\documentclass[numberedappendix]{emulateapj}
\usepackage{natbib}
\usepackage{graphicx}
\usepackage{afterpage}

\def\sun{\ifmmode\odot\else$\odot$\fi}
\def\micron{\hbox{~$\mu$m}}

\newcommand{\Hmol}{\hbox{H$_{2}$}}
\newcommand{\SIV}{\hbox{[\ion{S}{4}]10.51\micron}}
\newcommand{\Neii}{\hbox{[\ion{Ne}{2}]12.81\micron}}
\newcommand{\Neiii}{\hbox{[\ion{Ne}{3}]15.56\micron}}

\newcommand{\SIIIa}{\hbox{[\ion{S}{3}]18.71\micron}}
\newcommand{\SIIIb}{\hbox{[\ion{S}{3}]33.48\micron}}

\newcommand{\PAHseis}{\hbox{6.2\micron\ PAH}}
\newcommand{\PAHsiete}{\hbox{7.7\micron\ PAH}}
\newcommand{\PAHonce}{\hbox{11.3\micron\ PAH}}
\newcommand{\Hm}[1]{\hbox{H$_2$~S(#1)}}
\newcommand{\HII}{H\,{\sc ii}}

\def\Spitzer{\textit{Spitzer}}

\shorttitle{Local Luminous Infrared Galaxies. Spatially resolved observations with IRS}
\shortauthors{Pereira-Santaella et al.}

\begin{document}

\title{Local Luminous Infrared Galaxies. I. Spatially resolved observations with Spitzer/IRS $^{\star}$}
\author{Miguel Pereira-Santaella\altaffilmark{1}}
\author{Almudena Alonso-Herrero\altaffilmark{1,2}}
\author{George H. Rieke\altaffilmark{3}}
\author{Luis Colina\altaffilmark{1}}
\author{Tanio D\'iaz-Santos\altaffilmark{4,1}}
\author{J.-D. T. Smith\altaffilmark{5}}
\author{Pablo G. P\'erez-Gonz\'alez\altaffilmark{6,2}}
\author{Charles W. Engelbracht\altaffilmark{3}}

\email{pereira@damir.iem.csic.es}

\altaffiltext{$\star$}{This work is based on observations made with the Spitzer Space Telescope, which is operated by the Jet Propulsion Laboratory, California Institute of Technology under NASA contract 1407}
\altaffiltext{1}{Instituto de Estructura de la Materia, CSIC, Serrano 121, E-28006, Madrid, Spain}
\altaffiltext{2}{Associate Astronomer, Steward Observatory, University of Arizona, AZ 85721, USA}
\altaffiltext{3}{Steward Observatory, University of Arizona, 933 N. Cherry Avenue, Tucson, AZ 85721, USA}
\altaffiltext{4}{Actual address: University of Crete, Department of Physics, GR-71003, Heraklion, Greece}
\altaffiltext{5}{Ritter Astrophysical Research Center, University of Toledo, 2801 West Bancroft Street, Toledo, OH 43606, USA}
\altaffiltext{6}{Departamento de Astrof\'isica, Facultad de CC. F\'isicas, Universidad Complutense de Madrid, E-28040 Madrid, Spain}

\begin{abstract}
We present results from the \Spitzer/IRS spectral mapping observations of 15 local luminous infrared galaxies (LIRGs). In this paper we investigate the spatial variations of the mid-IR emission which includes: fine structure lines, molecular hydrogen lines, polycyclic aromatic features (PAHs), continuum emission and the 9.7\micron\ silicate feature. We also compare the nuclear and integrated spectra. 
We find that the star formation takes place in extended regions (several kpc) as probed by the PAH emission as well as the \Neii\ and \Neiii\ emissions. The behavior of the integrated PAH emission and 9.7\micron\ silicate feature is similar to that of local starburst galaxies.
We also find that the minima of the \Neiii\slash\Neii\ ratio tends to be located at the nuclei and its value is lower than that of \HII\ regions in our LIRGs and nearby galaxies. It is likely that increased densities in the nuclei of LIRGs are responsible for the smaller nuclear \Neiii\slash\Neii\ ratios. This includes the possibility that some of the most massive stars in the nuclei are still embedded in ultracompact \HII\ regions.
In a large fraction of our sample the \PAHonce\ emission appears more extended than the dust 5.5\micron\ continuum emission. We find a dependency of the \PAHonce\slash\PAHsiete\ and \Neii\slash\PAHonce\ ratios with the age of the stellar populations. Smaller and larger ratios respectively indicate recent star formation.
The estimated warm (300 K $<$ T $<$ 1000 K) molecular hydrogen masses are of the order of 10$^{8}$M$_\Sun$, which are similar to those found in ULIRGs, local starbursts and Seyfert galaxies.
Finally we find that the \Neii\ velocity fields for most of the LIRGs in our sample are compatible with a rotating disk at $\sim$kpc scales, and they are in a good agreement with H$\alpha$ velocity fields.

\end{abstract}
\keywords{galaxies: nuclei --- galaxies: structure --- galaxies: active --- galaxies: starburst --- infrared: galaxies}

\section{Introduction}\label{s:intro}

Infrared (IR) bright galaxies were first identified almost 40
years ago \citep{Rieke72}. The {\it IRAS} satellite detected a large
number of these galaxies in the local Universe
(see \citealt{Sanders96} for a review). Galaxies were classified 
according  to their IR ($8-1000\,\mu$m) luminosities into Luminous
Infrared Galaxies (LIRGs, 
$10^{11}L_{\sun}< L_{IR} <10^{12}L_{\sun}$) and Ultra Luminous
Infrared Galaxies (ULIRGs, $L_{IR} > 10^{12}L_{\sun}$).  Since then, an
increasing number of studies showed the important role of
(U)LIRGs at cosmological distances. LIRGs and ULIRGs are not common in
the local Universe, accounting for just $5\%$ and $<1\%$, respectively,
of the total IR emission of galaxies. However, at z$\sim$1, LIRGs
dominate the IR background and the co-moving star formation rate
density, while ULIRGs are dominant at z$\sim$2 \citep{LeFloch2005, PerezGonzalez2005, Caputi2007}.

Thanks to the high sensitivity of the Infrared Spectrograph (IRS) \citep{HouckIRS} on-board \Spitzer, we can, for the first time, study systematically the mid-IR properties of (U)LIRGs, both locally and at high-$z$. The majority of IRS local studies have, so far, focused on the properties of starburst galaxies \citep{Smith07, Brandl06} and ULIRGs \citep{Armus07, Farrah07}. The latter class of galaxies includes a large variety of physical and excitation conditions. For instance, most of these local ($z<0.3$) ULIRGs are powered predominantly by compact starburst events, although a substantial AGN contribution is present in about half of them \citep{Imanishi2007, Farrah07, Nardini2008}.
In the future the larger sample ($\sim$200) of local LIRGs of the Spitzer legacy program GOALS will enable a statistical study of this class of galaxies \citep{Armus09}.

Although ULIRGs at z$\sim$2 extend to even higher luminosities than local ones, their mid-IR spectra are similar to those of local starbursts, rather than to local ULIRGs \citep{Rigby2008, Farrah08}. A similar behavior is found for high-z submillimeter galaxies \citep{Pope2007, MenendezDelmestre2009}.
This behavior may indicate that star formation in IR-bright galaxies at z$\sim$2
is taking place over larger scales (a few kpc) than in local ULIRGs, where most of the IR
emission arises from very compact regions (sub-kiloparsec scales, 
see e.g., \citealt{Soifer2000, Farrah07}).

As a result, local LIRGs may provide important insights to the behavior of high-redshift ULIRGs. Their star formation is often distributed over a substantial fraction of the galaxy (e.g., \citealt{AAH06s}), and in any case is usually not embedded at the high optical depths characteristic of the nuclear star forming regions in local ULIRGs. To make good use of this analogy requires observations that map the local LIRGs to get an understanding of their global properties as well as to provide a better comparison with the observations of high-z galaxies, where the entire galaxy is encompassed in the IRS slit. The spectral mapping mode of \Spitzer/IRS is well suited for this purpose, since it can provide both the nuclear and integrated spectra of local galaxies, enabling study of the spatial distribution of the spectral features that make up the integrated spectrum.

This is the first in a series of papers studying the mid-IR
properties of local LIRGs. We present \Spitzer/IRS spectral mapping
observations of a representative sample of fifteen local LIRGs.
The first results and the goals of this program were presented by \citet{AAH09ASR}. 
In the present paper we focus on the analysis of the spatially resolved measurements.
The paper is organized as follows. In Section \ref{s:observations} we present the
sample and the observational details. Section \ref{s:analysis}
describes the analysis of the data. We examine the silicate feature in
Section \ref{s:silicate}. Sections \ref{s:lines}, \ref{s:PAH},
\ref{s:molecular_hydrogen} discuss the spatially resolved
  properties of the atomic fine structure lines,
PAHs, and molecular hydrogen emission, respectively. We explore the
velocity fields in Section \ref{s:velocity_fields}. Finally, in
Section \ref{s:conclusions} we summarize the main results. Throughout
this paper we assume a flat cosmology with $H_0 = 70$ km
s$^{-1}$Mpc$^{-1}$, $\Omega_M = 0.3$ and $\Omega_{\Lambda} = 0.7$. 

\section{Observations and data reduction}\label{s:observations}

\subsection{The sample}\label{ss:thesample}
The galaxies studied here were selected from the sample of 30 local LIRGs described in \citet{AAH06s}, which was drawn from the IRAS Revised Bright Galaxy Sample (RBGS, \citealt{SandersRBGS}).
The sample of \citet{AAH06s} is a volume limited sample, and the velocity range was selected
so that the Pa$\alpha$ emission line could be observed with the NICMOS F190N filter on-board the Hubble Space Telescope (HST). This restricted the sample to local LIRGs (40 Mpc $\lesssim$ d $\lesssim$ 75 Mpc).
From this sample we selected 15 galaxies (Table \ref{tbl_obs_map}) with extended emission in the NICMOS Pa$\alpha$ images to be mapped with the IRS spectrograph. This selection criterion does not introduce any bias in the luminosity, distance and nuclear activity distributions of the mapped galaxies as compared to the parent sample of \citet{AAH06s}.
The remaining galaxies in the sample have been observed using the IRS staring mode as part of different programs. The latter will be analyzed with the mapping data in a forthcoming paper \citetext{Pereira-Santaella et al., in preparation}. The interacting galaxy Arp~299 was studied in detail by \citet{AAH09Arp299} and will be included in the appropriate comparisons and figures in this paper.

Most of the nuclei of the LIRGs in our sample are classified as \HII\ region-like from optical spectroscopy, 3 galaxies have a Seyfert classification and 2 are classified as composite (intermediate between \HII\ and LINER, see \citealt{AAH09PMAS}). The infrared luminosity of the sample ranges from 10$^{11.0}$ to 10$^{11.8}$ L$_\sun$. The main properties of the galaxies are summarized in Table \ref{tbl_obs_map}.

\tabletypesize{\scriptsize}
\begin{deluxetable}{cccccc}
\tablewidth{0pt}
\tablecaption{\Spitzer/IRS Mapping Observation details\label{tbl_obs_det}}
\tablehead{
&  \colhead{Pixel size} & \colhead{Ramp} &  \# & Step size $\bot$ & Step size $\parallel$ \\
\colhead{IRS Module} & (arcsec) & (s) & \colhead{of cycles} & (arcsec) & (arcsec)
}
\startdata
SL1, SL2 & 1.85 & 14 & 2 & 1.8 & -\\
LL1, LL2 & 4.46 & 14 & 2 & 5.25 & -\\
SH & 2.26 & 30 & 2 & 2.35 & 5.65 \\
LH & 4.46 & 60 & 4 & 5.55 & -
\enddata
\end{deluxetable}

\subsection{Spectral Mapping}\label{ss:obs_mapping}
We obtained \Spitzer\ IRS spectral mapping observations through two
guaranteed time observation (GTO) programs P30577 and P40479 (PI:
G. H. Rieke). We used all four IRS modules: high 
resolution (R$\sim$600) modules with Short-High (SH; 9.9 - 19.6 \micron) and Long-High (LH;
18.7 - 37.2\micron), and low resolution (R$\sim$60 - 120) with Short-Low
(SL1; 7.4 - 14.5 \micron, SL2; 5.2 - 7.7\micron) and Long-Low (LL1;
19.5 - 38.0\micron, LL2; 14.0 - 21.3\micron). Maps were made by moving the telescope pointing position perpendicular to
the long axis of the slit with a step size half of the slit width. The
SL, LL and LH slits are long enough to cover the full region of interest
with just one slit length, but due to the smaller size of SH slit,
parallel steps were also needed. The parallel step size was half of
the slit height. This strategy produced maps with a redundancy per
pixel of 2 for SL, LL and LH, and 4 for SH. Table \ref{tbl_obs_det} details the integration time, pixel size and step size for each IRS module. The maps cover at least the central 20''$\times$20'' to 30''$\times$30'' of the galaxies.

Dedicated background observations are not needed for the low resolution modules (SL
and LL) because the separation on the sky between the two orders of
each module is larger than the galaxy size. That is, when order one
is on the galaxy, order two is observing the background and vice
versa. The SH and LH slits are too small to allow background
subtraction. We obtained dedicated background observations for LH by
observing a region 2 arcminutes away from the galaxy. SH sky
observations are available only for the galaxies observed under the
program P40479. For the galaxies without dedicated SH background
observations, we estimated the sky background emission by combining the
background spectra from the low resolution modules. Generally the sky
contribution is not very important for the galaxies in our
  sample, except for NGC~3110 and 
Zw~049.057. The sky background is bright at the position of these
galaxies and its subtraction is essential to match the fluxes from the 
different slits. 
Moreover, subtracting background observations is always useful because
it removes many of the rogue pixels and improves the final quality of
the spectra. This is especially important for the LH module 
since it contains a large number of such pixels. 

We  used Basic Calibrated Data (BCD), processed by \Spitzer\
pipeline version S15.3 (low resolution modules: SL and LL) and version
S17.2 (high resolution modules: SH and LH) to build data cubes with
CUBISM \citep{SmithCUBISM}. CUBISM is a tool that combines slit
observations to create spectral cubes. It also performs background
subtraction, flux calibration and estimates the statistical
uncertainty at each spectral wavelength.

\section{Analysis}\label{s:analysis}
\subsection{SL Spectral Maps}\label{ss:spectral_maps}
Using the SL cubes generated by CUBISM we constructed maps of the 5.5\micron\ continuum,
and  the most prominent aromatic features (hereafter PAH) in this module 
(\PAHseis, \PAHsiete\ and \PAHonce). To build them we used our own IDL procedures. First we extracted 1-D spectra from the cubes by averaging the spectra over a 2$\times$2 pixel aperture and
then by moving with a 1 pixel step until the whole field of view was covered. The 2$\times$2 pixel
extraction box allows us to increase the S/N of the 1-D spectra, with
only a slight loss of angular resolution (see below and Appendix \ref{apxPSF}).
Once we obtained the 1-D spectra, for each pixel we did the following.  
To create the 5.5\micron\ continuum map, we integrated the flux
between the rest frame 5.3\micron\ and 5.7\micron\ wavelengths (in
this section all wavelengths refer to rest frame wavelengths). For the
PAH maps, first we fitted the local continuum using a linear fit. The continuum pivots were set by averaging the spectrum over a $\sim$0.2\micron\ interval centered at 5.75\micron\ and
6.7\micron\ for the \PAHseis, at 7.2\micron\ and 8.2\micron\ for the
\PAHsiete, and at 10.6\micron\ and 11.8\micron\ for the
\PAHonce. After subtracting the continuum, we integrated the flux in
the wavelength ranges between 5.9 and 6.5\micron\, 7.3 and 7.9\micron\ and 10.8 and 11.6\micron\ for the
6.2, 7.7 and 11.3 \micron\ PAH features, respectively. We derived the flux
uncertainty of these maps with the statistical uncertainties provided by CUBISM. To
avoid noisy pixels, we required the feature peak to be at least above 1$\sigma$ to
plot that pixel in the map, where $\sigma$ is the continuum noise. The SL spectral maps of our sample are shown in Figure \ref{fig_map_sl}.

The angular resolution (FWHM) of the SL spectral maps is $\sim$4.5
arcsec. This is the resolution when we use 2$\times$2 boxes to construct the maps, as opposed to that measured from the standard star observations (see Appendix \ref{apxPSF}).
For the distances of our galaxies corresponds to $\sim$1 kpc (the exact
scale for each galaxy is given in Table \ref{tbl_obs_map}). In
Appendix \ref{apxPSF} we give full details on how the angular resolution of the spectral maps was derived.

We constructed maps of different PAH ratios by dividing the PAH maps obtained as described above. To avoid noisy pixels in the ratio maps we imposed further restrictions, and only those pixels with signal-to-noise ratio greater than 4 are plotted in the PAH ratio maps. Figures \ref{fig_pah11pah6_sl} and \ref{fig_pah11pah7_sl} show the \PAHonce\slash\PAHseis\ and \PAHonce\slash\PAHsiete\ ratios, respectively.

\subsection{SH Spectral Maps}\label{sh:spectral_maps}
We used the SH cubes to measure the brightest fine structure emission lines: \SIV, \Neii, \Neiii, \SIIIa\ and the molecular hydrogen lines, \Hm{2} at 12.3\micron\ and \Hm{1} at 17.0\micron. First we extracted 1-D spectra using the same method as for the low resolution cubes. Then we used the IDL package MPFIT \citep{MPFIT} to fit simultaneously a first-order polynomial to the continuum and a Gaussian to the line. The fit was over a wavelength range of $\sim$0.4\micron\ centered at the line wavelength.
The spectral resolution, R$\sim$600, is sufficiently high to allow us to separate the emission lines from the broad PAH bands. However, to improve the continuum determination, we used extra Gaussian profiles to fit the PAH features close to the emission lines (e.g., the 12.7\micron\ PAH feature next to the [NeII]). 
To avoid noisy pixels in the maps, we required: (1) a signal-to-noise ratio $>2$ in the continuum range; and (2) the line peak to be at least 1$\sigma$ above the continuum noise for the pixel to be shown in the corresponding map.

We also constructed an \PAHonce\ feature map by the same procedure as for SL cubes, but with the high resolution spectra. A 15\micron\ continuum was obtained by integrating the flux between 14.9\micron\ and 15.1\micron. We required a signal above 1$\sigma$ to plot a pixel in these two maps.

The SH spectral maps are shown in Figure \ref{fig_map_sh} together with the HST/NICMOS continuum-subtracted Pa$\alpha$ maps of \citet{AAH06s}.

We constructed \SIIIa\slash\Neii, \Neiii\slash\Neii\ and
\Neii\slash\PAHonce\ ratio maps by dividing the flux maps.
The variation of the PSF size with wavelength within each
module is negligible (see Appendix \ref{apxPSF}), and therefore we divided the flux maps
directly without attempting to correct for this small
effect (we only built ratio maps of lines observed with the SH module). We imposed stricter conditions to plot a pixel in a ratio map than those used for the line maps; only 
pixels with a flux measurement $>3\sigma$, for both features, were included in the
ratio maps. The maps of the SH line ratios are shown in Figure \ref{fig_co_sh}. The angular resolution of the SH maps is $\sim$6 arcsec. This corresponds to physical scales between $\sim$1 kpc, for the nearest galaxies, and $\sim$2 kpc for the most distant ones.

\subsection{Nuclear, Integrated and Extranuclear 1-D Spectra}\label{ss:1d_spectra}
We extracted 1-D spectra of the nuclei as well as of the integrated galaxy. 
To extract the nuclear spectra we used a 5.5''$\times$5.5'' square aperture centered at the nuclear coordinates (Table \ref{tbl_obs_map}). This aperture corresponds to physical scales between 1 and 2 kpc, depending on the galaxy. We set the orientation of the aperture aligned with the R.A. and Dec axis. We used the same aperture for both low and high resolution data cubes. 
To obtain the integrated spectrum we summed the flux of the cube. We excluded those pixels with the \PAHonce\ flux (15\micron\ continuum) below 1$\sigma$ to avoid adding too much noise to the low resolution (high resolution) integrated spectra. This threshold is the same as that required to plot a pixel in the \PAHonce\ and 15\micron\ continuum maps of Figure \ref{fig_map_sl} and Figure \ref{fig_map_sh} respectively. The physical scale of the integration area varies from galaxy to galaxy. The approximated physical sizes are listed in Tables \ref{tbl_lowres_map} and \ref{tbl_hires_map_integrated} for the low and high resolution maps respectively. They are of the order of 10$\times$8 kpc.
Although the extraction regions are slightly different for each module, the flux differences are only $\sim$10-20\%.

We plot the integrated spectra in Figure \ref{fig_spects}. The nuclear spectra will be presented in Pereira-Santaella et al. (in preparation), together with the nuclear spectra of the rest of the \citet{AAH06s} sample.

In Figure \ref{fig_spects_mapping} (right) we show a comparison of the nuclear and extranuclear low resolution spectra normalized at 14\micron\ for 4 of the galaxies. We selected these galaxies because in their spectral maps (left panel of Figure \ref{fig_spects_mapping}) it is possible to identify extranuclear \HII\ regions and extract their spectra. All the spectra, nuclear and extranuclear, show bright PAH features and fine structure lines. The variations of the spectral features from galaxy to galaxy and within a galaxy are relatively small, at least when compared to those observed in ULIRGs \citep{Armus07} or those observed in star-forming galaxies with different metallicities \citep{Engelbracht08}. That is, the comparison of the spectra shows that these LIRGs have a more or less uniform shallow silicate absorption and relatively high metallicity.

The PAH ratios, PAH equivalent width and silicate depth measured in the nuclear and integrated low resolution spectra are listed in Table \ref{tbl_lowres_map}. Table \ref{tbl_hires_map_integrated} and Table \ref{tbl_hires_map_nuclear} show respectively the integrated and nuclear fine structure line fluxes in the high resolution spectra.

\setcounter{figure}{5}
\begin{figure*}
\includegraphics[width=0.5\textwidth]{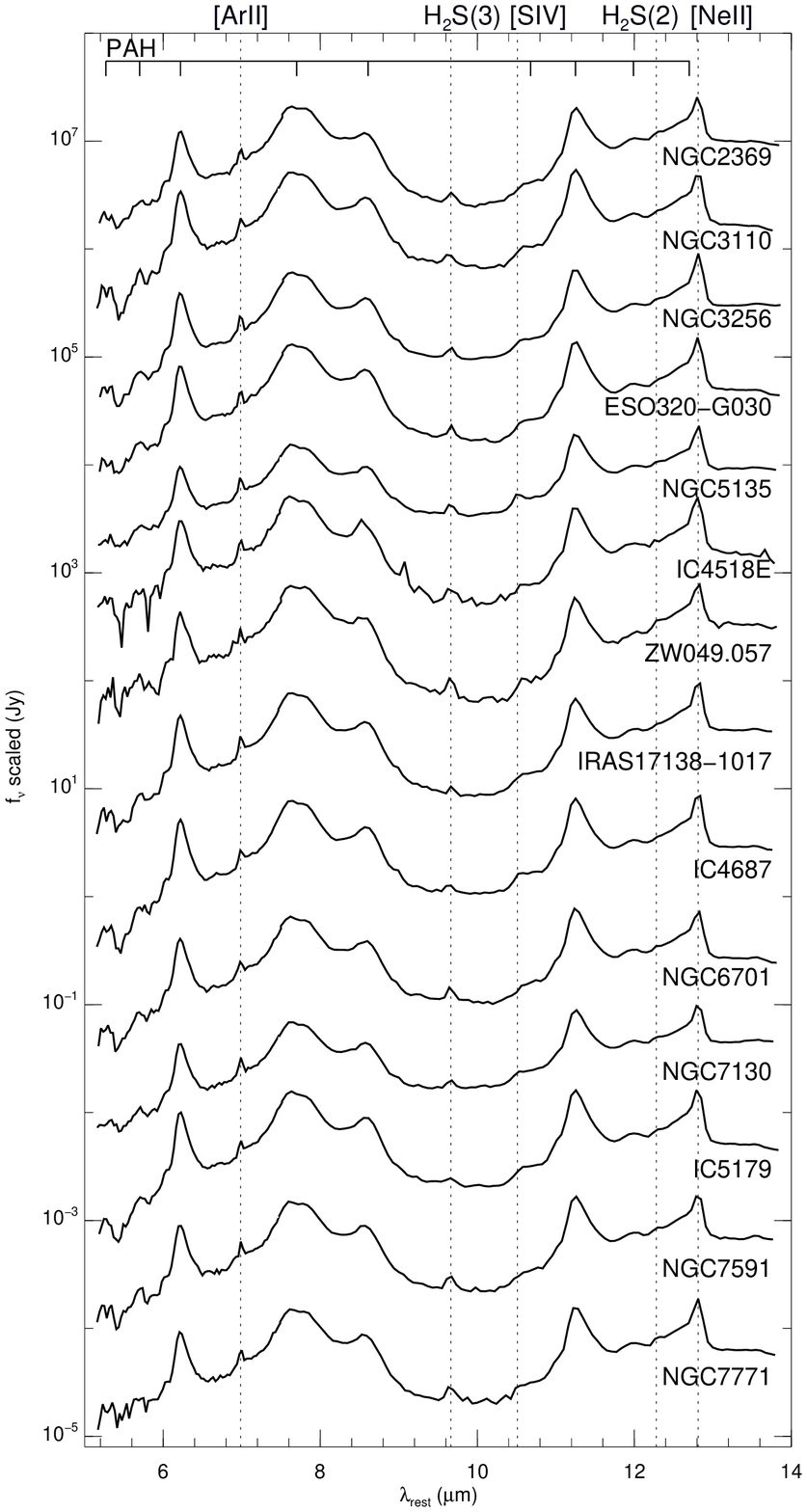}
\includegraphics[width=0.5\textwidth]{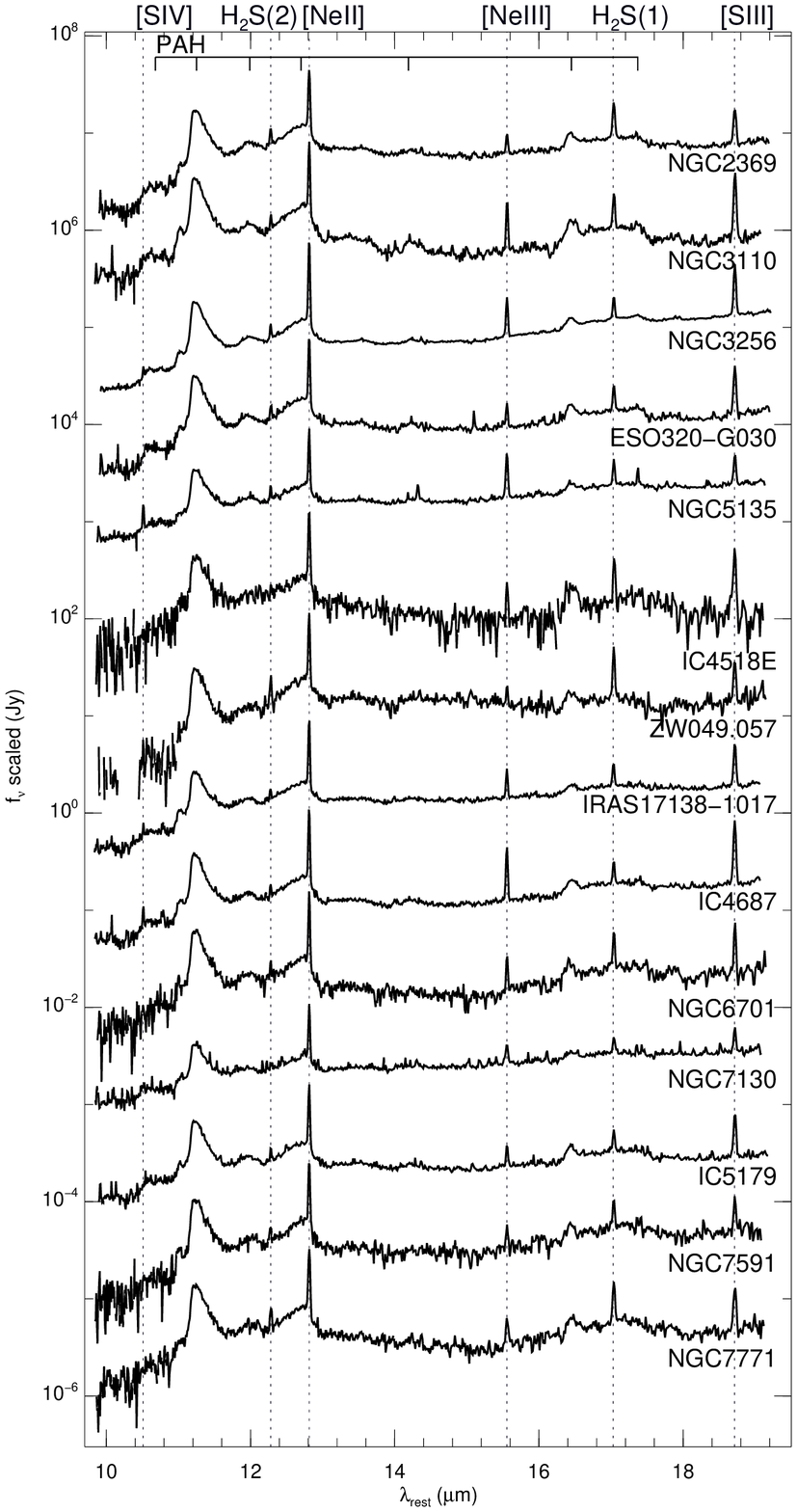}
\caption{Rest-frame SL (left panel) and SH (right panel) integrated spectra of the sample. The Arp~299 spectra was shown in \citet{AAH09Arp299}. The fluxes have been scaled for better comparison. The scale factors are, starting from the top galaxy, 2$\times 10^7$, 6$\times 10^6$, 1.2$\times 10^5$, 1.5$\times 10^5$, 2$\times 10^4$, 3$\times 10^4$, 7$\times 10^3$, 1$\times 10^2$, 8, 1, 1$\times 10^{-1}$, 1$\times 10^{-2}$, 5$\times 10^{-3}$, 2$\times 10^{-4}$  for the left panel and 1$\times 10^7$, 2$\times 10^6$, 2$\times 10^4$, 2$\times 10^3$, 5$\times 10^3$, 2$\times 10^2$, 2, 2$\times 10^{-1}$, 5$\times 10^{-2}$, 2$\times 10^{-3}$, 2$\times 10^{-4}$, 2$\times 10^{-4}$, 1$\times 10^{-5}$ for the right panel.}
\label{fig_spects}
\end{figure*}

\begin{figure*}
\includegraphics[width=0.42\textwidth]{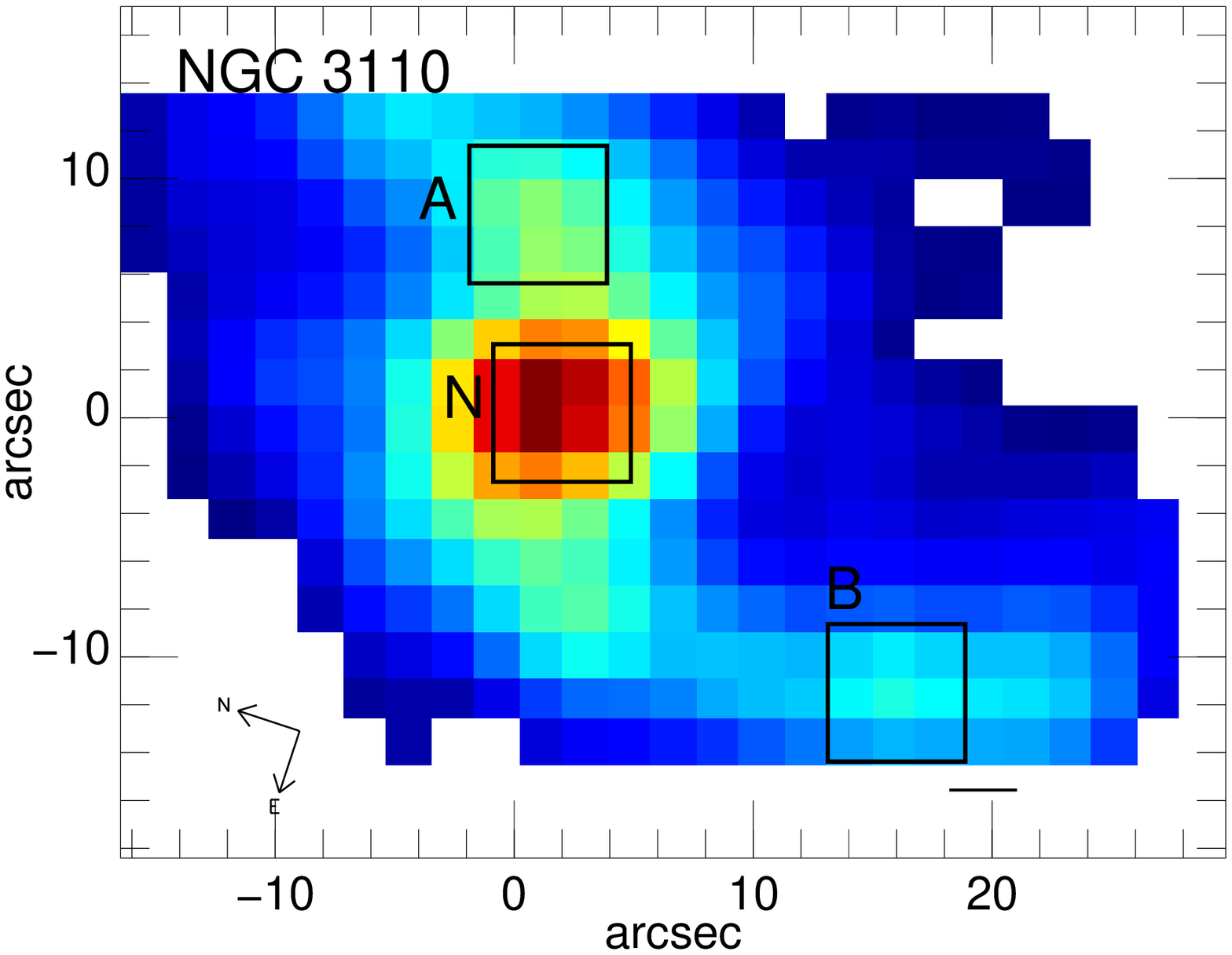}
\includegraphics[width=0.42\textwidth]{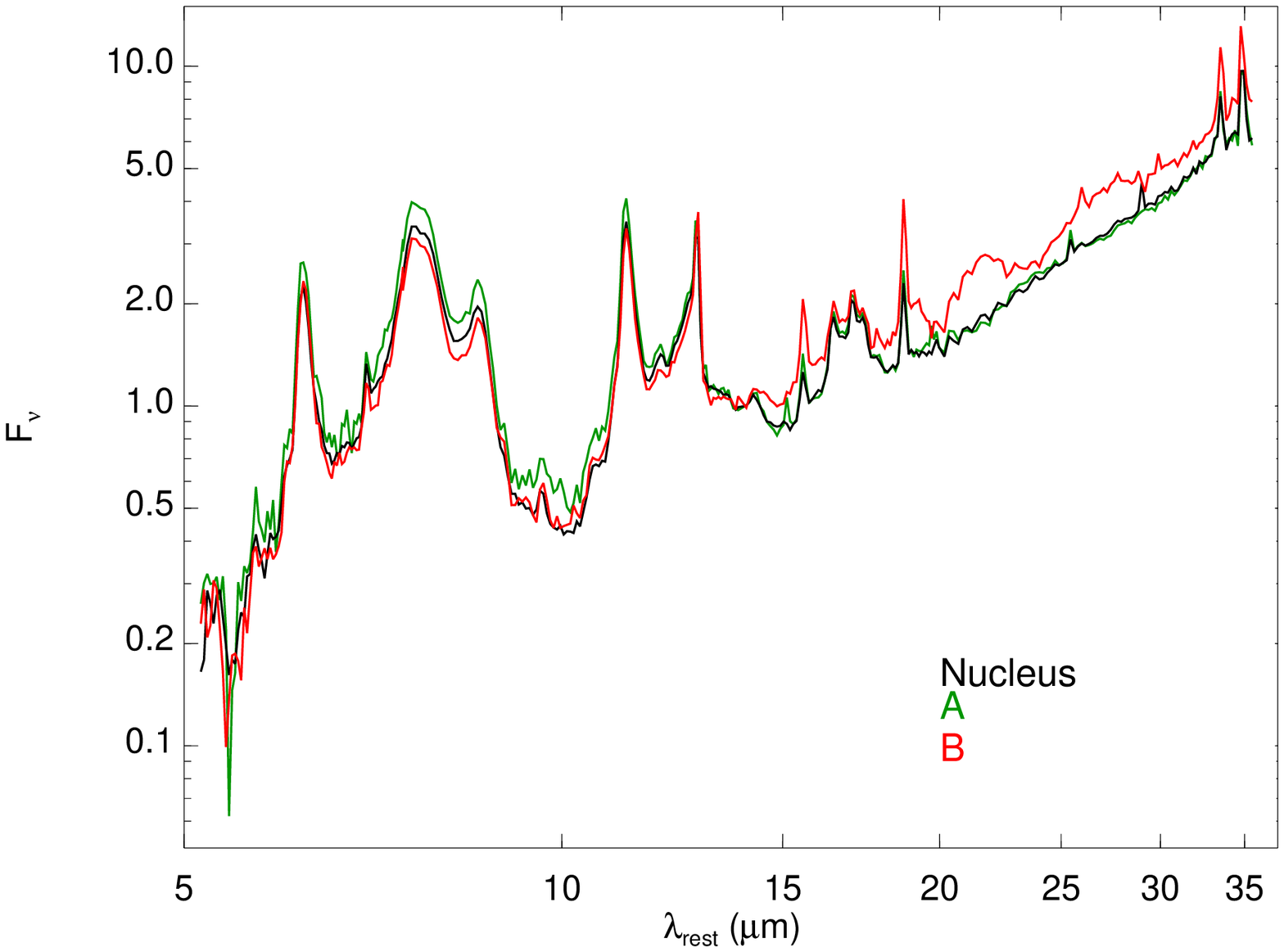}
\centering
\includegraphics[width=0.42\textwidth]{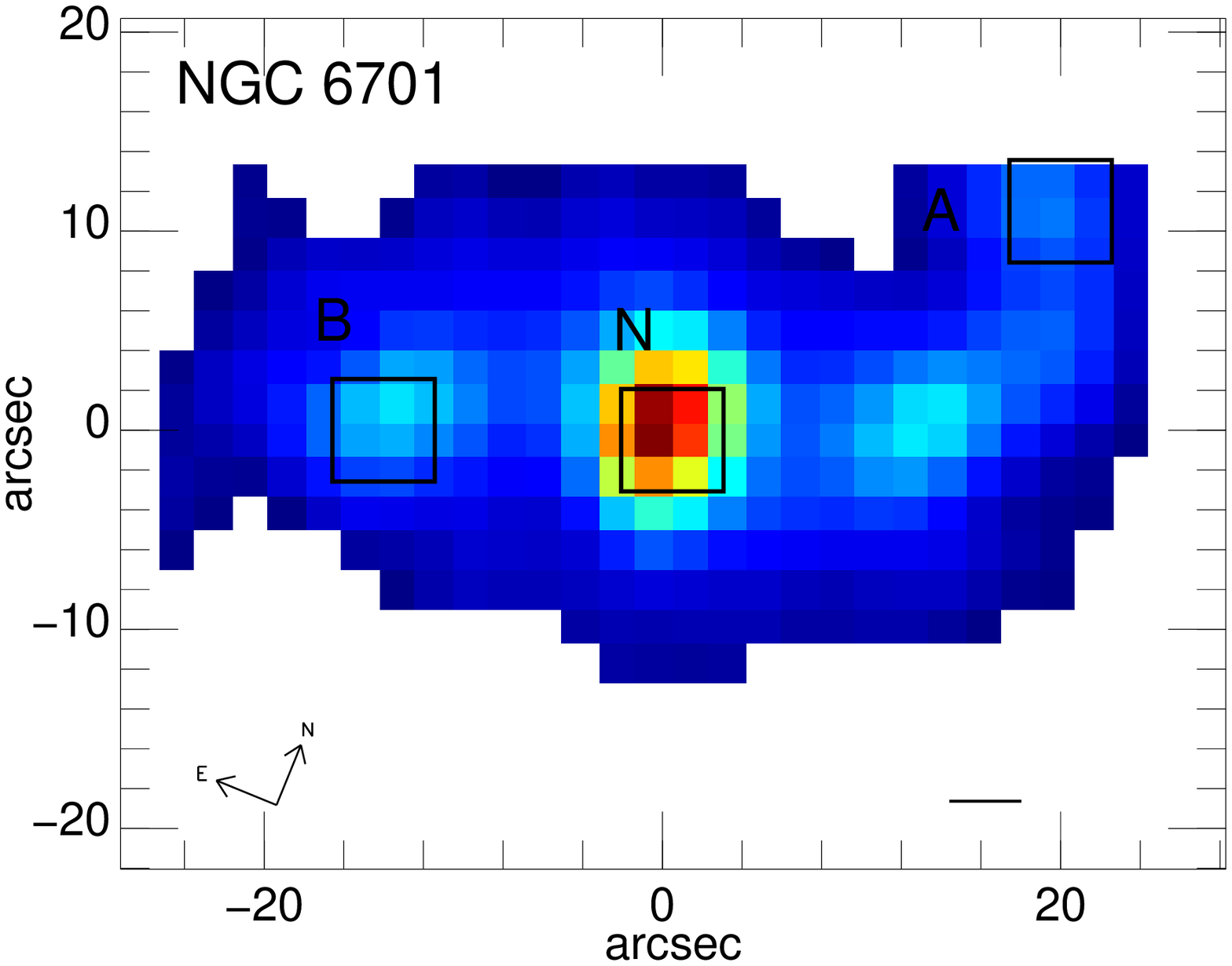}
\includegraphics[width=0.42\textwidth]{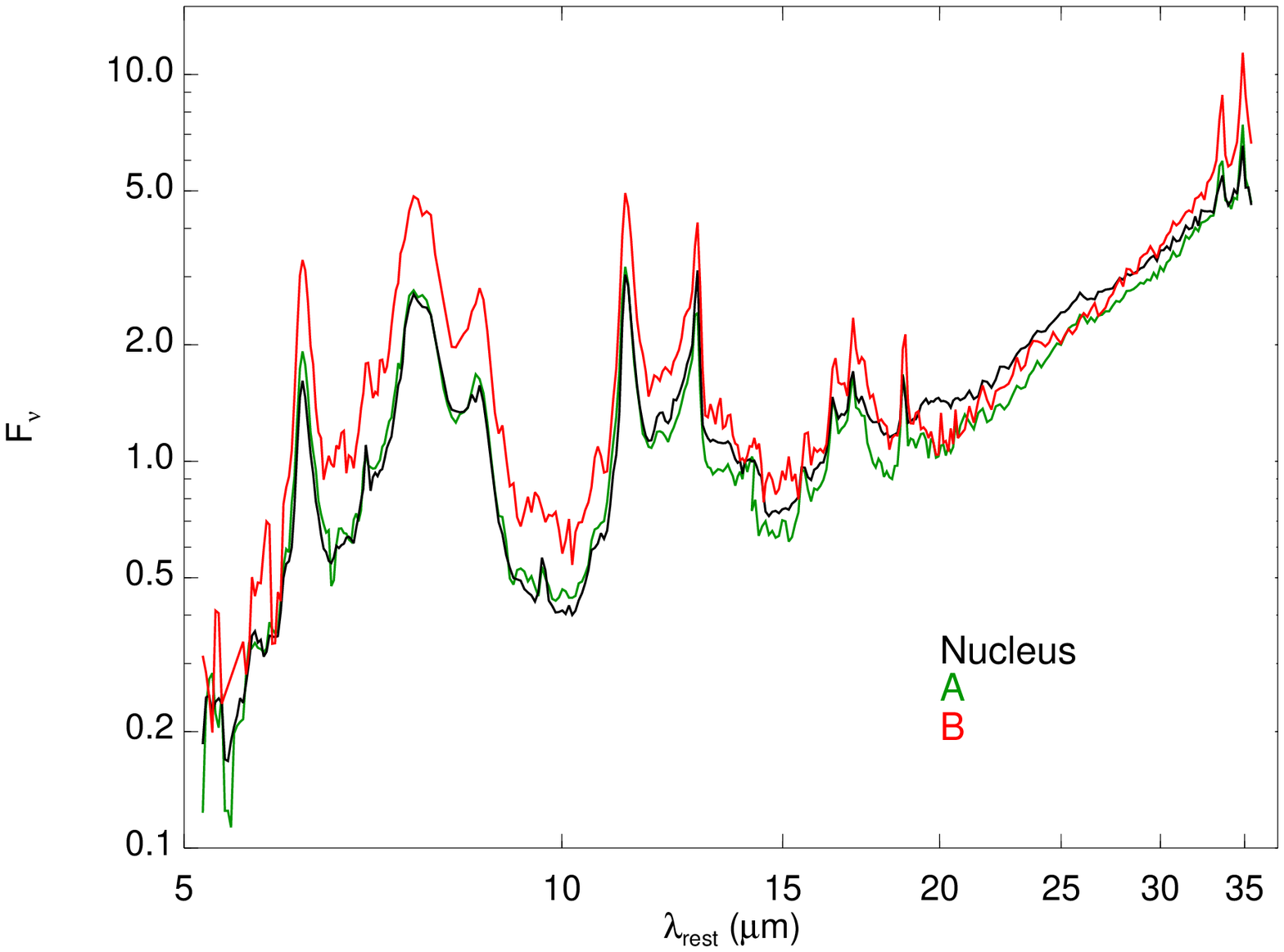}
\centering
\includegraphics[width=0.42\textwidth]{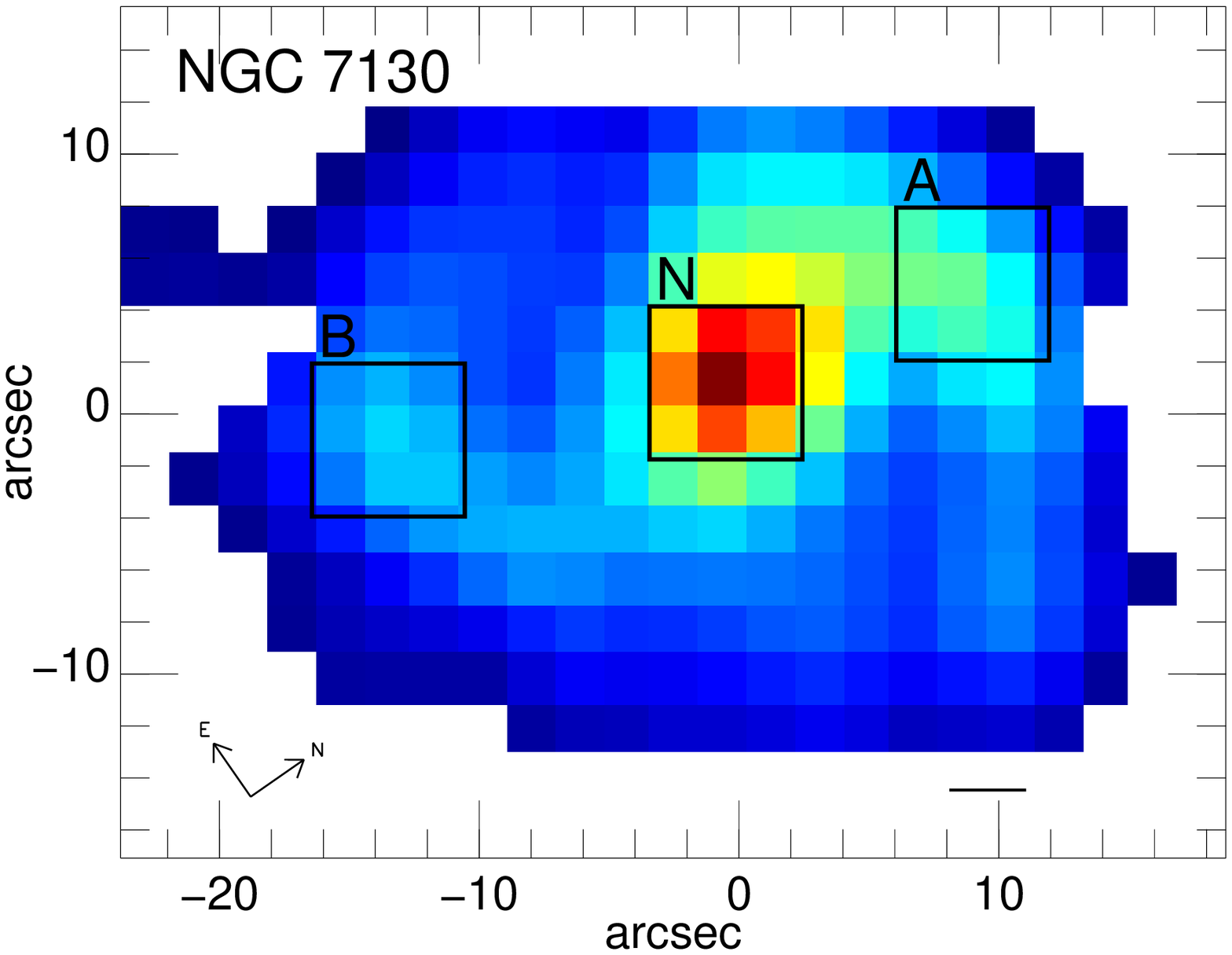}
\includegraphics[width=0.42\textwidth]{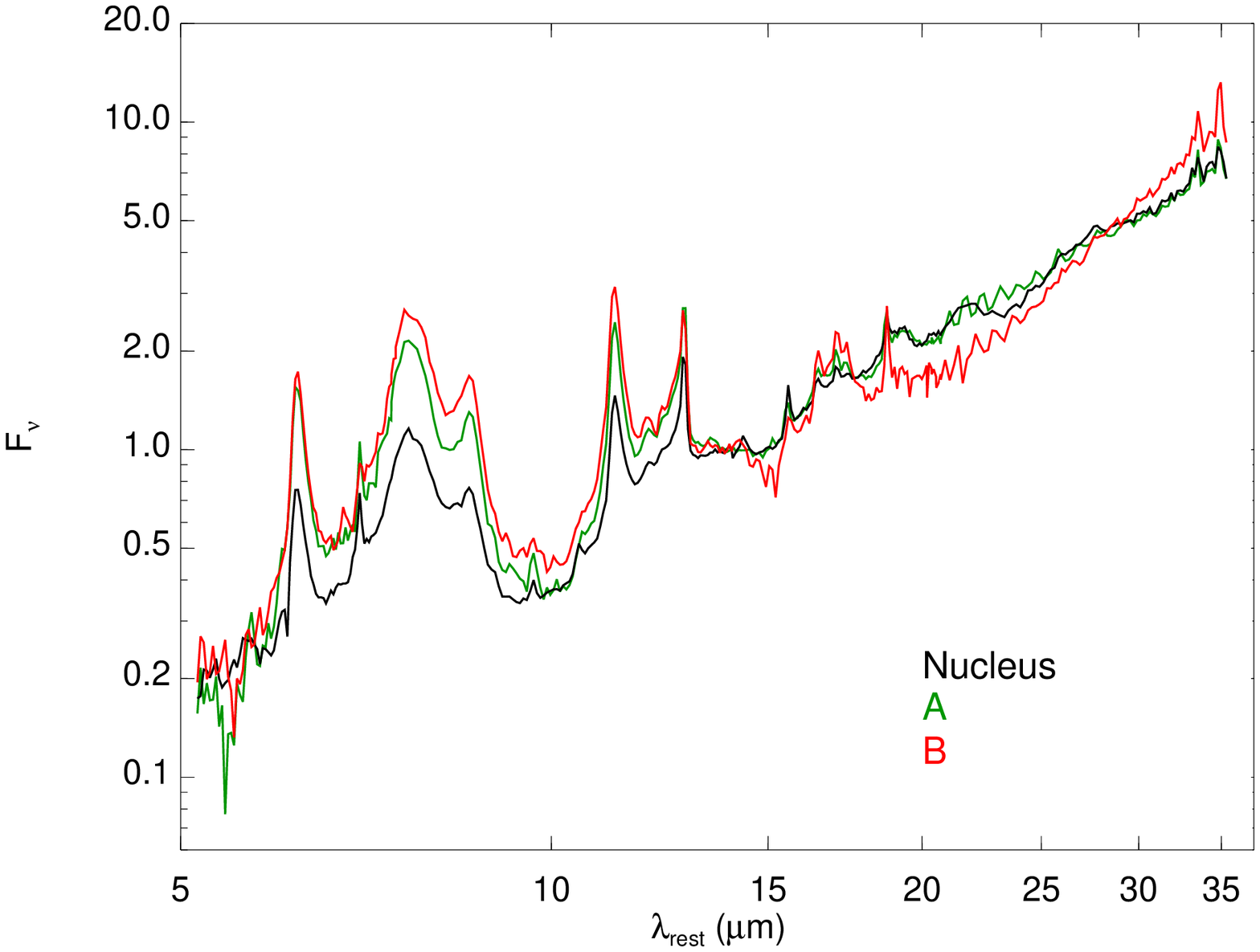}
\centering
\includegraphics[width=0.42\textwidth]{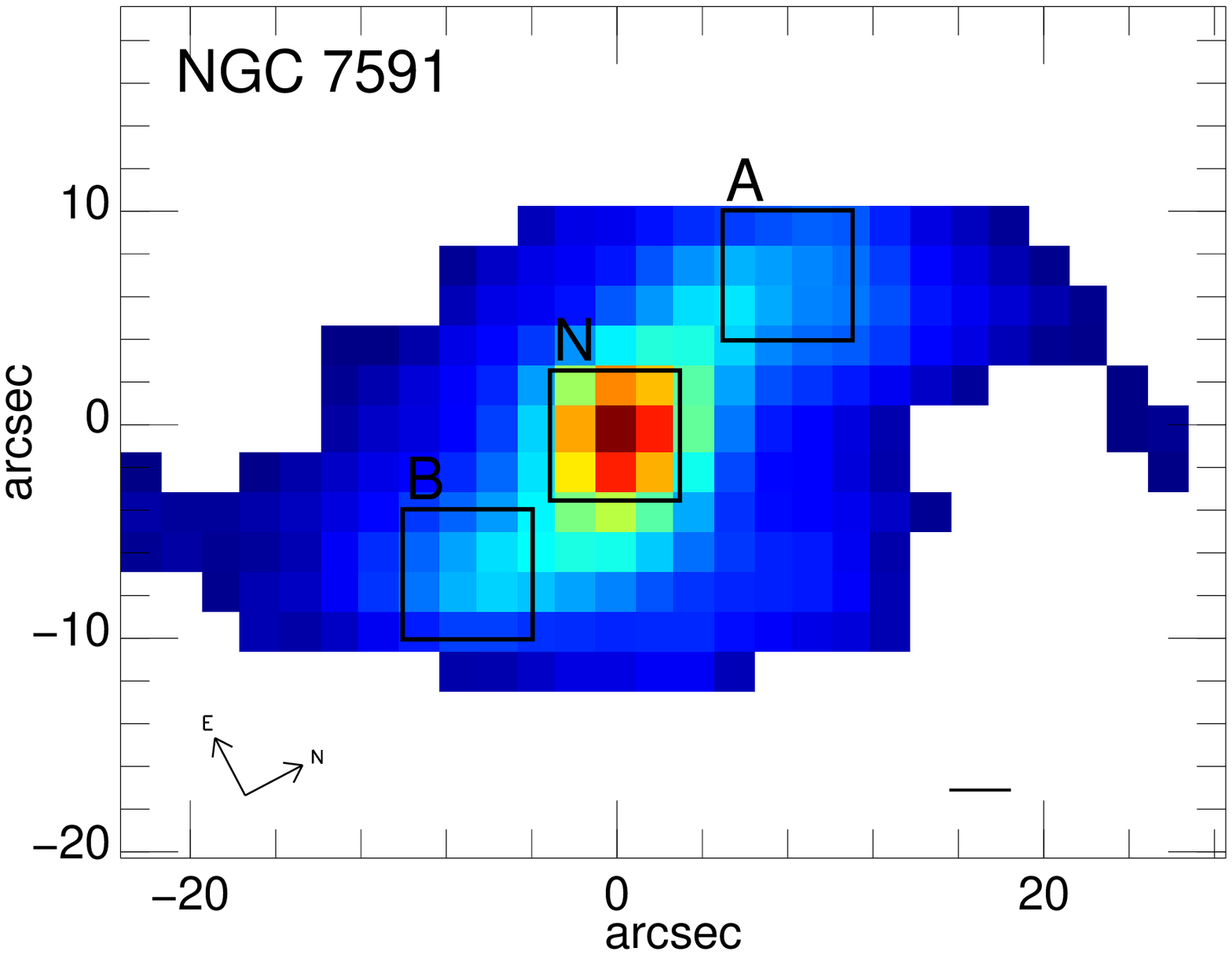}
\includegraphics[width=0.42\textwidth]{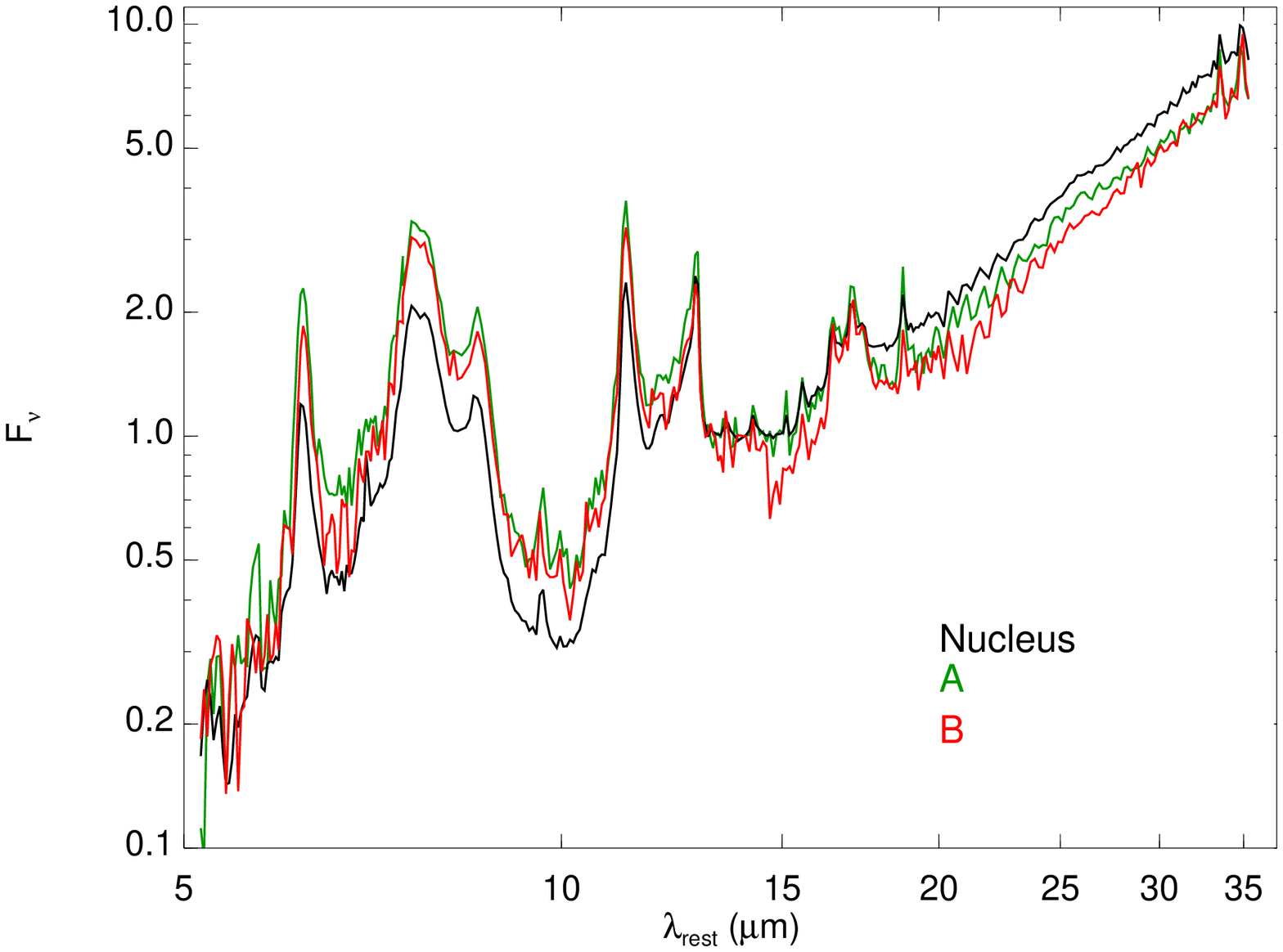}
\caption{Left: Spitzer/IRS SL spectral maps of the \PAHsiete\ feature. The image orientation is indicated on the maps for each galaxy. The scale represents 1 kpc. The black squares are the positions and sizes of the extraction apertures. Right: Low resolution spectra normalized at 14\micron.}
\label{fig_spects_mapping}
\end{figure*}

\subsection{Silicate Feature Maps}\label{ss:silicate_feature_maps}
Amorphous silicate grains present a broad feature centered at 9.7\micron\ that can be measured from the SL cubes. We used a method similar to that proposed by \citet{Spoon07} to measure the observed silicate strength. First, we extracted 1-D spectra from the cube as described in Section \ref{ss:spectral_maps}. For each spectrum, we estimated the continuum expected at 10\micron\ by fitting a power law through the feature-free continuum pivots at 5.5\micron\ and 13\micron. Then we compared the estimated continuum and the measured flux using the following expression: 
\begin{equation}
S_{\rm Si} = ln \frac{f_{obs}(10\micron)}{f_{cont}(10\micron)}
\end{equation}
where $S_{\rm Si}$ is the silicate strength. When the silicate strength is negative, the silicate feature is in absorption, whereas a positive silicate strength indicates that the feature is seen in emission. 

We estimated the flux at 10\micron\ instead of 9.7\micron\ (the maximum of the feature) to avoid a possible contamination by the molecular hydrogen emission line \Hm{3} at 9.67\micron. 

There are several factors that affect the accuracy of the silicate
strength measurement. The most important source of uncertainty is the
continuum estimate, because of the presence of broad PAH
bands in the spectra. We selected continuum pivots similar to those used by \citet{Spoon07}, so that we can make
meaningful comparisons with their results. 
The $S_{\rm Si}$ is equivalent to the optical depth of the silicate absorption ($\tau_{\rm Si}$) defined in \citet{Rieke85}. Alternatively, a more accurate way to measure the silicate optical depth is to fit all the features (PAHs, emission lines and dust continuum) present in the spectrum as PAHFIT does \citep{Smith07}. However, to avoid ambiguities between the PAH strength and the silicate absorption, the entire low resolution spectral range (5.2\micron\ -38\micron) is desirable. At longer wavelengths ($>$14.5\micron) the pixel size is about 2.5 times larger than at shorter wavelengths and since we are interested in the spatial distribution of the silicate feature we decided to use the first method to estimate the continuum. 
To compared these methods, first we fitted the integrated spectra using PAHFIT, assuming a foreground screen dust geometry to obtain the $\tau_{9.7}$, and then we compared this value with the measured $S_{\rm Si}$. We found that both quantities are well correlated.

Keeping in mind that the main source of uncertainty of the silicate
strength measurements is the continuum determination, we estimated
that the typical values of the statistical uncertainty
are $\sim$0.05 for the nuclei and less than $\sim$0.15 for
the extranuclear regions. The maps of the silicate feature strength
are presented in Figure \ref{fig_map_sil}.

\setcounter{figure}{8}
\begin{figure}
\center
\includegraphics[width=0.5\textwidth]{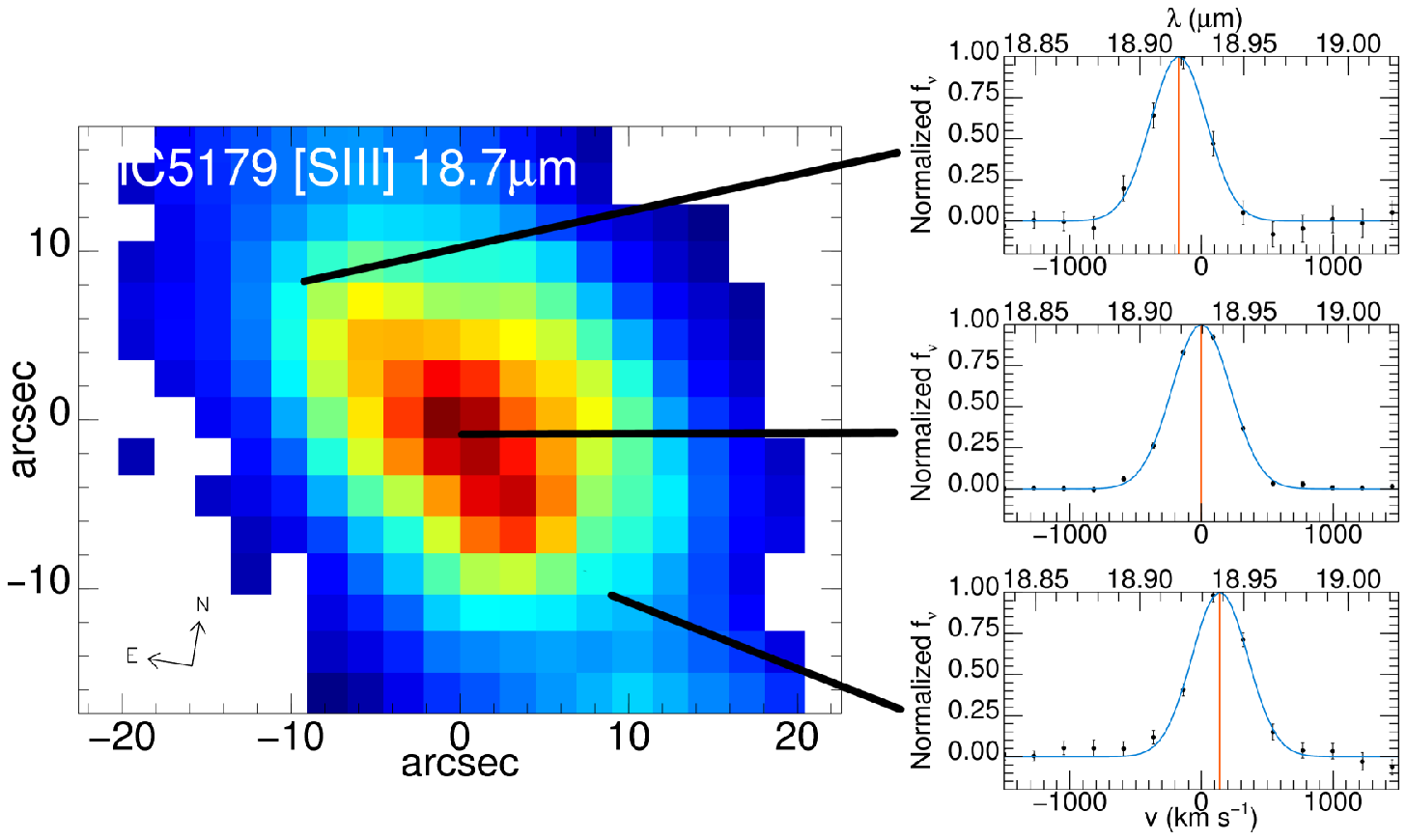}
\caption{The left panel is the map of the \SIIIa\ line flux of
  IC~5179. The right panels show the fit to this line for three
  selected regions of this galaxy.} 
\label{fig_velocity_field}
\end{figure}

\subsection{SH Velocity Fields}

The SH spectra were also used for rough determination of velocities. Figure \ref{fig_velocity_field} shows an example of line fitting for selected regions of
IC~5179 to illustrate the detection of variations in the line position, even with the relatively low spectral resolution of the SH module.

Since the spectral resolution, R$\sim$600, is lower than that of most kinematic studies, we tested the validity of the velocities with synthetic spectra. We created these spectra to simulate an unresolved spectral line as seen by the SH module and then we tried to recover the original line parameters. The equivalent width of the simulated lines was similar to the observed ones. We added noise to the spectra to match the continuum signal-to-noise ratio of the maps. The continuum signal-to-noise ratio, for most of the galaxies, is in the range $\sim$60-100 at the nuclei and $\sim$3-10 at the external regions.
To allow for the effect of the telescope pointing uncertainties, we distorted the wavelength scale by $\sim$5\% (estimated from the pointing information in the BCD file headers). We found that this effect limits the accuracy of the velocity determination to $\pm$ 10 km s$^{-1}$, independent of the signal-to-noise ratio of the spectra. 
For extended sources, which fill uniformly the slit, the wavelengths are more stable, and then the uncertainty due to the pointing inaccuracy would be lower. 
The galaxies studied here are not point sources, neither uniformly extended sources, thus this uncertainty estimation is an upper limit.
Including the uncertainty in the absolute wavelength calibration (equivalent to 10\% of a pixel), we found the minimum uncertainty to be $\pm$ 20 km s$^{-1}$. We therefore conclude that the uncertainty in the velocity field is in the range of 10 - 30 km s$^{-1}$, depending on the pointing accuracy achieved and the signal-to-noise ratio of an individual spectrum. Thus, we consider that variations in the velocity maps larger than $\sim$20 km s$^{-1}$ are likely to be real. 
Figure \ref{fig_v} shows the velocity fields derived for the \Neii\ and \Hm{1} emission lines; they range up to a total gradient of $\sim$200 km s$^{-1}$ and hence include a substantial amount of information about the sources. 

\section{The Silicate Feature}\label{s:silicate}

\subsection{Spatially Resolved Measurements}

LIRGs are known to contain highly obscured regions (A$_V \simeq
4-50\,$mag, see e.g., \citealt{Veilleux1995, Genzel1998, AAH2000, AAH06s}), usually coincident with
the nuclei of the galaxies. Assuming an extinction law and a dust distribution
geometry the silicate strength can be converted into a visual extinction.

The maps of the silicate feature of our sample are shown in Figure \ref{fig_map_sil}. In a large fraction of the systems (80\%) the most obscured regions appear to be coincident with the nuclei. In other cases, the nuclear regions appear to be slightly less obscured than the surrounding regions (e.g., NGC~7130) or they show a complicated morphology (e.g., IRAS~17138$-$1017, NGC~7771, IC~4687). We note that the spectra used
for measuring the strength of the silicate feature are averaged over
$\sim$1 kpc scales, and thus some of the real variations seen on smaller scales using ground-based mid-IR observations are smoothed out (e.g., the highly obscured southern nucleus of NGC~3256, see \citealt{Tanio2008}).

\setcounter{figure}{10}
\begin{figure}[h]
\centering
\includegraphics[width=0.45\textwidth]{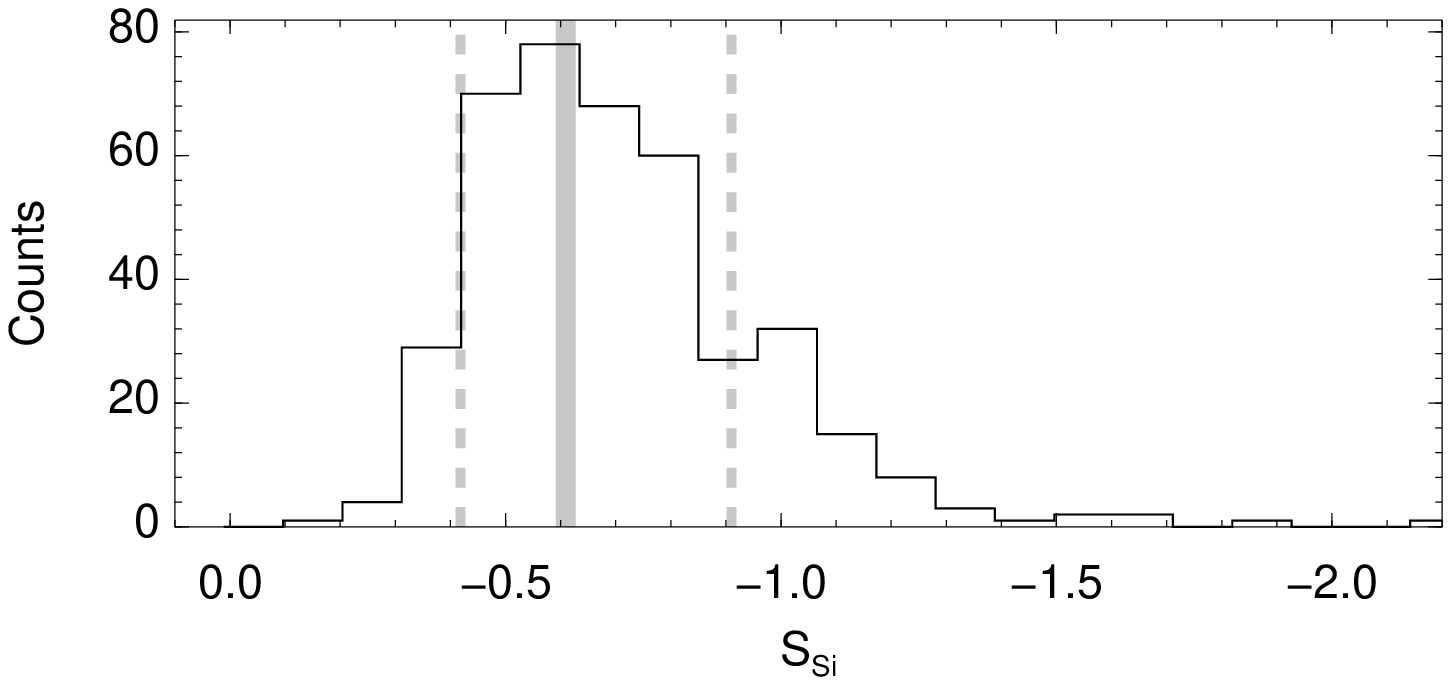}
\caption{Histogram of the spatially resolved measurements
  of the silicate strength for all galaxies
  measured in boxes of 2$\times$2 pixel. The solid grey line indicates
  the median silicate strength. The dashed grey lines define the range
  including 68\% of the data points. } 
\label{fig_ssi_histogram}
\end{figure}

Figure \ref{fig_ssi_histogram} shows the distribution of the silicate
strengths measured from the spatially resolved maps for all the
galaxies. They are in the range $\sim$0 to $\sim-2.0$, with a median of $-0.61$. 
The strengths of the majority of the LIRGs are moderate ($S_{\rm Si} \sim-0.4 \, {\rm to}  \, -0.9$), and intermediate
between those observed in starburst galaxies (\citealt{Brandl06}) and ULIRGs (\citealt{Spoon07}; \citealt{Sirocky08}).

If we assume a foreground screen of dust, the silicate absorption strength can be expressed in terms of the optical extinction (A$_V$) using the conversion factor A$_V/S_{\rm Si} = 16.6$ \citep{Rieke85}.  We then find that the median A$_V$ of the spatially resolved measurements is $\sim$10 mag. 
The most extincted region in our sample of LIRGs is the nuclear region of IC~694, also known as Arp~299-A (see also \citealt{AAH09Arp299}), with A$_V \sim$37 mag. The foreground screen assumption gives the minimum extinction compatible with the measurements. Most of the spatially resolved measurements (68\%) are in the range of -0.4 to -0.9. This range is much narrower than that observed in the nuclear regions of ULIRGs ($0 > S_{\rm Si} > -4$) \citep{Spoon07}. The similarity of the silicate absorption strengths among our sample members suggests that they may be in part a product of radiative transfer in dusty regions. We can approximate this possibility by assuming that the dust and emission sources are mixed. The resulting median A$_V$ is then $\sim$20 magnitudes, about 2 times greater than for the foreground screen geometry.

The nuclear and integrated values of the silicate strength are listed
in Table \ref{tbl_lowres_map}.  For all the galaxies, we find that the integrated strength of the silicate feature is smaller than, although comparable to, the nuclear one. The nuclear silicate strengths range from -0.4 to -1.9, although most of them are larger than -1.0. The integrated strength is $-0.3 < S_{\rm Si} < -1.2$, somewhat shallower than the nuclear strength.
The LIRGs with the largest variations between the nuclear and integrated silicate features are NGC~3256, Arp~299, NGC~7130, and NGC~7591.

\subsection{The $S_{\rm Si}$ vs. \PAHseis\ EW Diagram}

\begin{figure*}
\centering
\includegraphics[width=0.48\textwidth]{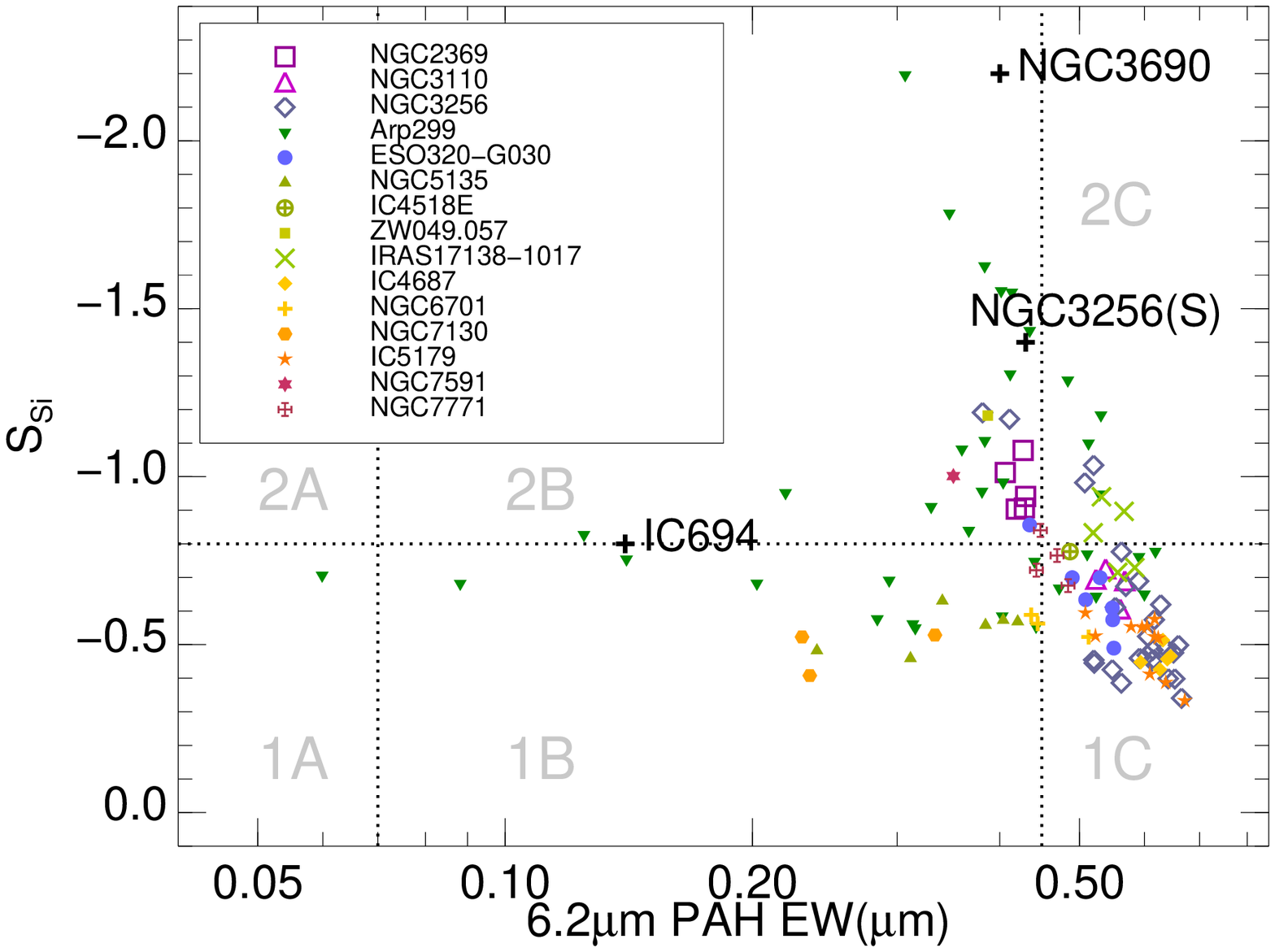}
\includegraphics[width=0.48\textwidth]{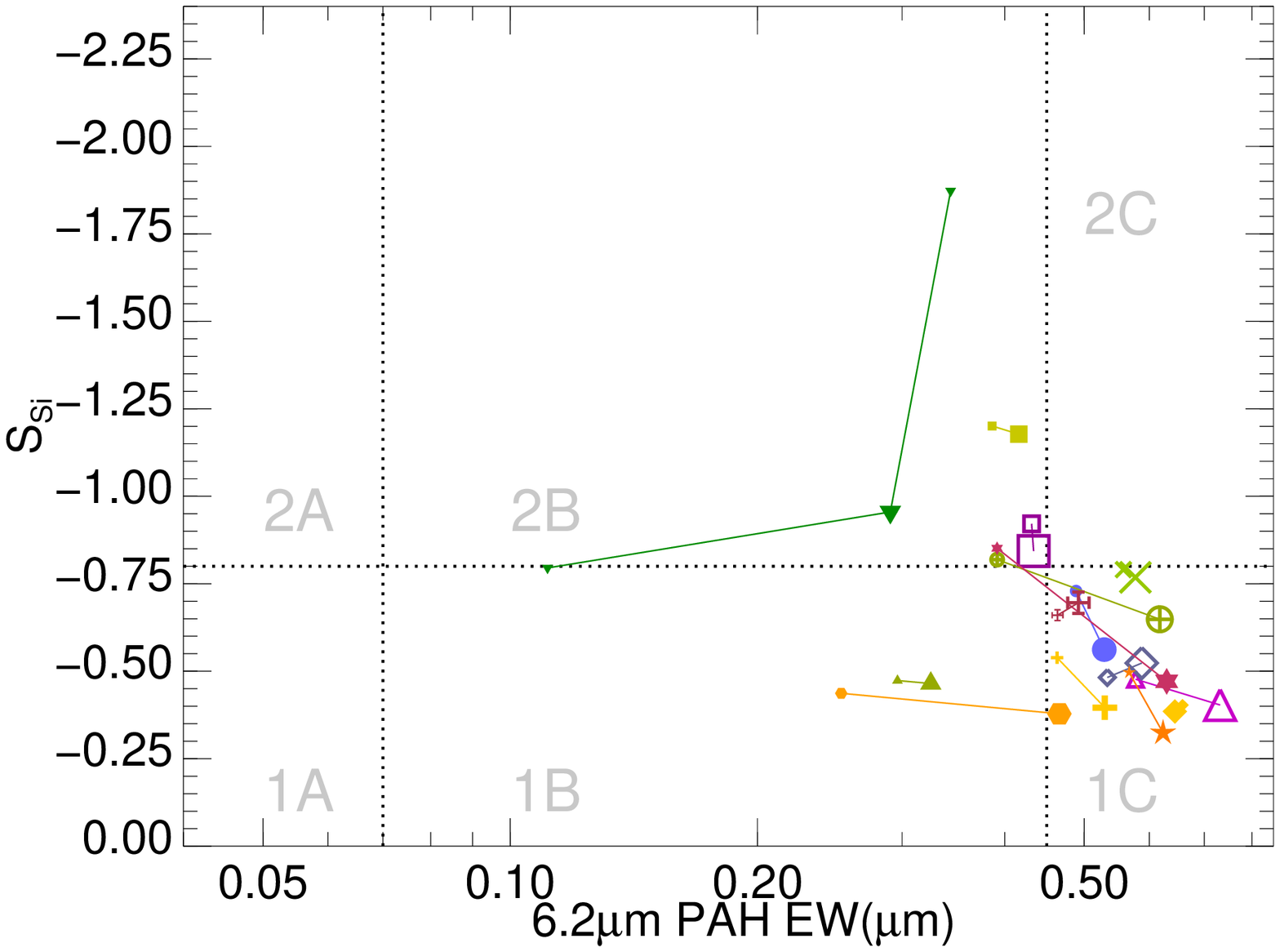} 
\caption{
  $S_{\rm Si}$ versus the \PAHseis\ equivalent width (EW) 
diagram, similar to that presented by \citet{Spoon07}.
\textit{Left:} Spatially resolved measurements for our sample of LIRGs
using a 2$\times$2 pixel box (3.7
  ''$\times$3.7 '' aperture). The black crosses are the two nuclei of Arp~299 (IC~694 and NGC~3690) and the southern nucleus of NGC~3256.
\textit{Right:} Nuclear (5.5
  ''$\times$5.5 '' aperture) and integrated values of each galaxy. The
  small symbols correspond to the nuclear value and the large symbols to
  the integrated galaxy. Nuclear and integrated are connected by a
  line to guide the eye. In both panels, following 
\citet{Spoon07}, the parameter space is divided into 6 classes out of
their 9 classes. Class
  1A is populated by unobscured AGNs. They have a featureless spectra with a
  shallow silicate feature. Class 1C corresponds to pure starbursts
  (PAH dominated spectra). The spectra of a galaxy in the class 1B is
  intermediate between the AGN spectra and the pure starburst. Classes
  2B and 2C have a deeper silicate absorption, so do 
classes 3A and 3B (not plotted here). The majority of 
  ULIRGs not dominated by an AGN lie in these
  classes. } 
\label{fig_pahew_ssi}
\end{figure*}

\citet{Spoon07} presented a diagram using the equivalent width (EW) of \PAHseis\ feature and the silicate strength to provide a general classification of infrared galaxies. ULIRGs appear distributed along two
branches. The horizontal branch is populated by unobscured AGNs (QSO and Sy1) in
the left-hand side of the diagram (class 1A) and by starburst galaxies
in the right-hand side
(class 1C) of the diagram. The diagonal branch goes from deeply
obscured nuclei (class 3A) to unobscured starbursts (class 1A). 

In Figure \ref{fig_pahew_ssi} we present a similar diagram for our sample of LIRGs.
The left panel shows the spatially resolved measurements of our sample of LIRGs. Most of them appear in this diagram at the region where both branches intercept. The nuclear and surrounding regions of galaxies classified as Seyferts using optical spectroscopy (NGC~5135: Sy2, NGC~7130: LINER/Sy and Arp~299-B1: Sy2, see Table \ref{tbl_obs_map}) are located in the 1B region (intermediate between AGN and starbursts), in agreement with their well-known composite nature. NGC~6701 and NGC~7591 are classified as composite objects which is likely to be a combination of an AGN and star-formation activity (see \citealt{AAH09PMAS}). Both nuclei appear quite close to region 1B. The two most obscured nuclei in the sample, Arp~299-A and the southern nucleus of NGC~3256, and their neighboring regions, are in the upper part of the diagonal branch populated by ULIRGs, although they do not reach the silicate strengths of the deeply embedded ULIRG nuclei.

The comparison between the nuclear and integrated values (right panel of Figure \ref{fig_pahew_ssi}) shows that the integrated spectra tend to have lower silicate strengths and larger \PAHseis\ EWs. 
We find that on average the integrated \PAHseis\ EW is $\sim$30\% larger than the nuclear values, whereas the integrated silicate strength is $\sim$15\% lower than that found in the nuclei.
That is, the integrated values move the galaxies to the pure starburst class region (1C) in this diagram. In the case of all the active galaxies of the sample, the nucleus is not sufficiently bright as to dominate the integrated mid-infrared spectrum and the dominant contribution from star formation, which is extended over several kpc (see \citealt{AAH06s, AAH09Arp299, AAH09PMAS}) makes the integrated spectrum look more starburst-like.

\section{Fine structure emission lines}\label{s:lines}

\subsection{Morphology}\label{ss:lines_morphology}
Although the limited spatial resolution ($\sim 1\,{\rm kpc}$)
of the SH maps does not allow us
to study the galaxy morphologies in great detail, it is possible to analyze
general trends. The exact spatial distribution of the emission lines depends on
the physical conditions and the age of the dominating stellar
population in each region. In Appendix \ref{apxDescription} we discuss the galaxies individually.

Figure \ref{fig_map_sh} shows the SH spectral  maps of the most prominent features, as well as those of the $15\micron$  continuum. The \Neii, \Neiii\ and \SIIIa\ fine structure lines have an overall morphology similar to that of the 15\micron\ continuum. We detect \SIV\ emission in only five systems. 
In two of them (NGC~7130 and NGC~5135), the \SIV\ emission mainly comes from the active nuclei \citep{Tanio09}. 
Our data do not allow us to determine whether there is a low surface brightness \SIV\ emission, associated with \HII\ regions, in these two galaxies due to the relatively low flux of this line.
In the other three cases, NGC~3256, IC~4687, and Arp~299 (for the last galaxy, see \citealt{AAH09Arp299}), the emission is more extended and we attribute it to the \HII\ regions seen in the Pa$\alpha$ and/or H$\alpha$ images. 

We find that the neon emission (\Neii\ and \Neiii) and the 15\micron\ continuum are spatially resolved. We calculated the ratio between the nuclear (circular aperture of radius 2 kpc; depending on the distance the aperture radius varies between 1.5 and 3 pixels)\footnote{The physical size of the nuclear aperture was chosen so it corresponds to the minimum spatial resolution of the most distant galaxy in the sample.} and the integrated neon line emission and 15\micron\ continuum emission for our galaxies, we also used the uncertainty maps to estimate the statistical uncertainty of these ratios (Table \ref{tbl_extended_ratio}). On average the uncertainty of the ratio is $\sim$6\%.
These are upper limits to the relative contribution of the nucleus to the total emission of the galaxy as in some cases the IRS maps do not cover the full extent of the galaxy. In general the emission arising from the central 2 kpc accounts for less than 50\% of the total emission for the neon lines and the continuum, indicating that the star formation is extended over several kpc. We note however that the LIRGs in this work were selected because they show extended Pa$\alpha$ emission (Section \ref{ss:thesample}), and thus, this result needs to be confirmed for a complete sample of LIRGs.

\begin{figure*}
\centering
\includegraphics[width=\textwidth]{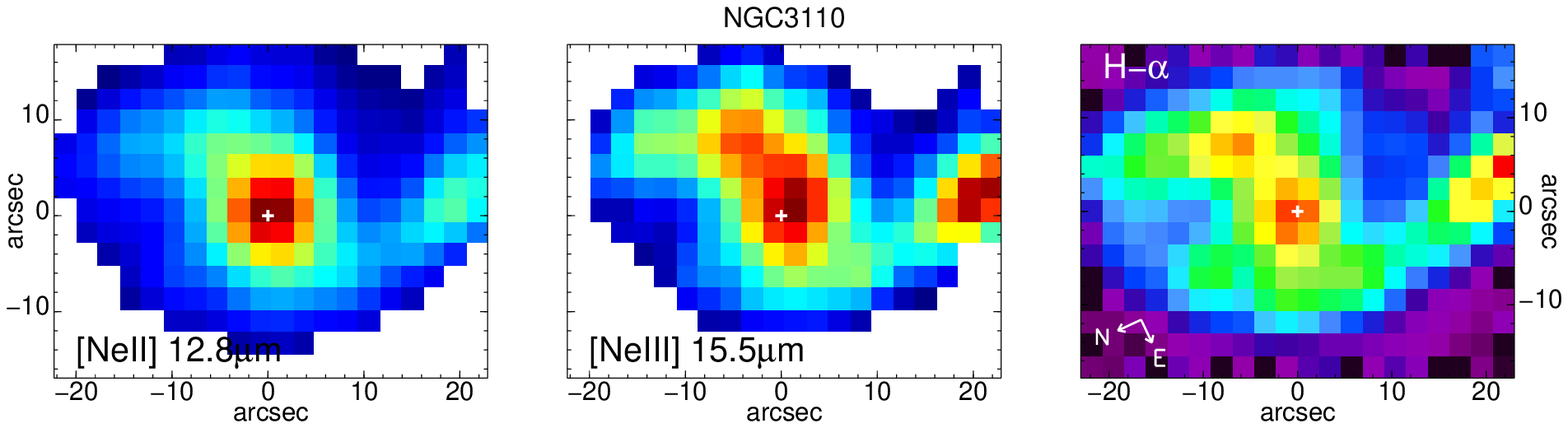}
\includegraphics[width=\textwidth]{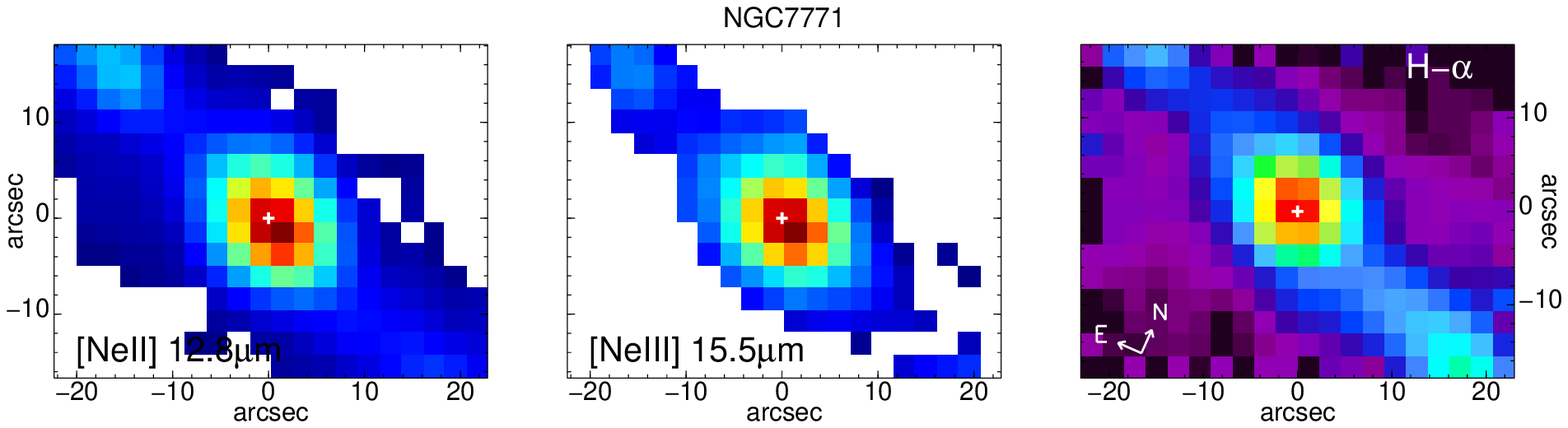}
\caption{Comparison of the \Neii\ and \Neiii\ maps with H$\alpha$ images from \citet{Hattori04} for NGC~3110 and NGC~7771. The image orientation for each galaxy is shown on the H$\alpha$ image. All the maps are shown in a square root scale.} 
\label{fig_Ne_Halpha}
\end{figure*}

The combined luminosities of the \Neii\ and \Neiii\ lines are good tracers of the star formation rate in galaxies \citep{Ho07}, so a good morphological correspondence is expected between them and
H$\alpha$. To further explore the behavior of the neon emission, Figure
\ref{fig_Ne_Halpha} compares the IRS neon spectral maps with H$\alpha$ images\footnote{The H$\alpha$ images  were degraded to the \Spitzer\slash IRS SH angular resolution, pixel size, and orientation.} of \cite{Hattori04} for NGC~3110 and NGC~7771.
The SH and H$\alpha$ maps of NGC~3110 show emission from the nuclear region and from \HII\ regions in the spiral arms. The \Neiii\ emission (also the \SIIIa\ emission) from these \HII\ regions is comparable to the emission arising from the nucleus, while \Neii\ is more concentrated around the nucleus. We note, however, that the H$\alpha$ image is not corrected for extinction. This is especially important in the nuclear region ($S_{\rm Si} \sim -0.48$, see Table \ref{tbl_lowres_map}, A$_V\sim 8\,$mag)  as it is the most extincted region in this galaxy. 
For NGC~7771, the \Neii\, \Neiii\ and H$\alpha$ emissions show similar morphologies,  both in the nuclear ring of star formation (seen in the Pa$\alpha$ image, Figure \ref{fig_map_sh}, but not resolved by the IRS spectral maps), and the extra-nuclear \HII\ regions, as well as in the diffuse regions between them. These two examples illustrate that at least from a qualitative point of view and on the scales probed here, the H$\alpha$, \Neii\ and \Neiii\ lines are tracing the same young ionizing stellar populations (see \citealt{Ho07} for a quantitative assessment).

\subsection{Spatially Resolved Line Ratios}\label{ss:line_ratios}

\subsubsection{The \Neiii\slash\Neii\ Line Ratios}\label{ss:ne3n2ratio}

\begin{figure}
\centering
\includegraphics[width=0.45\textwidth]{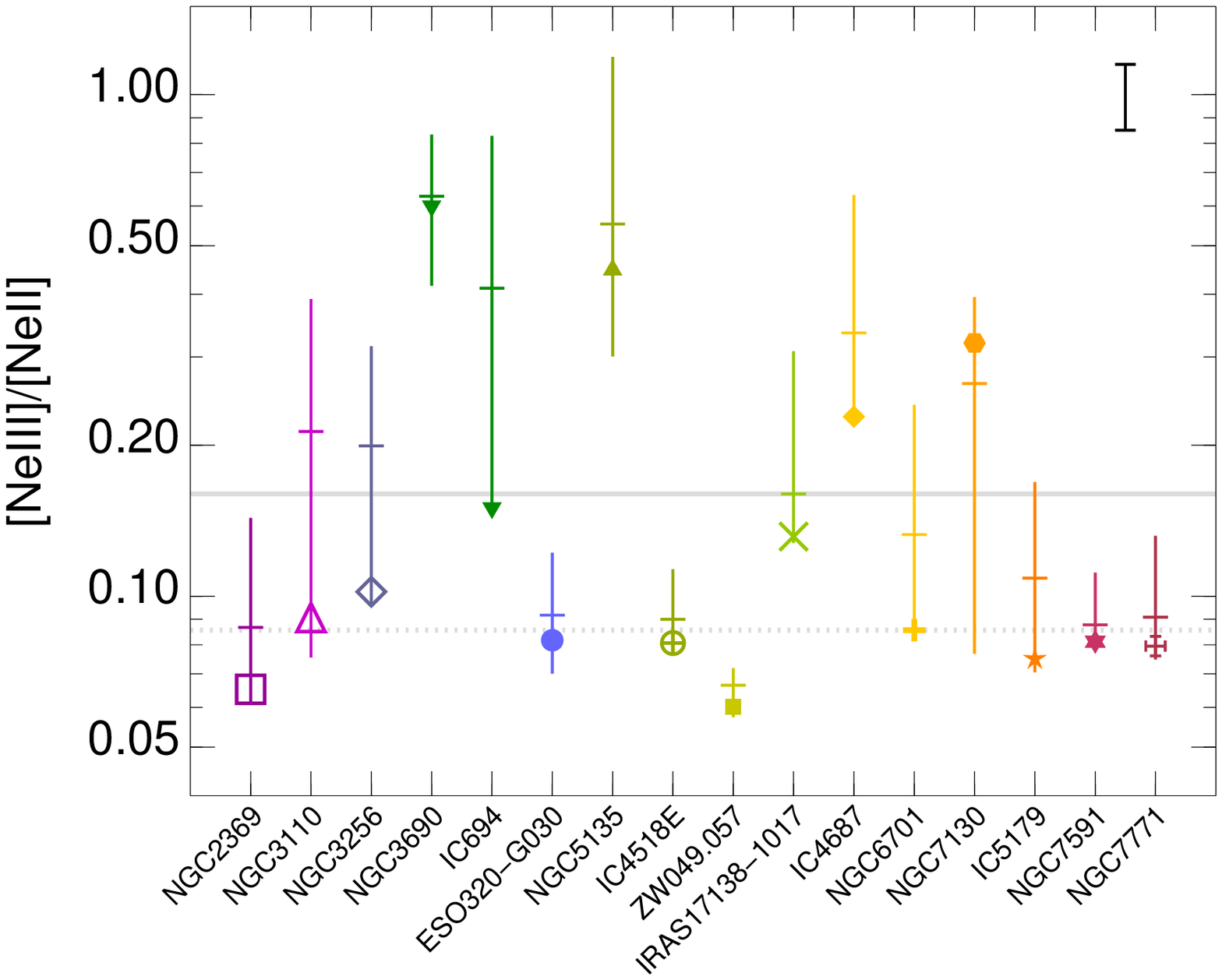} 
\caption{Range of the observed \Neiii\slash\Neii\ ratio from the spectral maps. 
The galaxy symbol (as in Figure \ref{fig_pahew_ssi}) indicates the value of the ratio in the
  nucleus. The horizontal mark is the median of the ratio of each
  galaxy. The solid gray line is the the median of the medians and the dotted line is the median of the nuclear ratios. The black line indicates the typical uncertainty of the ratios. } 
\label{fig_ne3ne2_gal}
\end{figure}

The spectral maps of the \Neiii\slash\Neii\ line ratios for our
sample of LIRGs are shown in Figure \ref{fig_co_sh}. 
In most cases the nuclei have the lowest \Neiii\slash\Neii\ ratios, while the circumnuclear regions and \HII\ regions tend to show higher ratios (see also Figure \ref{fig_ne3ne2_gal}).
With the exception of three cases with an active nucleus (NGC~3690, NGC~5135 and NGC~7130), the nuclear values of the \Neiii\slash\Neii\ ratio are all in the range of 0.06 - 0.10 (with three exceptions near 0.2, IC~694, IRAS17138-1017 and IC~4687 - see Table \ref{tbl_hires_map_nuclear}). Excluding the galaxies hosting an AGN, the median nuclear ratio is 0.08. The extra-nuclear values are between 0.06 and 1, while the integrated values (see Table \ref{tbl_hires_map_integrated}) are between 0.08 and 0.5. \citet{Thornley2000} find a similar range for the integrated ratios for an additional four LIRGs. This ratio ranges from 0.02 to 6 in starburst galaxies \citep{Verma03, Dale06, Dale2009}. For the \HII\ galaxies, the nuclear \Neiii\slash\Neii\ ratio is a factor $\sim$2-3 smaller than those measured in the extranuclear regions.

A similar behavior of increasing \Neiii\slash\Neii\ ratios with increasing galactocentric distances has been observed in the nearby galaxies NGC~253 \citep{Devost04} and M82 \citep{Beirao08}, although the physical scales probed are different ($\sim$80 pc in these two galaxies, $\sim$1 kpc in our sample of LIRGs). Galactic \HII\ regions follow the same trend, with \HII\ regions at increasing distances from the galactic nucleus showing larger \Neiii\slash\Neii\ ratios \citep{Giveon2002}.

Note that we did not correct this ratio for extinction. This correction however, only produces variations in the ratio of $5-15\%$ for the typical extinction of our sample ($S_{\rm Si}\sim-0.6$, that is A$_V\sim10$ mag). The correction (positive or negative) depends on the adopted extinction law (see \citealt{Farrah07}).

The strength of \Neiii\ relative to \Neii\ decreases with increasing metallicity, increasing age of the stellar population and increasing nebular density (e.g., \citealt{Snijders07}). In accordance with the luminosity- and mass-metallicity relations, LIRGs are characterized by uniformly relatively high metallicity (in many cases super-solar), with some indications of slightly (factor of two) reduced levels in their nuclei \citep{Rupke2008}. The \citet{Snijders07} models predict a $\sim$3 times larger \Neiii\slash\Neii\ ratio for a given age and density as the metallicity decreases from super-solar (2 Z$_\odot$) to solar. Thus, metallicity-related effects are unlikely to be the dominant cause of the relative increase of \Neiii\ strength outside the central regions of these galaxies. They may in fact to be overcome by some stronger mechanism. 

The trend with stellar population age is strong. The expected \Neiii\slash\Neii\ ratio of an instantaneous burst of star formation falls below $10^{-3}$ for ages $>$6 Myr \citep{Rigby2004}. 
\citet{Thornley2000} showed that the range of the \Neiii\slash\Neii\ ratio in starburst galaxies is compatible with models for bursts of star formation. However, from their models and those of \citet{Rigby2004}, the nuclear values for LIRGs are below the range that would be expected from such star-forming episodes. The observed nuclear ratios could only be explained in terms of the stellar population if: (1) the IMF is truncated at $\sim$30 M$_\odot$ in the LIRG nuclei (\citealt{Thornley2000}, Figure 6); or (2) the star formation rate in {\it all} the LIRG nuclei has decreased rapidly over the past 10-20 Myr (which seems contrived).

Another explanation is that due the high densities in the nuclear regions the \Neiii\slash\Neii\ ratio is suppressed, including the possibility that a fraction of the most massive stars are hidden in ultra-compact \HII\ regions \citep{Rigby2004}. The \Neiii\slash\Neii\ ratio decreases by approximately a factor of two if the density increases from 10$^3$ to 10$^5$ cm$^{-3}$ \citep{Snijders07}. We explore the density effects on the neon line ratio in the following section.

Before doing so, we point out that a few galaxies (NGC~3690, NGC~5135 and NGC~7130) have an opposite trend in that the \Neiii\slash\Neii\ ratio becomes larger in their nuclei. We attribute this behavior to the presence of active nuclei. These three galaxies are the only ones classified as Seyfert type and it is likely that the extra hard-ionizing radiation from the AGN is responsible for the high, three times larger than the median, \Neiii\slash\Neii\ nuclear ratio.

\subsubsection{The \SIIIa\slash\Neii\ Line Ratios}

The excitation potentials of the \Neii\ and \SIIIa\ emission lines are
21.6 eV and 23.3 eV, respectively. Therefore, the \SIIIa\slash\Neii\
ratio is almost insensitive to the hardness of the radiation field. On the other hand, this ratio is a good tracer of density in the range $10^4 - 10^6$ cm$^{-3}$, with lower \SIIIa\slash\Neii\ ratios indicating larger electron densities \citep{Snijders07}.

\begin{figure}
\centering
\includegraphics[width=0.45\textwidth]{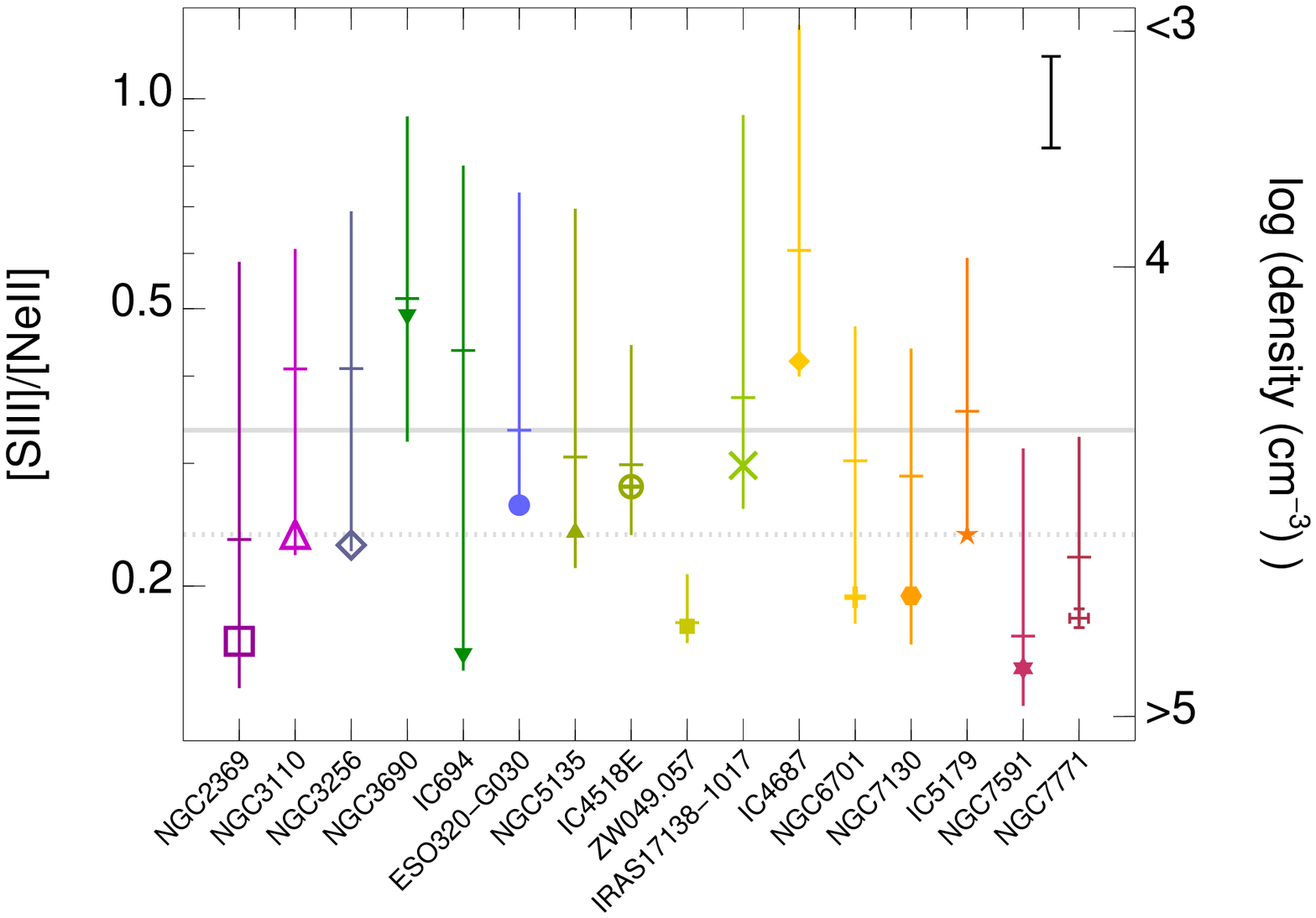} 
\caption{Range of the observed \SIIIa\slash\Neii\ ratio from the spectral maps. 
 The symbols are as in Figure \ref{fig_ne3ne2_gal}. The density label corresponds to the \SIIIa\slash\Neii\ ratio predicted by the \citet{Snijders07} models for solar metallicity, q = $1.6\times 10^8$ and age = 5 Myr. The \Neiii\slash\Neii\ ratio predicted using these parameters is $\sim$0.1-0.2 which is in agreement with the observed ratio.}
\label{fig_s3ne2_gal}
\end{figure}

Our spatially resolved measurements show that this ratio ranges from $\sim$0.1 to $\sim$1.25 (Figure \ref{fig_s3ne2_gal}). Assuming a 5 Myr old stellar population, an intermediate ionization parameter (q $=1.6\times10^8$) and solar metallicity (see caption of Figure \ref{fig_s3ne2_gal}), these ratios correspond to electron densities between $10^3$ and $10^5$ cm$^{-3}$ \citep{Snijders07}.
Like the \Neiii\slash\Neii\ ratio, for a given galaxy the lowest values of the \SIIIa\slash\Neii\ ratio occur in the nuclei (in most cases the nuclear value is between $\sim$0.13-0.25), while higher ratios are found around them (Figure \ref{fig_co_sh}). This indicates, as expected, that the nuclei are denser than the extranuclear \HII\ regions. According to this spatial distribution the integrated ratios ($\sim$0.2-0.5) are $\sim$30\% larger than the nuclear ratios.

\begin{figure*}
\centering
\includegraphics[width=0.45\textwidth]{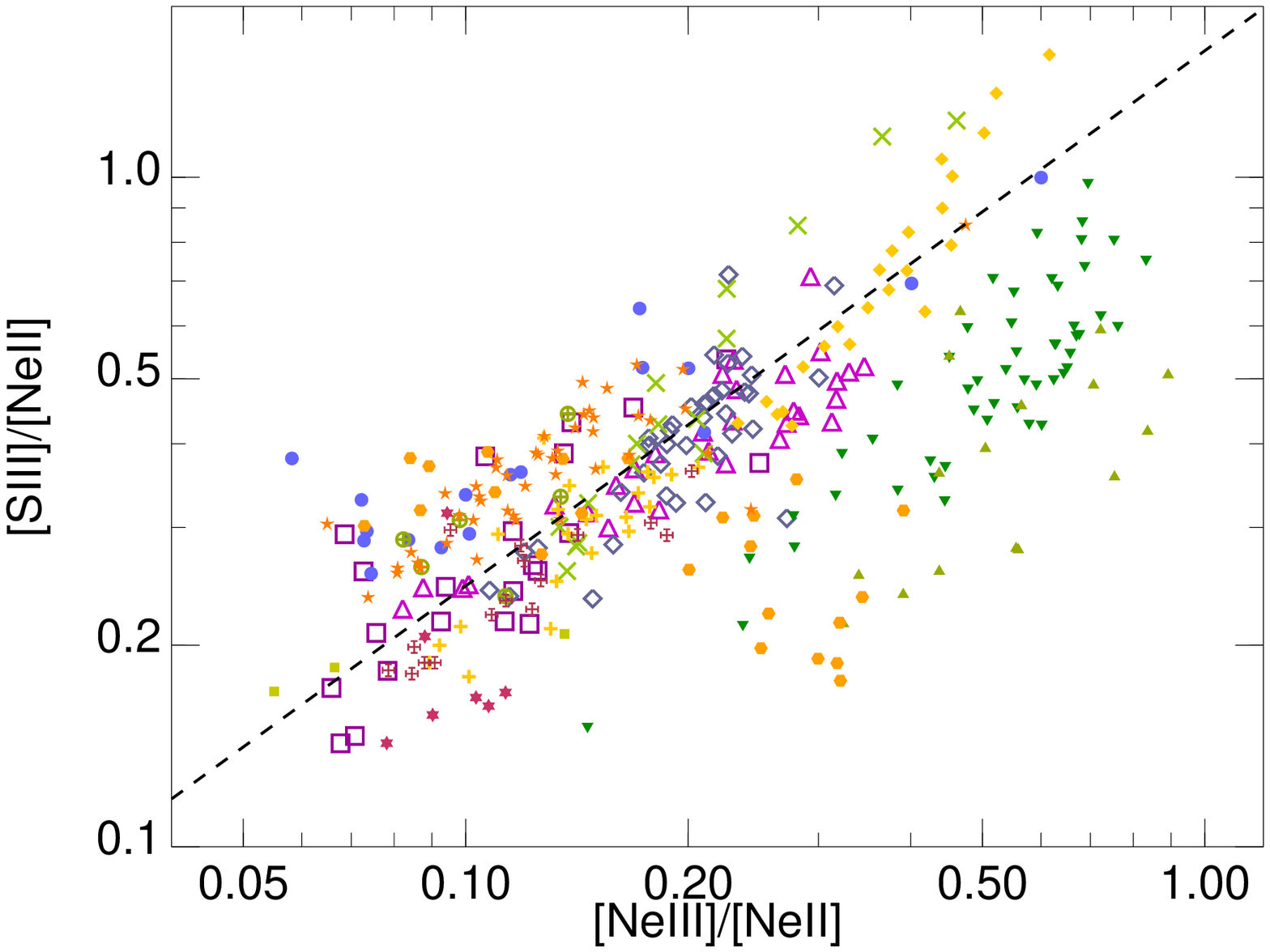} 
\includegraphics[width=0.45\textwidth]{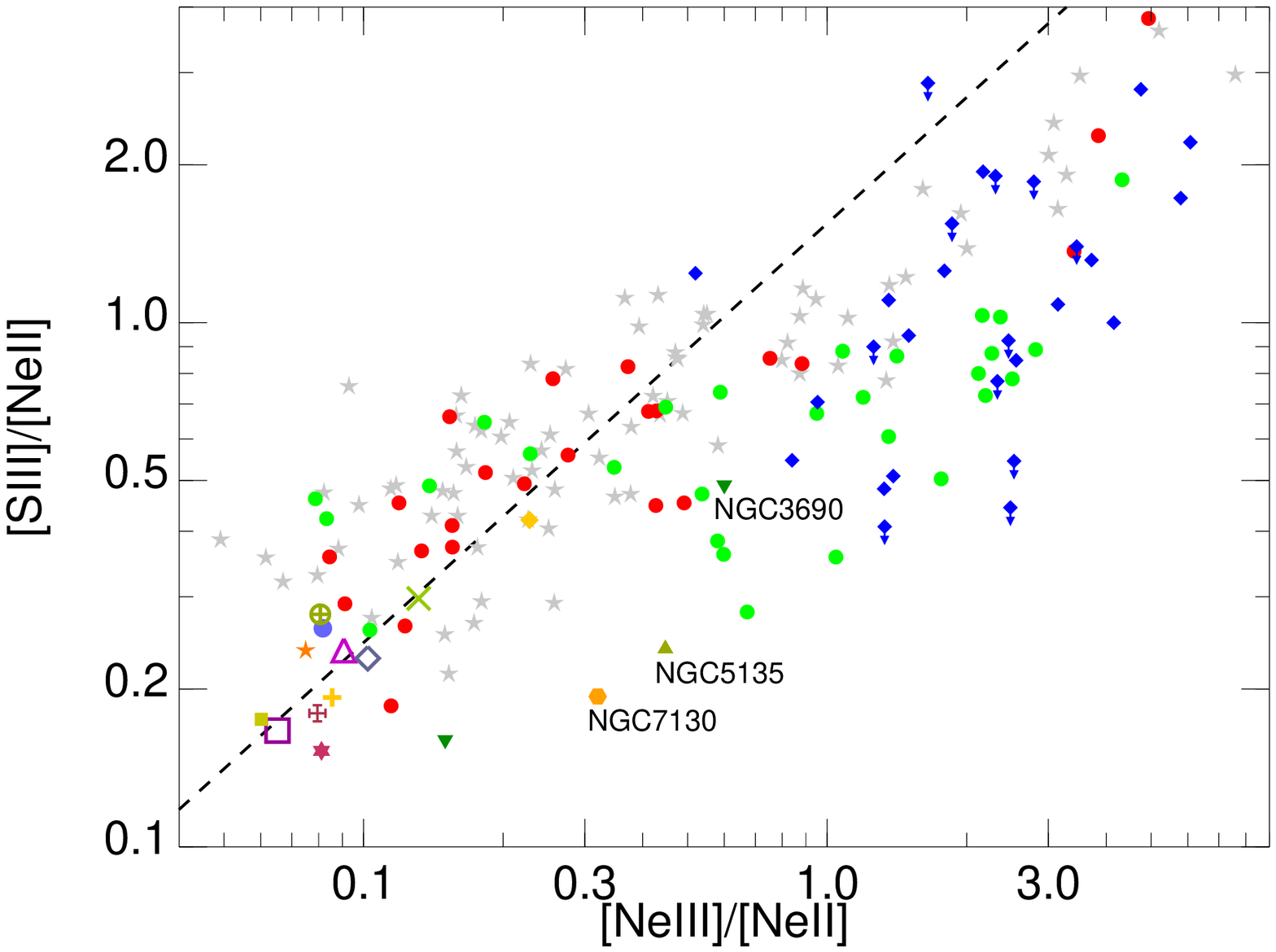} 
\caption{Left panel: \SIIIa\slash\Neii\ ratio vs. \Neiii\slash\Neii\ ratio from
  the spatially resolved maps (Each point corresponds to a resolution element including the nuclei). Right panel: Nuclear ratios of our sample of LIRGs. We can compare them with: \HII\ regions (grey stars) and star-forming galaxies (red circles) of \citet{Dale2009} , Seyfert galaxies of \citet{Tommasin08} (green circles) and quasars of \citet{Veilleux2009} (blue diamonds). The LIRG symbols are as in Figure \ref{fig_ne3ne2_gal}. The black dashed line is the best fit to the spatially resolved data of the LIRGs excluding those galaxies containing an active nucleus which lie below the correlation. We did not apply an extinction correction.} 
\label{fig_s3ne2_ne3ne2}
\end{figure*}

The left panel of Figure \ref{fig_s3ne2_ne3ne2} shows the \SIIIa\slash\Neii\ ratio
versus the \Neiii\slash\Neii\ ratio from our spatially resolved measurements. As
can be seen from this Figure, there is a tight correlation between the two line ratios if we exclude those galaxies hosting an active nucleus (NGC~3690, NGC~5135 and NGC~7130). The best linear fit to the data (excluding the AGNs) is:
\begin{equation}
log(y) = 0.19\pm 0.02 + (0.81\pm 0.03)\times log(x)
\end{equation}
Where y is the \SIIIa\slash\Neii\ ratio and x the \Neiii\slash\Neii\ ratio.
In the right panel of Figure \ref{fig_s3ne2_ne3ne2} we plot the nuclear ratios of our LIRGs together with those of extranuclear \HII\ regions and nuclei of nearby galaxies \citep{Dale2009}, Seyfert galaxies \citep{Tommasin08} and quasars \citep{Veilleux2009}.
Both diagrams show that, in general, for a given density (as traced by the \SIIIa\slash\Neii\ ratio) the AGNs show larger \Neiii\slash\Neii\ ratios than starburst galaxies and \HII\ regions. The most straightforward explanation is that the ionizing photons from the active nucleus increase the radiation hardness (the \Neiii\slash\Neii\ ratio). However some of the Seyfert galaxies and quasars are in the diagram close to \HII\ regions. These may be galaxies harboring an AGN as well as star formation. Moreover the position of the AGNs and starbursts in Figure \ref{fig_s3ne2_ne3ne2} is consistent with models \citep{Groves2004, Dopita2006}.

\subsubsection{The \SIV\ to \Neiii\ and to \SIIIa\ Line Ratios}\label{ss:s4ne3ratio}

\begin{figure}[!bt]
\centering
\includegraphics[width=0.45\textwidth]{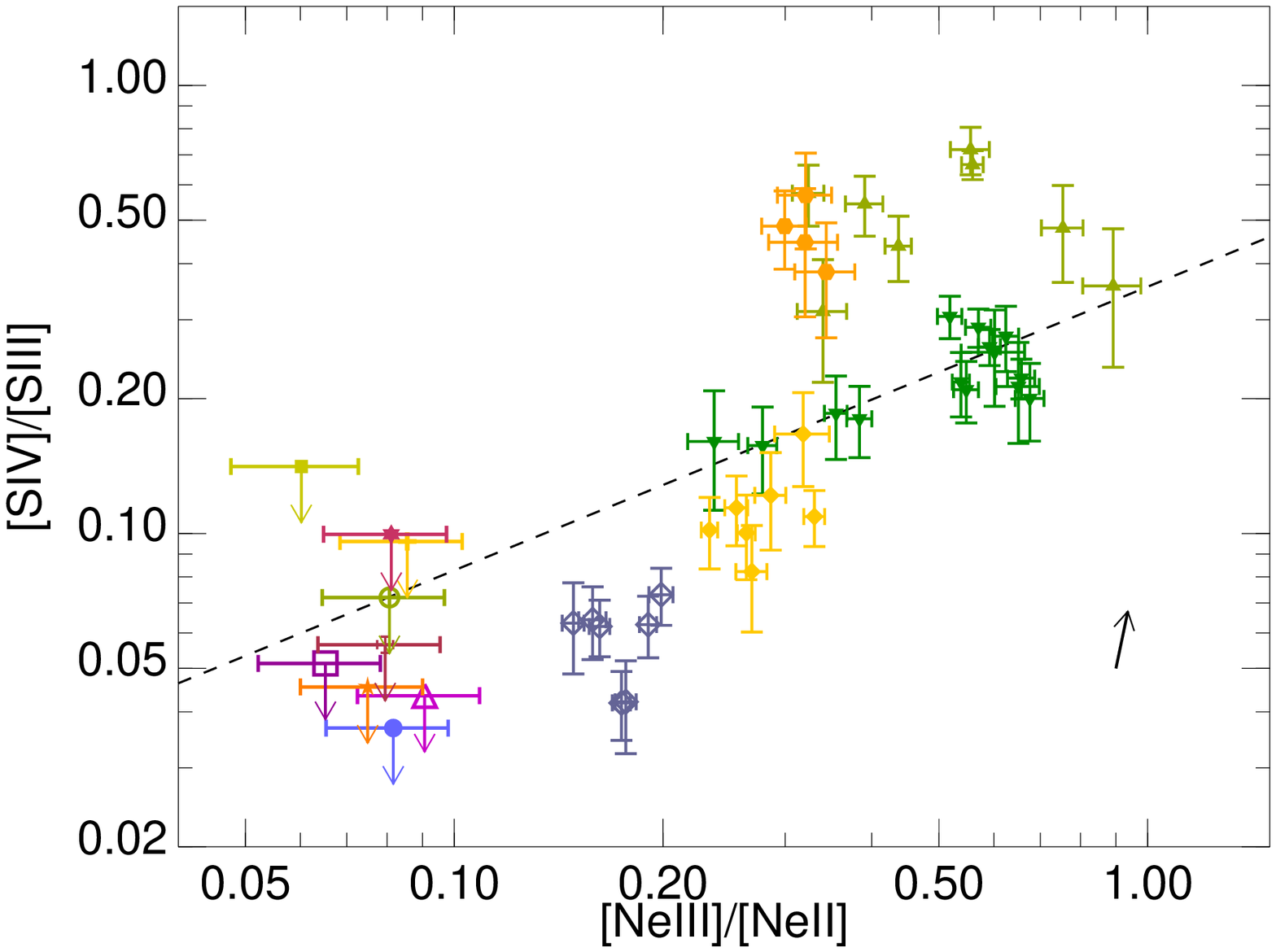} 
\caption{\SIV\slash\SIIIa\ ratio vs. \Neiii\slash\Neii\ ratio from the spatially resolved maps. We did not apply an extinction correction. The galaxy symbols are as in Figure \ref{fig_ne3ne2_gal}. When \SIV\ is not detected we represent the upper limit of the nuclear \SIV\slash\SIIIa\ ratio. The dashed line is the  correlation found by \citet{Gordon08} for M101 \HII\ regions and starburst galaxies. The black arrow indicates the extinction correction of A$_V$ = 10 mag, assuming a screen geometry. The orange and light green points above the correlation corresponds to galaxies harboring an AGN (see Section \ref{ss:s4ne3ratio}).}
\label{fig_ne3ne2_s4s3}
\end{figure}

The \SIV\ and \Neiii\ lines arise from ions with a similar excitation potential, 35 eV
and 41 eV, respectively. While the \Neiii\ line is detected in
all galaxies, the \SIV\ line is detected in just six galaxies
(40\% of the sample). If the \SIV\ emission arises from silicate dust embedded
regions, it can be heavily affected by extinction as it is inside the
broad 9.7\micron\ silicate feature. This seems
to be the case in ULIRGs \citep{Farrah07}. However, the depth of the silicate features
 in our sample of LIRGs are moderate, and generally much lower than in
ULIRGs probably indicating lower extinction (see Section \ref{s:silicate}).

Using the photoionization models of \cite{Snijders07}, we can infer the expected range of \SIV\slash\Neiii\ ratios.
The values for the range of densities, ionization parameters and stellar population ages covered by the models, excluding the most extreme cases, are compatible with the upper limits we measure. 
Moreover, we do not find any correlation between the \SIV\slash\Neiii\ ratio and the observed silicate strength ($S_{\rm Si}$). Thus it is likely that the non-detection of the \SIV\ line is because its flux is under the detection limit.

The \SIV\slash\SIIIa\ and the \Neiii\slash\Neii\ ratios  are well correlated and can be used to estimate the hardness of the radiation field \citep{Verma03, Dale06, Gordon08}. Figure \ref{fig_ne3ne2_s4s3} shows that there is a good correlation between  the \SIV\slash\SIIIa\ and \Neiii\slash\Neii\ ratios of the spatially resolved measurements for our star-forming LIRGs, and indeed, they approximately follow the correlation found by \cite{Gordon08} for M101 \HII\ regions and starburst galaxies. The upper limits to the \SIV\slash\SIIIa\ ratio are also consistent with this trend.

Two of the galaxies classified as active, NGC~5135 and NGC~7130, lie above the
\cite{Gordon08} correlation. A similar trend was found by \citet{Dale06} for  
AGN in the \SIV\slash\SIIIb\ versus the \Neiii\slash\Neii\ diagram. 
A possible explanation is that extra \SIV\ emission is produced by the active nucleus. High spatial resolution ground-based observations of these two LIRGs indicate that most of the \SIV\ emission comes from the nucleus, whereas this line is not detected in the surrounding star-forming regions of these galaxies \citep{Tanio09}. For the other two galaxies NGC~3690 (classified as Seyfert), and NGC~7591 (classified as composite, intermediate between \HII\ and LINER) the \SIV\ line is more likely to be affected by extinction, as both have nuclear $S_{\rm Si} \sim -0.8$, and thus the observed \SIV\slash\SIIIa\ ratios are lower limits. 

\section{PAH Features}\label{s:PAH}
 
Models predict that the relative strength of the PAH bands depends on the ionization state \citep{Draine01, Galliano2006}. The \PAHseis\ and \PAHsiete\ bands dominate the emission of ionized PAH, whereas the \PAHonce\ is more intense for neutral PAHs. More complete models of PAH feature behavior have been published recently by \citet{Galliano2008}, who explore the effects of the radiation field hardness and intensity as well as the size distribution of the PAH carriers.

\subsection{Morphology}\label{ss:pah_morphology}

Figure \ref{fig_map_sl} shows flux maps of three brightest
PAH features present in the SL module at 6.2\micron, 7.7\micron\ and 11.3\micron, together with the 5.5\micron\ continuum maps. 
Table \ref{tbl_extended_ratio} shows the ratio between the flux arising from the central 2 kpc and the total emission. The \PAHonce\ is clearly more extended than the 5.5\micron\ continuum in 60\% of the galaxies, arising also from more diffuse regions (see also \citealt{Tanio09}). However, the nuclear contribution to the integrated \PAHseis\ emission is comparable to that at the 5.5\micron\ continuum, and only in 2 cases (NGC~7130 and NGC~7591) the \PAHseis\ emission is more extended than the continuum. When comparing both PAHs, the \PAHonce\ emission is more extended than the \PAHseis\ emission in about 40\% of the galaxies, while the opposite is never the case.

A direct comparison of the PAH ratios with fine structure line ratios
is not possible because the low spectral resolution (R$\sim$60-120)
is not sufficient to separate the emission lines from the
PAHs. Alternatively we used high resolution (R$\sim$600) data to
measure the fine structure lines and the \PAHonce\ (Figure \ref{fig_map_sh}).

In general, the \PAHonce\ emission appears to be more extended than the \Neii\ line. The PAH emission is also more extended than the \Neii$+$\Neiii\ emission which, as discussed in Section \ref{s:lines}, traces recent star formation.
PAHs have also been used to measure the star formation, since they are a proxy, although not perfect, for the total infrared luminosity \citep{Peeters2004, Smith07}. From this comparison, it is clear that the \PAHonce\ emission does not trace the same stars as the \Neii\ emission line. In fact, \citet{Peeters2004} found that the \PAHseis\ is not a good tracer of massive star formation (O stars) as seems to be the case of the \PAHonce.

\subsection{Extinction Effects on the PAH Ratios}\label{ss:pah_extinction}
 As discussed above, models
\citep{Draine01, Galliano2006} predict that neutral PAHs show larger
\PAHonce\ to \PAHseis\ and \PAHonce\ to \PAHsiete\
ratios than ionized PAHs. Figures \ref{fig_pah11pah6_sl} and
\ref{fig_pah11pah7_sl} show the SL 
spectral maps of the 
 \PAHonce\slash\PAHseis\ and \PAHonce\slash\PAHsiete\ 
ratios, respectively. The \PAHonce\slash\PAHseis\ ratio spans a factor of 2,
from $\sim$0.45 to $\sim$1.15 and the \PAHonce\slash\PAHsiete\ ratio
range is $\sim0.3 - 0.9$. These ranges are similar to those found in
starburst galaxies \citep{Brandl06}.

\begin{figure*}
\centering
\includegraphics[width=0.45\textwidth]{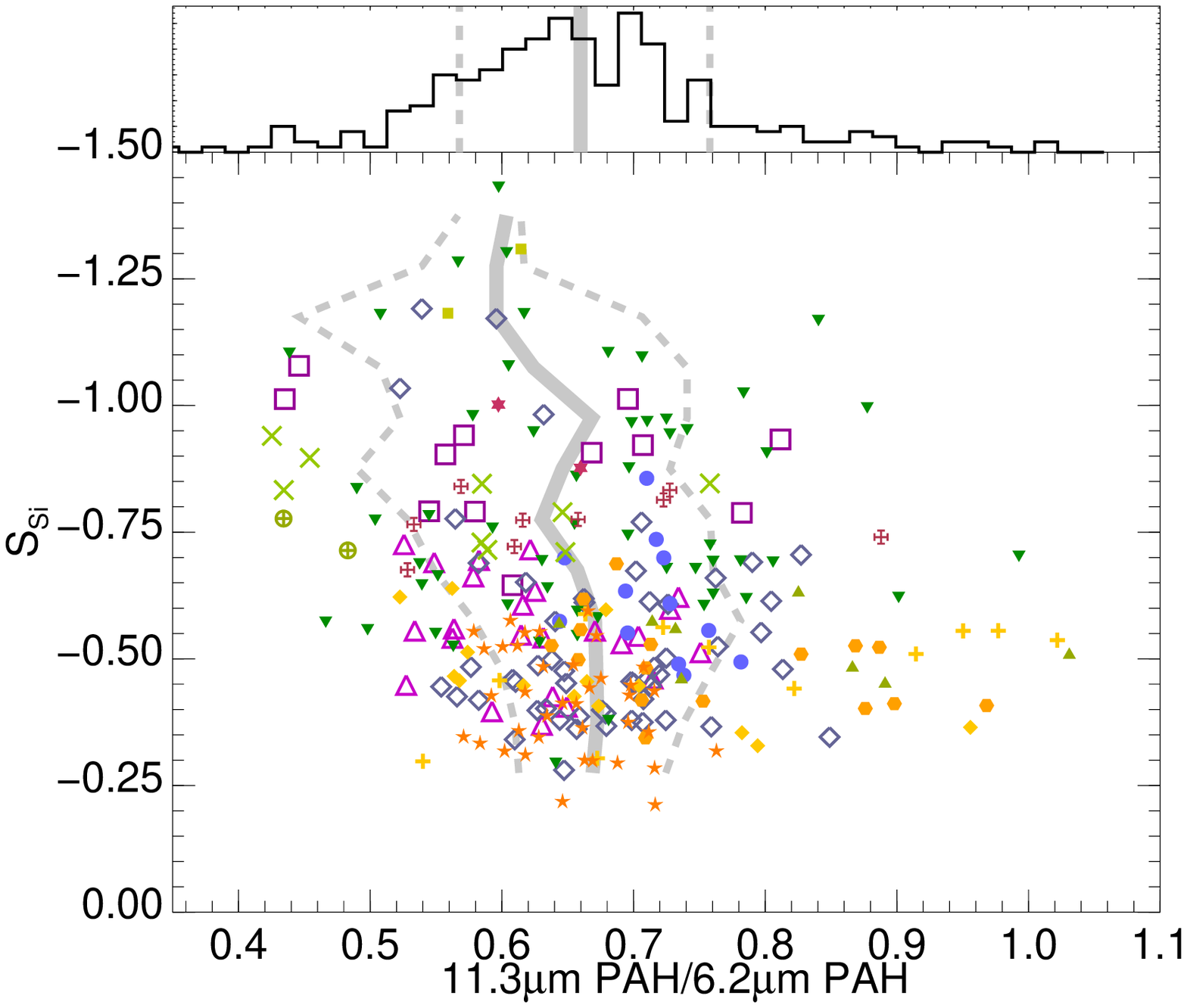}
\includegraphics[width=0.45\textwidth]{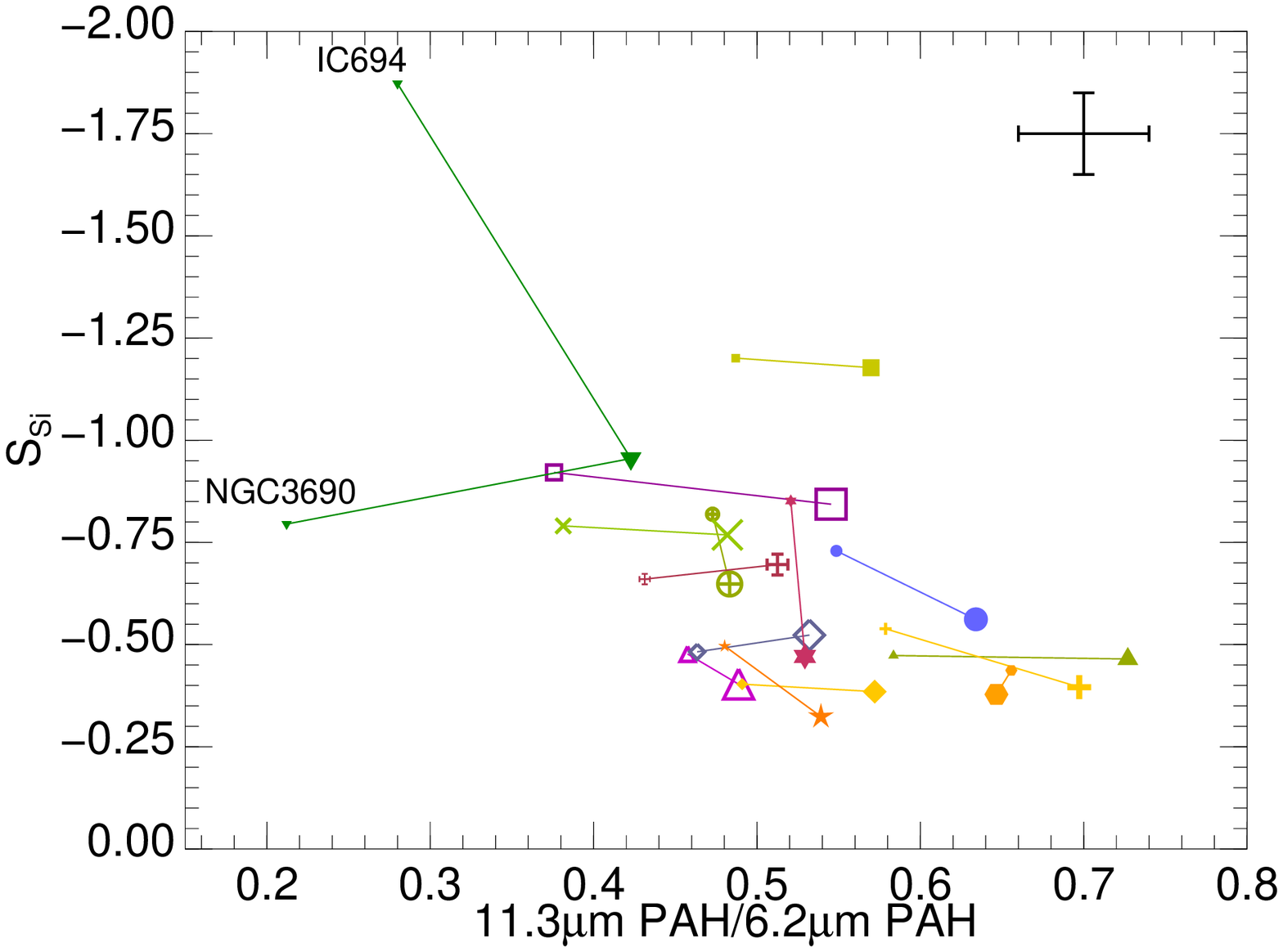}
\caption{\PAHonce\slash\PAHseis\ ratio vs. silicate strength ($S_{\rm
    Si}$). \textit{Left}:  Spatially resolved measurements 
using a 2$\times$2 pixel box. The
  symbols are as in Figure \ref{fig_ne3ne2_gal}. The solid
  gray line is the median and the dashed gray lines encircle the
  region containing 68\% of the data points. These are calculated in
  intervals of 0.25 of the $S_{\rm Si}$. The top histogram is the
  distribution of the the \PAHonce\slash\PAHseis\ ratio for the whole
  sample of LIRGs, in gray are marked the median and the
  1-$\sigma$. \textit{Right}: Same as left panel, but comparing
    the nuclear (small symbols) and
  integrated (large symbols) values. For each galaxy the 
values are connected by a line to guide the eye. 
}
\label{fig_pahratio_ssi}
\end{figure*}

The \PAHonce\ is affected by the broad 9.7\micron\ silicate feature,
thus in highly obscured regions the 
the \PAHonce\slash\PAHseis\ and the \PAHonce\slash\PAHsiete\ ratios can be
underestimated. An example of this is the southern nucleus of NGC~3256,
one of the most obscured regions ($S_{\rm Si}$
= -1.4) in the sample. Its \PAHonce\slash\PAHseis\ ratio is
lower than that in the surrounding regions. 

To explore the effects of the 9.7\micron\ silicate 
feature on the PAH ratios, we compare
the \PAHonce\slash\PAHseis\ ratio and $S_{\rm Si}$ for the spatially resolved 
measurements of our LIRGs in Figure~\ref{fig_pahratio_ssi} (left
panel). A similar figure was presented 
by \citet{Brandl06} for the nuclear regions of starburst
  galaxies. They found that the extinction can
change the relative strength of the PAHs by up to
a factor of 2. We find, however, that the median value of the
\PAHonce\slash\PAHseis\ ratio is approximately constant 
for silicate strengths weaker than 
$\sim-1.0$, and only deeper silicate absorptions seem to affect the
\PAHonce\slash\PAHseis\ ratio. Some caution is needed here
because there are very few regions in our sample with 
silicate strengths stronger than $-1.0$. 

As an alternative approach, we corrected these PAH ratios for extinction using the spatially resolved measurements of the $S_{\rm Si}$. We assumed a foreground screen dust geometry and the extinction law of \citet{Smith07}. 
For a typical value of the silicate strength in our sample of $S_{\rm Si}=-0.5$,
this extinction reduces the \PAHonce\ to \PAHseis\ and to \PAHsiete\ ratios
by $\sim15\%$. If we increase the silicate strength up to $-1.0$ the
reduction is $\sim30\%$. These are much less than the factor of $\sim$2
variations in the \PAHonce\slash\PAHseis\ ratio observed in some galaxies
(e.g., NGC~3256, NGC~3110, NGC~6701), at an almost constant value of 
$S_{\rm Si}$. Thus, we conclude that the observed variations in the
PAH ratios are real, and do not significantly depend on the extinction, at least for regions in our sample of LIRGs
with relatively shallow silicate absorptions ($S_{\rm Si} >-1)$.

\subsection{Nuclear vs. Integrated PAH Ratios}

The right panel of Figure \ref{fig_pahratio_ssi} compares the
\PAHonce\slash\PAHseis\ ratio vs $S_{\rm Si}$ values of the nuclear
spectra and the integrated spectra. The general trend is that the integrated
spectra have shallower silicate absorptions ($\sim15\%$ lower
$S_{\rm Si}$, on average) and slightly larger ($\sim20\%$, 
on average) \PAHonce\slash\PAHseis\ ratios. 

\begin{figure}
\centering
\includegraphics[width=0.45\textwidth]{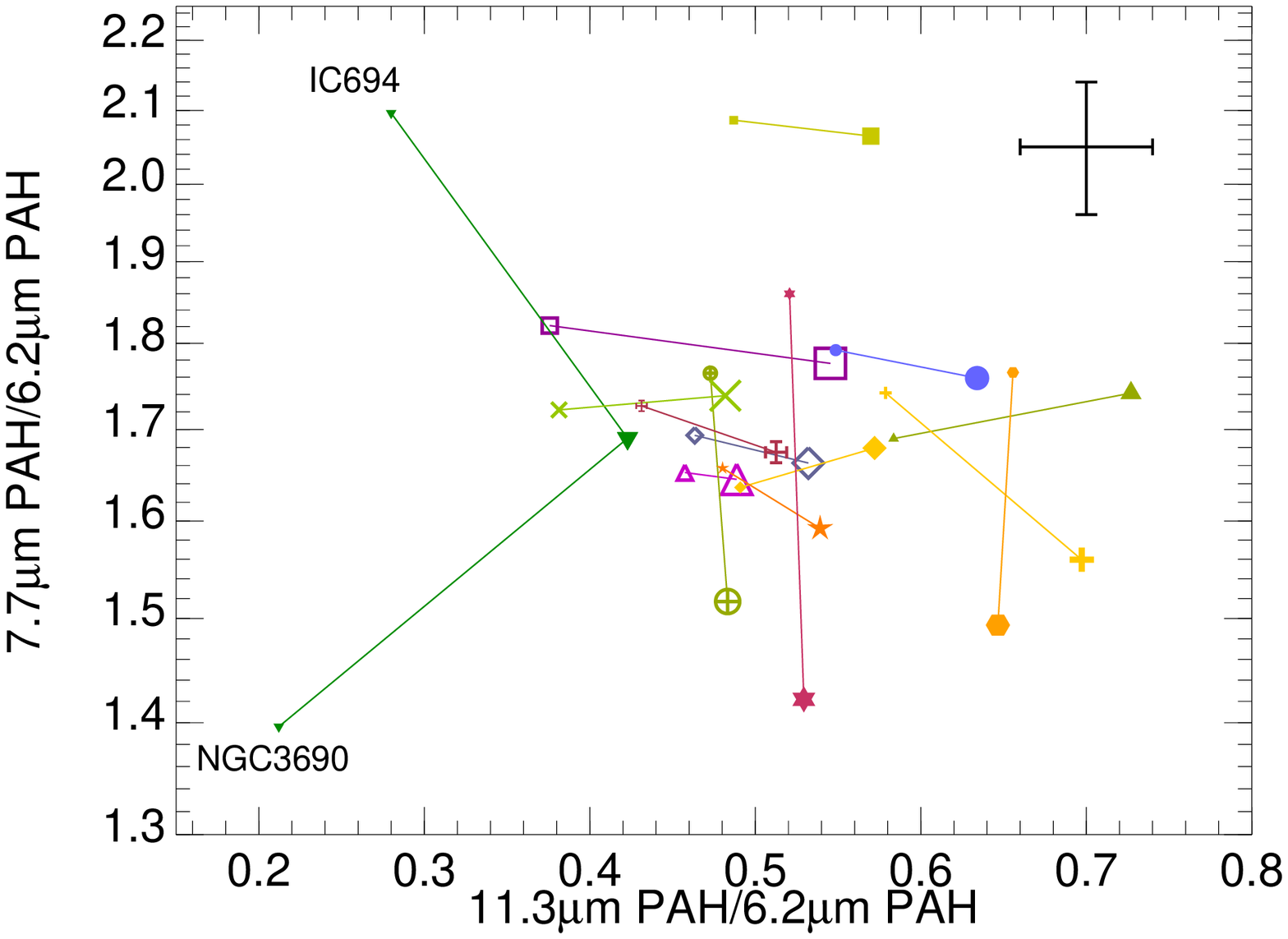}
\caption{ \PAHonce\slash\PAHseis\ ratio
    vs. \PAHsiete\slash\PAHseis\ ratio for nuclear and integrated
      values. Symbols are as in the right panel of Figure
    \ref{fig_pahratio_ssi}.} 
\label{fig_pahratios}
\end{figure}

In Figure \ref{fig_pahratios} we represent the \PAHonce\slash\PAHseis\ ratio versus \PAHsiete\slash\PAHseis\ ratio, comparing the nuclear and the integrated values of each galaxy. The distribution of the values of the ratio is approximately uniform.
However it is interesting to note two clearly different behaviors. The majority of the galaxies show a small variation of the \PAHsiete\slash\PAHseis\ ratio and a somewhat larger \PAHonce\slash\PAHseis\ ratio in the integrated spectrum. This behavior is  expected, since the \PAHsiete\ and \PAHseis\ are related to ionized PAHs and the \PAHonce\ to neutral PAHs. The ionized PAHs emission is concentrated in the nuclear regions, but the there is a substantial amount of the \PAHonce\ emission arising from diffuse regions throughout the galaxies (see Section \ref{ss:pah_morphology}). This explains the larger integrated \PAHonce\slash\PAHseis\ ratios of all the LIRGs (see Table \ref{tbl_lowres_map}).

Some galaxies, however, show a completely different behavior, with an almost constant 
\PAHonce\slash\PAHseis\ ratio but a lower \PAHsiete\slash\PAHseis\
ratio in the integrated spectrum. The most extreme cases of this
behavior are NGC~7130 and NGC~7591. \citet{Galliano2008} showed
that the \PAHsiete\slash\PAHseis\ ratio is very
sensitive to the size distribution of the PAHs, with smaller PAH molecules
producing lower \PAHsiete\slash\PAHseis\ ratios. 
One possibility is that smaller PAH molecules are destroyed in harsher environments, such as those of Seyfert and LINER nuclei. The destruction of small PAHs might suppress the nuclear \PAHseis\ emission. This may lead to the different behavior observed in the AGN/LINER nuclei in Figure \ref{fig_pahratios}. 
Moreover, the absolute value of the ratio seems to give little information about the nature, starburst or AGN, of the nuclei and we can only distinguish them by the relative variation between nuclear and integrated ratios. 

\begin{figure}
\centering
\includegraphics[width=0.45\textwidth]{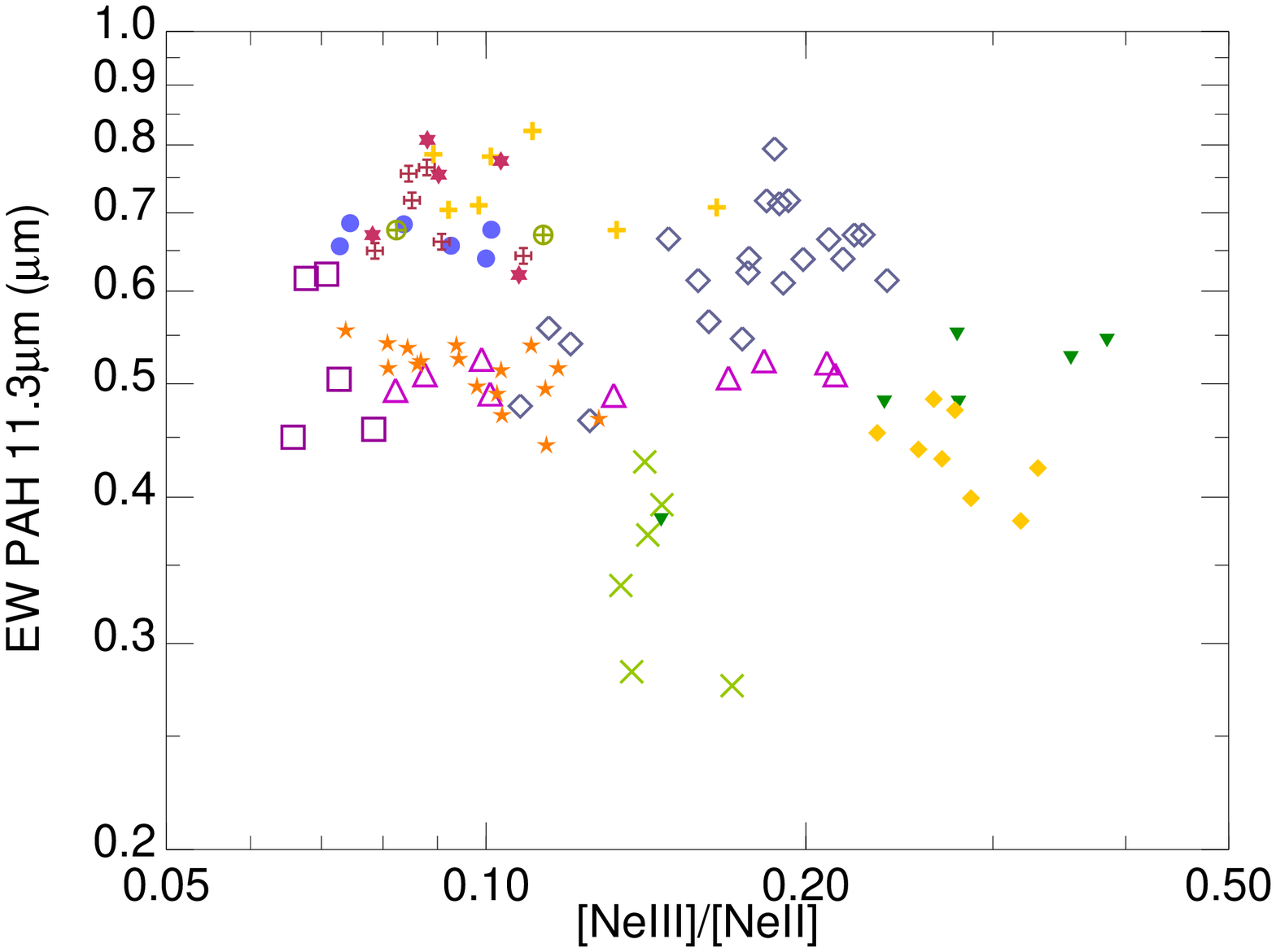}
\caption{Spatially resolved measurements of the
EW of the \PAHonce\ versus the
  \Neiii\slash\Neii\ ratio for those galaxies not classified as
  active. Galaxy symbols as in Figure
  \ref{fig_ne3ne2_gal}. } 
\label{fig_pah113ew_ne3ne2}
\end{figure}

\subsection{The Effects of the Radiation Field}\label{ss:pah_radiation}
Since the \Neiii\slash\Neii\ ratio is a proxy for the hardness of the radiation field, we can compare the spatial variations of this ratio with those of the \PAHonce\slash\PAHseis\ to ascertain if the radiation field has an effect on the PAH ratios. A comparison between Figures \ref{fig_pah11pah6_sl} and \ref{fig_co_sh} indicates that the \PAHonce\slash\PAHseis\ ratio does not decrease necessarily in regions of enhanced \Neiii\slash\Neii\ ratios.
 
On the other hand, the morphology of the \PAHonce\slash\PAHsiete\ ratio (Figure \ref{fig_pah11pah7_sl}) resembles that of the \Neiii\slash\Neii\ ratio. However this resemblance is the opposite of what one would expect. 
Both ratios present their minima in the nuclear regions for most of the galaxies. Thus while the observed nuclear \Neiii\slash\Neii\ ratio would indicate lower radiation hardness in the nuclei, the PAH ratio would suggest a larger ionization state. Again, it is likely that the typical radiation field hardness observed in our sample of LIRGs is not sufficiently large as to affect the PAH emission, as is the case of the low metallicity galaxies of \citet{Wu06}.
Another possible interpretation is that the PAHs and the neon emission are related to different physical processes that due to our limited spatial resolution we cannot separate. 
As a result, the bulk of the \PAHsiete\ emission may be produced in star-forming regions, while the \PAHonce\ emission also comes from diffuse regions. Thus, low values of the \PAHonce\slash\PAHsiete\ ratio trace \HII\ regions, and larger values indicate less recent star formation. The latter interpretation is similar to that of \citet{Galliano2008}. They find that the lowest values of the \PAHonce\slash\PAHsiete\ ratio are associated with the brightest PAH emission. However, in their galaxies, the \PAHonce\slash\PAHseis\ ratio follows the same trend, whereas this is not the case for the LIRGs studied here. Alternatively, as discussed in the previous section, the size of the grains may affect this ratio. If the smaller grains are favored in and around the nucleus, the \PAHsiete\ emission would be increased respect the \PAHonce\ emission. As consequence the \PAHonce\slash\PAHsiete\ ratio would be lower in the nucleus.

\citet{Brandl06} studied the dependency of the EW of the PAHs with the radiation hardness for a sample of high metallicity starburst galaxies. \citet{Wu06} carried out a similar study, but for low metallicity galaxies. Whereas \citet{Brandl06} found that this dependency does not exist, \citet{Wu06} found the opposite, that is, the harsher radiation of low metallicity galaxies field lowers the PAH EW. Eventually \citet{Gordon08} showed that the reason for the discrepancy is the different range of the radiation field hardness in high and low metallicity galaxies. In particular, \citet{Gordon08} found that for ionizations above a certain threshold there is a correlation between the radiation hardness and the PAH EW, but for lower ionizations this dependency disappears.
We show in Figure \ref{fig_pah113ew_ne3ne2} the EW of the \PAHonce\ and the \Neiii\slash\Neii\ ratio of those galaxies  not classified as AGN. The ionization and metallicity of our LIRGs are similar to those galaxies studied by \citet{Brandl06} and, similarly to their findings, there is no correlation between these parameters in our sample of LIRGs.

\subsection{Dependency on the Age of the Stellar Population}
Since the \Neii\ emission traces young stellar populations ($<$10 Myr), and since the \PAHonce\ feature can also be excited by older populations (see \citealt{Peeters2004} and Section \ref{ss:pah_morphology}), the \Neii\slash\PAHonce\ ratio can give us an indication of age variations throughout the galaxies \citep{Tanio09}. As can be see from Figure \ref{fig_co_sh}, the \Neii\slash\PAHonce\ ratio ranges from $\sim$0.1 to $\sim$0.6 for our sample of LIRGs.
In some LIRGs (e.g., NGC~6701, NGC~7130, NGC~7771, IC~4687) the \Neii\slash\PAHonce\ ratio appears enhanced in the high surface brightness \HII\ regions (see the HST/NICMOS Pa$\alpha$ images in Figure \ref{fig_map_sh}) with respect to regions of more diffuse emission. 
In other galaxies the interpretation might not be as straightforward. 
In some cases the IRS  angular resolution is not sufficiently high to isolate the circumnuclear \HII\ regions (e.g., NGC~5135), and the age differences may be smoothed out.

\section{Molecular hydrogen}\label{s:molecular_hydrogen}

We constructed maps of the molecular hydrogen emission using the pure rotational lines \Hm{1}\ at 17.0\micron\ and the \Hm{2}\ at 12.3\micron\ (Figure \ref{fig_map_sh}). These lines trace warm (300 K $<$ T $<$ 1000 K) molecular gas in the interstellar medium. 
The \Hm{1}\ morphology appears, generally, different from that of the fine structure lines, which are associated with the ionized gas medium. There are also differences in the velocity fields  that will be discussed in Section \ref{s:velocity_fields}.
Using the same procedure as described in Section \ref{ss:lines_morphology} we find that the \Hm{1}\ emission is more extended than the 15\micron\ continuum in $\sim$60\% of the sample. In most of these galaxies the size of the \Hm{1} emitting region is larger than that of the fine structure lines (see Table \ref{tbl_extended_ratio}).

\subsection{Temperatures, Column Densities and Masses}
We extracted the nuclear spectra of all galaxies using a 13.4''$\times$13.4'' aperture. This aperture corresponds to physical scales between 3 and 5 kpc depending on the galaxy. This aperture size is different from that used in the other Sections because now we are limited by the lowest spatial resolution of the LH module, where the \Hm{0} line at 28.2\micron\ lies.
The fluxes of the \Hmol\ lines measured from these spectra are listed in Table \ref{tbl_molecular_hydrogen_fluxes}.

The temperature of the warm molecular hydrogen was calculated as described by \citet{Roussel07}. We assumed an ortho-to-para ratio $=3$ \citep{Rigopoulou02, Roussel07}, and thus this assumption may introduce some uncertainty as it is used in the temperature calculations.
The column densities and masses are highly dependent on the \Hmol\ lines used in these calculations. It is preferable to use the lines \Hm{0} and \Hm{1} because they come from the lowest rotational levels of the hydrogen molecule, and therefore they trace the molecular hydrogen gas at the lowest temperatures which dominates determination of the masses \citep{Roussel07}. However the \Hm{0} line is not detected in all galaxies, and to allow meaningful comparisons we estimated the column densities and masses with the \Hm{1} and \Hm{2} lines.
The results are presented in Table \ref{tbl_molecular_hydrogen_temp}. The masses estimated from the \Hm{1} and \Hm{0} lines are in the range 0.4 to 3 $\times 10^8$M$_\sun$. These values are comparable to those found in ULIRGs \citep{Higdon2006}, local starbursts and Seyfert galaxies \citep{Rigopoulou02, Tommasin08}.
It should be noted that these masses correspond to a beam size of 13.4''$\times$13.4'' which does not include all the molecular hydrogen emission. Then the mass values listed in Table \ref{tbl_molecular_hydrogen_temp} should be considered as lower limits to the total \Hmol\ mass of the galaxy.
For instance the integrated fluxes of the \Hm{1} line (Table \ref{tbl_hires_map_integrated}) are a factor of between two and three larger than those used in the mass calculations (Table \ref{tbl_molecular_hydrogen_fluxes}). Thus assuming that the physical conditions of the \Hmol\ in the outer parts of the galaxy are similar to those in the inner $\sim$3-5 kpc, the total mass would be at least a factor of $\sim$2-3 larger than the values listed in Table \ref{tbl_molecular_hydrogen_temp}.

\subsection{Excitation Mechanism and Relation to the PAH Emission}
Studies of the molecular hydrogen emission of galactic PDRs indicate that the formation of \Hmol\ occurs in the surfaces of PAHs (see e.g., \citealt{Habart03, Velusamy08}) as both the PAH carriers and molecular hydrogen can be excited by UV radiation. Thus the \Hmol\ and the PAH emission morphologies are correlated in these regions. A similar correlation is seen in external galaxies \citep{Rigopoulou02, Roussel07}. However, \Hmol\ can be excited by other mechanisms, such as X-rays (e.g., \citealt{Lepp1983}) and shock fronts (e.g., \citealt{Hollenbach1989}).

\begin{figure}
\centering
\includegraphics[width=0.45\textwidth]{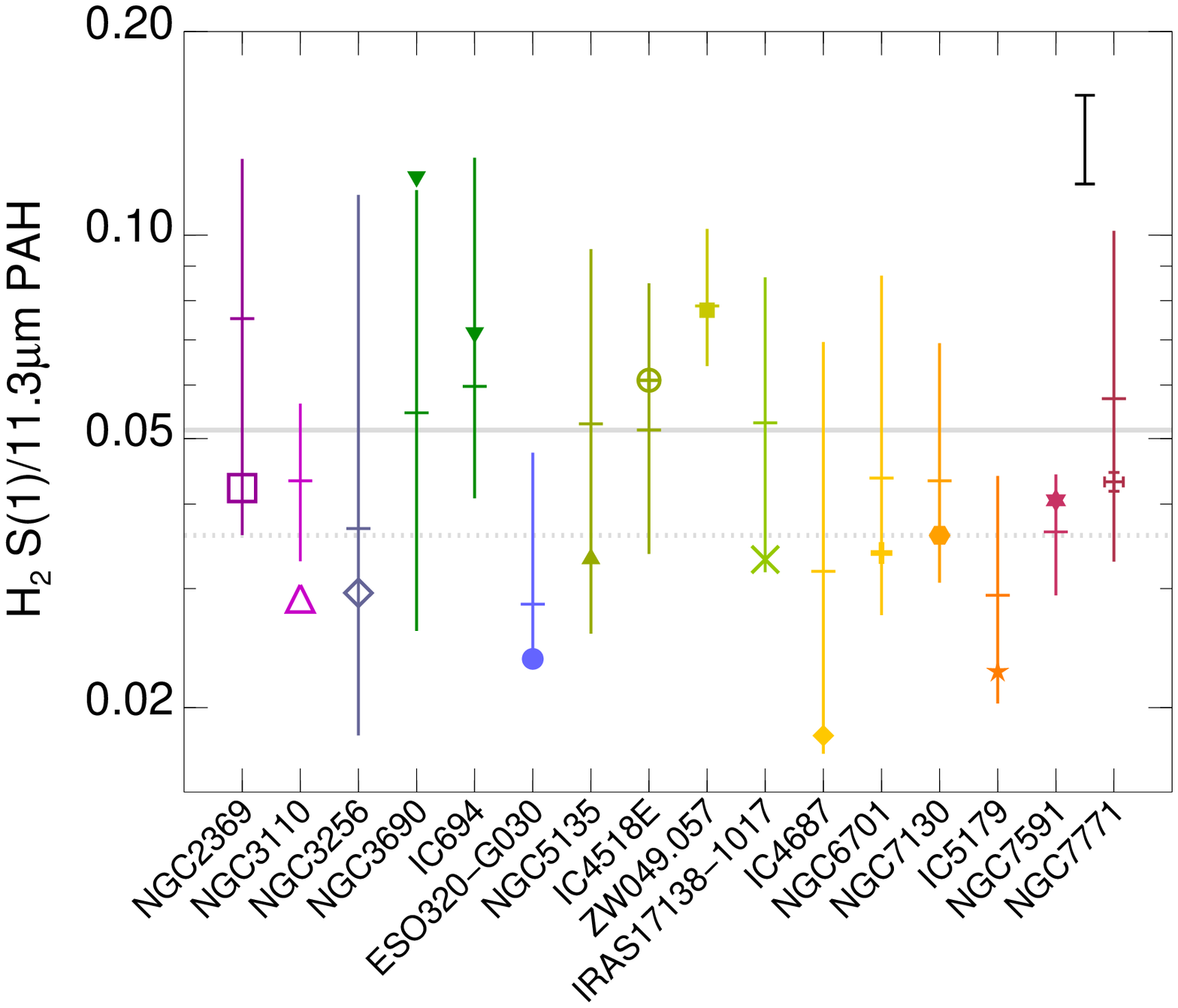} 
\caption{Range of the observed \Hm{1}\slash\PAHonce\ ratio for
  each galaxy. Symbols are as in Figure
  \ref{fig_ne3ne2_gal}. } 
\label{fig_H2S1pah11_gal}
\end{figure}

\begin{figure}
\centering
\includegraphics[width=0.45\textwidth]{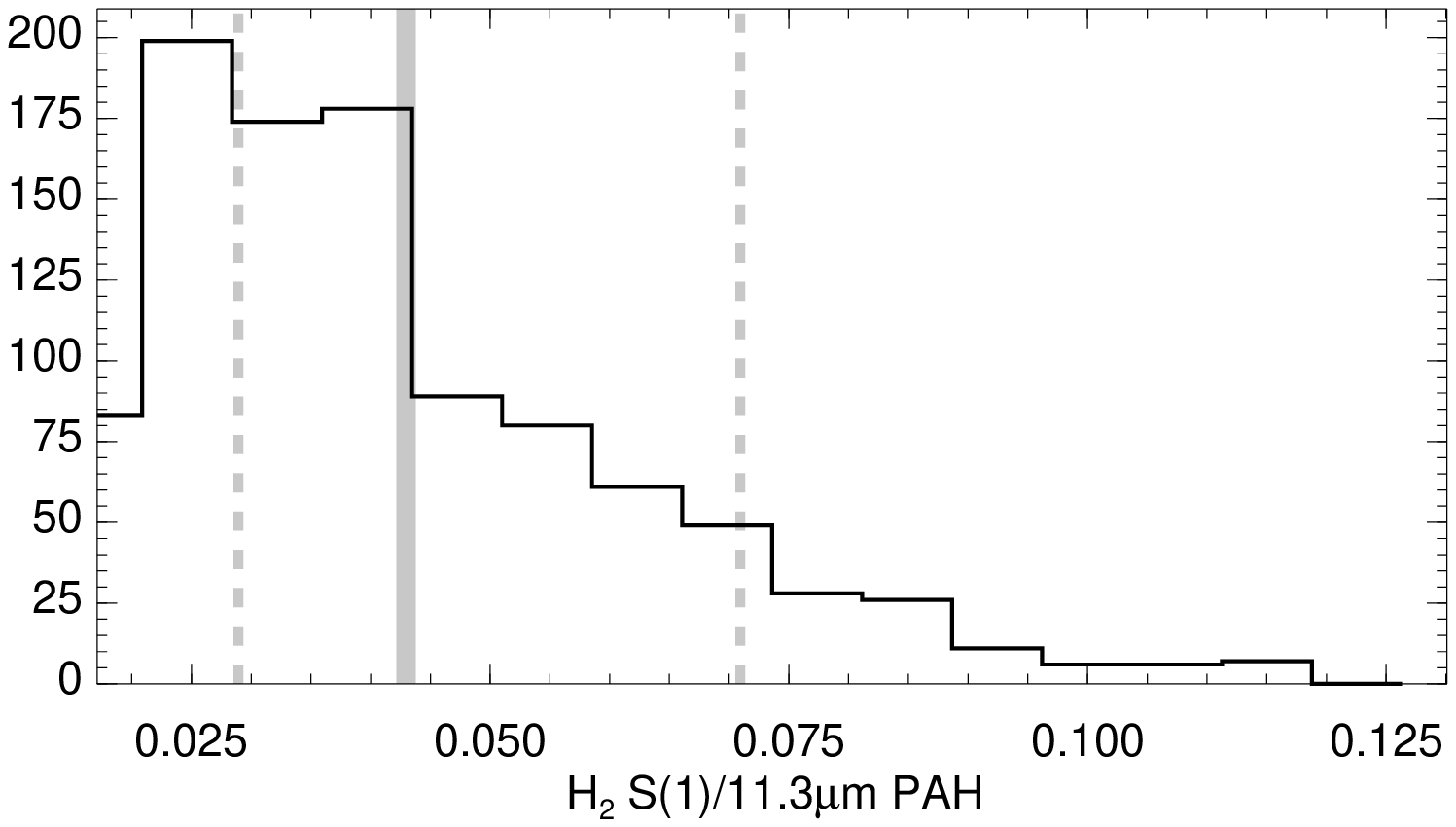} 
\caption{Distribution of the observed  \Hm{1}\slash\PAHonce\ ratio for all the galaxies  from spatially resolved measurements. The lines are as in Figure \ref{fig_ssi_histogram}}
\label{fig_H2S1pah11_histogram}
\end{figure}

\begin{figure*}
\centering
\includegraphics[width=\textwidth]{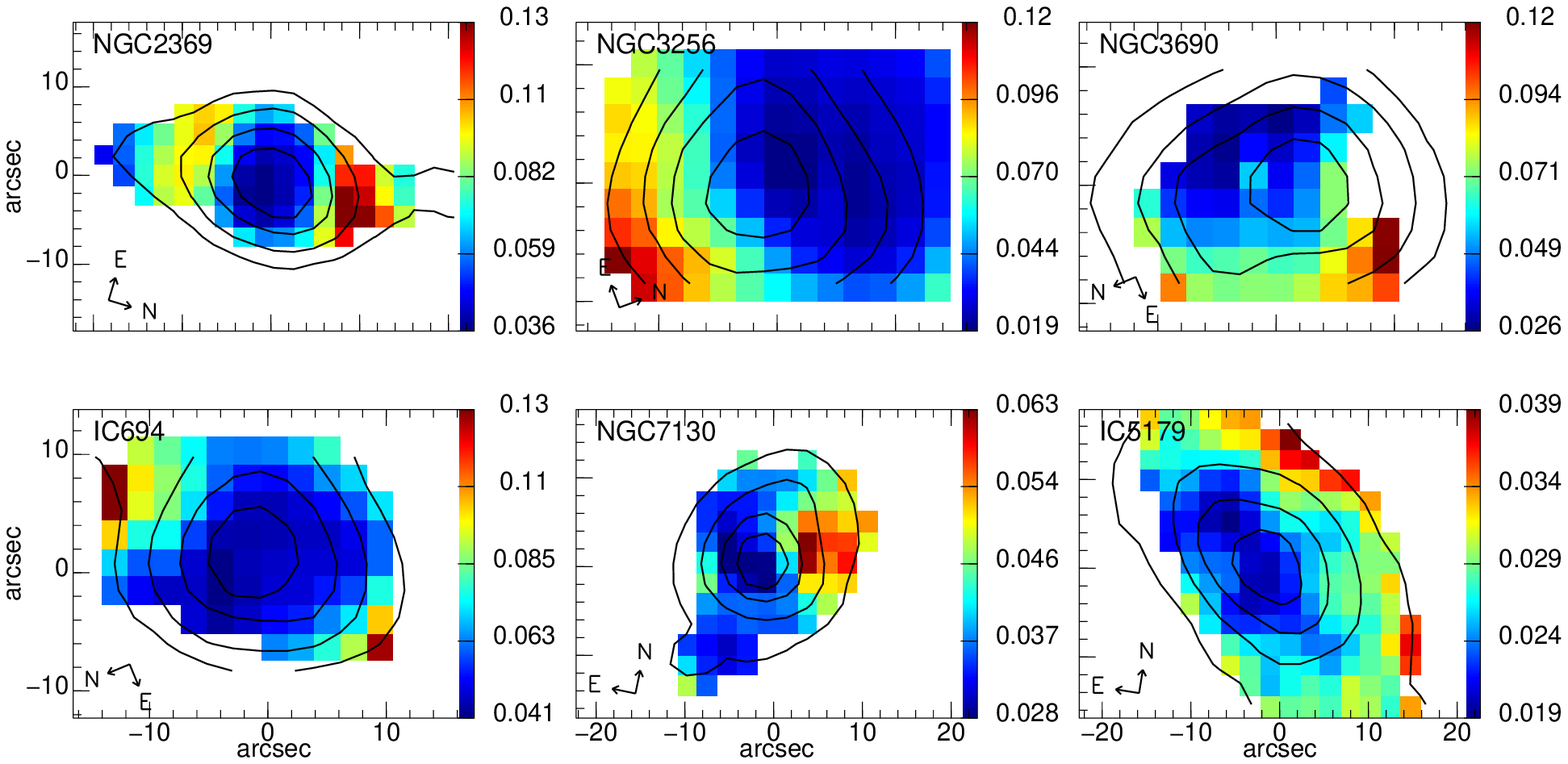} 
\caption{Maps of the observed \Hm{1}\slash\PAHonce\ ratio for selected galaxies. For reference we represent the 15\micron\ continuum contours. }
\label{fig_H2S1pah11_maps}
\end{figure*}

To explore the morphological differences we constructed maps of the \Hm{1}\slash\PAHonce\ ratio. The range of this ratio for each galaxy is shown in Figure \ref{fig_H2S1pah11_gal}. This ratio varies by a factor of 10 in our sample of LIRGs, although, as can be seen in the Figure \ref{fig_H2S1pah11_histogram}, 68\% of the values are in a  narrower range, from 0.028 to 0.073. Figure \ref{fig_H2S1pah11_maps} shows the ratio maps of those galaxies with the most noticeable morphological characteristics. 

The \PAHonce\ feature falls close to the the 9.7\micron\ silicate absorption and, although the \Hm{1}\ is also close to the silicate absorption at 18\micron, extinction variations could increase artificially the \Hm{1}\slash\PAHonce\ ratio. We used the extinction curve adopted by \citet{Smith07} to quantify this effect. If we assume a mixed model for the dust distribution, the ratio can be increased by up to a factor of $\sim$1.4, for the typical silicate absorption observed in our sample of LIRGs. Which is much lower than the observed variation. Choosing a screen geometry, to increase the ratio by a factor of 2 we would need an optical extinction of A$_V\sim$100 mag,  which is much larger than our  estimates (Section \ref{s:silicate}). Therefore, the effect of the extinction is small compared to the observed variations of the \Hm{1}\slash\PAHonce\ ratio. 

We can try to distinguish among the different excitation mechanisms by comparing the \PAHonce\ morphology with that of the \Hmol\ emission (Figure \ref{fig_map_sh}). It is clear that the morphologies are different. The \Hm{1}\ emission is more extended than the PAH emission, although the maxima of both emissions are located in the nucleus for most of the galaxies. The only exception is NGC~3256 where the \Hm{1}\ peaks at the southern nucleus while the PAH maximum is located at northern nucleus. We note, however, that the southern nucleus of this galaxy is very obscured (see Figure~\ref{fig_map_sil}, and \citealt{Tanio09}), and the \PAHonce\ map is not corrected for extinction. This suggests that the warm \Hmol\ in LIRGs is produced not only in PDRs, but that the other mechanisms also play a role. 

We attribute the increased values of the \Hm{1}\slash\PAHonce\ ratio, not to a deficiency of PAH emission, but rather to an excess emission by \Hmol, excited by X-ray or shocks.
For instance, in the merger system NGC~3256, the largest value of the ratio occurs where the
components of the merger are colliding. 
The \Hmol\ line ratios are inconsistent with X-ray excitation \citep{Shaw2005}. Thus, it is likely that there is some contribution from shocks to the \Hmol, as appears also to be the case for some regions in NGC~3690 and IC~694 \citep{AAH09Arp299}. 
In many cases, the hypothesis of shock excitation is supported by large velocity dispersions (see \citealt{GarciaMarin06} for NGC~3690/IC~694 and the following section for additional examples).

The galaxy IC~5179 is interesting because its velocity field (see next section) is highly
ordered, so strong shocks are unlikely, and its X-ray luminosity is low. Thus the \Hmol\ emission in this galaxy is likely to be produced in PDRs.
The small variations (a factor of 2) of the \Hm{1}\slash\PAHonce\ ratio are probably due to differences in the extinction or differences in the \PAHonce\ relative strength with respect to the total PAH luminosity from region to region. For the rest of the galaxies, the morphology of \Hm{1}\slash\PAHonce\
ratio does not allow us to determine whether the \Hmol\ is produced in PDRs or is excited by shock fronts or X-ray.

\section{Velocity Fields}\label{s:velocity_fields}

Figure \ref{fig_v} shows the velocity fields of the \Neii\ and \Hm{1} emission lines in the SH module. They cover $\sim$30''$\times$30'', which corresponds to physical scales of $\sim$6-11 kpc, depending on the galaxy.
Most of the \Neii\ velocity fields are consistent with a rotating disk, although some of them show more complex morphologies (NGC~3110, NGC~3256, IC~694, NGC~6701 and NGC~7130). Only two of the galaxies (NGC~3256 and IC~694) with perturbed velocity fields are part of merger systems. NGC~3110 and NGC~7130 are paired with other galaxies. However not all the galaxies in groups present disturbed velocity fields (e.g., NGC~7771, IC~4518E), at least on the physical scales ($\sim$kpc) probed by the IRS spectral maps.

\begin{figure*}
\centering
\includegraphics[width=\textwidth]{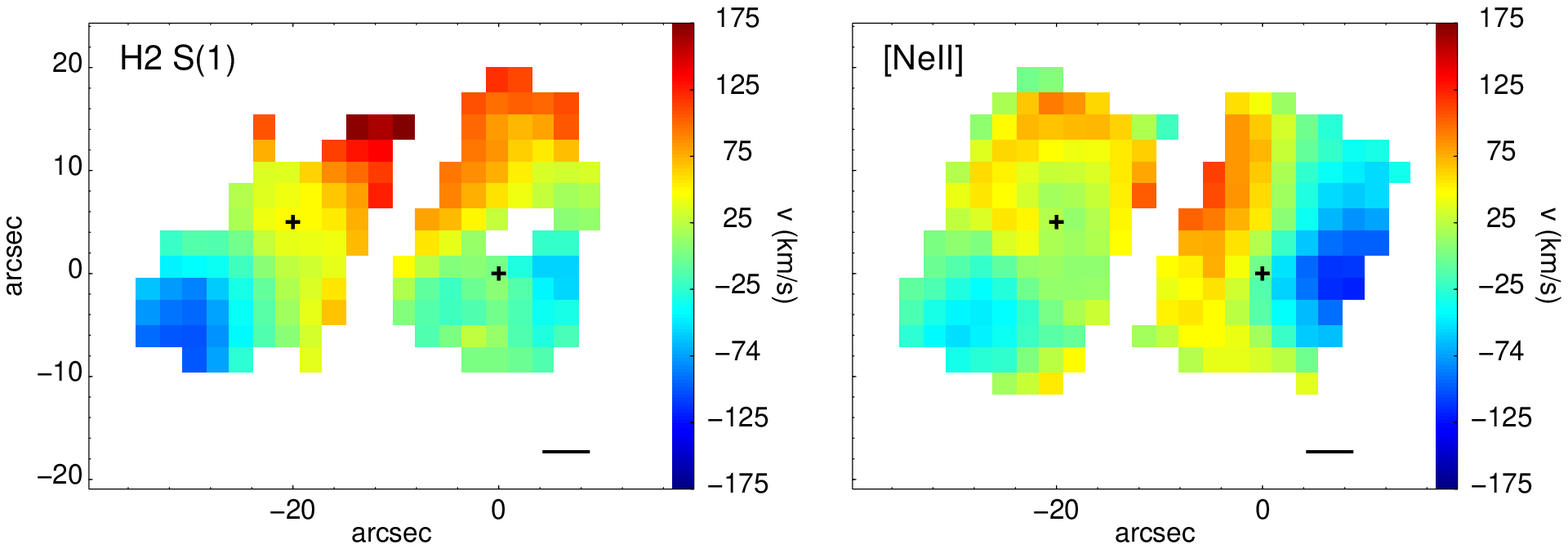} 
\caption{Arp 299 (NGC~3690 and IC~694) velocity fields of the molecular gas traced by the \Hm{1} line (left) and the ionized gas traced by the \Neii\ line (right). The velocities are referred to the the nucleus B1 of Arp~299. The black crosses mark the nuclei A and B1 of Arp~299. The scale represents 1 kpc. North is up and east to the left. }
\label{fig_arp299_velocity_field}
\end{figure*}

Figure \ref{fig_arp299_velocity_field} shows the velocity field of the interacting system Arp~299  (IC~694 and NGC~3690) rotated to the usual north up east to the left orientation for easy comparison with other works. The \Neii\ velocity field, which traces the ionized gas, agrees well with that measured from H$\alpha$ (see \citealt{GarciaMarin06}). The peak-to-peak variation of the velocity field is $\sim$250 km s$^{-1}$ also comparable to that measured from H$\alpha$.
The warm molecular gas (300 K $<$ T $<$ 1000 K) velocity field, traced by the \Hm{1} line, it is similar to the \Neii\ field  (Figure \ref{fig_arp299_velocity_field}), as both rotates in the same sense, although there are differences.
The molecular hydrogen velocity field does not present the irregular structure of the ionized gas and the neutral gas (see \citealt{GarciaMarin06}). However, the \Hm{1} velocity field resembles that produced by two disks rotating in opposite sense. This suggests that before the interaction both galaxies were spirals \citep{Augarde1985}. Moreover the \Hm{1}\ velocity field shows evident similarities with the CO velocity field (see \citealt{Aalto1997, Casoli1999}).

H$\alpha$ velocity fields are available for NGC~3256 \citep{Lipari2000}, NGC~6701 and NGC~7771  \citep{AAH09PMAS} and NGC~7591 \citep{Amram1994}.
The \Neii\ velocity field of NGC~3256 appears perturbed due the merger nature of this galaxy. It is consistent with the H$\alpha$ velocity field in shape and amplitude.
The H$\alpha$ velocity maps of NGC~7771 and NGC~6701 are slightly smaller than those presented here. NGC~7771 has and ordered velocity field compatible with a rotating disk whereas NGC~6701 presents a complex morphology both in H$\alpha$ and mid-IR emission lines. The structure and the peak-to-peak velocity amplitude ($\sim$300 km s$^{-1}$ for NGC~7771 and $\sim$150 km s$^{-1}$ for NGC~6701) of the \Neii\ velocity field are in good agreement with those of the H$\alpha$ velocity field.
As in the previous cases, the H$\alpha$ velocity field of NGC~7591 is also similar to that of the \Neii.

\section{Conclusions}\label{s:conclusions}

We presented the analysis of \Spitzer/IRS mapping observations of 15 local LIRGs with extended Pa$\alpha$ emission. We studied the spatial distribution of the mid-IR spectral features and compared the nuclear and integrated spectra. We calculated fine structure line and PAH ratios which trace the physical conditions in the star-forming regions. The main results are the following:

\begin{itemize}

\item We used the 9.7\micron\ silicate feature strength vs. \PAHseis\ equivalent width diagram to classify the activity of the LIRGs. There is a good agreement with the optical classification and this diagram. Most of the LIRGs populate the pure starburst class and only the nuclei classified as Seyfert from optical spectroscopy appear in the diagram in the AGN/SB class.
The integrated values show, in general, larger \PAHseis\ EW and shallower silicate absorption than the nuclear spectra. That is, the extended star formation partially masks the nuclear activity, resulting in a starburst-like integrated spectra.
The silicate feature strength, nuclear and integrated, of our sample of local LIRGs ($S_{\rm Si} \sim-0.4 \, {\rm to}  \, -0.9$) is small compared to that found in ULIRGs ($S_{\rm Si}$ up to -4).

\item We constructed maps of the spatial distribution of the \Neiii\slash\Neii, which traces the radiation field hardness, and the \SIIIa\slash\Neii\ that varies with the electron density. In general, the minimum of both ratios is located at the nucleus. The nuclear \Neiii\slash\Neii\ ratio is below the expected range derived by star-formation models. A possible explanation is that due the high densities in the nuclear regions this ratio is suppressed, including the possibilities that the most massive stars are either missing of buried in ultracompact \HII\ regions. Alternatively the star formation rate may have decreased rapidly in all nuclei over the last 10-20 Myr.

\item We find a positive correlation between the \Neiii\slash\Neii\ and \SIIIa\slash\Neii\ ratios for star forming regions in our sample of LIRGs and other starburst galaxies. On the other hand, AGNs, for a given \SIIIa\slash\Neii\ ratio, show a systematically larger \Neiii\slash\Neii\ ratio. The few starburst galaxies where we detect the \SIV\ emission line follow the correlation between the \SIV\slash\SIIIa\ and \Neiii\slash\Neii\ observed in \HII\ regions in other galaxies. We observe larger \SIV\slash\SIIIa\ ratios for those galaxies harboring an AGN.

\item We find that the \PAHonce\ emission is more extended than that of the 5.5\micron\ continuum. However, the ratio of the nuclear (2 kpc) \PAHseis\ emission with respect to the integrated emission is comparable to that of the 5.5\micron\ continuum in most cases.
We find no correlation between the \Neiii\slash\Neii\ ratio and the \PAHonce\ EW, thus the effect of the radiation field hardness in the PAH emission, for the \Neiii\slash\Neii\ ratio range in our sample of LIRGs, may not be important. 
We propose that the \PAHonce\slash\PAHsiete\ ratio depends on the age of the stellar population. While the \PAHsiete\ comes from young star-forming regions, the \PAHonce\ arises also from diffuse medium, thus the \PAHonce\slash\PAHsiete\ is lower in young \HII\ regions. We also found that large \Neii\slash\PAHonce\ ratios may indicate recent star formation.

\item We explored the variation of the PAH ratios across the galaxies. In most cases these variations are real, that is, they are not due to extinction). In general the integrated \PAHonce\slash\PAHseis\ ratios are larger than the nuclear values, probably indicating that the integrated emission includes more neutral conditions. The nuclear and integrated  \PAHsiete\slash\PAHseis\ ratios are almost constant, except in the nuclei of galaxies classified as AGN where this ratio is higher. Since the \PAHseis\ emission is more associated to small molecules, the increased \PAHsiete\slash\PAHseis\ ratio in AGN could be explained if the small PAH molecules are more easily destroyed in the harsh environments of active nuclei.

\item Using the \Hm{0} at 28.2\micron\ and the \Hm{1} at 17.0\micron\ lines integrated over 13''$\times$13'' (3 to 5 kpc) we estimated the mass of the warm (T$\sim$300 K) molecular hydrogen. It ranges from 0.4 to 2.6$\times$10$^8$M$_\sun$, and they are similar to those of ULIRGs, local starbursts and Seyfert galaxies. However these masses are lower limits since it is only included the \Hmol\ emission from the central few kpc of the galaxies and the total mass of warm molecular hydrogen is likely to be, at least, a factor of 2 larger.

\item{The similarity between the PAHs and molecular hydrogen morphologies suggests that the main excitation mechanism of the latter is the UV radiation too. However there are some regions with an excess of \Hmol\ emission with respect to the PAH emission, thus the other mechanisms should contribute noticeably to the \Hmol\ emission. Some of these regions are associated to interacting systems, where large scale shocks may also play a role in exciting the molecular hydrogen.}

\item Despite the modest spectral resolution of the SH module, we show that useful velocity information can be obtained from the SH spectra. For most of the galaxies and on the physical scales probed by the IRS spectra ($\sim$kpc) the velocity fields are comparable with that produced by a rotating disk.
For the galaxies with available H$\alpha$ velocity fields, the \Neii\ velocity fields are in good agreement with those of the H$\alpha$, both in shape and peak-to-peak velocities.

\end{itemize}

\section*{Acknowledgements}
The authors would like to thank Takashi Hattori, M. Ag\'undez, B. Groves and Dimitra Rigopoulou for their help and enlightening discussions, as well as to Miwa Block for producing some of the data cubes. We thank the anonymous referee for the useful comments and suggestions, which significantly improved this paper.

MP-S acknowledges support from the CSIC under grant JAE-Predoc-2007. Support for this work was provided by NASA through contract 1255094 issued by JPL/California Institute of Technology. 
MP-S, AA-H, LC, and TD-S acknowledge support from the Spanish Plan Nacional del Espacio under grant ESP2007-65475-C02-01.
AA-H also acknowledges support for this work from the Spanish Ministry of Science and Innovation through Proyecto Intramural Especial under grant number 200850I003.
TDS acknowledges support from the Consejo Superior de Investigaciones Cient\'{\i}ficas under grant I3P-BPD-2004.
PGP-G acknowledges support from the Ram\'on y Cajal Program financed by the Spanish Government and the European Union, and from the Spanish Programa Nacional de Astronom\'{\i}a y Astrof\'{\i}sica under grants AYA 2006--02358 and AYA 2006--15698--C02--02.
This research has made use of the NASA/IPAC Extragalactic Database (NED), which is operated by the Jet Propulsion Laboratory, California Institute of Technology, under contract with the National Aeronautics and Space Administration.

\begin{turnpage}
\setcounter{table}{0}
\begin{deluxetable*}{llcccccccc}
\tabletypesize{\scriptsize}
\tablewidth{0pt}
\tablecaption{Log of Spitzer/IRS Spectral Mapping Observations\label{tbl_obs_map}}
\tablehead{
 &  & \colhead{RA\tablenotemark{a}} & \colhead{Dec\tablenotemark{a}}  & \colhead{v$_{hel}$\tablenotemark{b}} &
 \colhead{Dist.\tablenotemark{c}} & \colhead{Scale} & & Spect. &  \\[0.5ex]
\colhead{Target} & \colhead{IRAS Name} & \colhead{(J2000.0)} & \colhead{(J2000.0)} & \colhead{(km s$^{-1}$)} & \colhead{(Mpc)}  & \colhead{(pc arcsec$^{-1}$)} &\colhead{L$_{IR}$\tablenotemark{d}} & class\tablenotemark{e} &
\colhead{AOR\tablenotemark{f}} }

\startdata
NGC~2369  & IRAS F07160-6215 & 07 16 37.75  & -62 20 37.5  & 3196 & 46.0 & 220 & 11.1 & - & 17659392 \\
NGC~3110  & IRAS F10015-0614 & 10 04 02.12  & -06 28 29.1  & 5014 & 72.5 & 350 & 11.2 & \HII\ & 17659648 \\
NGC~3256  & IRAS F10257-4339 & 10 27 51.28  & -43 54 13.5  & 2790 & 40.1 & 190 & 11.6 & \HII\ & 17659904 \\
NGC~3690\tablenotemark{*}  & IRAS F11257+5850 & 11 28 30.80  & +58 33 44.0  & 3121 & 44.9 & 220 & 11.8 & Sy2 & 17660160  \\
IC~694\tablenotemark{*}    &  & 11 28 33.80  & +58 33 46.5  & & & & & \HII\ & 17660416  \\
ESO320-G030 & IRAS F11506-3851 & 11 53 11.72  & -39 07 48.7  & 3038 & 43.7 & 210 & 11.2 & \HII\ & 17660672 \\
NGC5135  & IRAS F13229-2934 & 13 25 44.06  & -29 50 01.2  & 4074 & 58.8 & 280 & 11.2 & Sy2 & 17660928 \\
IC4518E  & IRAS F14544-4255 & 14 57 44.80  & -43 07 53.0  & 4584 & 64.5 & 320 & 11.2 & - & 21927936 \\
Zw049.057 & IRAS F15107+0724 & 15 13 13.12  & +07 13 31.7  & 3858 & 55.6 & 270 & 11.2 & \HII\ & 17661184 \\
IRAS17138-1017 & IRAS F17138-1017 & 17 16 35.80  & -10 20 39.4  & 5161 & 74.7 & 360 & 11.4 & \HII\ & 17661440 \\
IC4687   & IRAS F18093-5744 & 18 13 39.83  & -57 43 31.2  & 5105 & 73.8 & 360 & 11.5 & \HII\ & 17665024 \\
NGC6701  & IRAS F18425+6036 & 18 43 12.46  & +60 39 12.0  & 3896 & 56.2 & 270 & 11.0 & Composite & 21928448 \\
NGC7130  & IRAS F21453-3511 & 21 48 19.40  & -34 57 04.9  & 4837 & 69.9 & 340 & 11.4 & LINER/Sy2 & 17661696 \\
IC5179   & IRAS F22132-3705 & 22 16 09.09  & -36 50 37.5  & 3363 & 48.4 & 230 & 11.2 & \HII\ & 17661952 \\
NGC7591  & IRAS F23157+0618 & 23 18 16.28  & +06 35 08.9  & 4907 & 70.9 & 340 & 11.0 & Composite & 21928960 \\
NGC7771  & IRAS F23488+1949 & 23 51 24.88  & +20 06 42.6  & 4302 & 62.1 & 300 & 11.4 & \HII\ & 21929472
\enddata
\tablenotetext{*}{NGC~3690 and IC~694 are the two components, western and eastern respectively, of the interacting system Arp~299}
\tablenotetext{a}{Coordinates of the nucleus of the galaxy.}
\tablenotetext{b}{Heliocentric velocity from the IRS nuclear spectrum of each galaxy.}
\tablenotetext{c}{Distance calculated using the measured heliocentric velocity and our cosmology.}
\tablenotetext{d}{Logarithm of the L(8-1000 \micron), as defined in \citet{Sanders96}, in solar luminosities ($3.826\times10^{26}$ W). It is calculated using the IRAS fluxes from \citet{SandersRBGS} and from \citet{Surace2004}, the measured redshift, and our cosmology. }
\tablenotetext{e}{Classification of the nuclear activity from optical spectroscopy. Composite indicates intermediate LINER/\HII\ classification. The references for the nuclear classification are given in \citet{AAH06s}, except for NGC~3690 \citep{GarciaMarin06}, NGC~6701 and NGC~7591 \citep{AAH09PMAS}.}
\tablenotetext{f}{Astronomical Observation Request key.}
\end{deluxetable*}

\begin{deluxetable*}{lccccccccccccc}
\tabletypesize{\scriptsize}
\tablewidth{0pt}
\tablecaption{Low resolution (SL) spectra\label{tbl_lowres_map}}
\tablehead{ & \multicolumn{6}{c}{Nuclear} & & \multicolumn{6}{c}{Integrated} \\[0.1ex]
\cline{2-7} \cline{9-14}\\[-1.5ex]
	& & & \multicolumn{2}{c}{PAH EW\tablenotemark{b}} & & & & &
	& \multicolumn{2}{c}{PAH EW\tablenotemark{b}} \\[0.1ex]
	\cline{4-5} \cline{11-12}\\[-1.5ex]
\colhead{Name} 
	& \colhead{Size\tablenotemark{a}} & 
	\colhead{S$_{\rm Si}$} 
	& \colhead{6.2\micron} &  \colhead{11.3\micron} &
 		\colhead{\case{\PAHsiete}{\PAHseis}} & \colhead{\case{\PAHonce}{\PAHseis}} & &
	 \colhead{Size\tablenotemark{c}} & \colhead{S$_{\rm Si}$}  
	& \colhead{6.2\micron} &  \colhead{11.3\micron} &
	 \colhead{\case{\PAHsiete}{\PAHseis}} & \colhead{\case{\PAHonce}{\PAHseis}} \\[-1.5ex]
}
\startdata
NGC2369 & 1.2 & -0.92 & 0.43~$\pm$~0.04 & 0.33~$\pm$~0.02 & 1.8~$\pm$~0.0 & 0.38~$\pm$~0.01 & & 8.8$\times$6.2&-0.84 & 0.43~$\pm$~0.08 & 0.40~$\pm$~0.03 & 1.8~$\pm$~0.0 & 0.55~$\pm$~0.01 \\
NGC3110 & 2.0 & -0.48 & 0.58~$\pm$~0.11 & 0.48~$\pm$~0.05 & 1.7~$\pm$~0.0 & 0.46~$\pm$~0.01 & & 15.4$\times$9.8&-0.40 & 0.73~$\pm$~1.02 & 0.53~$\pm$~0.05 & 1.6~$\pm$~0.3 & 0.49~$\pm$~0.06 \\
NGC3256 & 1.1 & -0.48 & 0.53~$\pm$~0.02 & 0.33~$\pm$~0.01 & 1.7~$\pm$~0.0 & 0.46~$\pm$~0.01 & & 7.6$\times$3.8&-0.52 & 0.59~$\pm$~0.02 & 0.43~$\pm$~0.01 & 1.7~$\pm$~0.0 & 0.53~$\pm$~0.01 \\
NGC3690\tablenotemark{*} & 1.2 & -0.80 & 0.11~$\pm$~0.01 & 0.045~$\pm$~0.001 & 1.4~$\pm$~0.1 & 0.21~$\pm$~0.01 & & 8.3$\times$5.7&-0.96 & 0.29~$\pm$~0.01 & 0.21~$\pm$~0.01 & 1.7~$\pm$~0.1 & 0.42~$\pm$~0.03 \\
IC694\tablenotemark{*} & 1.2 & -1.9 & 0.34~$\pm$~0.01 & 0.23~$\pm$~0.01 & 2.1~$\pm$~0.1 & 0.28~$\pm$~0.02 & & 8.3$\times$5.7 &-0.96 & 0.29~$\pm$~0.01 & 0.21~$\pm$~0.01 & 1.7~$\pm$~0.1 & 0.42~$\pm$~0.03 \\
ESO320-G030 & 1.2 & -0.73 & 0.49~$\pm$~0.06 & 0.50~$\pm$~0.03 & 1.8~$\pm$~0.0 & 0.55~$\pm$~0.01 & & 5.9$\times$4.6 &-0.56 & 0.53~$\pm$~0.10 & 0.53~$\pm$~0.04 & 1.8~$\pm$~0.0 & 0.63~$\pm$~0.01 \\
NGC5135 & 1.6 & -0.47 & 0.30~$\pm$~0.02 & 0.31~$\pm$~0.02 & 1.7~$\pm$~0.0 & 0.58~$\pm$~0.01 & & 10$\times$5.9&-0.46 & 0.33~$\pm$~0.05 & 0.35~$\pm$~0.03 & 1.7~$\pm$~0.0 & 0.73~$\pm$~0.02 \\
IC4518E & 1.8 & -0.82 & 0.39~$\pm$~0.28 & 0.44~$\pm$~0.17 & 1.8~$\pm$~0.2 & 0.47~$\pm$~0.05 & & 6.4$\times$6.4 &-0.65 & 0.62~$\pm$~0.39 & 0.53~$\pm$~0.16 & 1.5~$\pm$~0.1 & 0.48~$\pm$~0.03 \\
ZW049.057 & 1.5 & -1.2 & 0.39~$\pm$~0.15 & 0.48~$\pm$~0.16 & 2.1~$\pm$~0.1 & 0.49~$\pm$~0.03 & & 4.9$\times$3.8 &-1.2 & 0.42~$\pm$~0.30 & 0.47~$\pm$~0.19 & 2.1~$\pm$~0.2 & 0.57~$\pm$~0.06 \\
IRAS17138-1017 & 2.0 & -0.79 & 0.56~$\pm$~0.06 & 0.40~$\pm$~0.03 & 1.7~$\pm$~0.0 & 0.38~$\pm$~0.01 & & 10$\times$4.3 &-0.77 & 0.58~$\pm$~0.10 & 0.43~$\pm$~0.04 & 1.7~$\pm$~0.0 & 0.48~$\pm$~0.01 \\
IC4687 & 2.0 & -0.40 & 0.66~$\pm$~0.08 & 0.55~$\pm$~0.03 & 1.6~$\pm$~0.0 & 0.49~$\pm$~0.01 & & 13$\times$5.8 &-0.38 & 0.64~$\pm$~0.11 & 0.54~$\pm$~0.04 & 1.7~$\pm$~0.0 & 0.57~$\pm$~0.01 \\
NGC6701 & 1.5 & -0.54 & 0.46~$\pm$~0.07 & 0.45~$\pm$~0.05 & 1.7~$\pm$~0.0 & 0.58~$\pm$~0.01 & & 13$\times$6.5 &-0.40 & 0.53~$\pm$~0.12 & 0.52~$\pm$~0.05 & 1.6~$\pm$~0.0 & 0.70~$\pm$~0.02 \\
NGC7130 & 1.9 & -0.44 & 0.25~$\pm$~0.04 & 0.23~$\pm$~0.04 & 1.8~$\pm$~0.1 & 0.66~$\pm$~0.02 & & 12$\times$8.2 &-0.38 & 0.47~$\pm$~0.10 & 0.37~$\pm$~0.05 & 1.5~$\pm$~0.0 & 0.65~$\pm$~0.02 \\
IC5179 & 1.3 & -0.50 & 0.57~$\pm$~0.08 & 0.46~$\pm$~0.03 & 1.7~$\pm$~0.0 & 0.48~$\pm$~0.01 & & 10$\times$5.5 &-0.32 & 0.62~$\pm$~0.08 & 0.51~$\pm$~0.03 & 1.6~$\pm$~0.0 & 0.54~$\pm$~0.01 \\
NGC7591 & 1.9 & -0.85 & 0.39~$\pm$~0.12 & 0.35~$\pm$~0.06 & 1.9~$\pm$~0.1 & 0.52~$\pm$~0.02 & & 16$\times$8.8 &-0.47 & 0.63~$\pm$~0.31 & 0.43~$\pm$~0.08 & 1.4~$\pm$~0.1 & 0.53~$\pm$~0.03 \\
NGC7771 & 1.7 & -0.66 & 0.46~$\pm$~0.06 & 0.38~$\pm$~0.04 & 1.7~$\pm$~0.0 & 0.43~$\pm$~0.01 & & 11$\times$7.8 &-0.70 & 0.49~$\pm$~0.07 & 0.41~$\pm$~0.03 & 1.7~$\pm$~0.0 & 0.51~$\pm$~0.01
\enddata
\tablenotetext{*}{NGC~3690 and IC~694 are part of the interacting system Arp~299.}
\tablenotetext{a}{Projected square size of the aperture in kpc. The aperture size is 5.5''$\times$5.5'' for all the galaxies.}
\tablenotetext{b}{Equivalent width units are \micron.}
\tablenotetext{c}{Approximated projected square size of the aperture in kpc. See Section \ref{ss:1d_spectra} for details.}
\end{deluxetable*}

\end{turnpage}

\begin{deluxetable*}{lccccccccccccc}
\tabletypesize{\scriptsize}
\tablewidth{0pt}
\tablecaption{High resolution (SH) spectra. Integrated spectra\label{tbl_hires_map_integrated}}
\tablehead{
 	& & \colhead{[\ion{S}{4}]} & \colhead{[\ion{Ne}{2}]} 
	& \colhead{[\ion{Ne}{3}]}
	& \colhead{[\ion{S}{3}]} & \colhead{\Hm{2}} & \colhead{\Hm{1}} & 
	\\[0.5ex]
	\colhead{Name} & \colhead{Size\tablenotemark{a}} 
	& 10.51$\mu$m & 12.81$\mu$m & 15.55$\mu$m & 18.71$\mu$m &
	12.28$\mu$m & 17.04$\mu$m & \colhead{\PAHonce}
}
\startdata
NGC2369 & 9.7$\times$6.2& $<$5.5 & 194.~$\pm$~11. & 15.~$\pm$~3. & 44.~$\pm$~4. & 19.~$\pm$~6. & 51.~$\pm$~2. & 798.~$\pm$~34. \\
NGC3110 &15$\times$8.4& $<$6.9 & 164.~$\pm$~14. & 30.~$\pm$~3. & 59.~$\pm$~5. & $<$5.1 & 33.~$\pm$~3. & 877.~$\pm$~29. \\
NGC3256 &6.0$\times$3.8& 15.~$\pm$~3. & 1229.~$\pm$~48. & 189.~$\pm$~5. & 395.~$\pm$~8. & 65.~$\pm$~6. & 124.~$\pm$~3. & 2998.~$\pm$~30. \\
ESO320-G030 &6.3$\times$4.2& $<$4.8 & 152.~$\pm$~7. & 12.~$\pm$~2. & 47.~$\pm$~5. & 13.~$\pm$~4. & 21.~$\pm$~2. & 677.~$\pm$~29. \\
NGC5135 &9.0$\times$5.0& 19.~$\pm$~2. & 141.~$\pm$~12. & 66.~$\pm$~3. & 39.~$\pm$~3. & 14.~$\pm$~4. & 32.~$\pm$~2. & 593.~$\pm$~24. \\
IC4518E &8.0$\times$6.4& $<$1.6 & 11.~$\pm$~1. & 0.90~$\pm$~0.32 & 3.7~$\pm$~1.0 & $<$1.0 & 2.1~$\pm$~0.5 & 34.~$\pm$~3. \\
ZW049.057 &4.8$\times$4.3& $<$4.3 & 23.~$\pm$~3. & $<$2.2 & $<$1.4 & $<$6.1 & 7.9~$\pm$~1.0 & 89.~$\pm$~24. \\
IRAS17138-1017 &11.5$\times$5.8& $<$3.9 & 159.~$\pm$~9. & 25.~$\pm$~1. & 54.~$\pm$~3. & 8.3~$\pm$~3.7 & 24.~$\pm$~2. & 505.~$\pm$~22. \\
IC4687 &11$\times$7.2& 10.~$\pm$~2. & 198.~$\pm$~10. & 57.~$\pm$~2. & 103.~$\pm$~3. & 8.9~$\pm$~5.5 & 22.~$\pm$~1. & 750.~$\pm$~23. \\
NGC6701 &8.1$\times$5.4& $<$14. & 117.~$\pm$~9. & 16.~$\pm$~3. & 32.~$\pm$~7. & $<$8.8 & 26.~$\pm$~3. & 600.~$\pm$~14. \\
NGC7130 &16$\times$10& $<$9.6 & 158.~$\pm$~14. & 36.~$\pm$~4. & 39.~$\pm$~7. & $<$7.0 & 29.~$\pm$~5. & 643.~$\pm$~29. \\
IC5179 &10$\times$6.9& $<$9.2 & 290.~$\pm$~17. & 28.~$\pm$~3. & 91.~$\pm$~5. & 19.~$\pm$~5. & 35.~$\pm$~6. & 1334.~$\pm$~29. \\
NGC7591 &8.8$\times$6.8& $<$5.8 & 54.~$\pm$~5. & 4.4~$\pm$~2.5 & 10.~$\pm$~3. & $<$3.6 & 10.~$\pm$~4. & 266.~$\pm$~13. \\
NGC7771 &14$\times$6& $<$12. & 148.~$\pm$~11. & $<$9.2 & 41.~$\pm$~9. & 20.~$\pm$~10. & 41.~$\pm$~4. & 675.~$\pm$~23.
\enddata
\tablecomments{Fluxes are expressed in units of $10^{-17}Wm^{-2}$. The fluxes of Arp~299 (NGC~3690/IC~694) are given in \citet{AAH09Arp299}.}
\tablenotetext{a}{Approximated projected square size of the aperture in kpc. See Section \ref{ss:1d_spectra} for details.}
\end{deluxetable*}

\begin{deluxetable*}{lccccccccccccc}
\tabletypesize{\scriptsize}
\tablewidth{0pt}
\tablecaption{High resolution (SH) spectra. Nuclear spectra\label{tbl_hires_map_nuclear}}
\tablehead{
 	& & \colhead{[\ion{S}{4}]} & \colhead{[\ion{Ne}{2}]} 
	& \colhead{[\ion{Ne}{3}]}
	& \colhead{[\ion{S}{3}]} & \colhead{\Hm{2}} & \colhead{\Hm{1}} & 
	\\[0.5ex]
	\colhead{Name} & \colhead{Size\tablenotemark{a}} 
	& 10.51$\mu$m & 12.81$\mu$m & 15.55$\mu$m & 18.71$\mu$m &
	12.28$\mu$m & 17.04$\mu$m & \colhead{\PAHonce}
}
\startdata
NGC2369 & 1.2 & $<$0.44 & 52.~$\pm$~3. & 3.4~$\pm$~0.2 & 8.6~$\pm$~0.2 & 3.1~$\pm$~0.2 & 5.5~$\pm$~0.2 & 138.~$\pm$~3. \\
NGC3110 & 1.9 & $<$0.27 & 26.~$\pm$~1. & 2.4~$\pm$~0.1 & 6.2~$\pm$~0.2 & 1.3~$\pm$~0.3 & 2.8~$\pm$~0.1 & 99.~$\pm$~2. \\
NGC3256 & 1.1 & 1.8~$\pm$~0.3 & 260.~$\pm$~14. & 27.~$\pm$~1. & 60.~$\pm$~1. & 8.0~$\pm$~0.8 & 14.~$\pm$~1. & 489.~$\pm$~5. \\
ESO320-G030 & 1.2 & $<$0.39 & 41.~$\pm$~2. & 3.3~$\pm$~0.1 & 11.~$\pm$~0.1 & 2.8~$\pm$~0.3 & 4.0~$\pm$~0.1 & 172.~$\pm$~3. \\
NGC5135 & 1.6 & 8.0~$\pm$~0.2 & 51.~$\pm$~3. & 23.~$\pm$~1. & 12.~$\pm$~0.1 & 3.2~$\pm$~0.3 & 4.9~$\pm$~0.2 & 166.~$\pm$~3. \\
IC4518E & 1.8 & $<$0.082 & 1.1~$\pm$~0.1 & 0.13~$\pm$~0.02 & 0.24~$\pm$~0.08 & 0.13~$\pm$~0.05 & 0.34~$\pm$~0.05 & 5.4~$\pm$~0.5 \\
ZW049.057 & 1.5 & $<$0.26 & 10.~$\pm$~1. & 0.63~$\pm$~0.07 & 1.8~$\pm$~0.2 & 1.4~$\pm$~0.3 & 2.2~$\pm$~0.1 & 32.~$\pm$~3. \\
IRAS17138-1017 & 2.0 & 1.0~$\pm$~0.2 & 64.~$\pm$~3. & 8.5~$\pm$~0.2 & 19.~$\pm$~1. & 2.1~$\pm$~0.6 & 4.0~$\pm$~0.4 & 135.~$\pm$~3. \\
IC4687 & 2.0 & 2.3~$\pm$~0.2 & 55.~$\pm$~2. & 12.~$\pm$~0.1 & 23.~$\pm$~0.1 & 2.0~$\pm$~0.3 & 3.2~$\pm$~0.2 & 189.~$\pm$~3. \\
NGC6701 & 1.5 & $<$0.49 & 27.~$\pm$~2. & 2.3~$\pm$~0.1 & 5.1~$\pm$~0.2 & 1.9~$\pm$~0.2 & 3.4~$\pm$~0.2 & 103.~$\pm$~2. \\
NGC7130 & 1.9 & 3.6~$\pm$~0.4 & 38.~$\pm$~3. & 12.~$\pm$~1. & 7.4~$\pm$~0.6 & 1.7~$\pm$~0.8 & 2.9~$\pm$~0.3 & 93.~$\pm$~3. \\
IC5179 & 1.3 & $<$0.39 & 36.~$\pm$~2. & 2.7~$\pm$~0.1 & 8.5~$\pm$~0.3 & 1.4~$\pm$~0.1 & 2.6~$\pm$~0.1 & 120.~$\pm$~3. \\
NGC7591 & 1.9 & $<$0.24 & 16.~$\pm$~1. & 1.3~$\pm$~0.1 & 2.4~$\pm$~0.2 & 1.1~$\pm$~0.2 & 1.8~$\pm$~0.2 & 48.~$\pm$~2. \\
NGC7771 & 1.7 & $<$0.38 & 37.~$\pm$~2. & 3.0~$\pm$~0.1 & 6.7~$\pm$~0.4 & 3.0~$\pm$~0.5 & 4.8~$\pm$~0.2 & 115.~$\pm$~4. \enddata
\tablecomments{Fluxes are expressed in units of $10^{-17}Wm^{-2}$. The fluxes of Arp~299 (NGC~3690/IC~694) are given in \citet{AAH09Arp299}.}
\tablenotetext{a}{Projected square size of the aperture in kpc. The aperture size is 5.5''$\times$5.5'' for all the galaxies. }
\end{deluxetable*}

\begin{deluxetable*}{lcccccccccccccc}
\tablewidth{0pt}
\tabletypesize{\small}
\tablecaption{Nuclear flux\slash\ Integrated flux\label{tbl_extended_ratio}}
\tablehead{ & \multicolumn{3}{c}{Central 2.0 kpc\slash\ Integrated flux} & & & \multicolumn{4}{c}{Central 2.0 kpc\slash\ Integrated flux} \\[0.1ex]
\cline{2-5} \cline{7-11}\\[-1.5ex]
	& \colhead{5.5\micron} & \colhead{\PAHseis} & \colhead{\PAHonce} & &  
	& \colhead{15\micron} & \colhead{\Neii} &  \colhead{\Neiii} & \colhead{\Hm{1}} \\[-0.4ex]
\colhead{Name}   & \colhead{continuum} & & & & & \colhead{continuum} \\[-3.0ex]
}
\startdata
NGC2369 & 0.41 $\pm$ 0.02 & 0.41  $\pm$ 0.01 & 0.32  $\pm$ 0.01 & & & 0.39  $\pm$ 0.02 & 0.36  $\pm$ 0.02 & 0.36  $\pm$ 0.02 & 0.20  $\pm$ 0.02 & \\
NGC3110 & 0.13 $\pm$ 0.05 & 0.12 $\pm$ 0.02 & 0.10 $\pm$ 0.01 & & & 0.13 $\pm$ 0.01 & 0.07 $\pm$ 0.03 & 0.12  $\pm$ 0.02 & 0.08 $\pm$ 0.03  & \\
NGC3256 & 0.38  $\pm$ 0.01 & 0.36  $\pm$ 0.01 & 0.34  $\pm$ 0.01 & & & 0.39  $\pm$ 0.01 & 0.31  $\pm$ 0.01 & 0.41  $\pm$ 0.01 & 0.20  $\pm$ 0.03& \\
ESO320-G030 & 0.57  $\pm$ 0.02 & 0.54  $\pm$ 0.01 & 0.49  $\pm$ 0.01 & & & 0.48  $\pm$ 0.01 & 0.49  $\pm$ 0.02 & 0.40  $\pm$ 0.01 & 0.38 $\pm$ 0.02  & \\
NGC5135  & 0.45  $\pm$ 0.01 & 0.42  $\pm$ 0.01 & 0.35  $\pm$ 0.01 & & & 0.39 $\pm$ 0.01 & 0.36  $\pm$ 0.01 & 0.28  $\pm$ 0.01 & 0.19 $\pm$ 0.03  & \\
IC4518E  & 0.21  $\pm$ 0.10 & 0.18  $\pm$ 0.10 & 0.15  $\pm$ 0.03 & & & 0.14  $\pm$ 0.02 & - & 0.15  $\pm$ 0.01 & 0.19  $\pm$ 0.06 & \\
ZW049.057  & 0.53 $\pm$ 0.08 & - & 0.50  $\pm$ 0.02 & & & 0.52  $\pm$ 0.02 & - & 0.47 $\pm$ 0.02 & -  & \\
IRAS17138-1017  & 0.31 $\pm$ 0.03 & 0.31 $\pm$ 0.02 & 0.25  $\pm$ 0.01 & & & 0.31  $\pm$ 0.01 & 0.27  $\pm$ 0.01 & 0.19  $\pm$ 0.01 & 0.15  $\pm$ 0.05 & \\
IC4687  & 0.24  $\pm$ 0.03 & 0.25  $\pm$ 0.01 & 0.23  $\pm$ 0.01 & & & 0.19  $\pm$ 0.01 & 0.17  $\pm$ 0.01 & 0.15  $\pm$ 0.01 & 0.13  $\pm$ 0.05 & \\
NGC6701  & 0.25  $\pm$ 0.03 & 0.23  $\pm$ 0.02 & 0.20  $\pm$ 0.02 & & & 0.22  $\pm$ 0.01 & 0.16  $\pm$ 0.02 & 0.24  $\pm$ 0.01 & 0.15  $\pm$ 0.03 & \\
NGC7130  & 0.28  $\pm$ 0.02 & 0.13  $\pm$ 0.03 & 0.14  $\pm$ 0.02 & & & 0.14  $\pm$ 0.01 & 0.21  $\pm$ 0.03 & 0.10  $\pm$ 0.01 & 0.09  $\pm$ 0.05 & \\
IC5179  & 0.23  $\pm$ 0.03 & 0.20  $\pm$ 0.01 & 0.18  $\pm$ 0.01 & & & 0.19  $\pm$ 0.01 & 0.16  $\pm$ 0.02 & 0.11  $\pm$ 0.01 & 0.13  $\pm$ 0.03 & \\
NGC7591  & 0.22  $\pm$ 0.05 & 0.14  $\pm$ 0.05 & 0.13  $\pm$ 0.02 & & & 0.22 $\pm$ 0.01 & 0.23  $\pm$ 0.03 & 0.23  $\pm$ 0.01 & 0.17  $\pm$ 0.04 & \\
NGC7771  & 0.24  $\pm$ 0.02 & 0.28  $\pm$ 0.02 & 0.23  $\pm$ 0.01 & & & 0.24  $\pm$ 0.01 & 0.20  $\pm$ 0.03 & 0.23  $\pm$ 0.01 & 0.12  $\pm$ 0.04&
\enddata
\end{deluxetable*}

\begin{deluxetable*}{lcccccccccccccc}
\tablewidth{0pt}
\tablecaption{Molecular hydrogen emission line fluxes\label{tbl_molecular_hydrogen_fluxes}}
\tablehead{  &  & \colhead{\Hm{2}} & \colhead{\Hm{1}} & \colhead{\Hm{0}} \\
      \colhead{Name} & \colhead{Size\tablenotemark{a}} & 12.27\micron & 17.03\micron & 28.22\micron
}
\startdata
NGC2369 & 2.9 & 10.~$\pm$~1. & 19.~$\pm$~2. & 3.5~$\pm$~0.5 \\
NGC3110 & 4.6 & 4.~$\pm$~1. & 10.~$\pm$~0.9 & 1.6~$\pm$~0.2 \\
NGC3256 & 2.7 & 34.~$\pm$~3. & 61.~$\pm$~5. & 9.~$\pm$~3. \\
NGC3690 & 2.9 & 11.~$\pm$~4. & 22.~$\pm$~1. & 3.8~$\pm$~0.6 \\
IC694 & 2.9 & 20.~$\pm$~3. & 29.~$\pm$~1. & 5.7~$\pm$~0.9 \\
ESO320-G030 & 2.9 & 8.1~$\pm$~1.2 & 13.~$\pm$~1. & 1.7~$\pm$~0.7 \\
NGC5135 & 3.9 & 9.3~$\pm$~1.4 & 17.~$\pm$~1. & 1.3~$\pm$~0.5 \\
IC4518E & 4.4 & 0.3~$\pm$~0.1 & 1.1~$\pm$~0.2 & 1.0~$\pm$~0.1 \\
ZW049.057 & 3.7 & 3.3~$\pm$~0.8 & 5.8~$\pm$~0.4 & $<$~0.6 \\
IRAS17138-1017 & 4.9 & 6.3~$\pm$~1.3 & 14.~$\pm$~1. & 1.9~$\pm$~0.5 \\
IC4687 & 4.9 & 6.5~$\pm$~1.1 & 12.~$\pm$~1. & 1.8~$\pm$~0.4 \\
NGC6701 & 3.7 & 4.6~$\pm$~0.8 & 10.~$\pm$~1. & 1.9~$\pm$~0.9 \\
NGC7130 & 4.6 & 4.2~$\pm$~1.7 & 9.9~$\pm$~0.7 & 1.4~$\pm$~0.3 \\
IC5179 & 3.2 & 6.3~$\pm$~0.8 & 11.2~$\pm$~0.8 & 1.4~$\pm$~0.2 \\
NGC7591 & 4.6 & 3.0~$\pm$~1.0 & 5.5~$\pm$~0.6 & 1.0~$\pm$~0.3 \\
NGC7771 & 4.1 & 8.7~$\pm$~1.2 & 16.~$\pm$~1. & 3.5~$\pm$~1.8
\enddata
\tablecomments{Fluxes are expressed in units of $10^{-17}Wm^{-2}$. }
\tablenotetext{a}{Projected square size of the aperture in kpc. The aperture size is 13.4''$\times$13.4'' for all the galaxies.}
\end{deluxetable*}

\begin{deluxetable*}{lcccccccccccccc}
\tablewidth{0pt}
\tabletypesize{\small}
\tablecaption{Derived warm molecular hydrogen temperatures, column densities and masses.\label{tbl_molecular_hydrogen_temp}}
\tablehead{  & \colhead{T(S(0)-S(1))} & \colhead{N$_0$} & \colhead{M$_0$} & \colhead{T(S(1)-S(2))} & \colhead{N$_1$} & \colhead{M$_1$} \\
      \colhead{Name} & \colhead{(K)} & \colhead{($\times 10^{20}$~cm$^{-2}$)} & \colhead{($\times 10^8$~M$_\sun$)} & \colhead{(K)} & \colhead{($\times 10^{20}$~cm$^{-2}$)} & \colhead{($\times 10^8$~M$_\sun$)}
}
\startdata

NGC2369 & 168 & 8.8 & 1.3 & 348 & 5.1 & 0.73 \\
NGC3110 & 178 & 3.7 & 1.3 & 305 & 2.4 & 0.86 \\
NGC3256 & 181 & 19. & 2.1 & 369 & 12. & 1.3 \\
NGC3690 & 172 & 8.9 & 1.2 & 351 & 5.3 & 0.72 \\
IC694 & 163 & 15. & 2.0 & 415 & 7.8 & 1.1 \\
ESO320-G030 & 189 & 3.4 & 0.43 & 390 & 2.2 & 0.28 \\
NGC5135 & 236 & 1.9 & 0.44 & 361 & 1.5 & 0.36 \\
IC4518E & 108 & 9.2 & 2.7 & 252 & 3.0 & 0.87 \\
ZW049.057 & $>$205 & $<$1.0 & $<$0.22 & 373 & 0.74 & 0.16 \\
IRAS17138-1017 & 185 & 4.1 & 1.5 & 324 & 2.7 & 1.0 \\
IC4687 & 178 & 4.1 & 1.5 & 361 & 2.5 & 0.92 \\
NGC6701 & 168 & 4.7 & 1.0 & 323 & 2.8 & 0.61 \\
NGC7130 & 183 & 3.0 & 0.97 & 320 & 2.0 & 0.65 \\
IC5179 & 194 & 2.6 & 0.40 & 369 & 1.7 & 0.27 \\
NGC7591 & 167 & 2.5 & 0.85 & 362 & 1.4 & 0.48 \\
NGC7771 & 157 & 10.0 & 2.6 & 363 & 5.2 & 1.3 
\enddata
\tablecomments{N$_0$ and M$_0$ are the column density and mass derived from the \Hmol\ lines S(0) and S(1). N$_1$ and M$_1$ correspond to the column density and mass derived from the S(1) and S(2) lines.}
\end{deluxetable*}

\setcounter{figure}{0}
\begin{figure*}[!p]
\includegraphics[width=\textwidth]{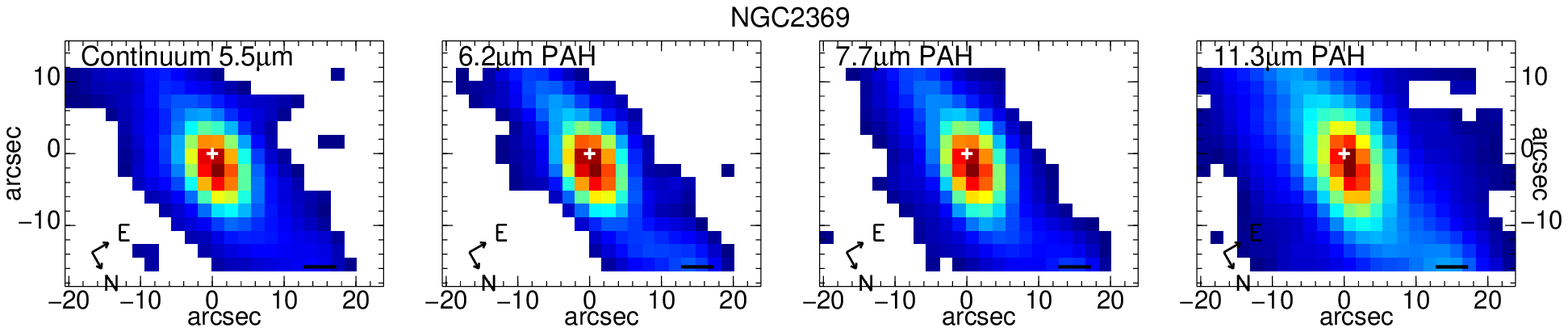}
\includegraphics[width=\textwidth]{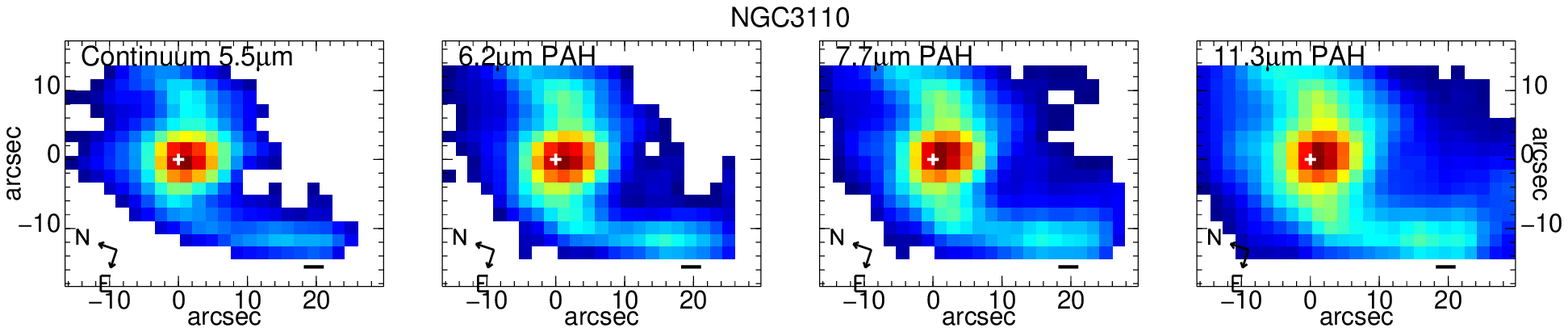}
\includegraphics[width=\textwidth]{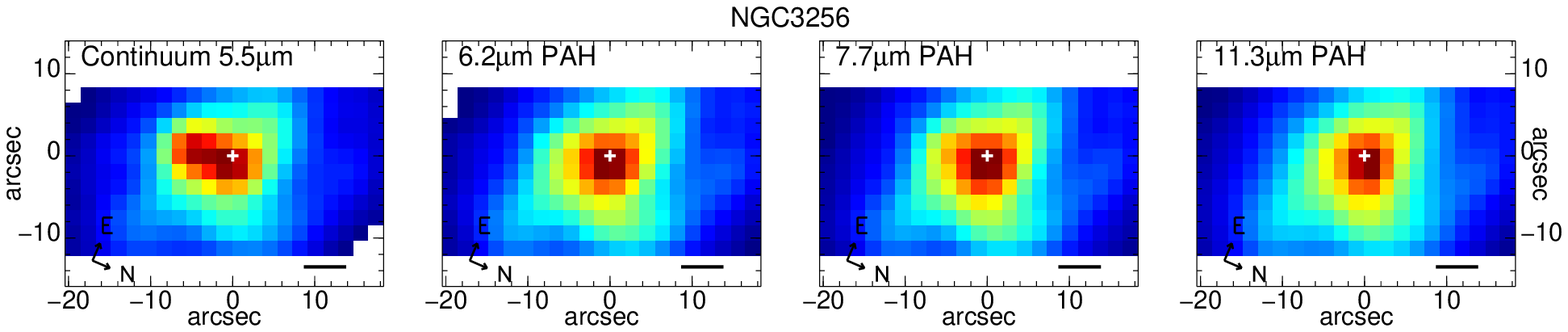}
\includegraphics[width=\textwidth]{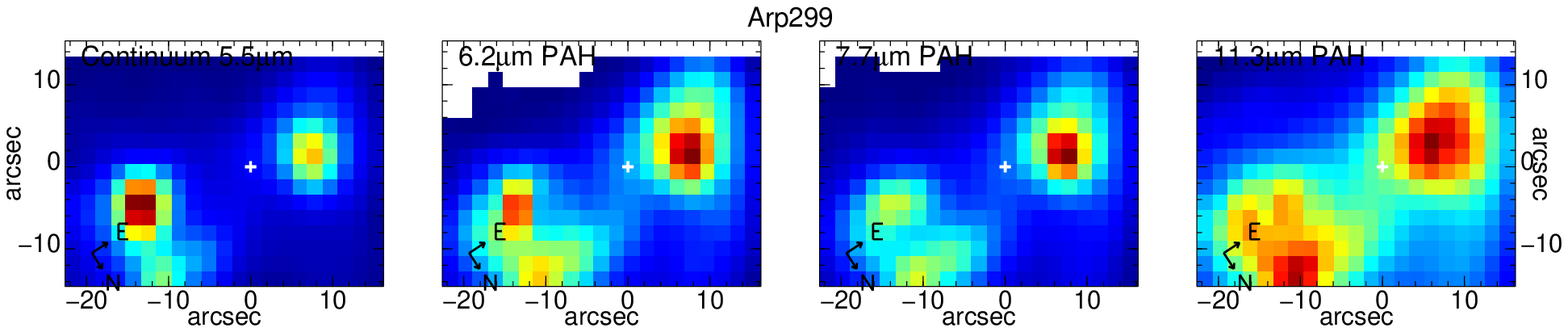}
\includegraphics[width=\textwidth]{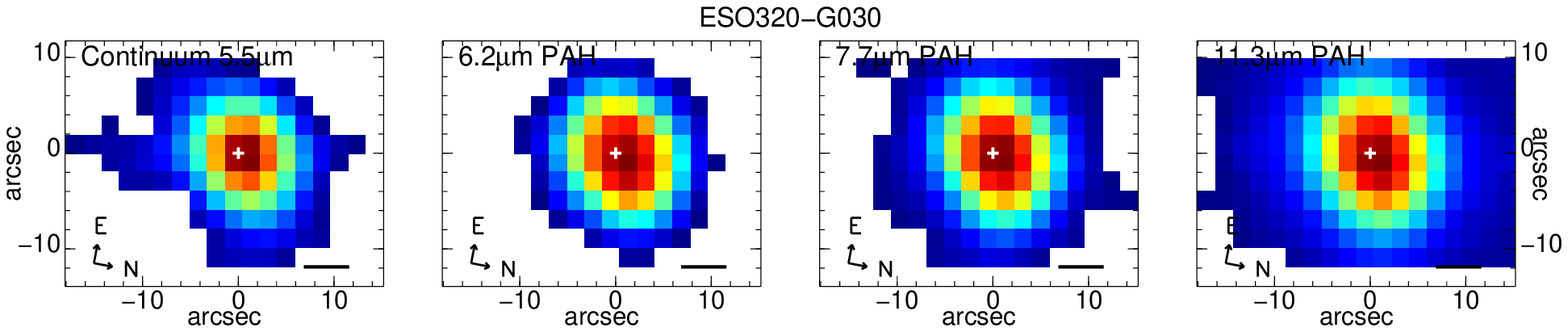}
\includegraphics[width=\textwidth]{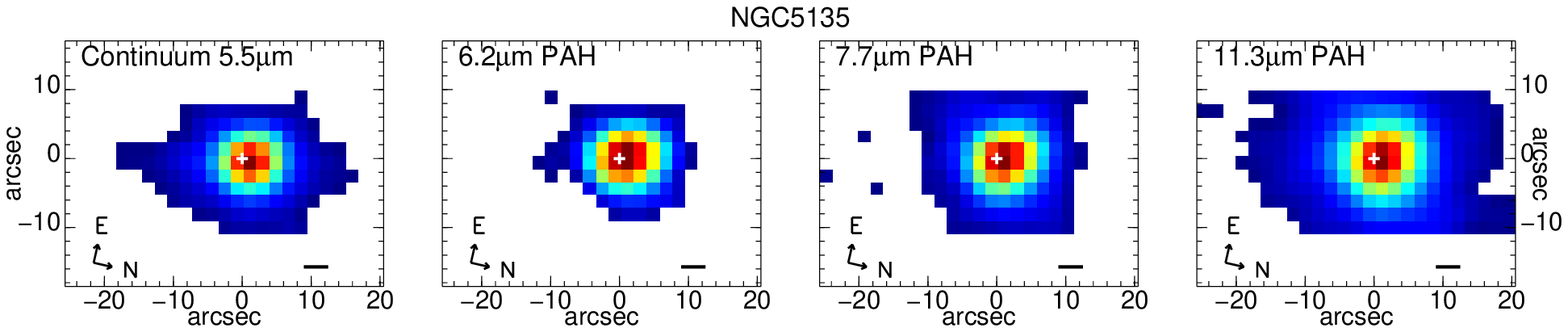}
\caption{\Spitzer/IRS SL spectral maps of the 5.5\micron\ continuum, the \PAHseis, the \PAHsiete\ and the \PAHonce. The white cross marks the coordinates of the nucleus as listed in Table \ref{tbl_obs_map}. The image orientation is indicated on the maps for each galaxy. The scale represents 1 kpc. The maps are shown in a square root scale.}
\label{fig_map_sl}
\end{figure*}

\begin{figure*}
\addtocounter{figure}{-1}
\includegraphics[width=\textwidth]{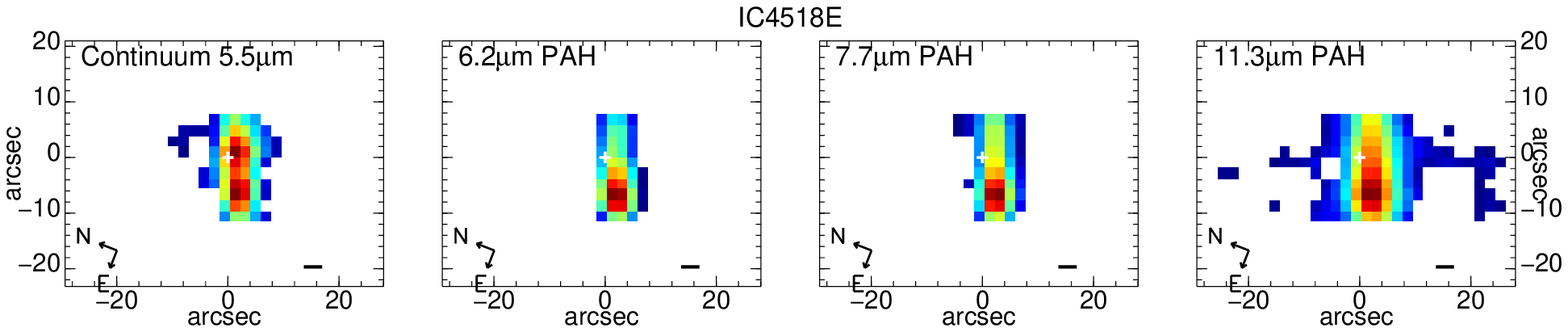}
\includegraphics[width=\textwidth]{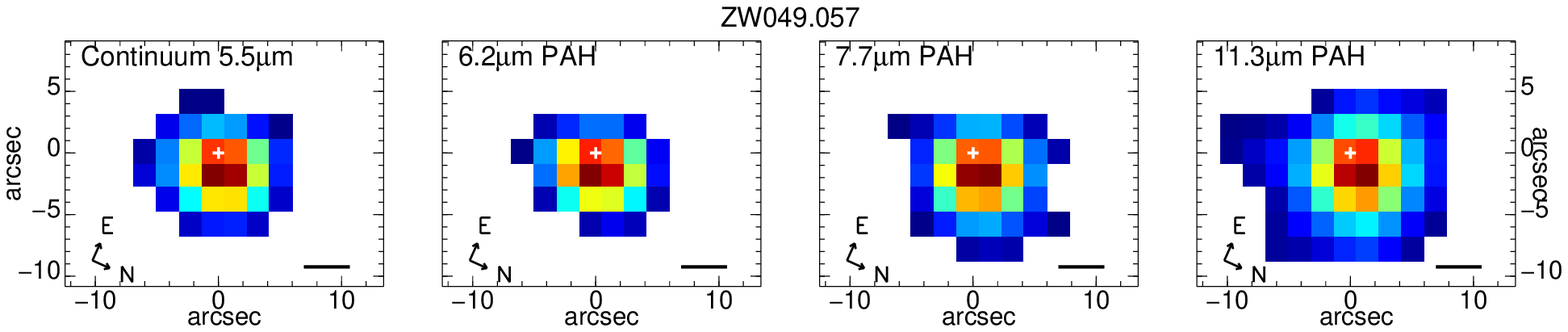}
\includegraphics[width=\textwidth]{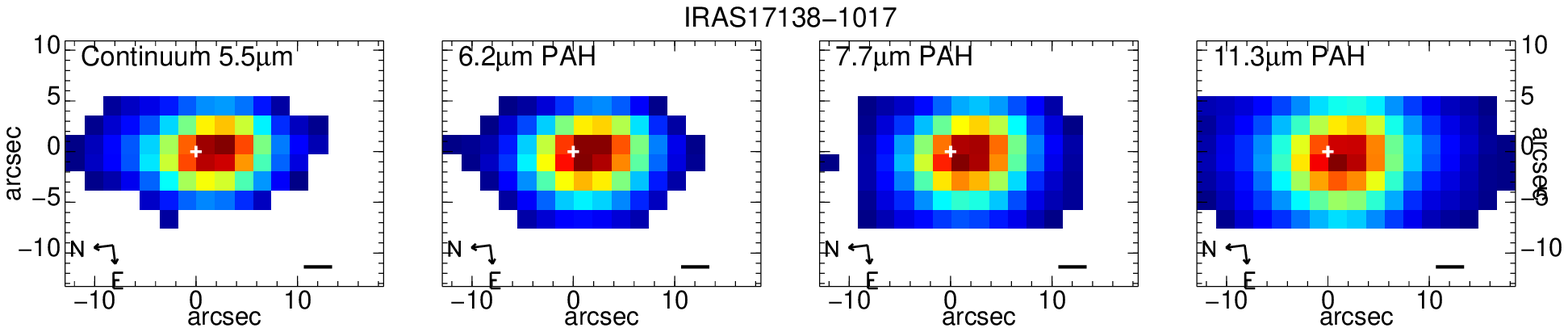}
\includegraphics[width=\textwidth]{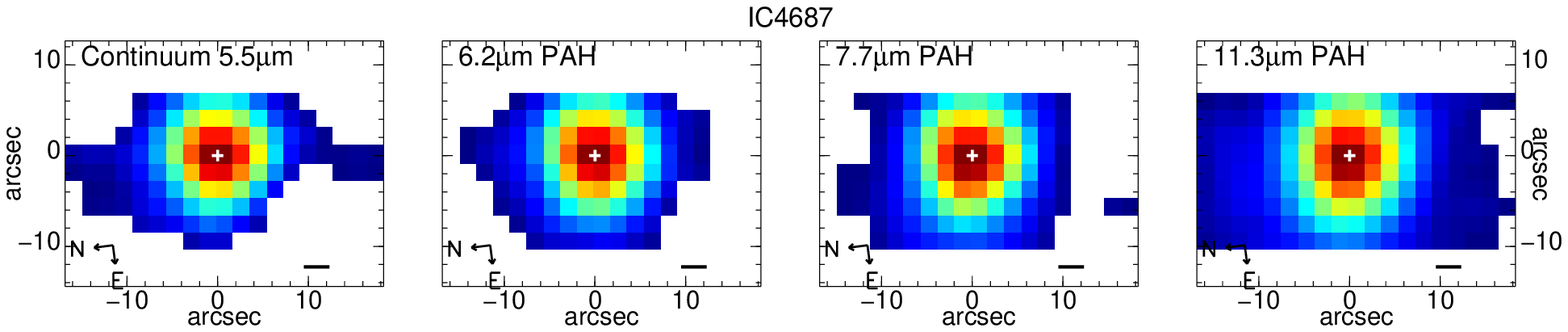}
\includegraphics[width=\textwidth]{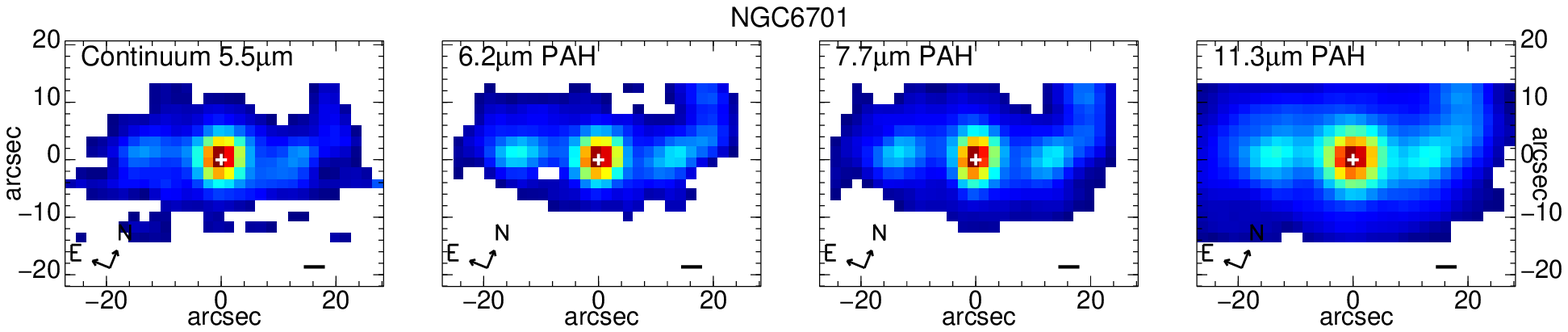}
\includegraphics[width=\textwidth]{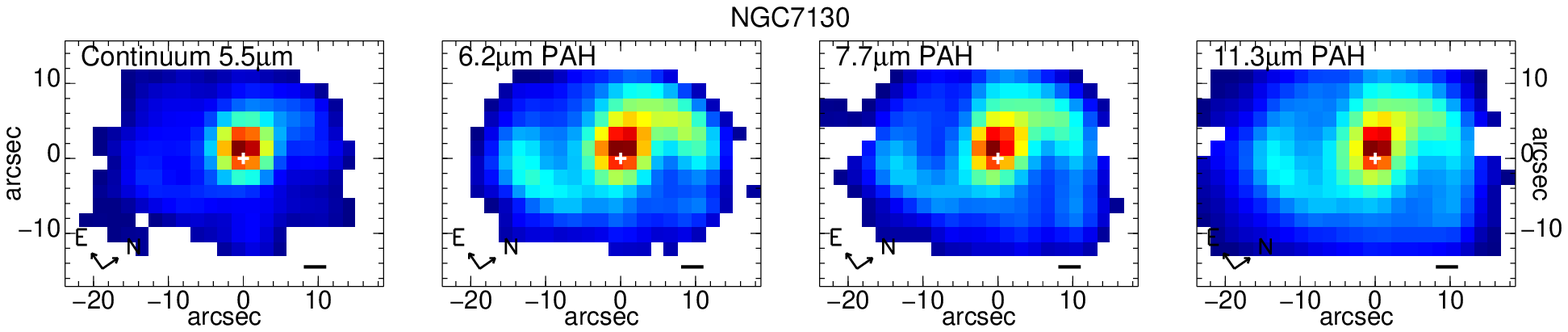}
\caption{Continued.}
\end{figure*}

\begin{figure*}
\addtocounter{figure}{-1}
\includegraphics[width=\textwidth]{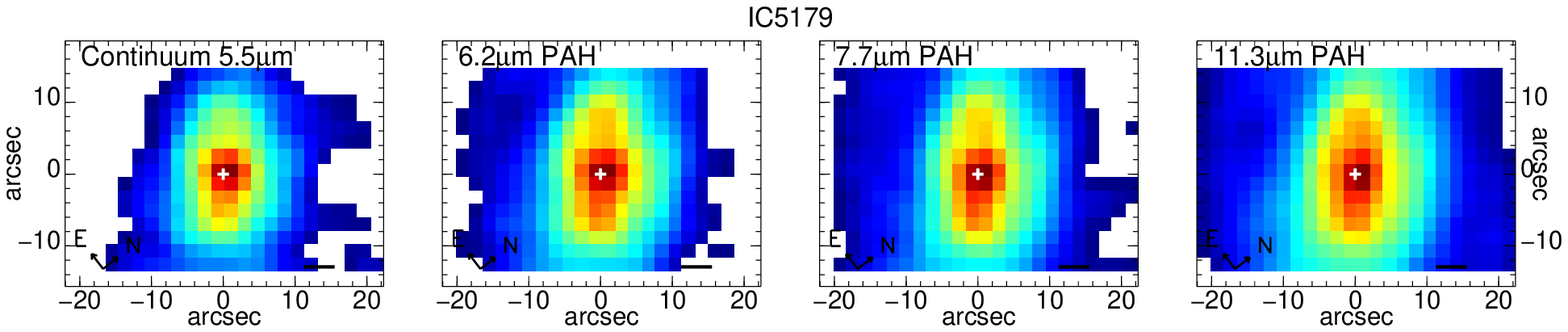}
\includegraphics[width=\textwidth]{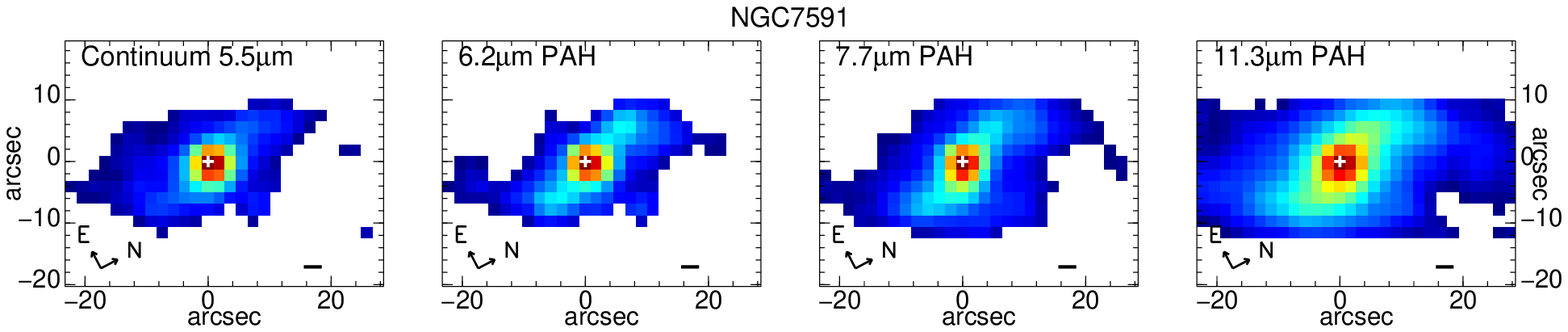}
\includegraphics[width=\textwidth]{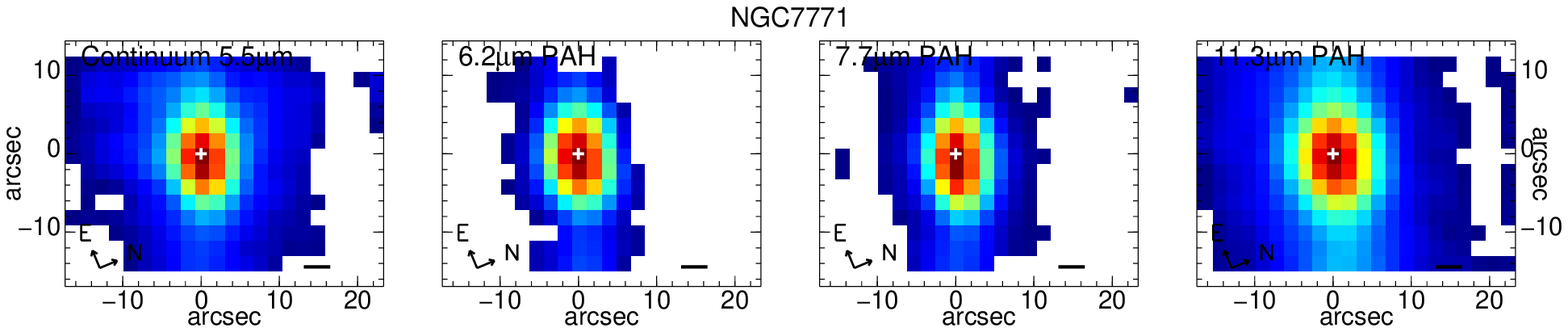}
\caption{Continued.}
\end{figure*}
\afterpage{\clearpage}

\begin{figure*}[!p]
\includegraphics[width=0.33\textwidth]{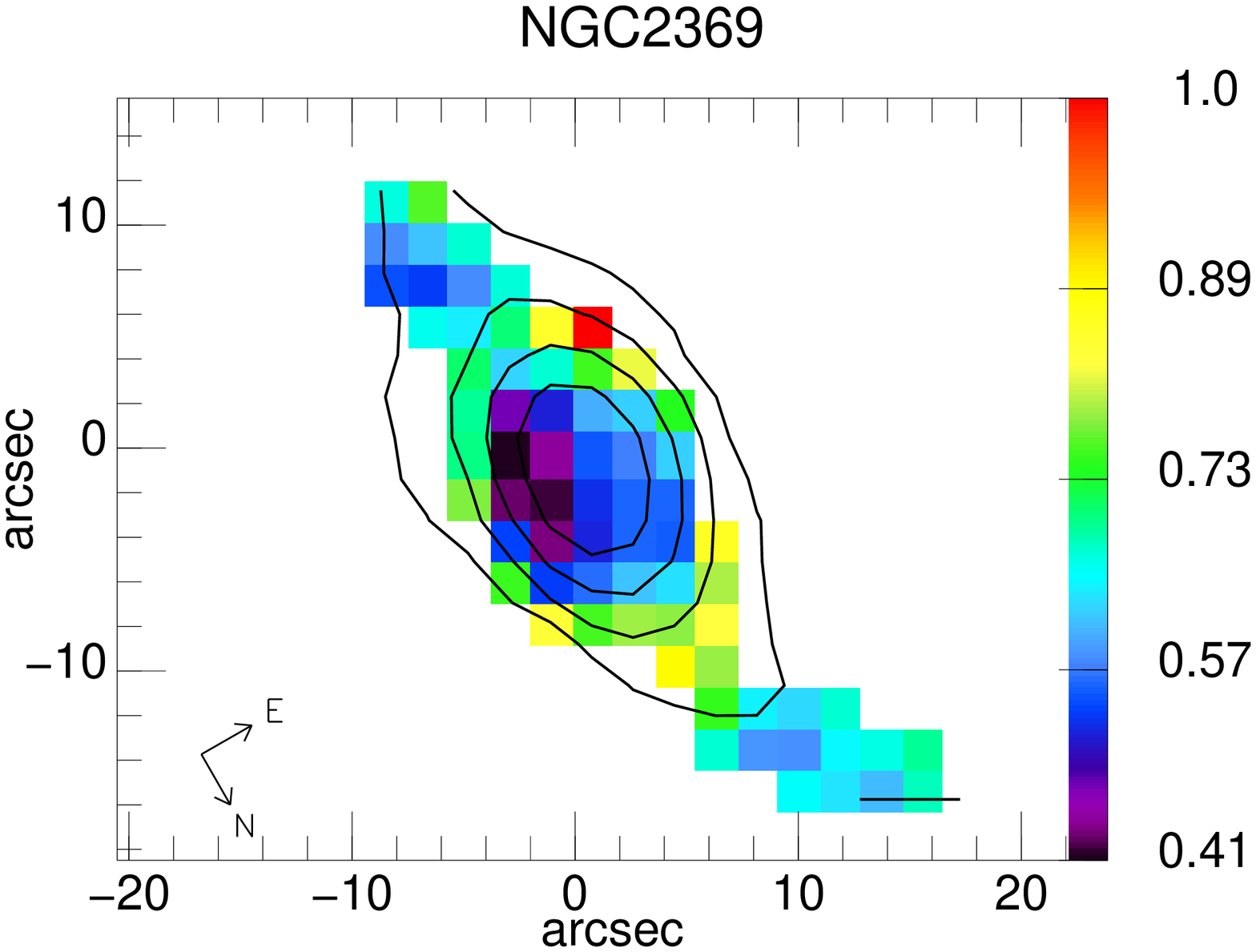}
\includegraphics[width=0.33\textwidth]{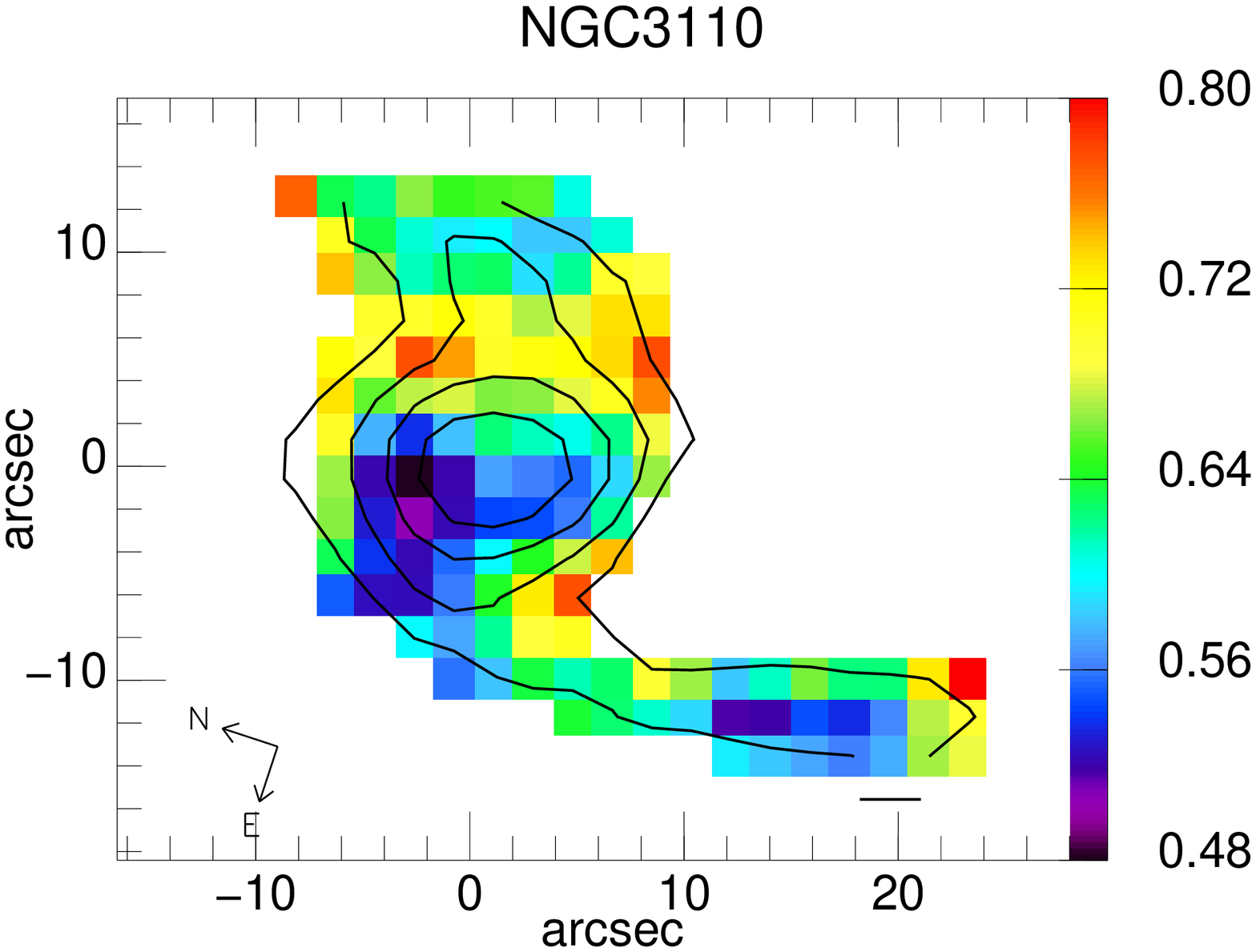}
\includegraphics[width=0.33\textwidth]{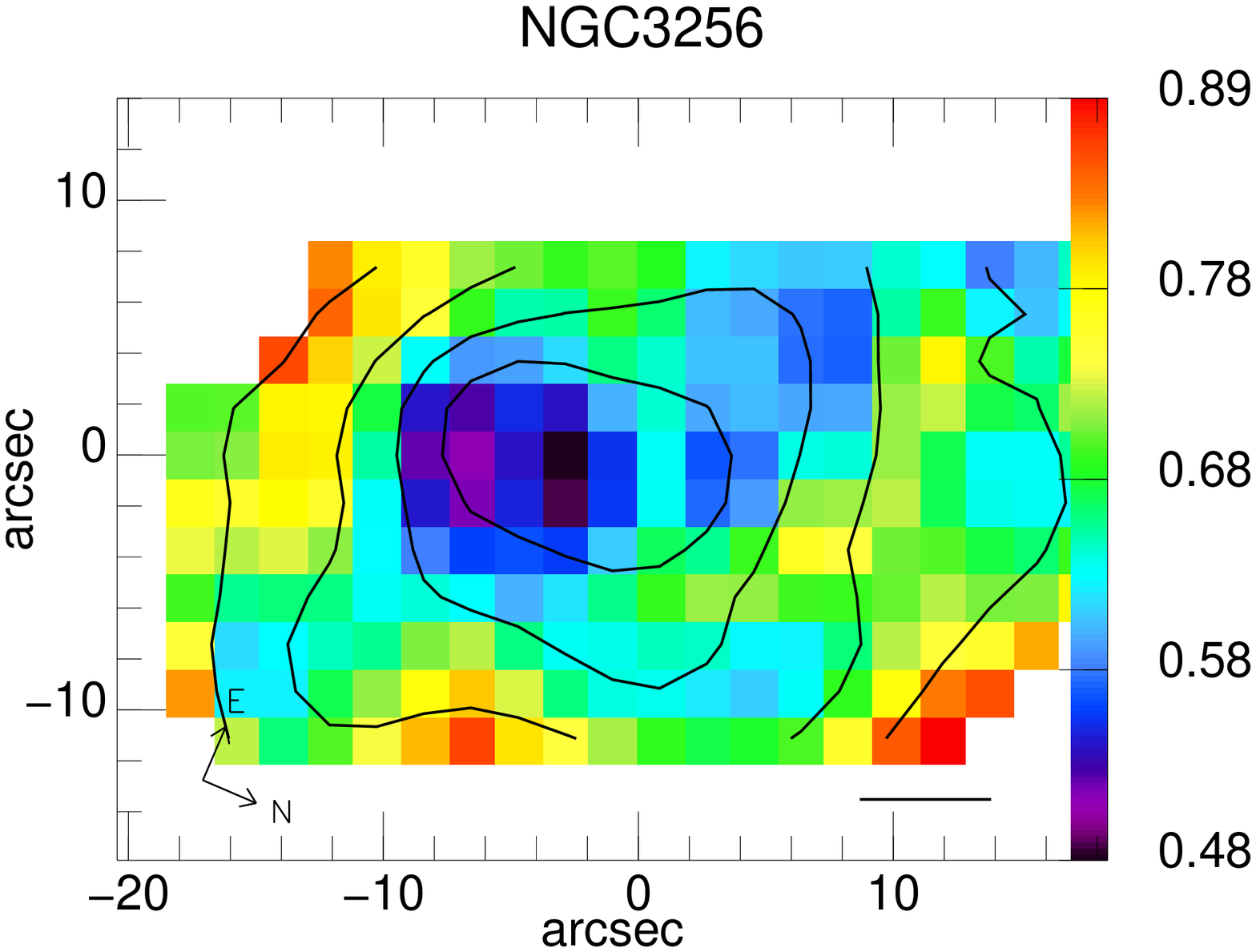}
\includegraphics[width=0.33\textwidth]{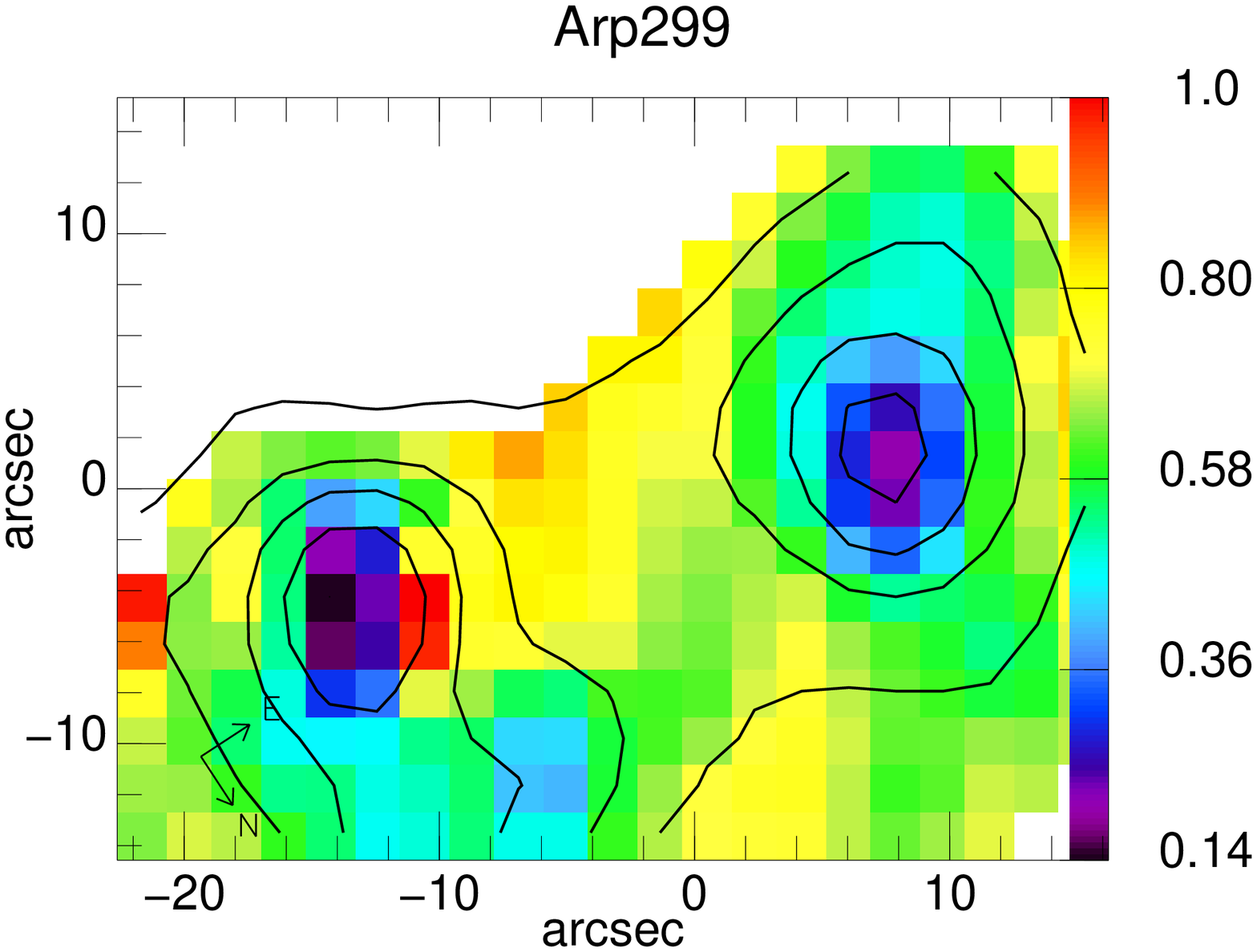}
\includegraphics[width=0.33\textwidth]{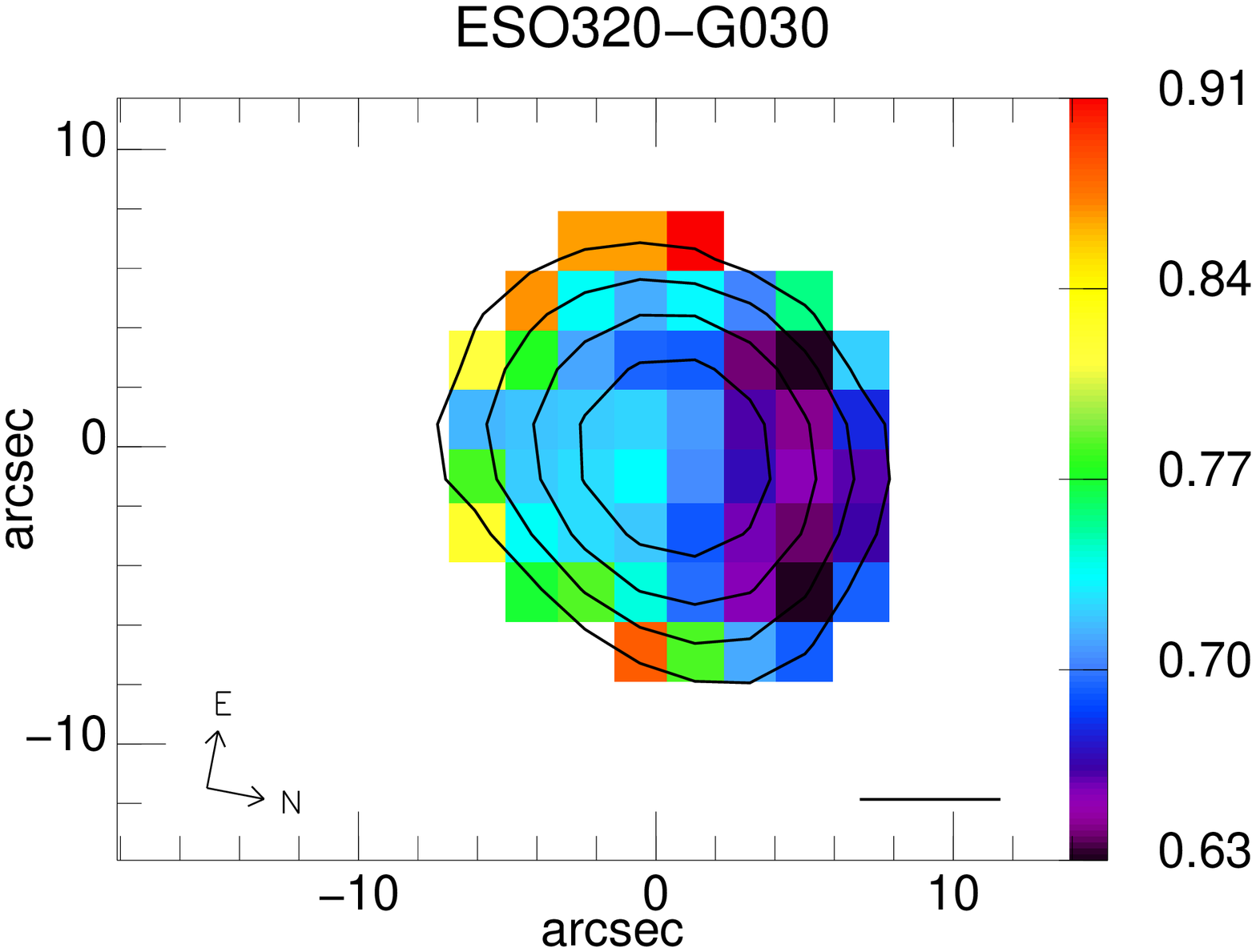}
\includegraphics[width=0.33\textwidth]{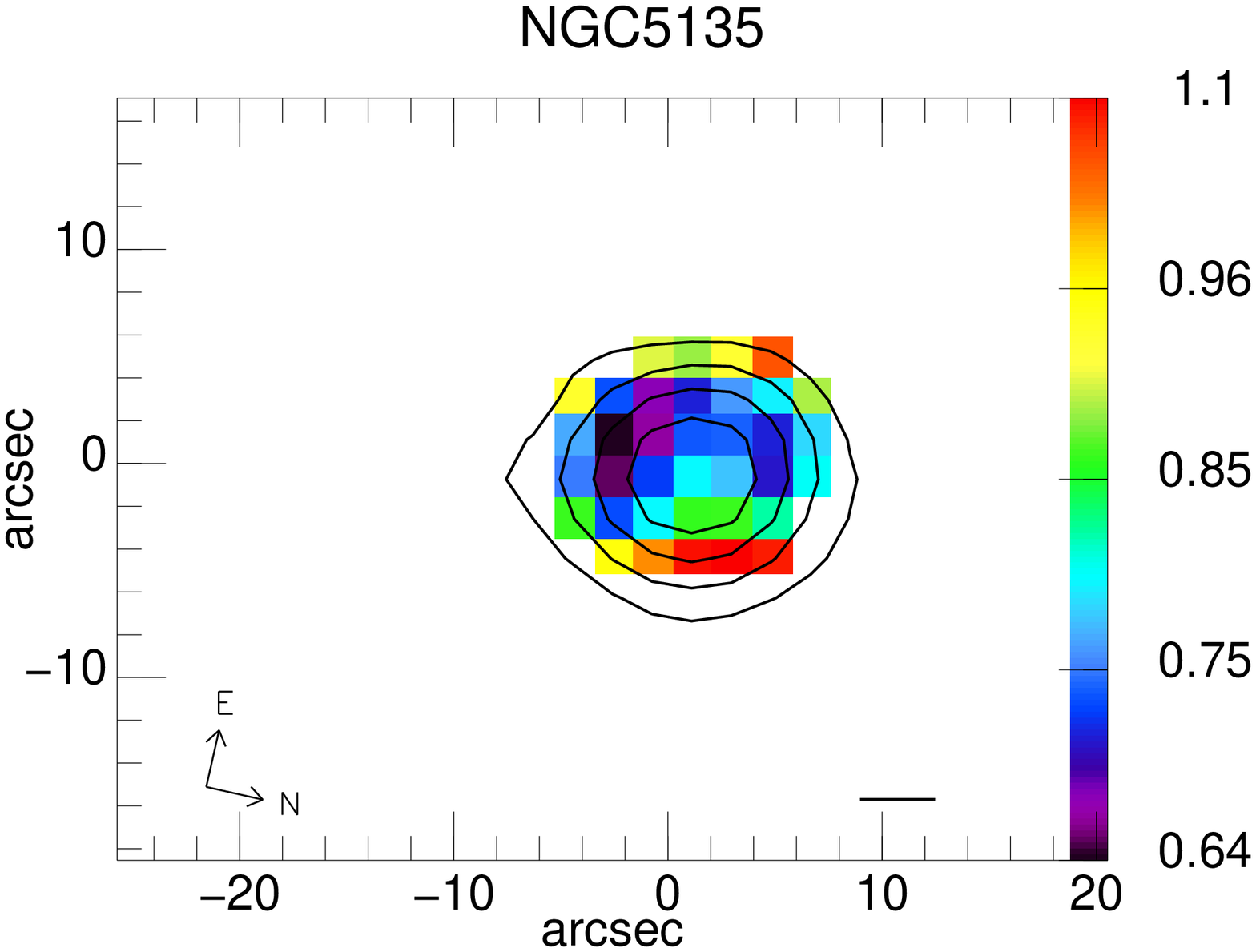}
\includegraphics[width=0.33\textwidth]{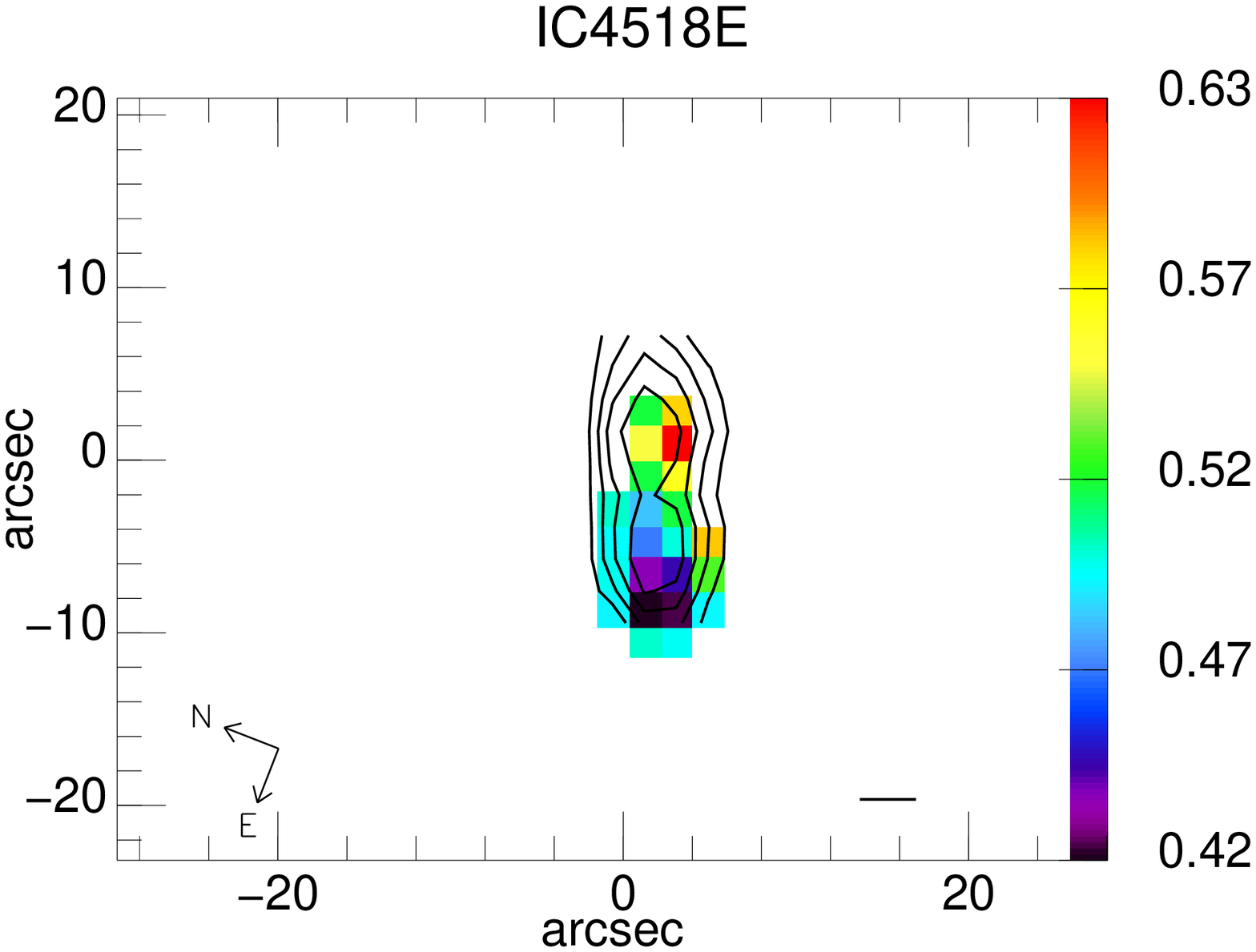}
\includegraphics[width=0.33\textwidth]{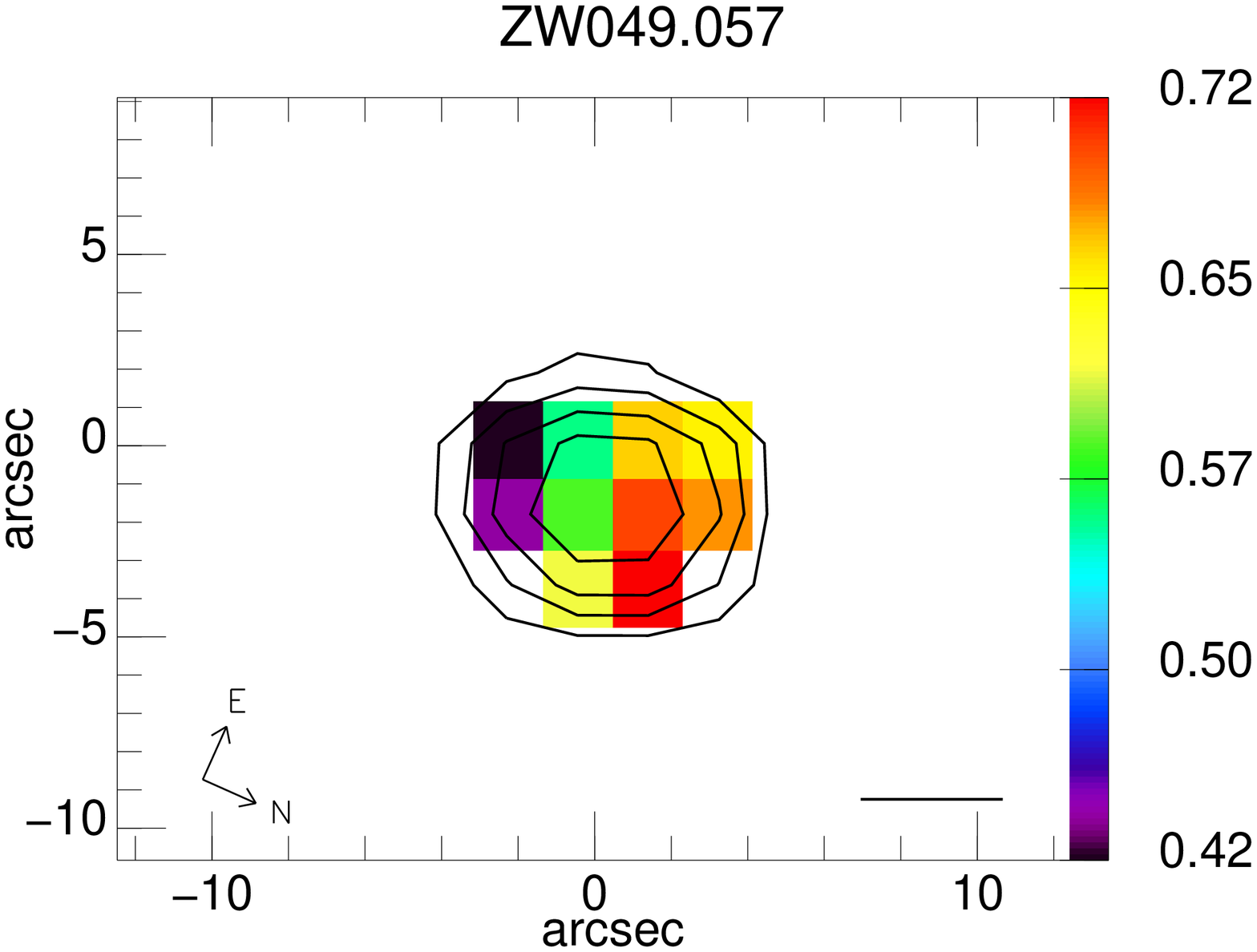}
\includegraphics[width=0.33\textwidth]{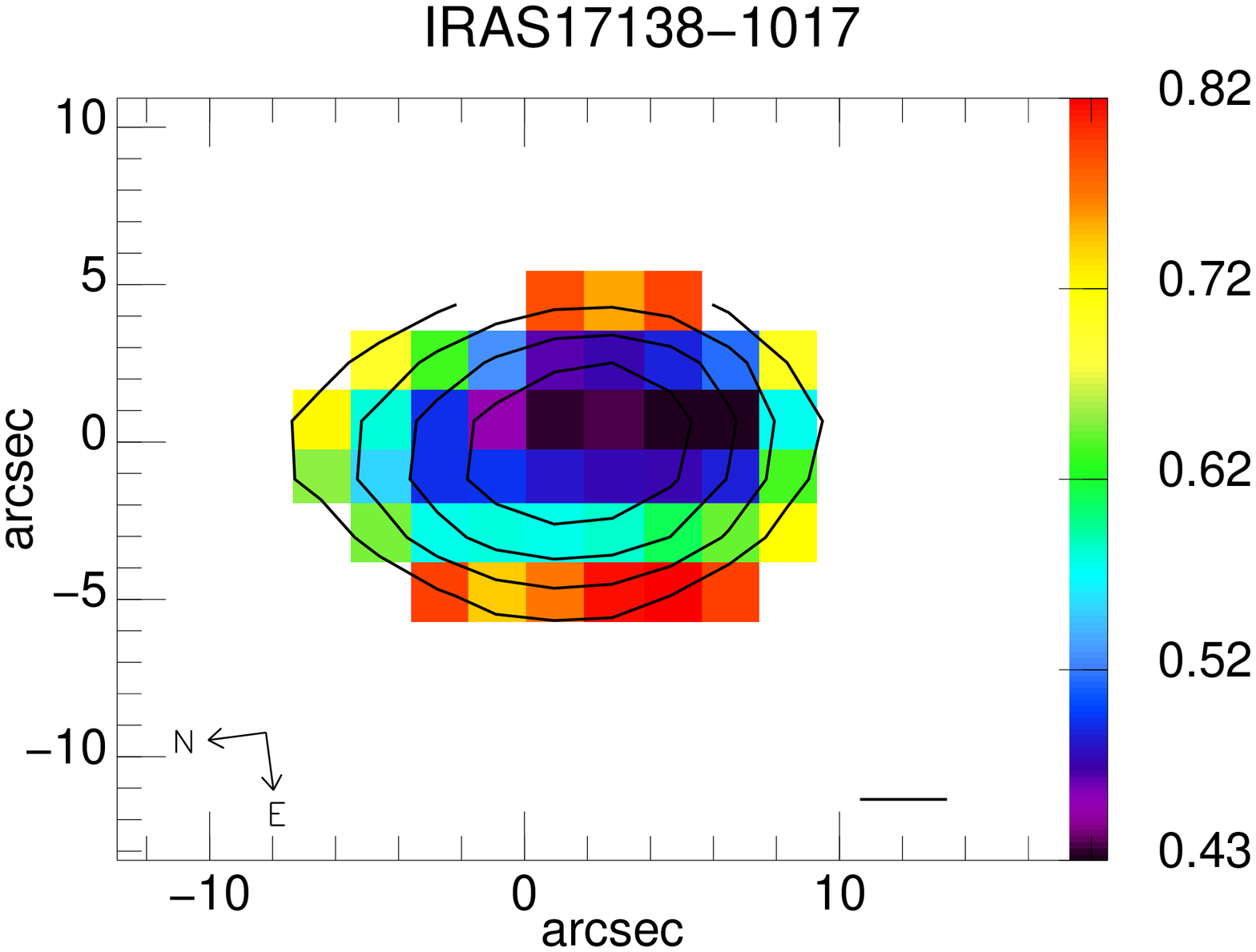}
\includegraphics[width=0.33\textwidth]{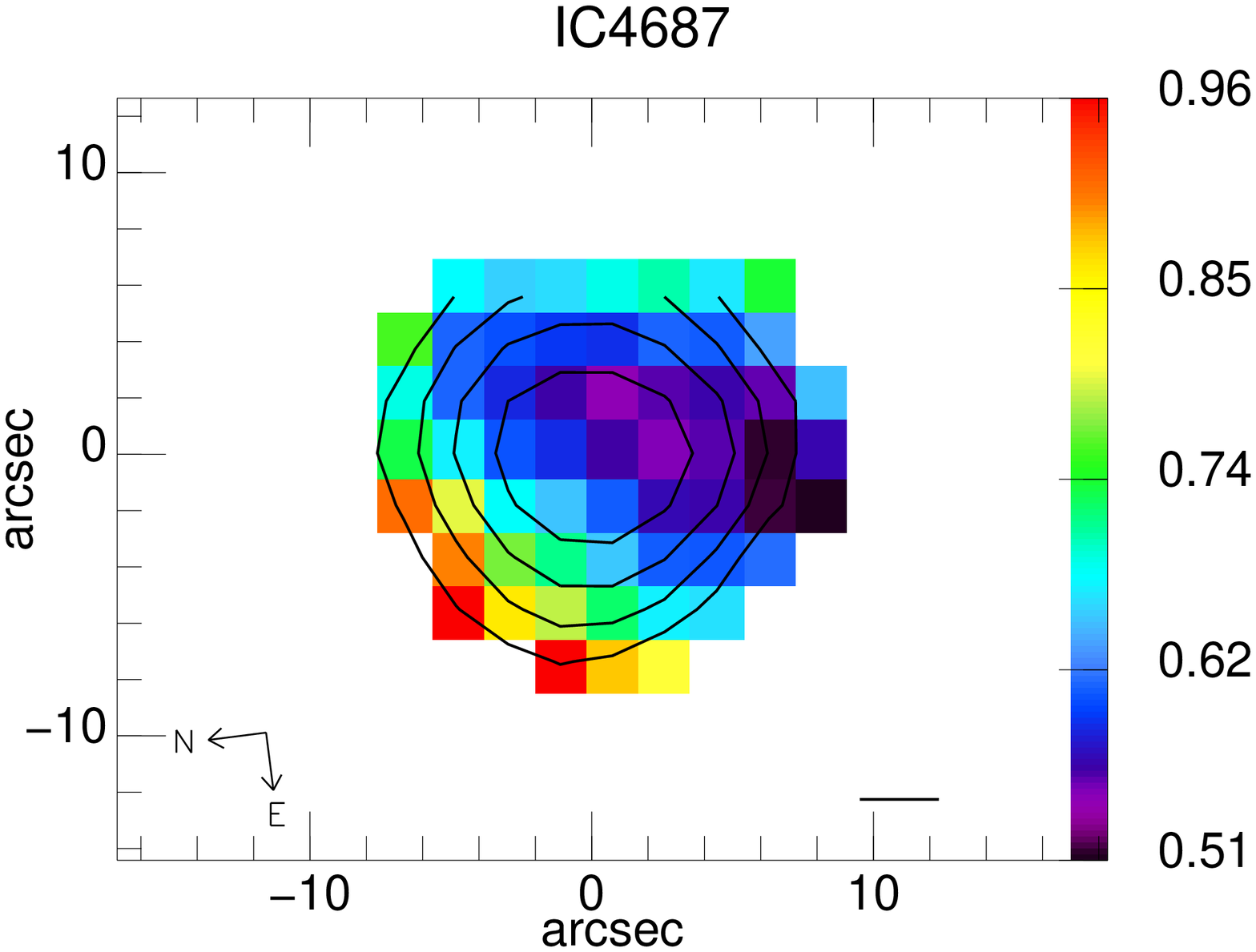}
\includegraphics[width=0.33\textwidth]{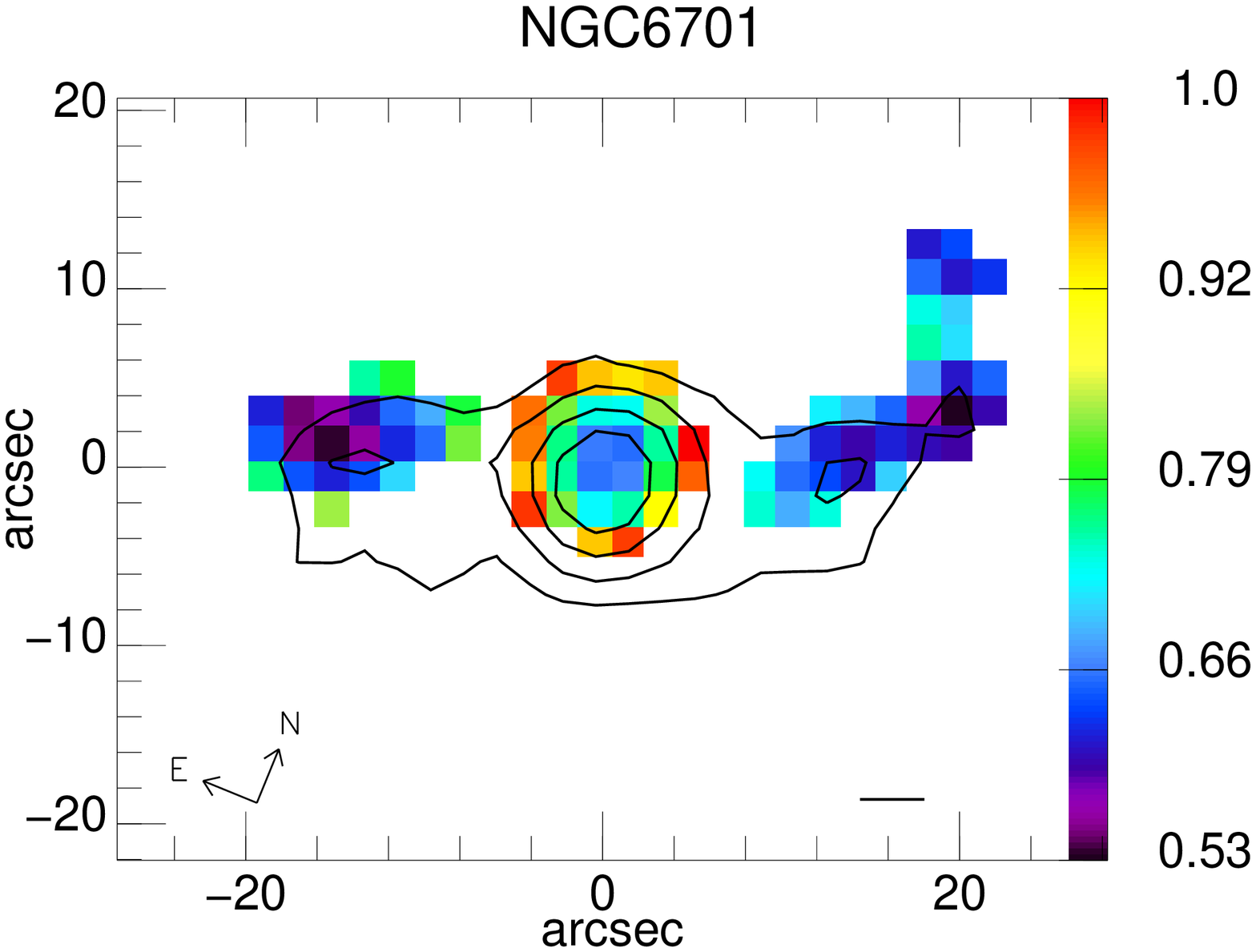}
\includegraphics[width=0.33\textwidth]{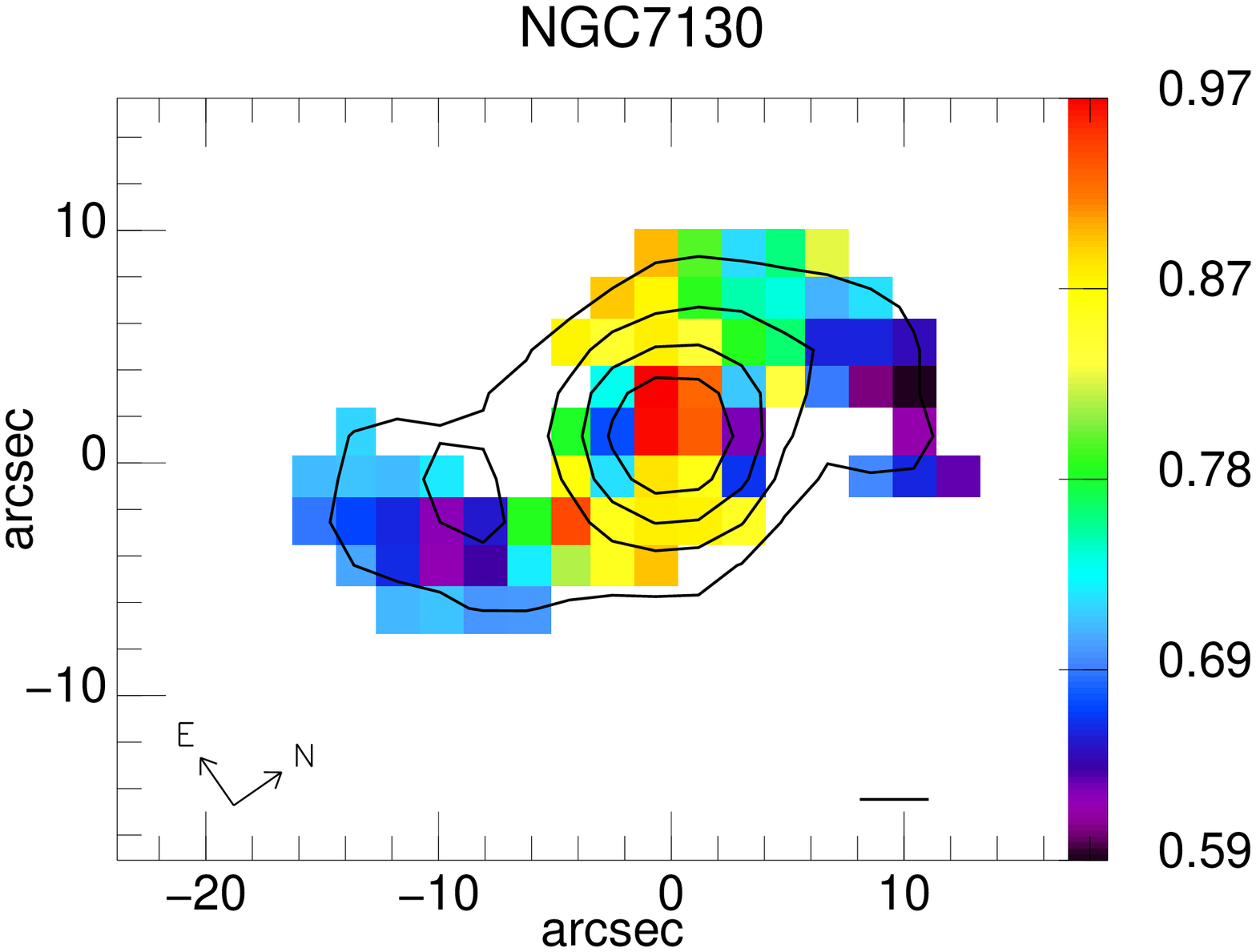}
\includegraphics[width=0.33\textwidth]{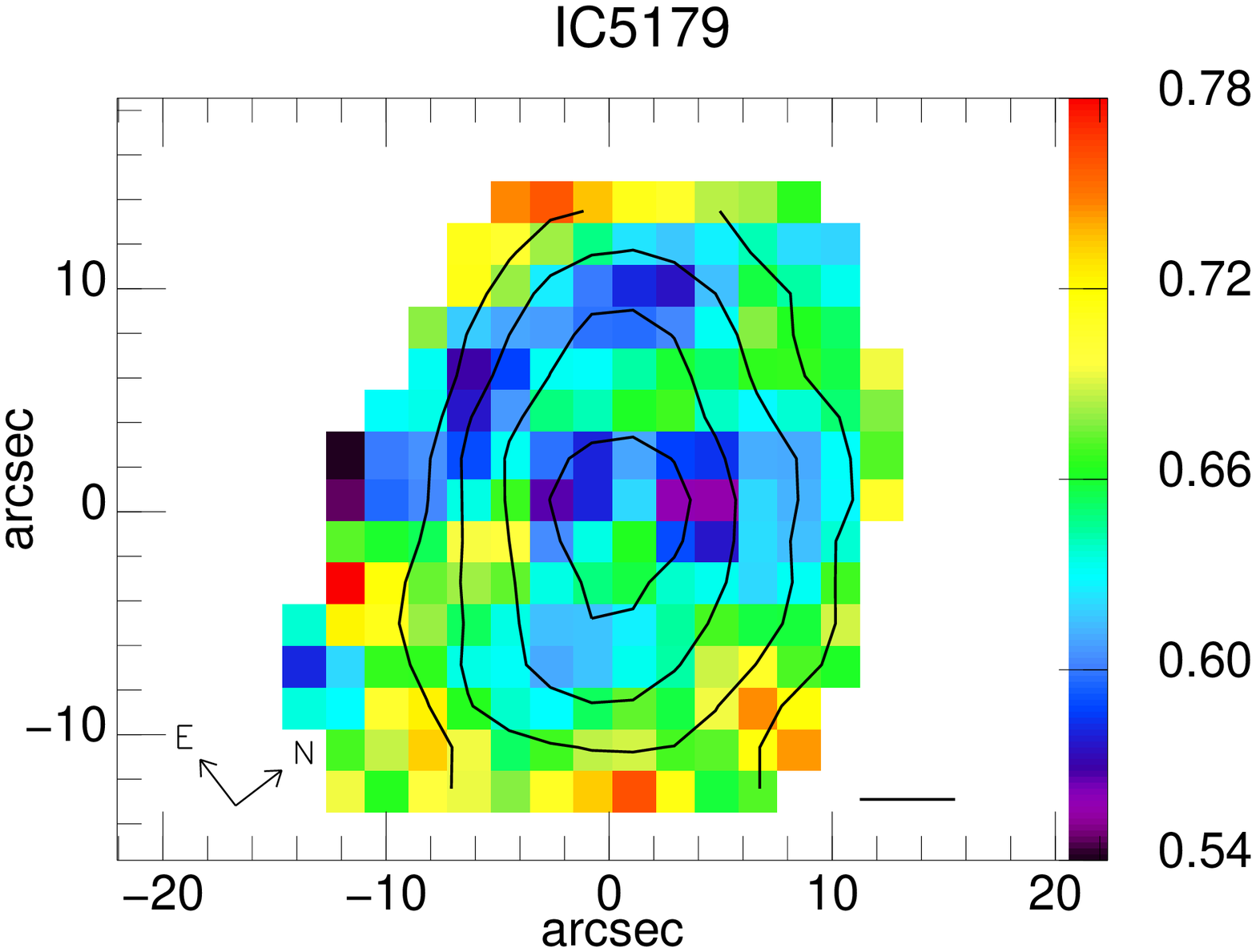}
\includegraphics[width=0.33\textwidth]{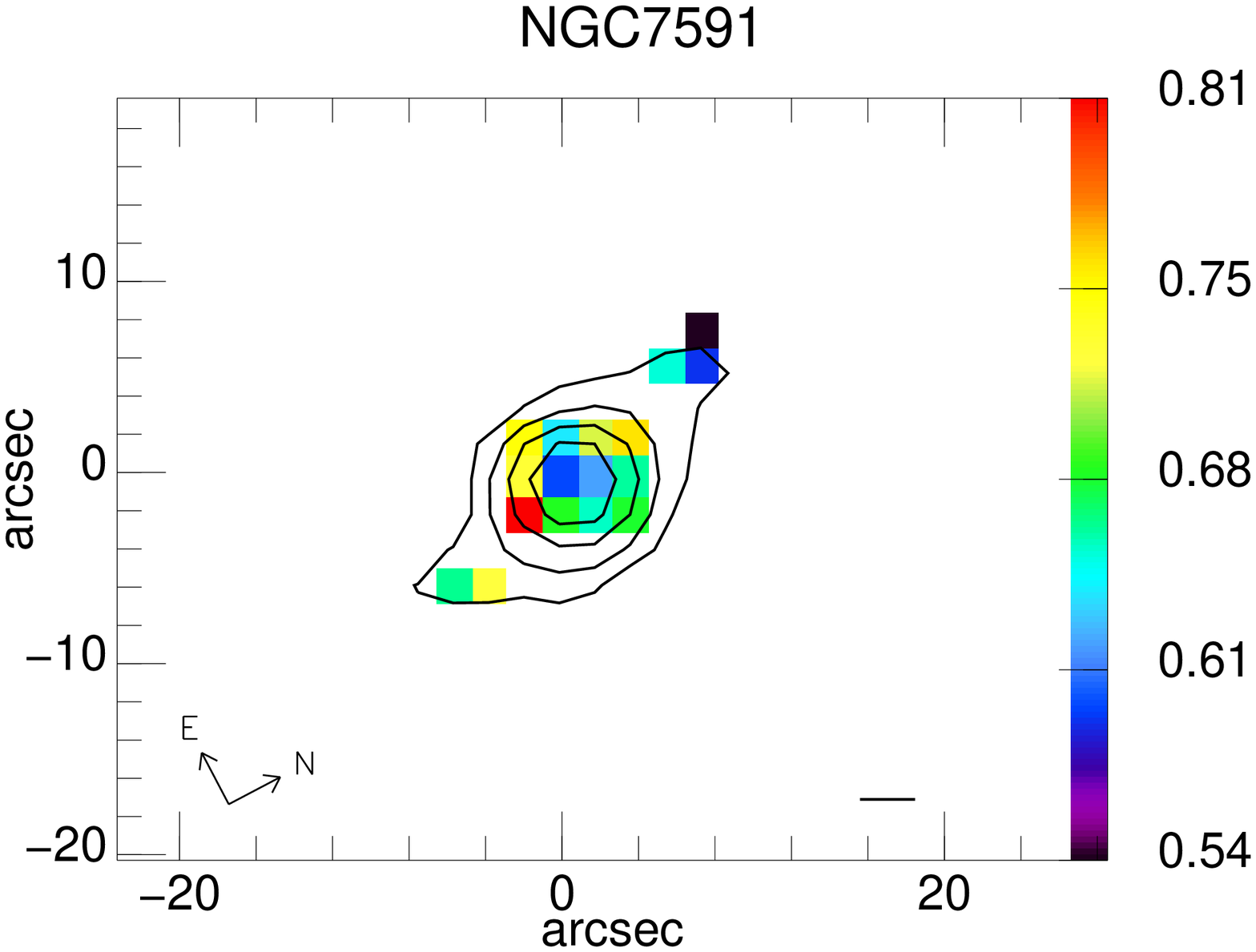}
\includegraphics[width=0.33\textwidth]{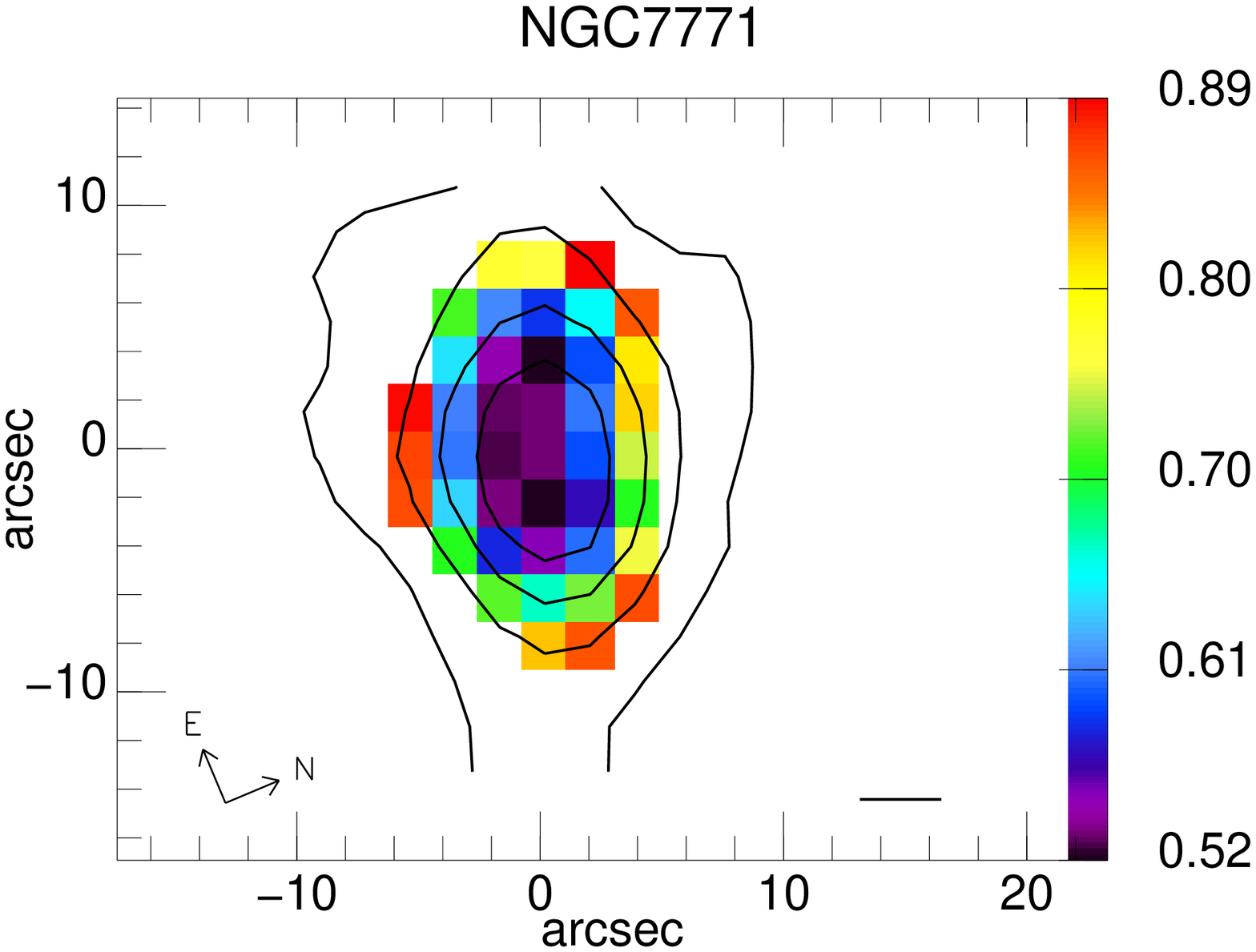}
\caption{Spitzer/IRS SL spectral maps of the \PAHonce\slash\PAHseis\ ratio. The 5.5\micron\ continuum contours, in a logarithm scale, are displayed to guide the eye. The image orientation is indicated on the maps for each galaxy. The scale represents 1 kpc. The ratio maps are shown in a linear scale.}
\label{fig_pah11pah6_sl}
\end{figure*}

\begin{figure*}[!p]
\includegraphics[width=0.33\textwidth]{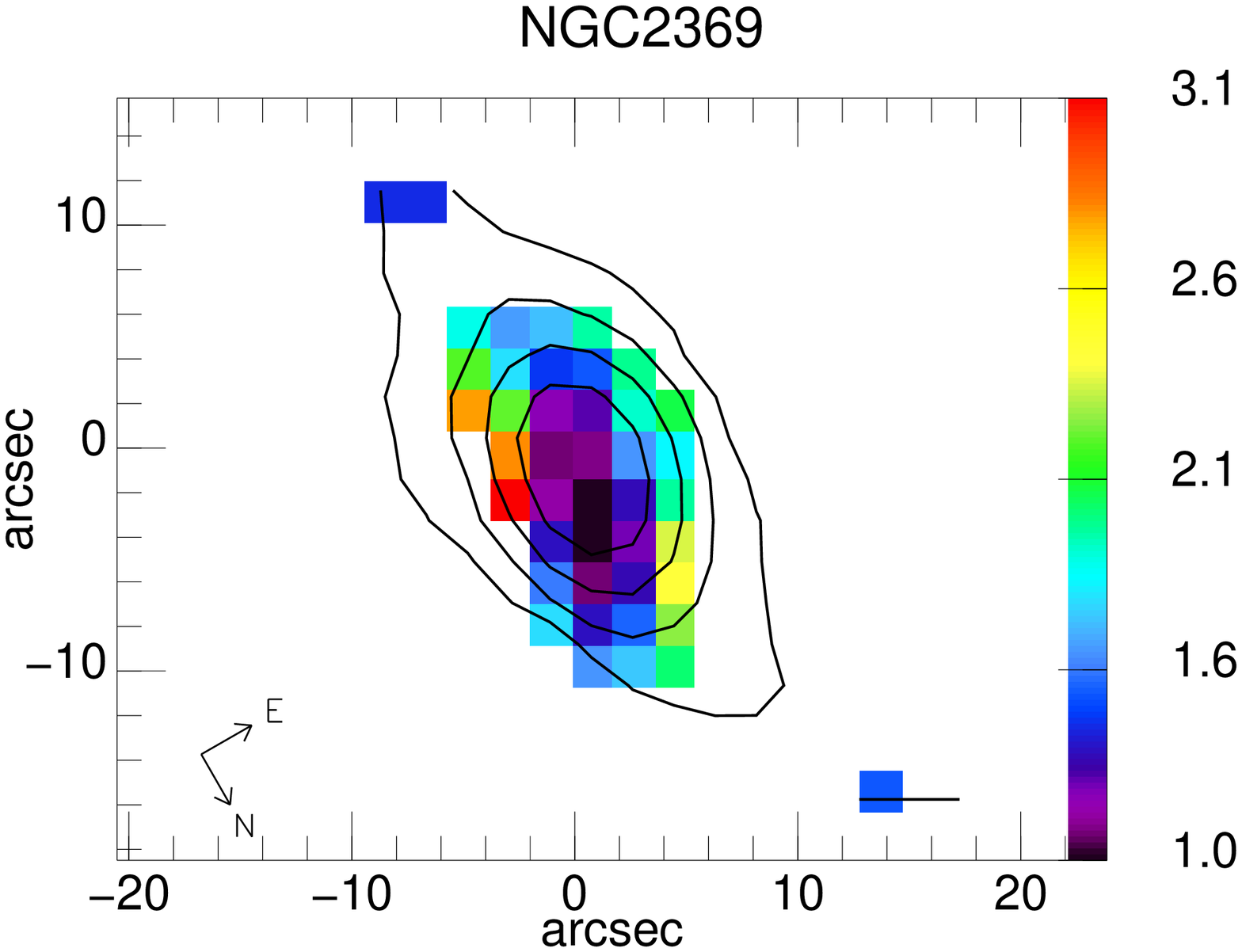}
\includegraphics[width=0.33\textwidth]{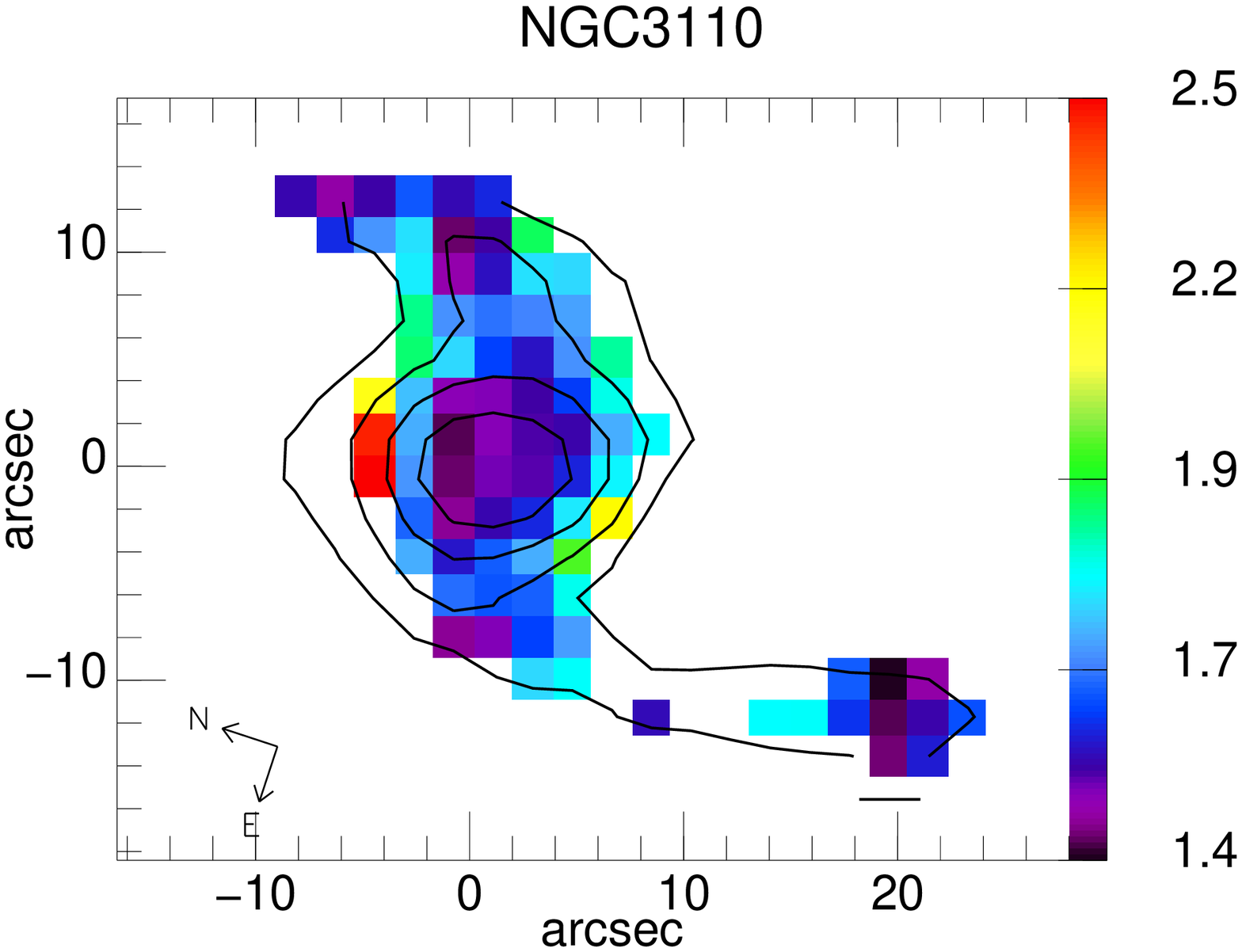}
\includegraphics[width=0.33\textwidth]{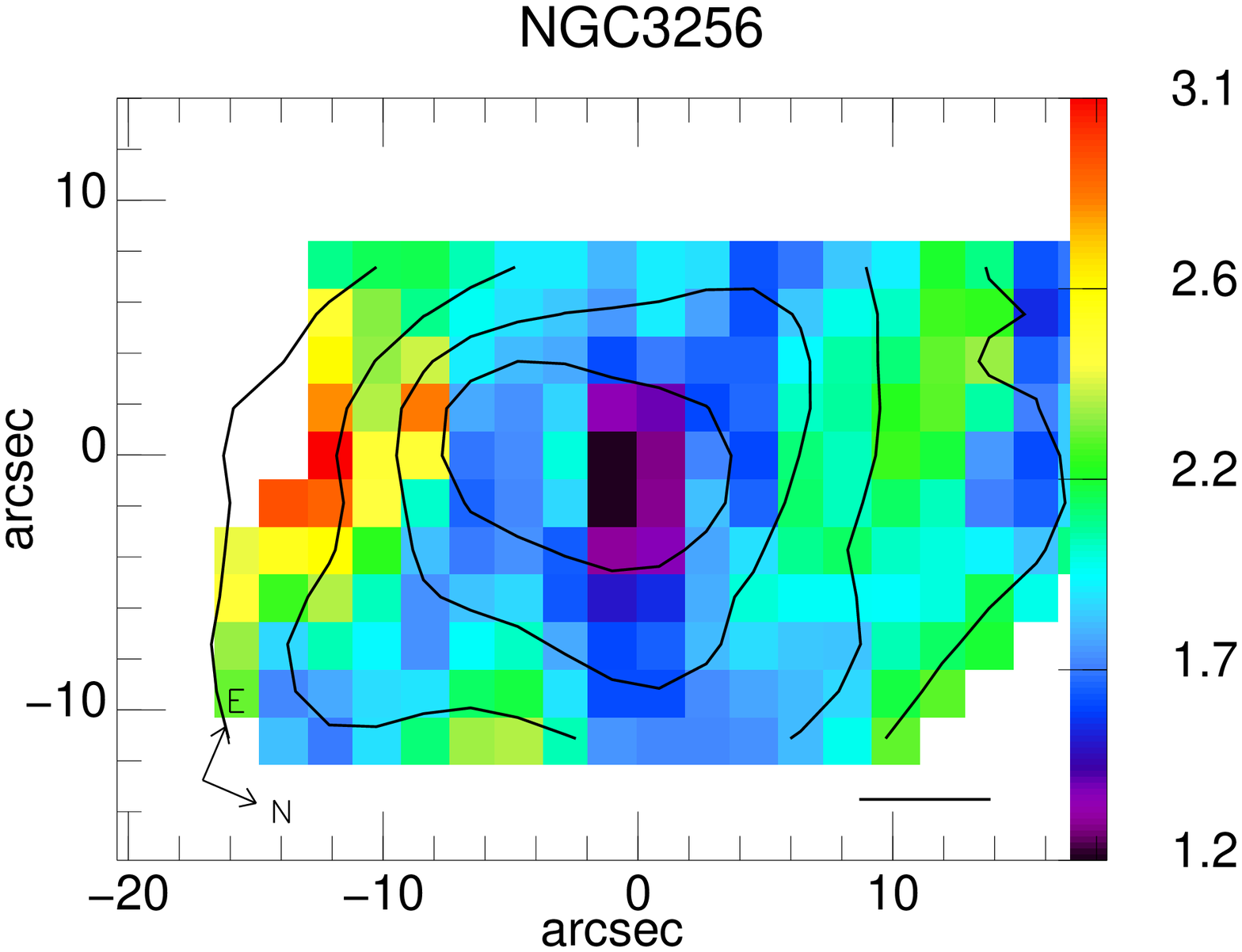}
\includegraphics[width=0.33\textwidth]{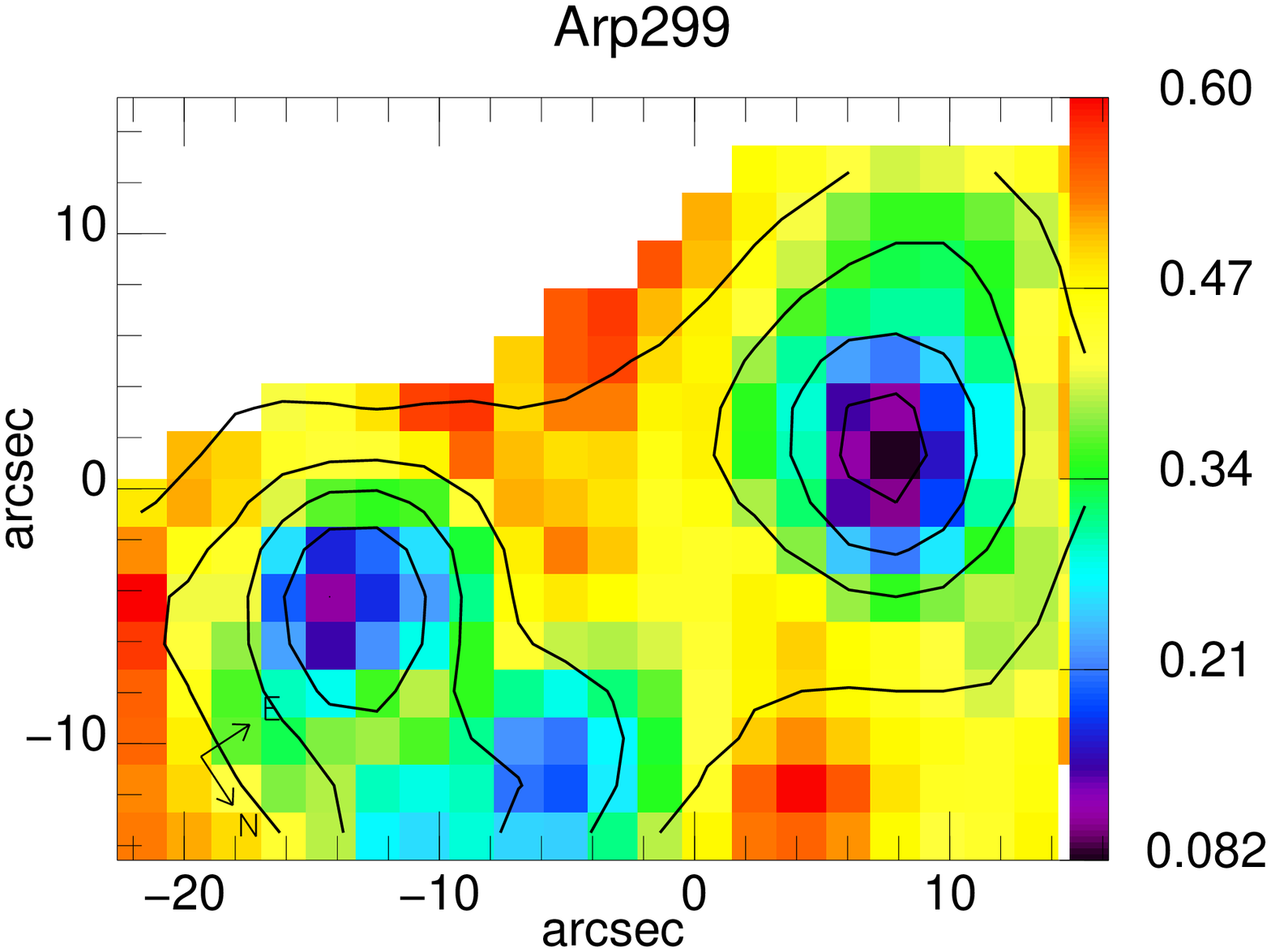}
\includegraphics[width=0.33\textwidth]{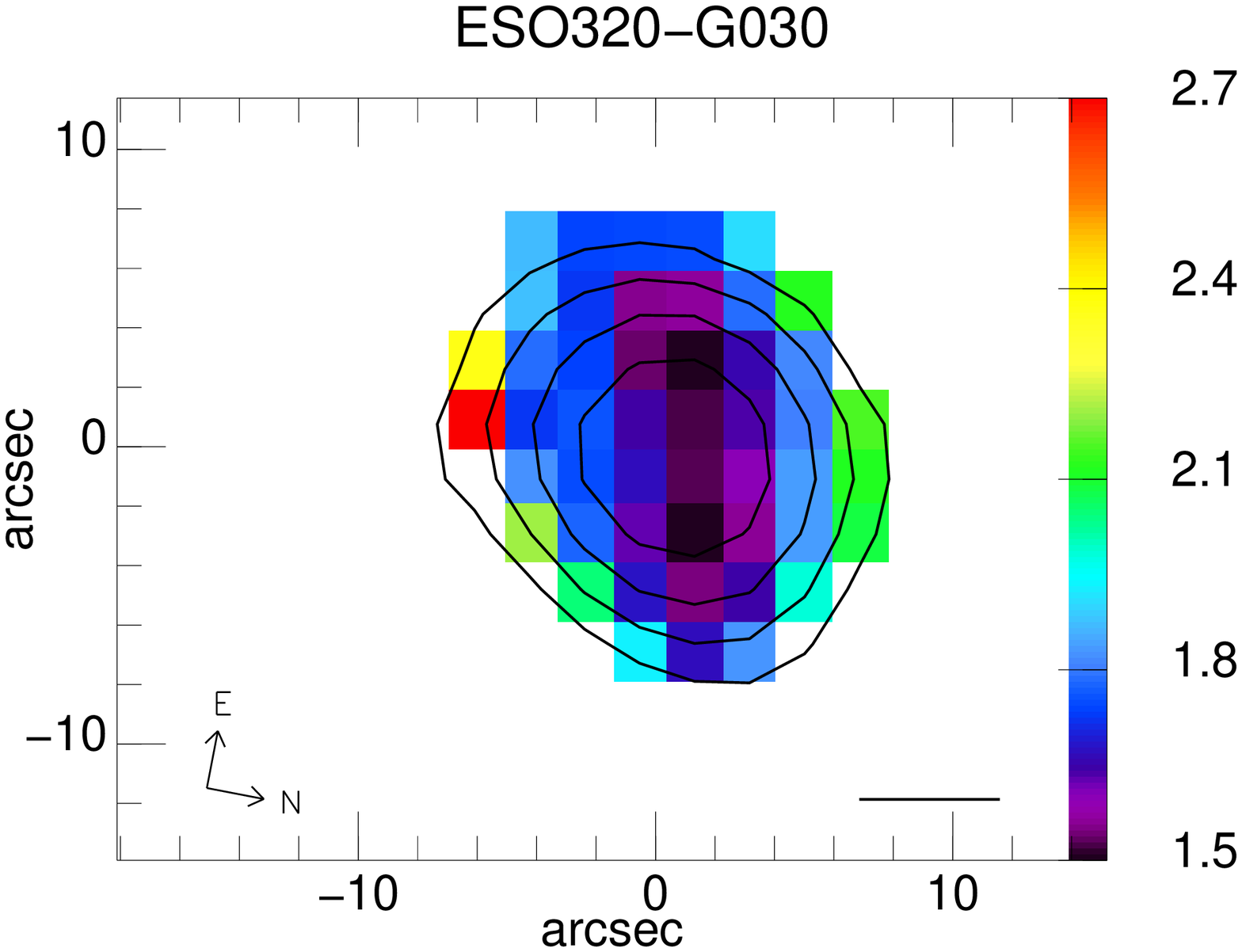}
\includegraphics[width=0.33\textwidth]{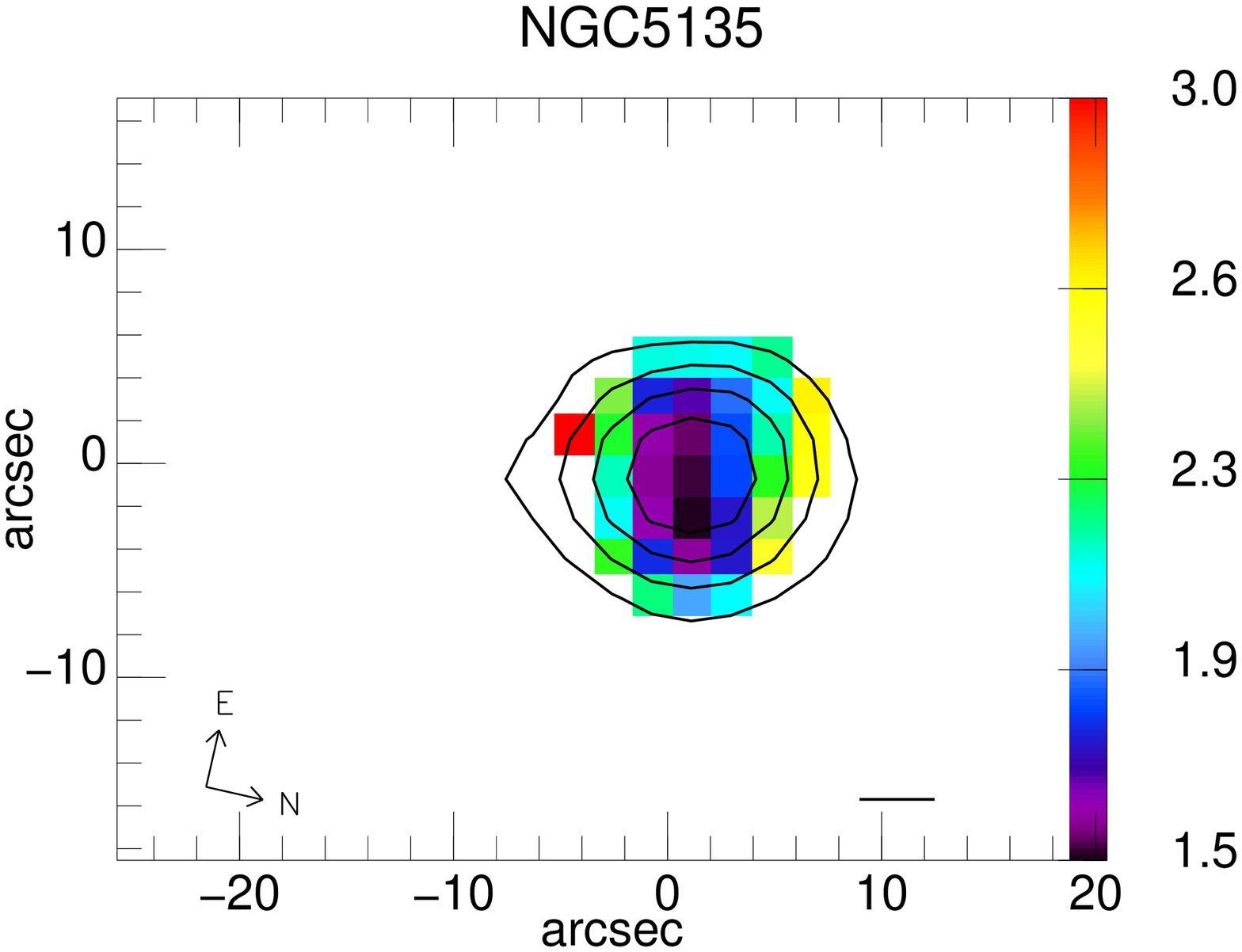}
\includegraphics[width=0.33\textwidth]{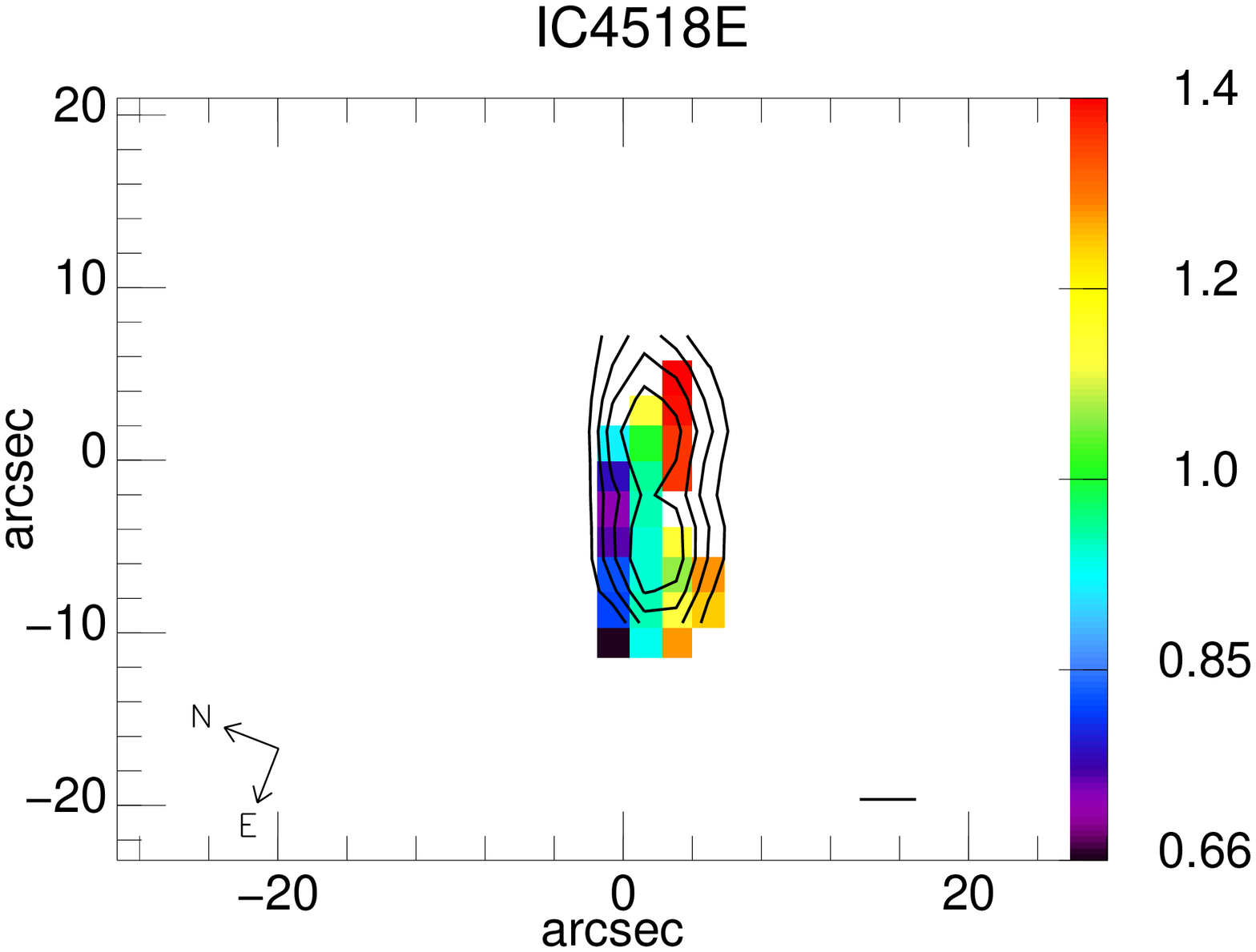}
\includegraphics[width=0.33\textwidth]{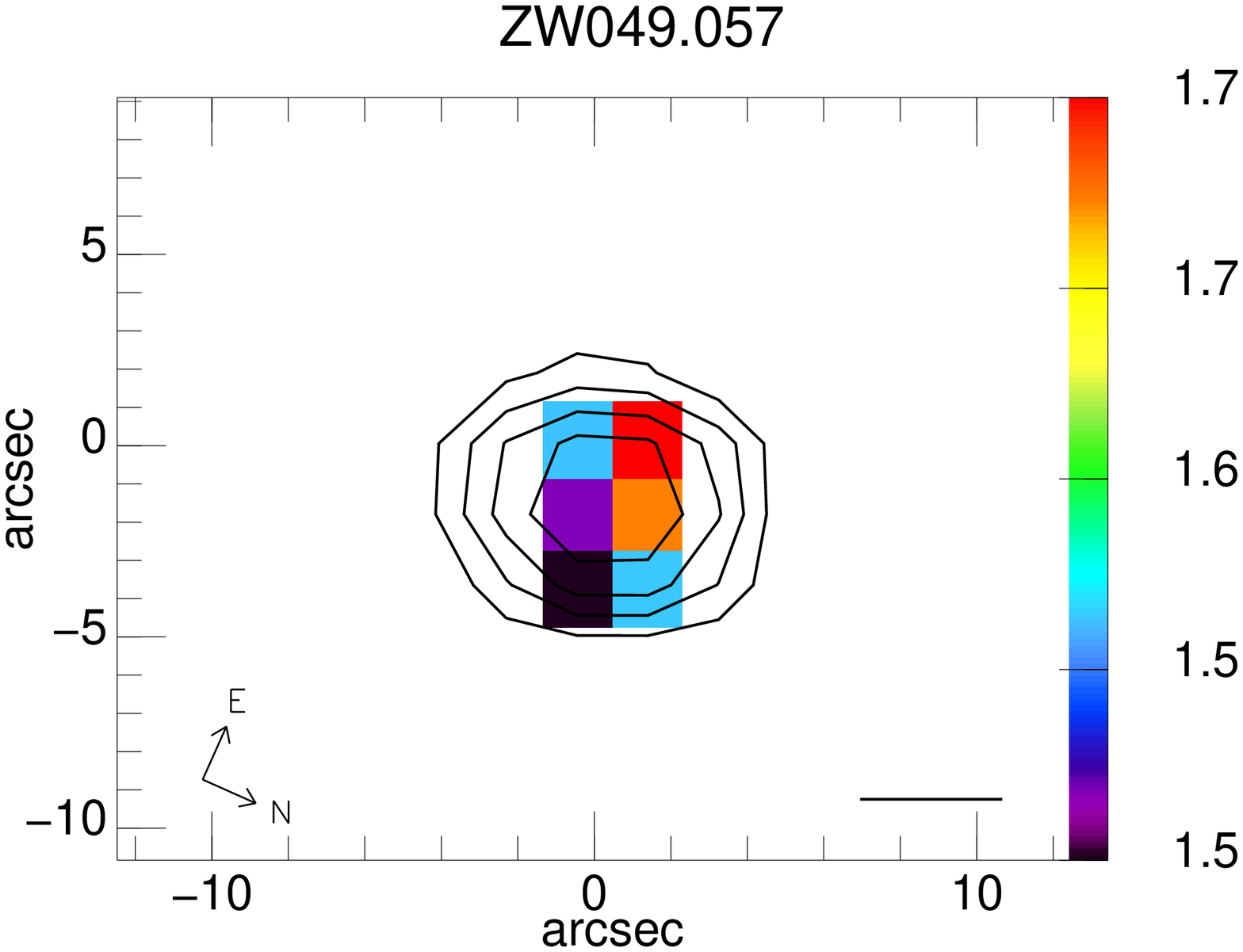}
\includegraphics[width=0.33\textwidth]{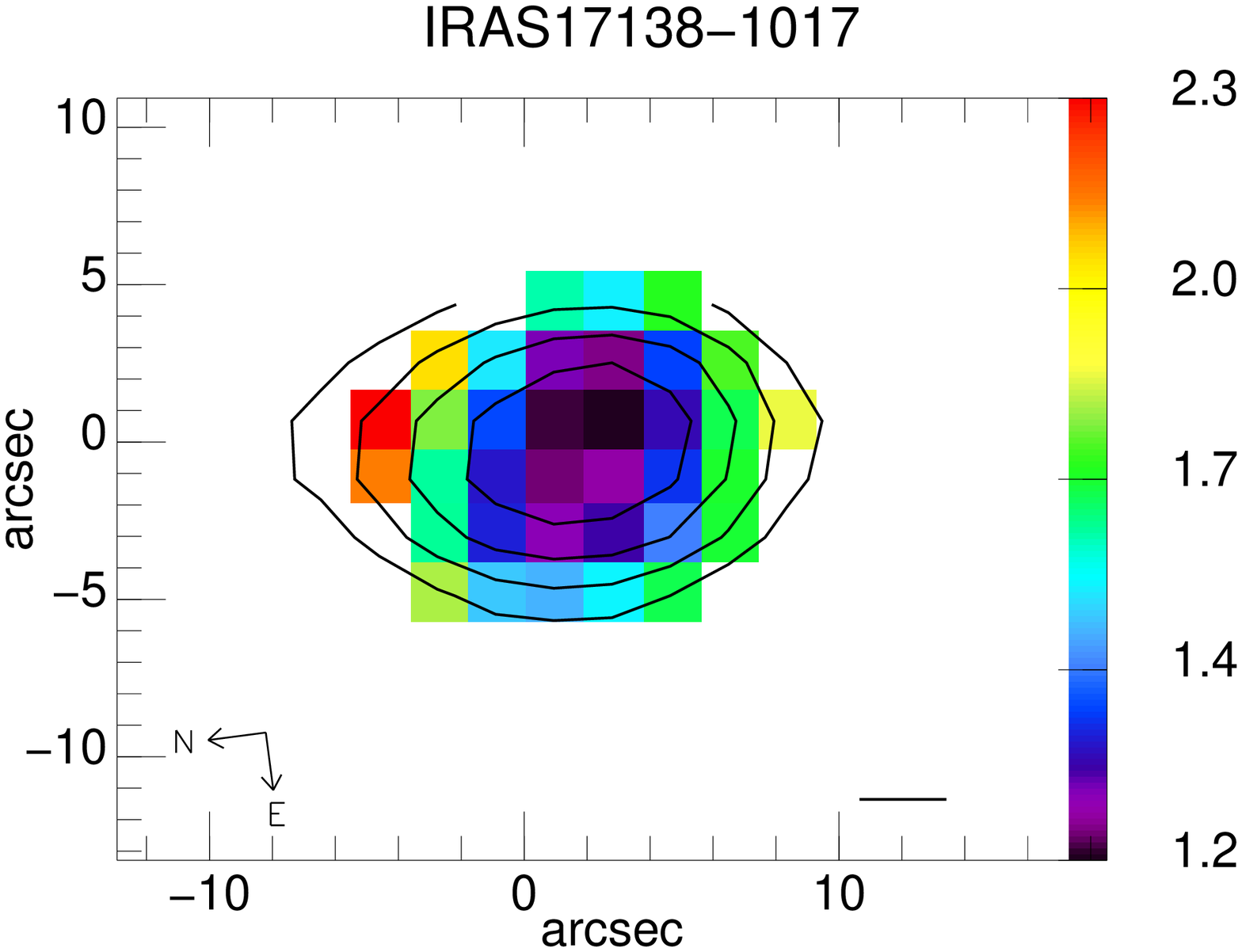}
\includegraphics[width=0.33\textwidth]{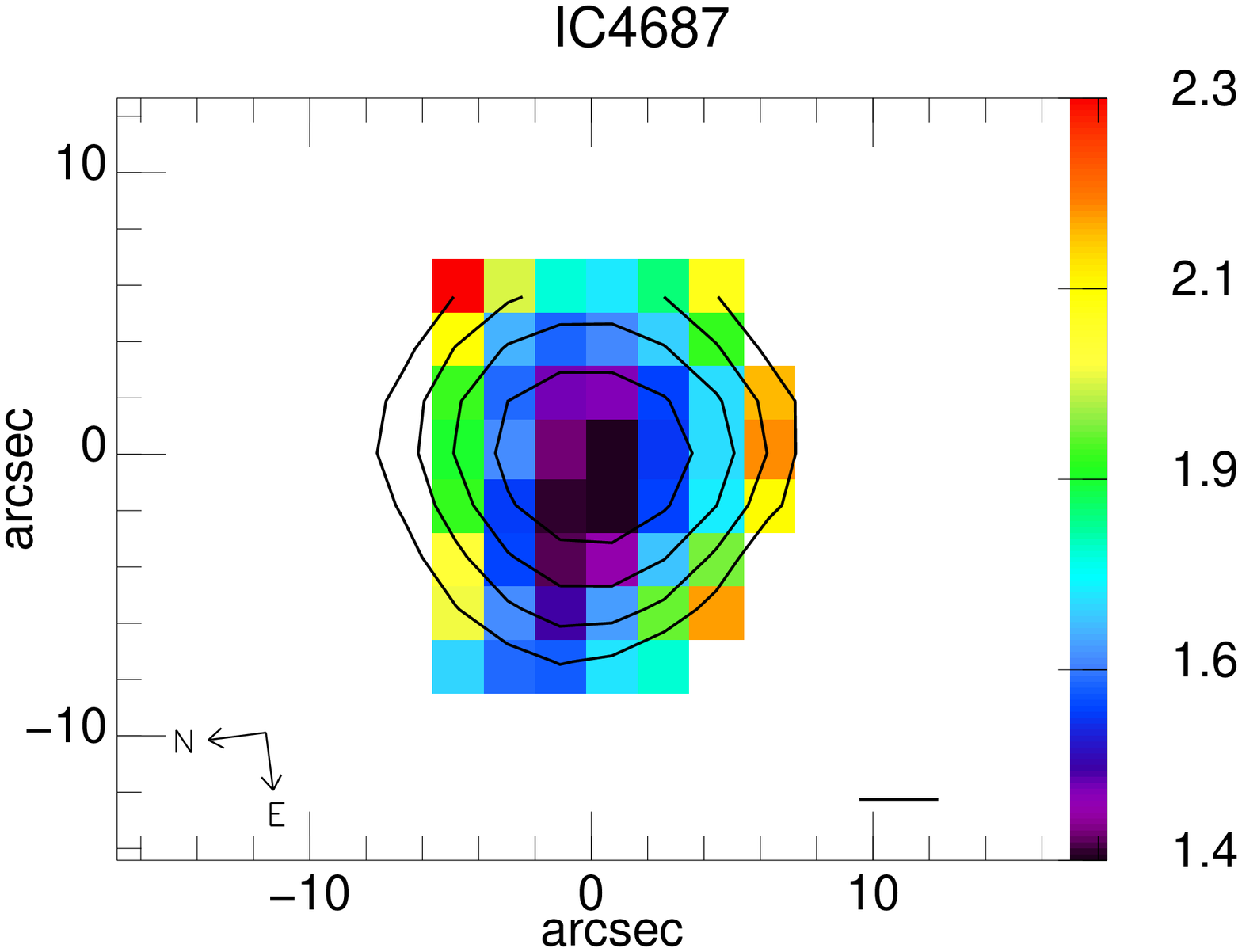}
\includegraphics[width=0.33\textwidth]{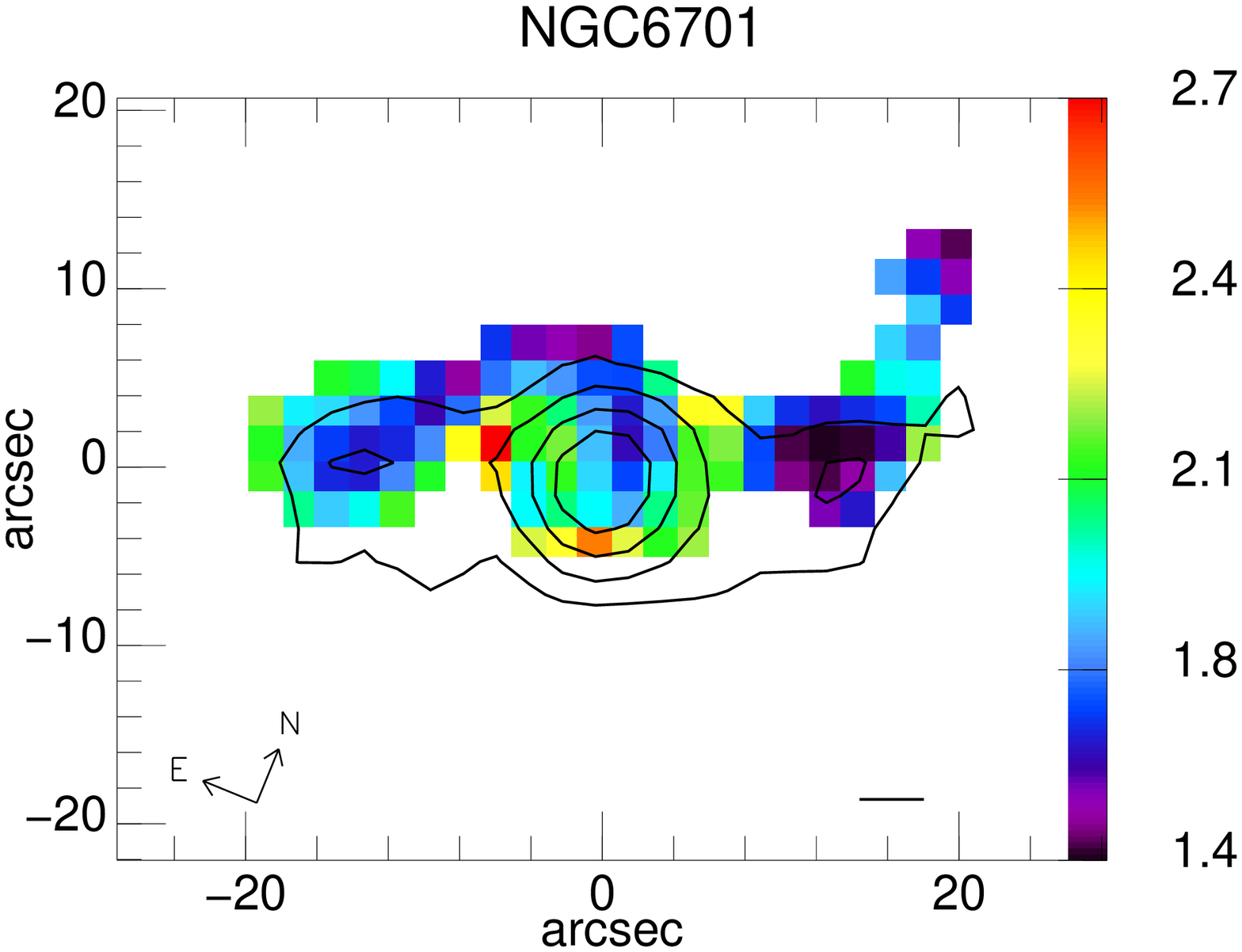}
\includegraphics[width=0.33\textwidth]{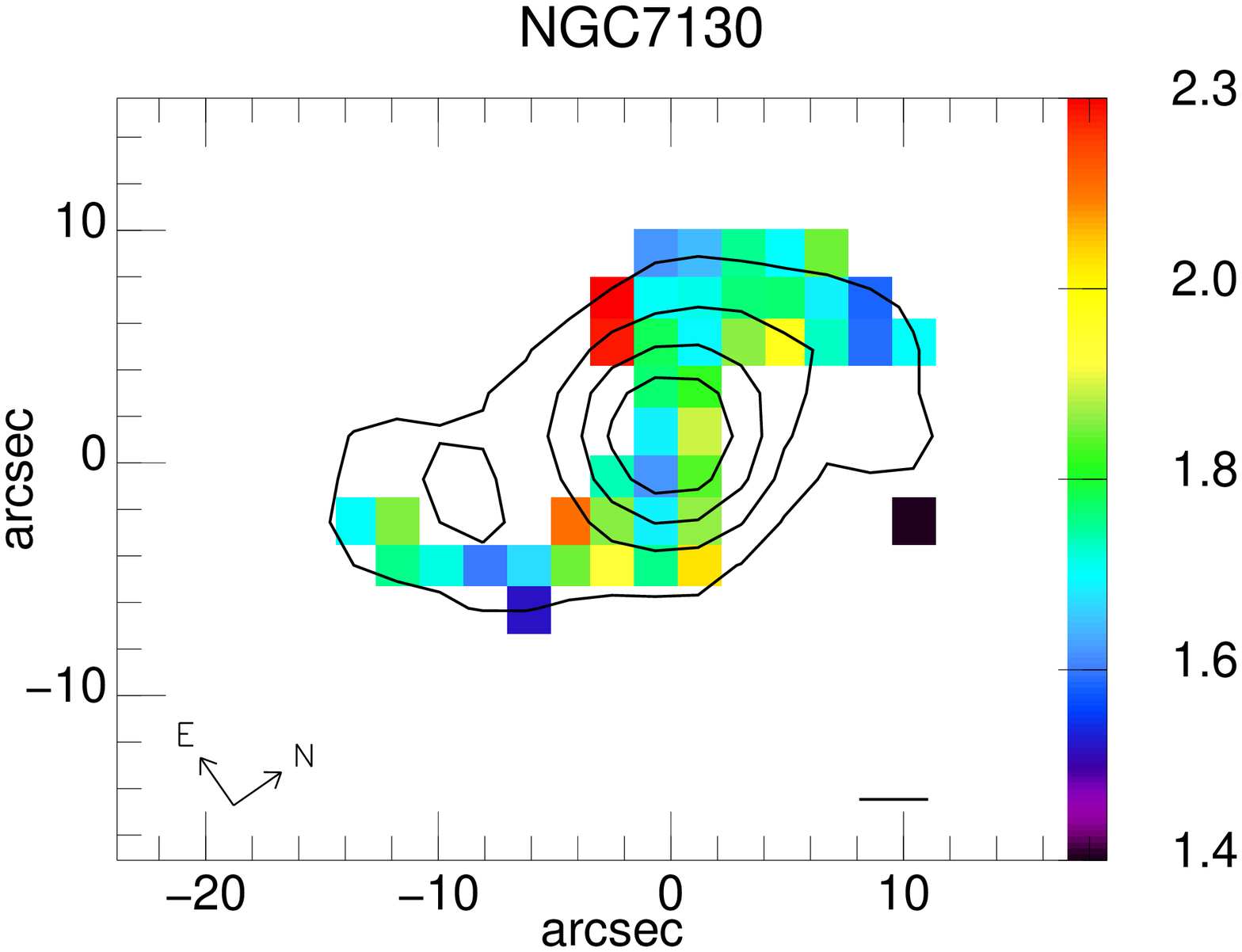}
\includegraphics[width=0.33\textwidth]{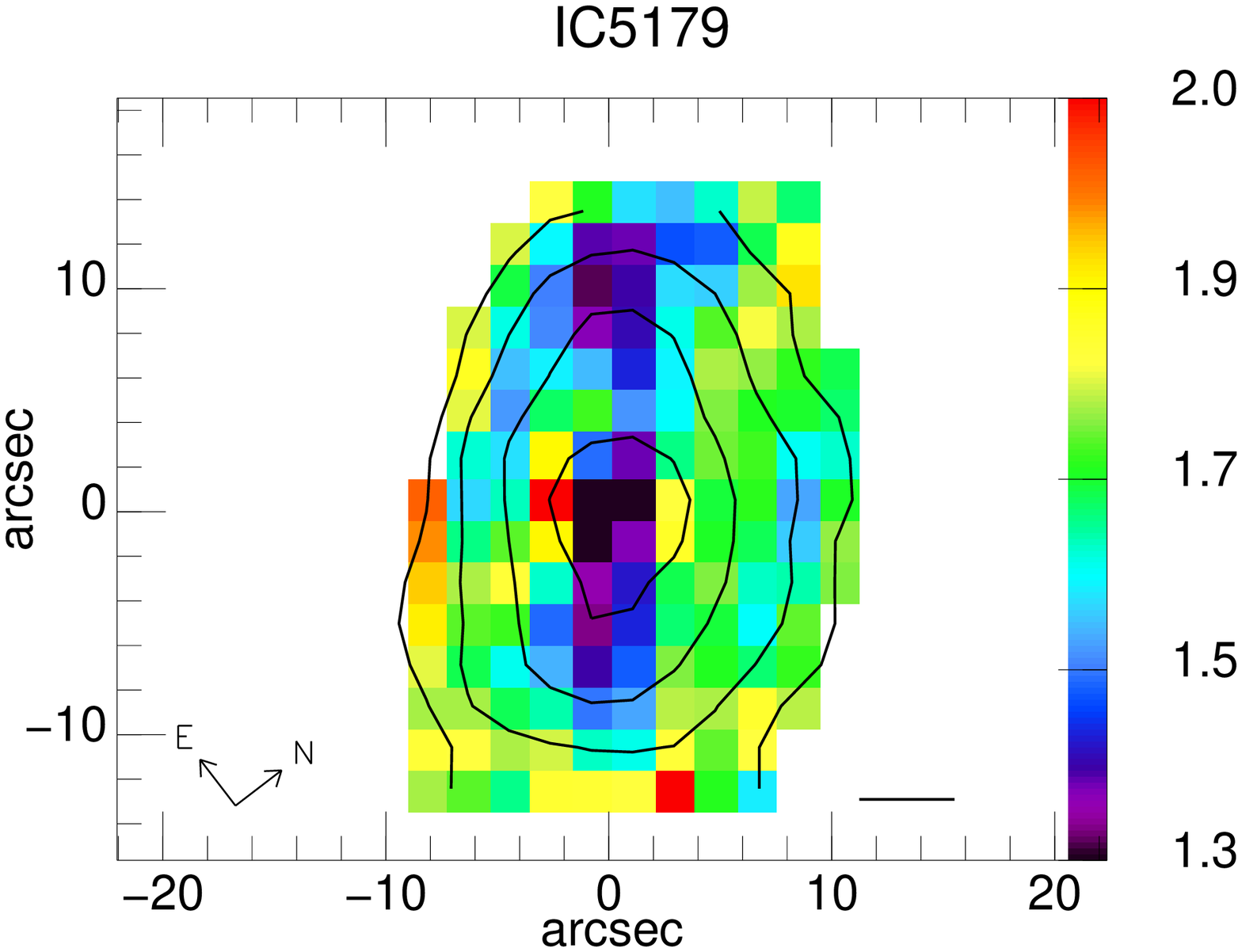}
\includegraphics[width=0.33\textwidth]{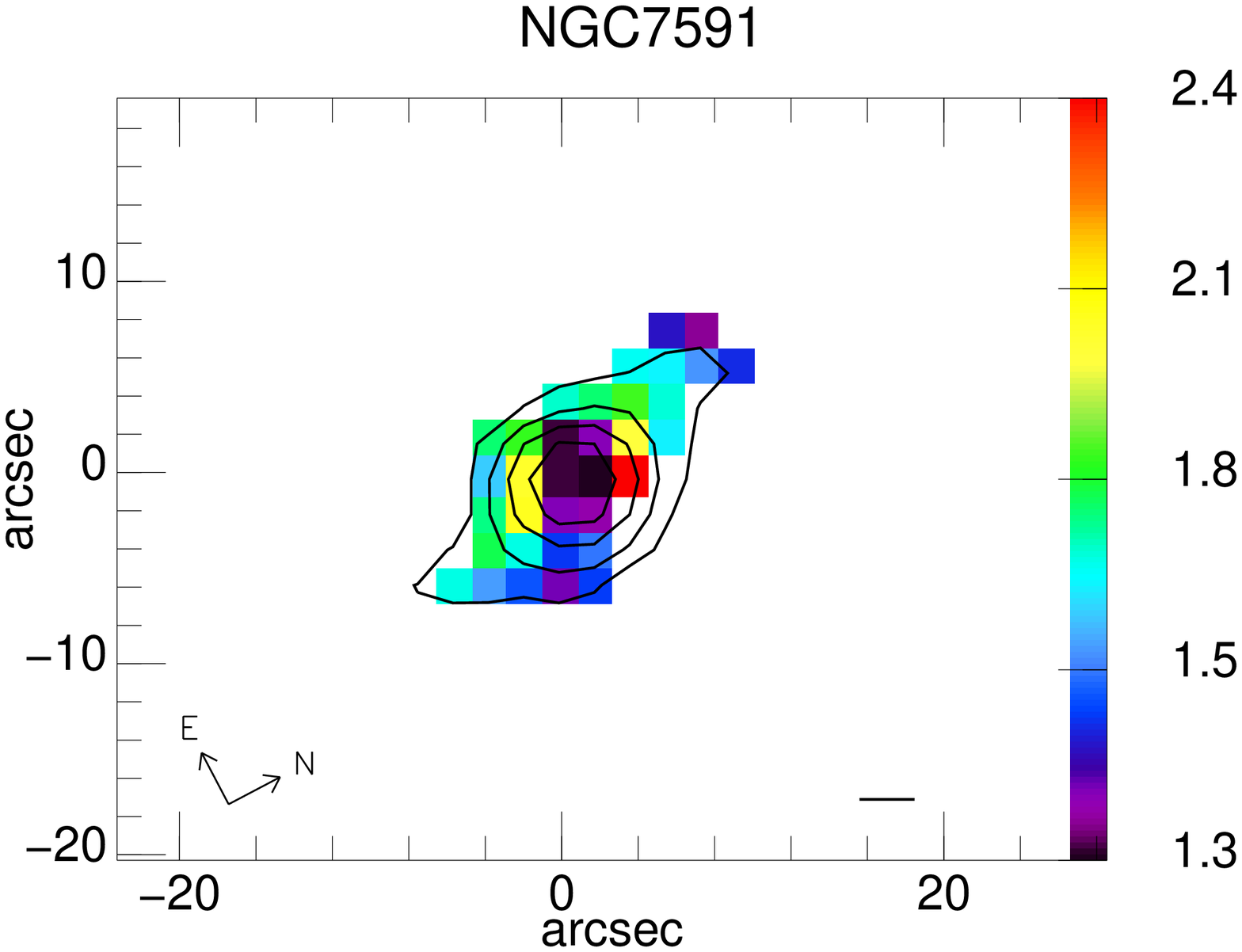}
\includegraphics[width=0.33\textwidth]{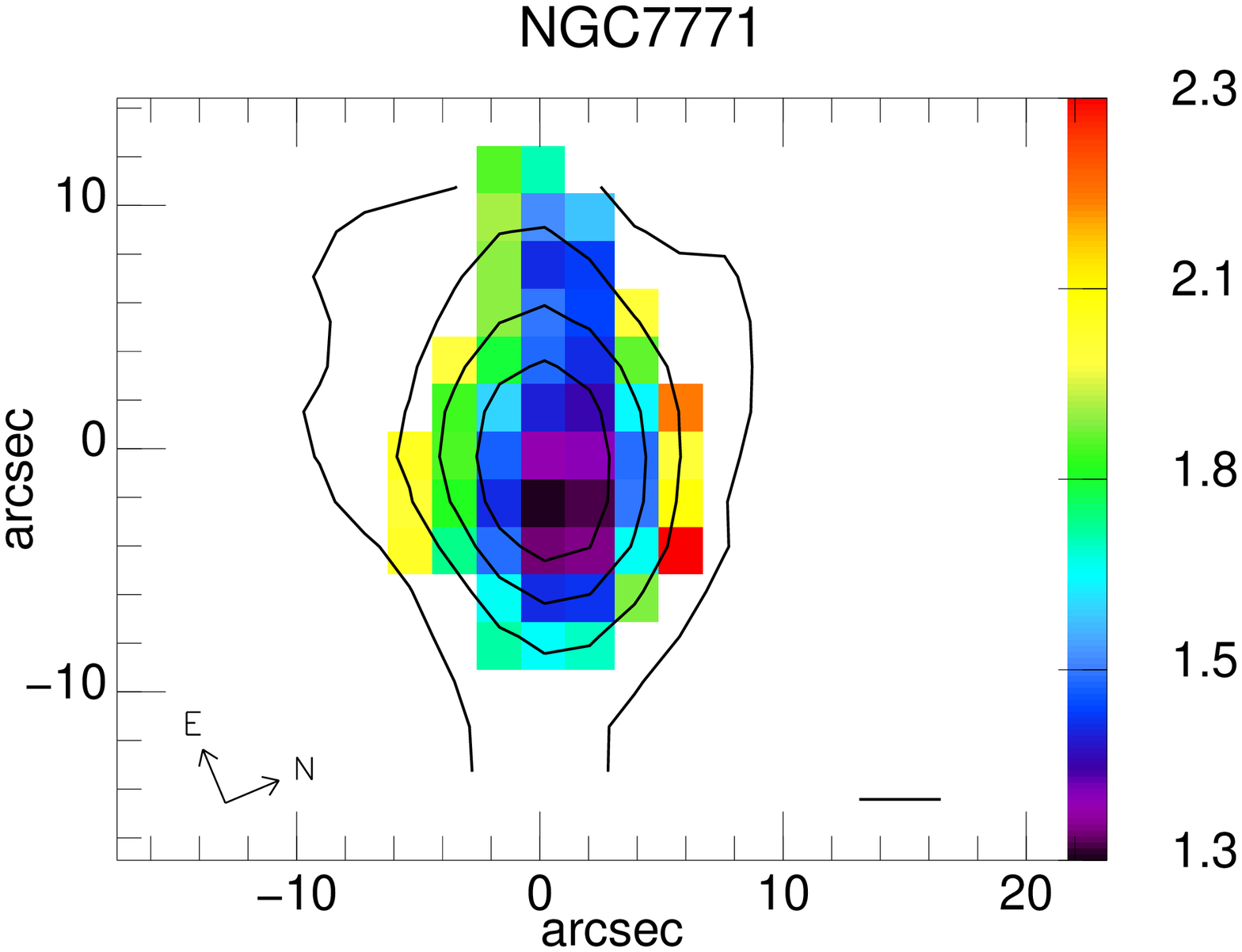}
\caption{Spitzer/IRS SL spectral maps of the \PAHonce\slash\PAHsiete\ ratio. The 5.5\micron\ continuum contours, in a logarithm scale, are displayed to guide the eye. The image orientation is indicated on the maps for each galaxy. The scale represents 1 kpc. The ratio maps are shown in a linear scale.}
\label{fig_pah11pah7_sl}
\end{figure*}


\begin{figure*}[!p]
\center
\includegraphics[width=0.9\textwidth]{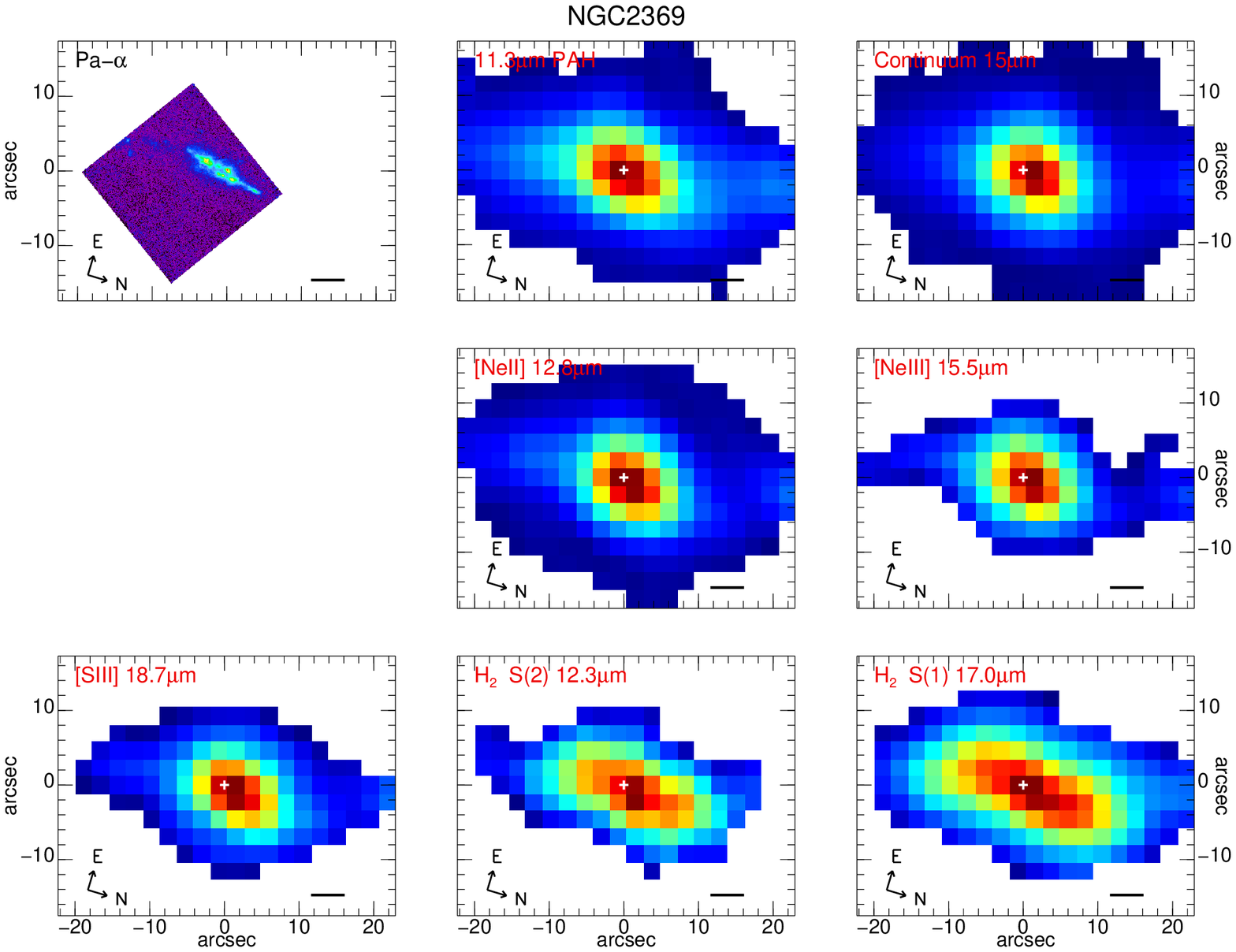}
\includegraphics[width=0.9\textwidth]{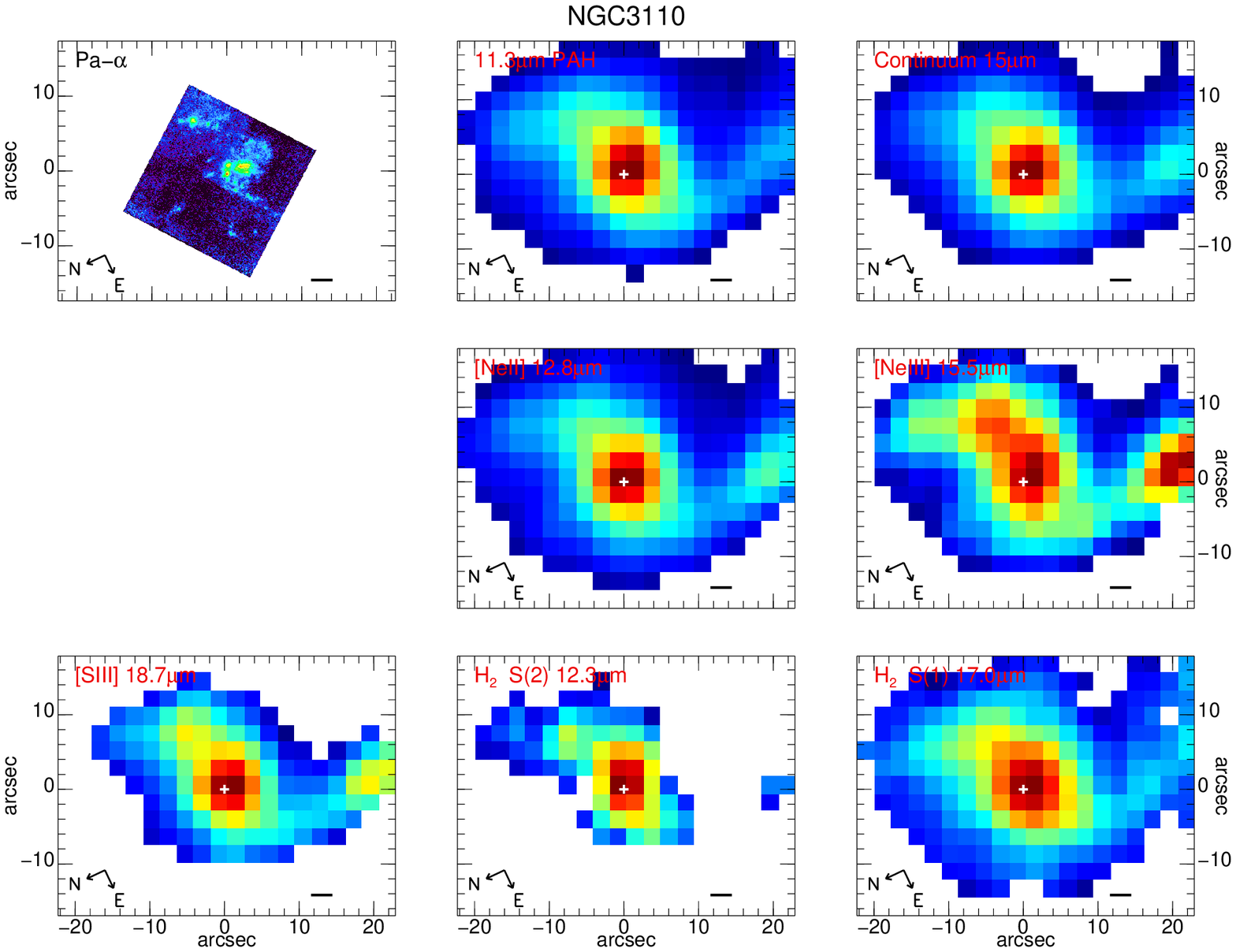}
\caption{Spitzer/IRS SH spectral maps of the \PAHonce\ feature, the 15\micron\ continuum, the fine structure line emission of \SIV, \Neii, \Neiii, \SIIIa\ and the molecular hydrogen lines \Hm{2} at 12.3\micron\ and \Hm{1} at 17.0\micron. The white cross marks the coordinates of the nucleus as listed in Table \ref{tbl_obs_map}. The image orientation is indicated on the maps for each galaxy. The scale represents 1 kpc. The maps are shown in a square root scale.}
\label{fig_map_sh}
\end{figure*}

\begin{figure*}
\addtocounter{figure}{-1}
\includegraphics[width=0.9\textwidth]{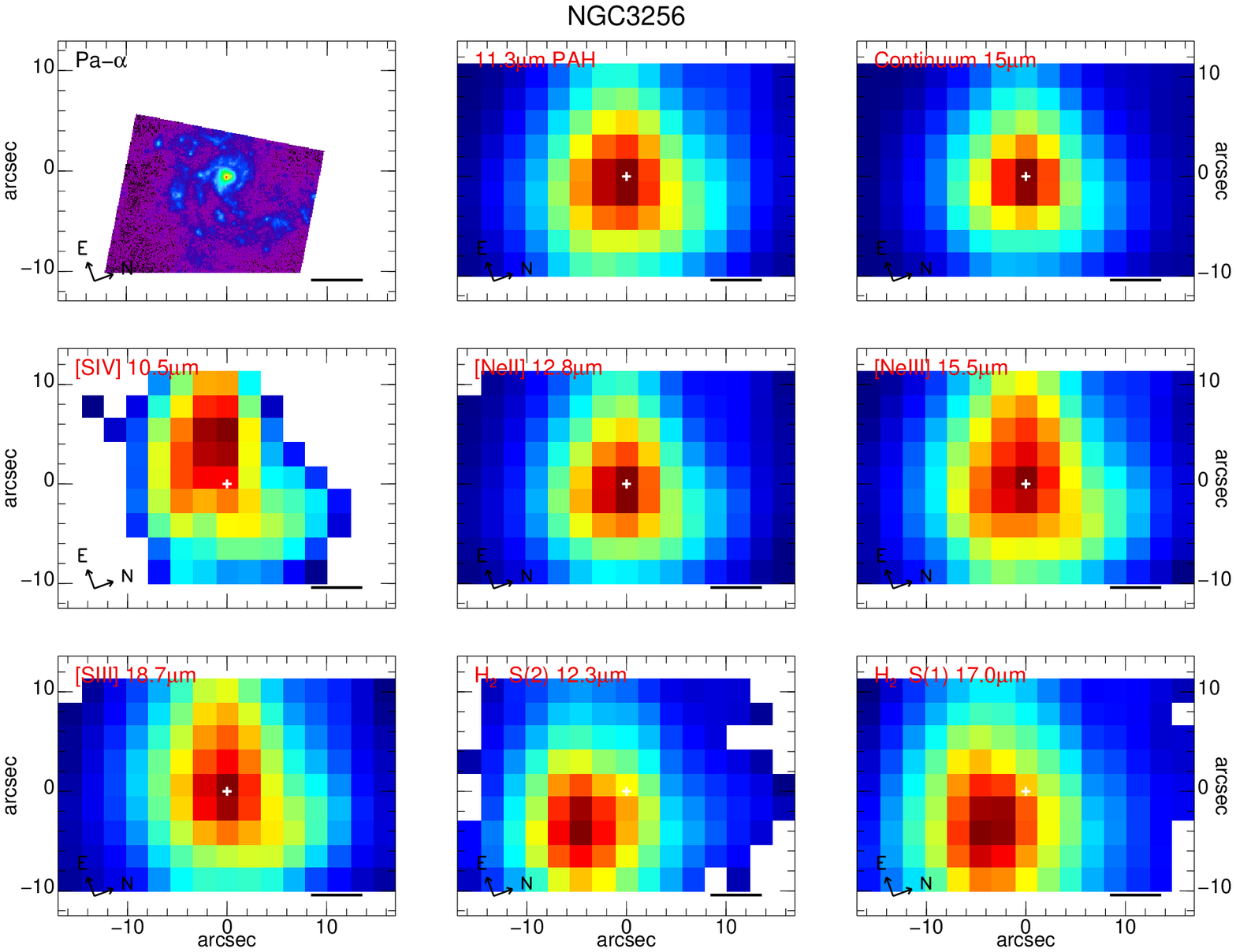}
\includegraphics[width=0.9\textwidth]{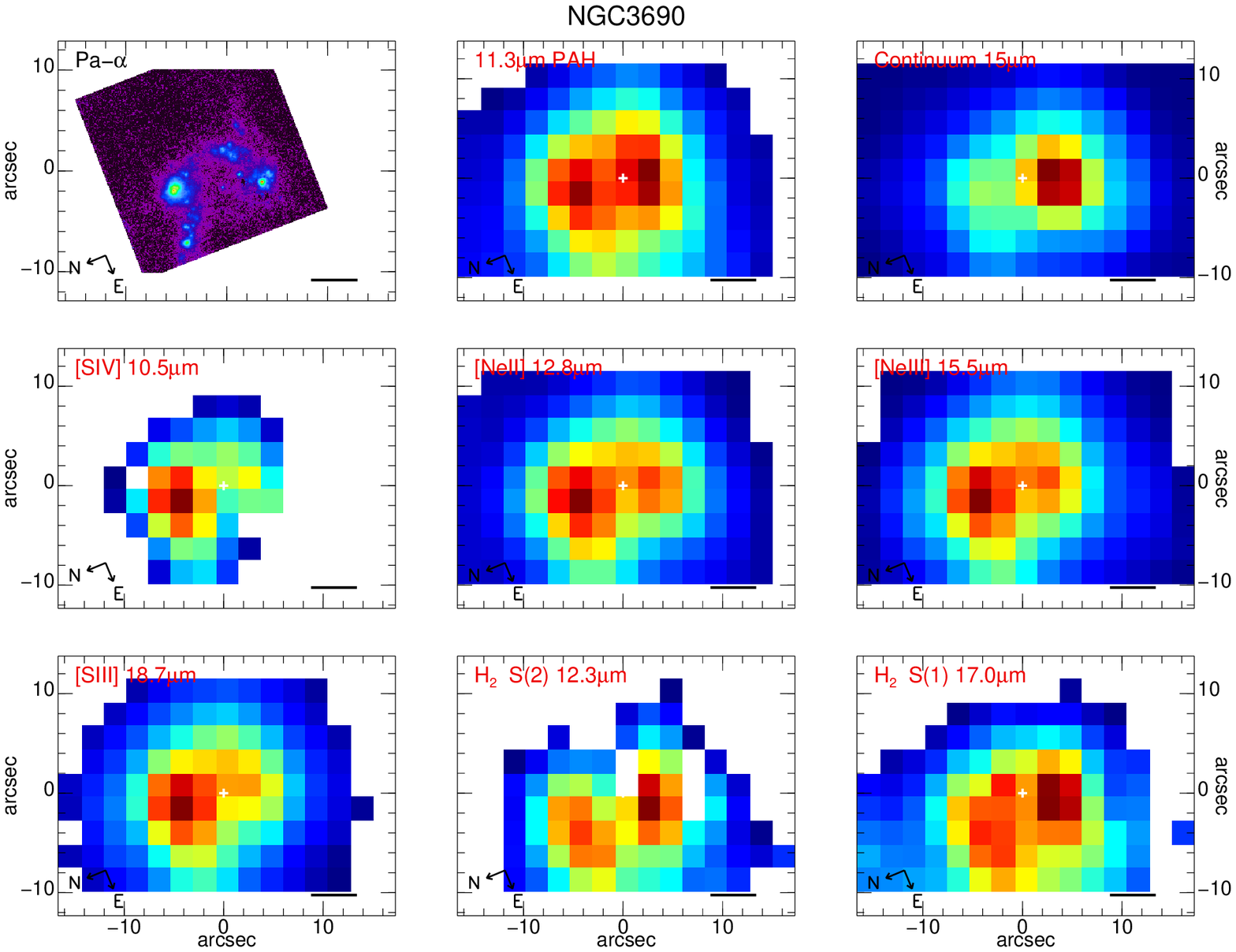}
\caption{Continued.}
\end{figure*}

\begin{figure*}
\addtocounter{figure}{-1}
\includegraphics[width=0.9\textwidth]{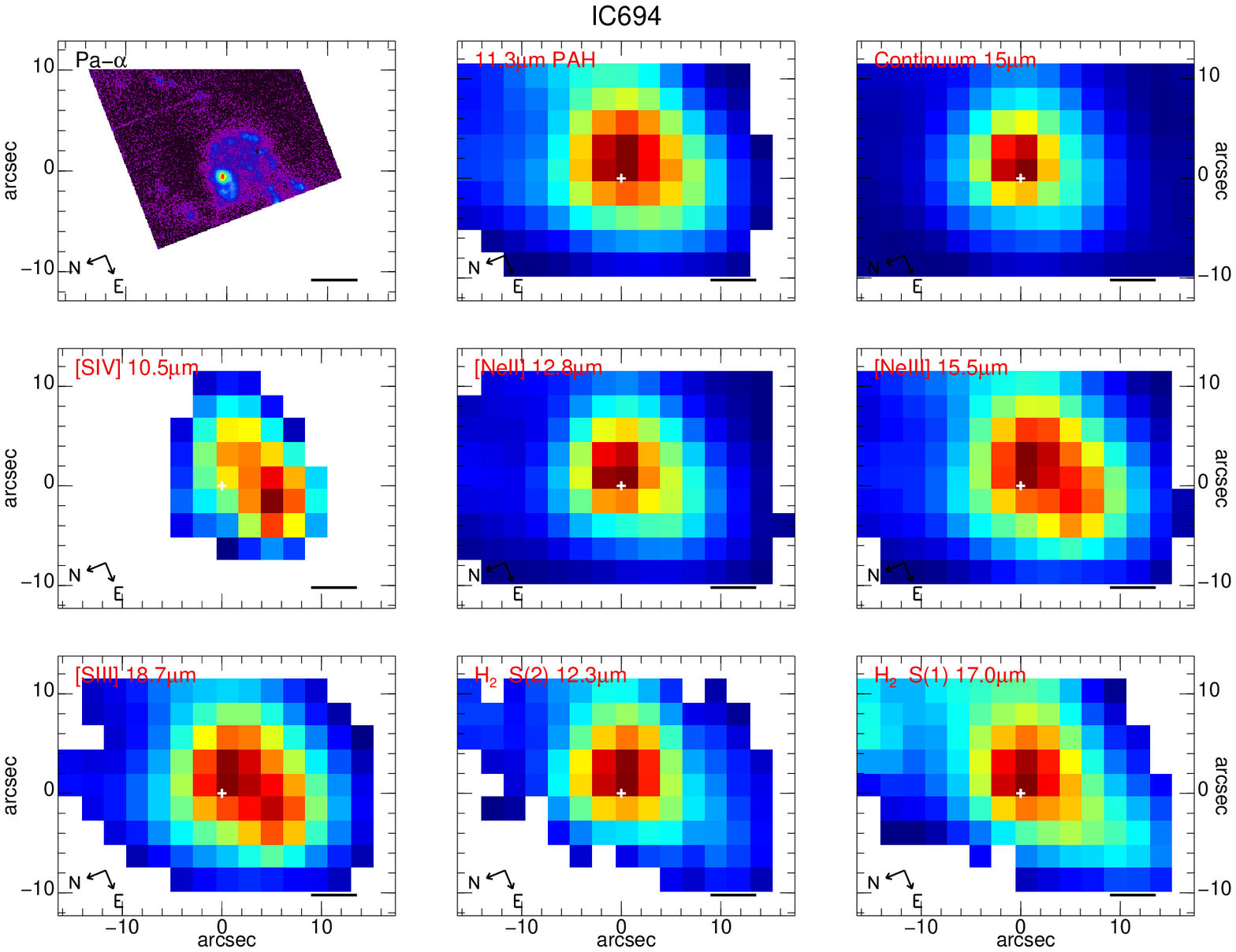}
\includegraphics[width=0.9\textwidth]{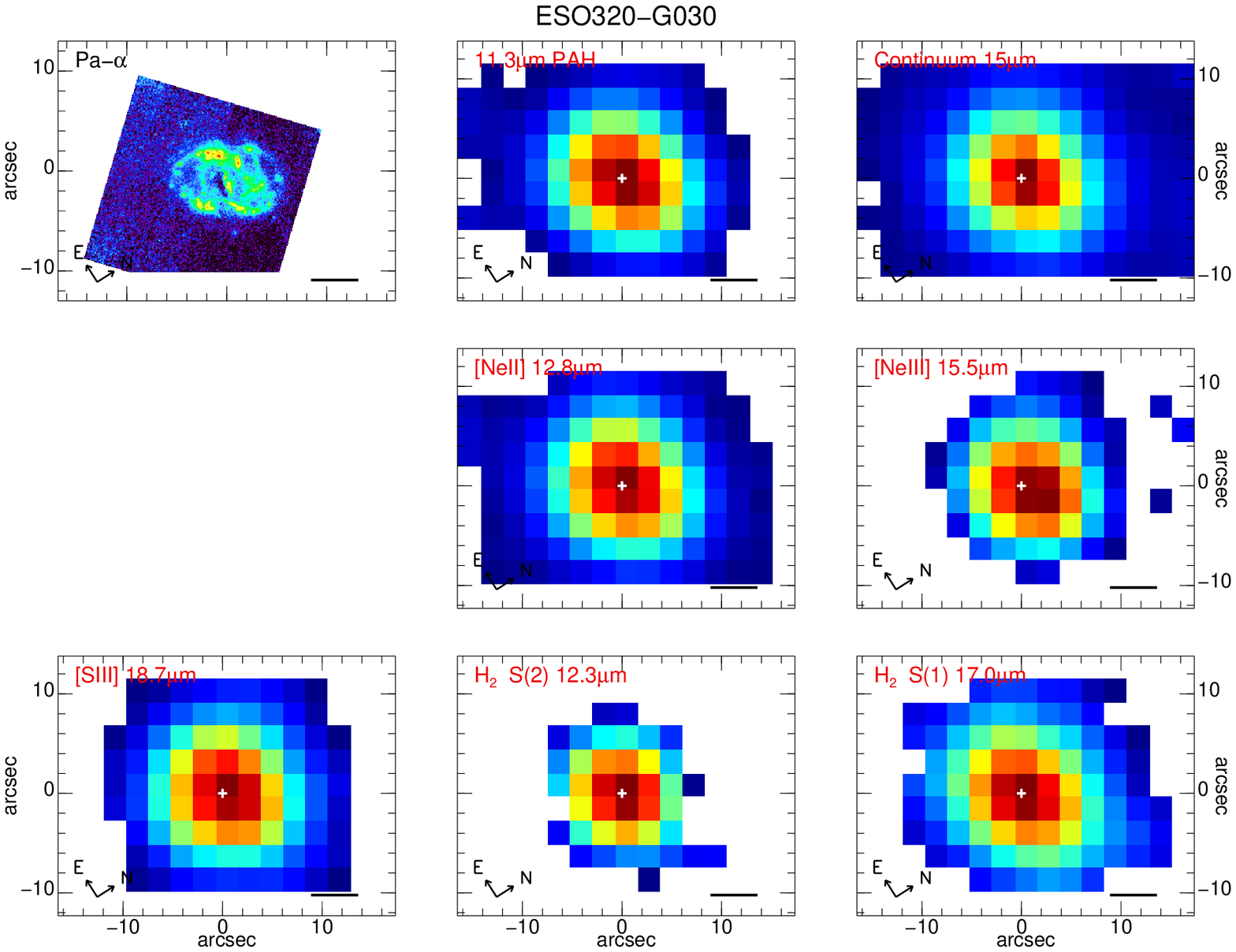}
\caption{Continued.}
\end{figure*}

\begin{figure*}
\addtocounter{figure}{-1}
\includegraphics[width=0.9\textwidth]{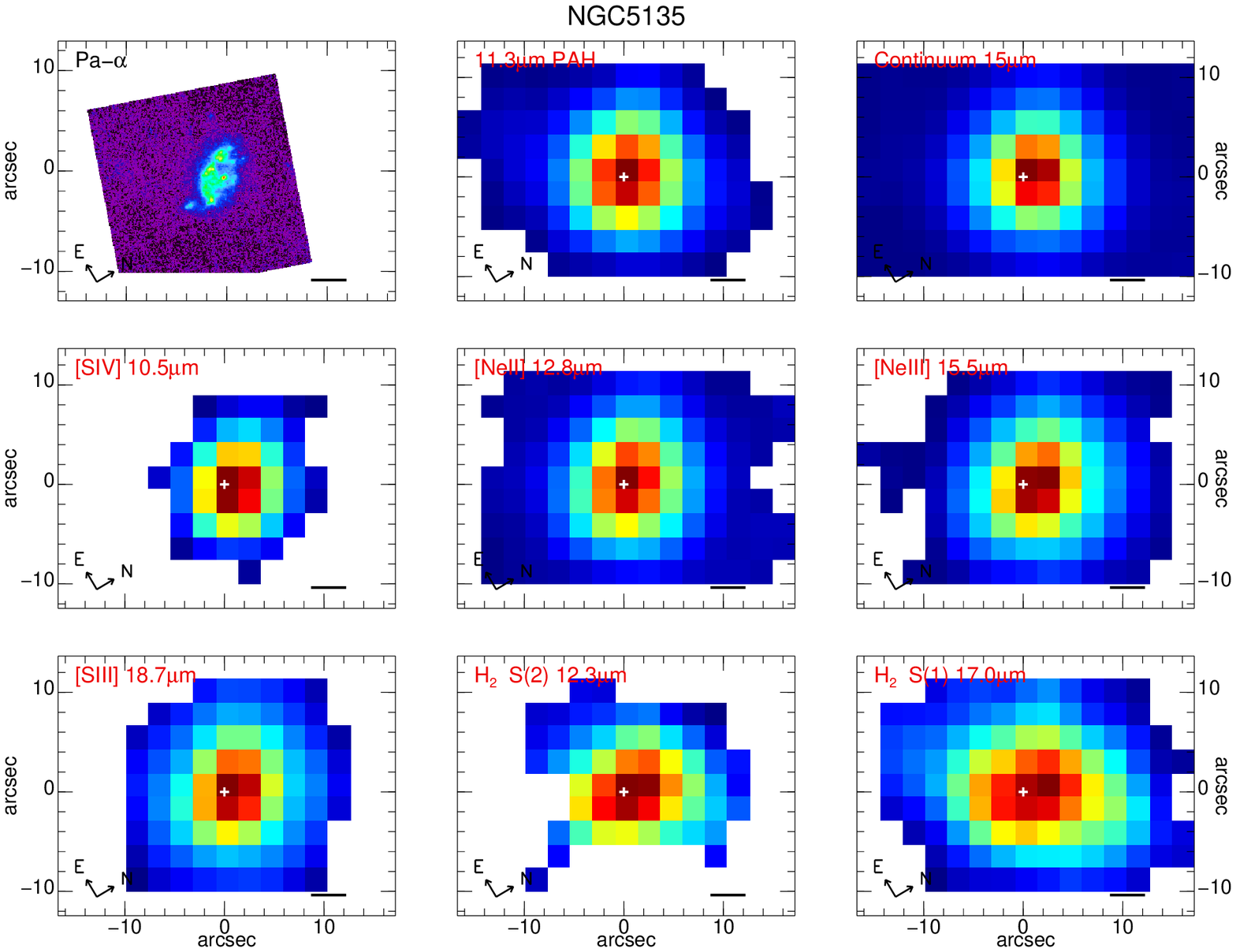}
\includegraphics[width=0.9\textwidth]{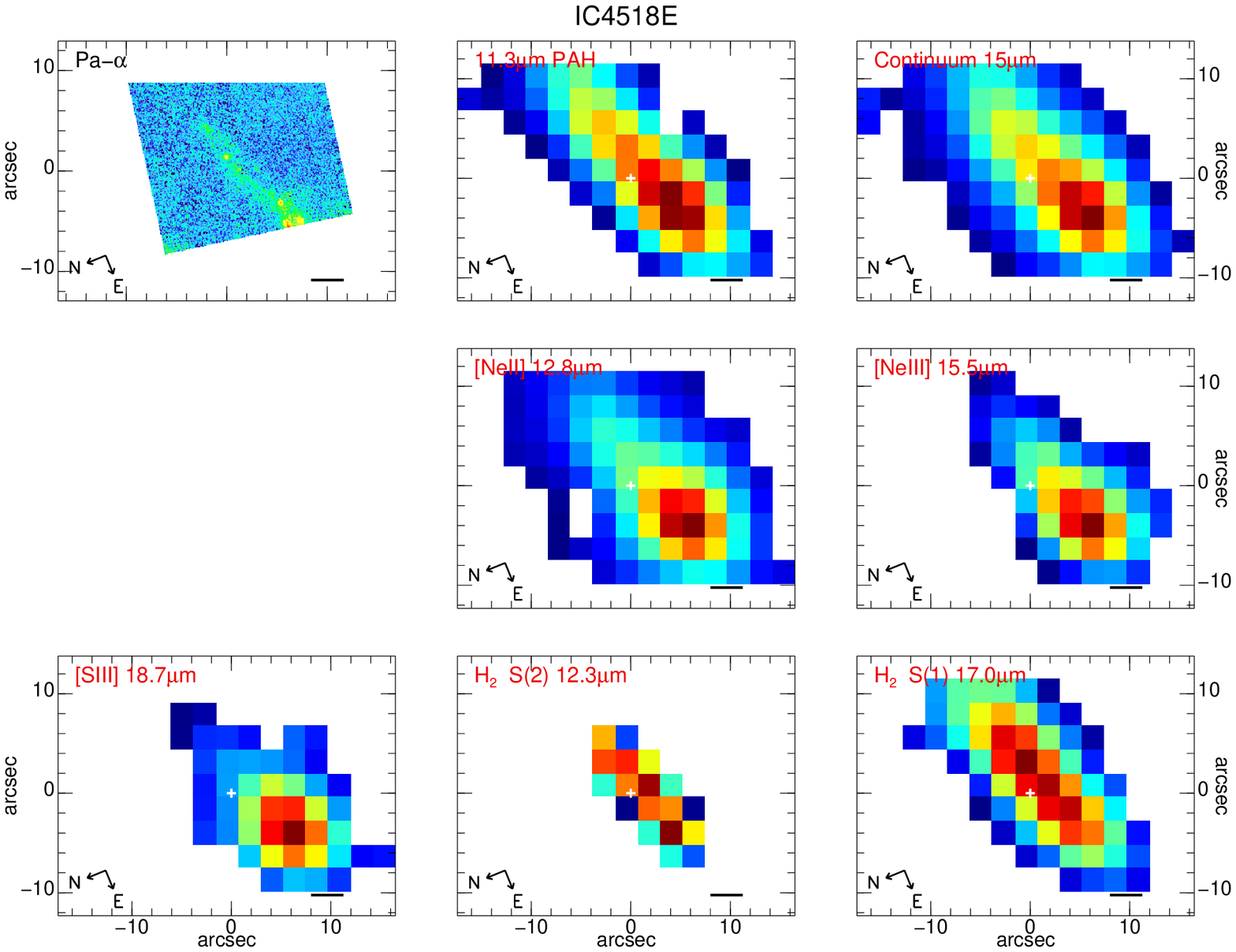}
\caption{Continued.}
\end{figure*}

\begin{figure*}
\addtocounter{figure}{-1}
\includegraphics[width=0.9\textwidth]{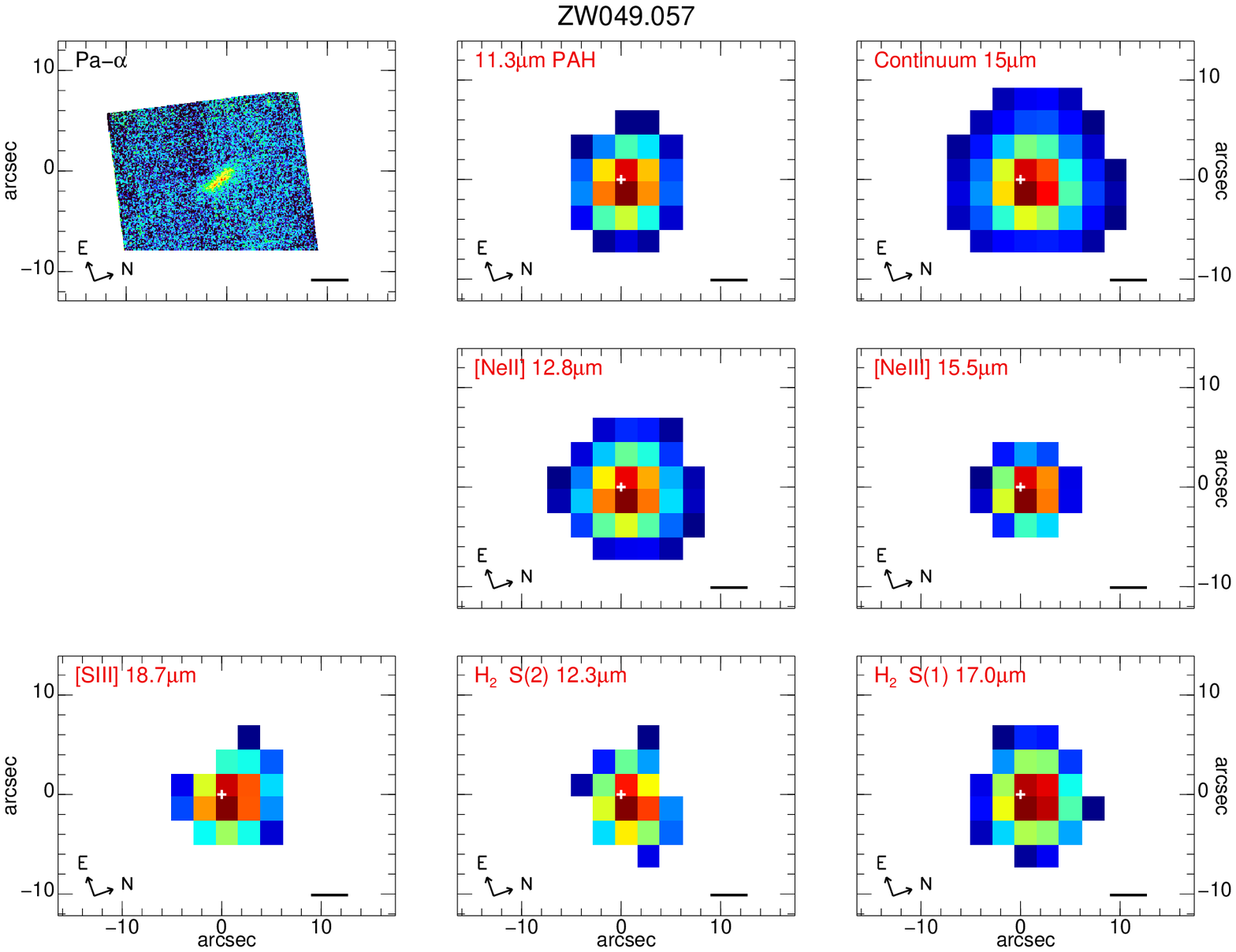}
\includegraphics[width=0.9\textwidth]{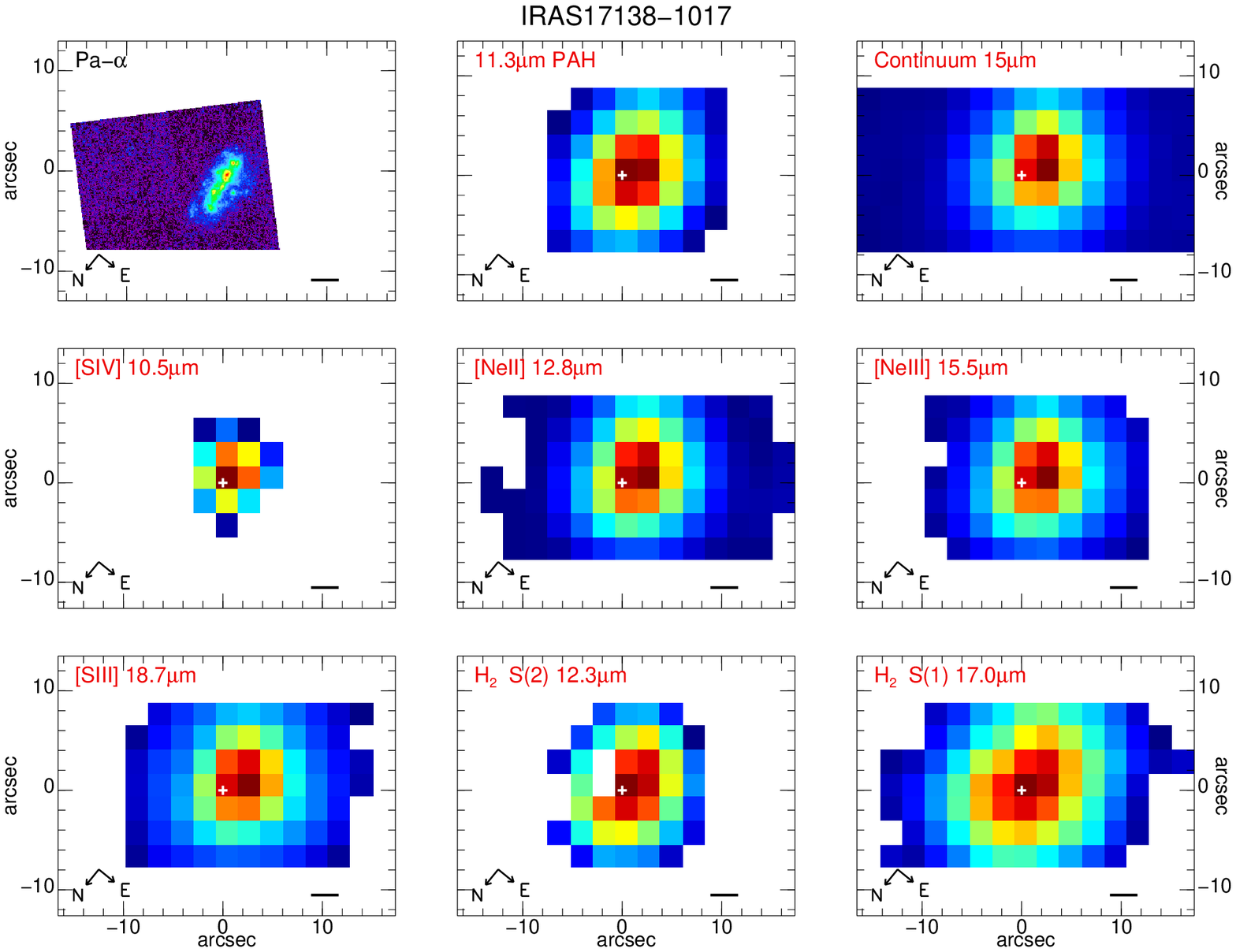}
\caption{Continued.}
\end{figure*}

\begin{figure*}
\includegraphics[width=0.9\textwidth]{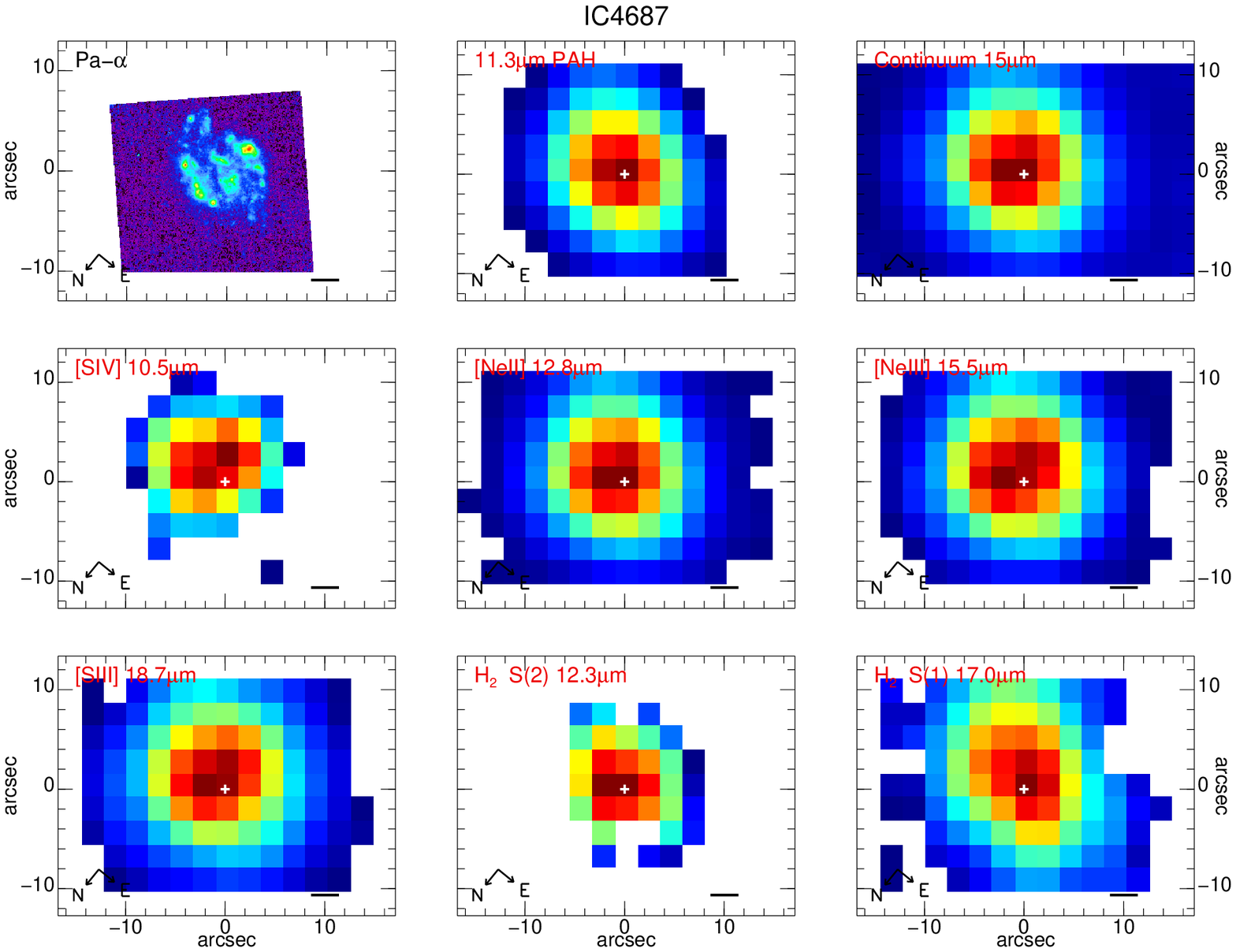}
\includegraphics[width=0.9\textwidth]{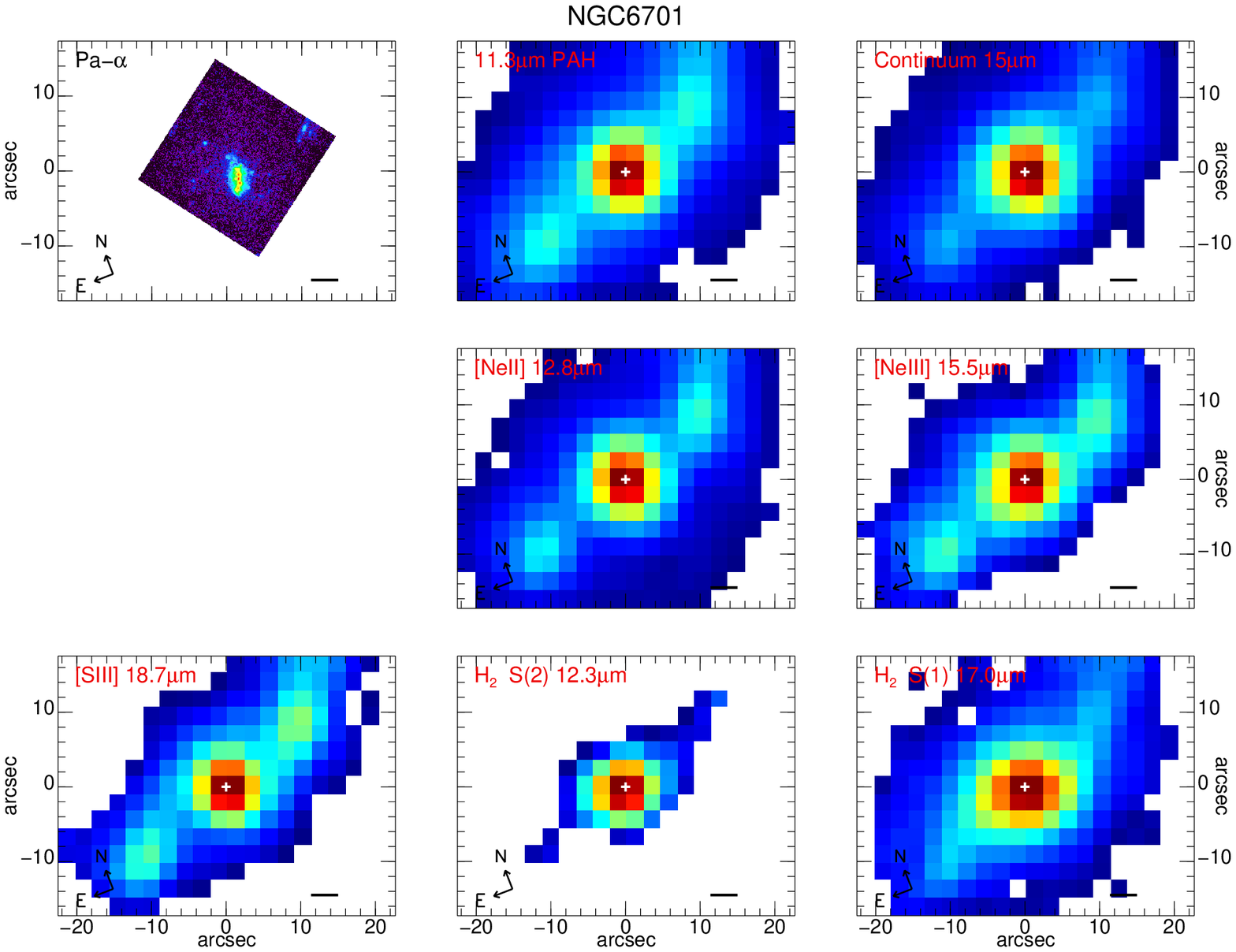}
\addtocounter{figure}{-1}
\caption{Continued.}
\end{figure*}

\begin{figure*}
\includegraphics[width=0.9\textwidth]{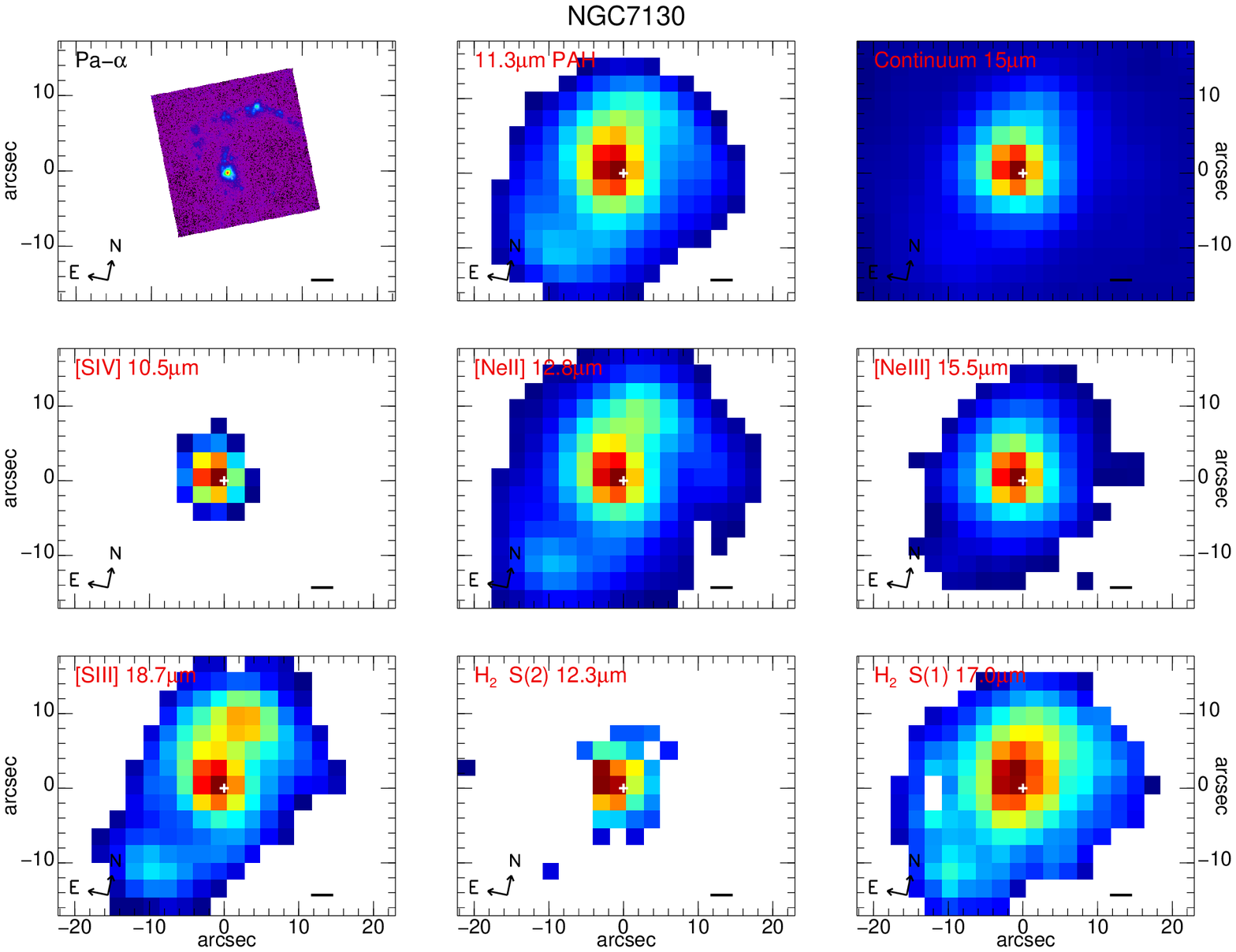}
\includegraphics[width=0.9\textwidth]{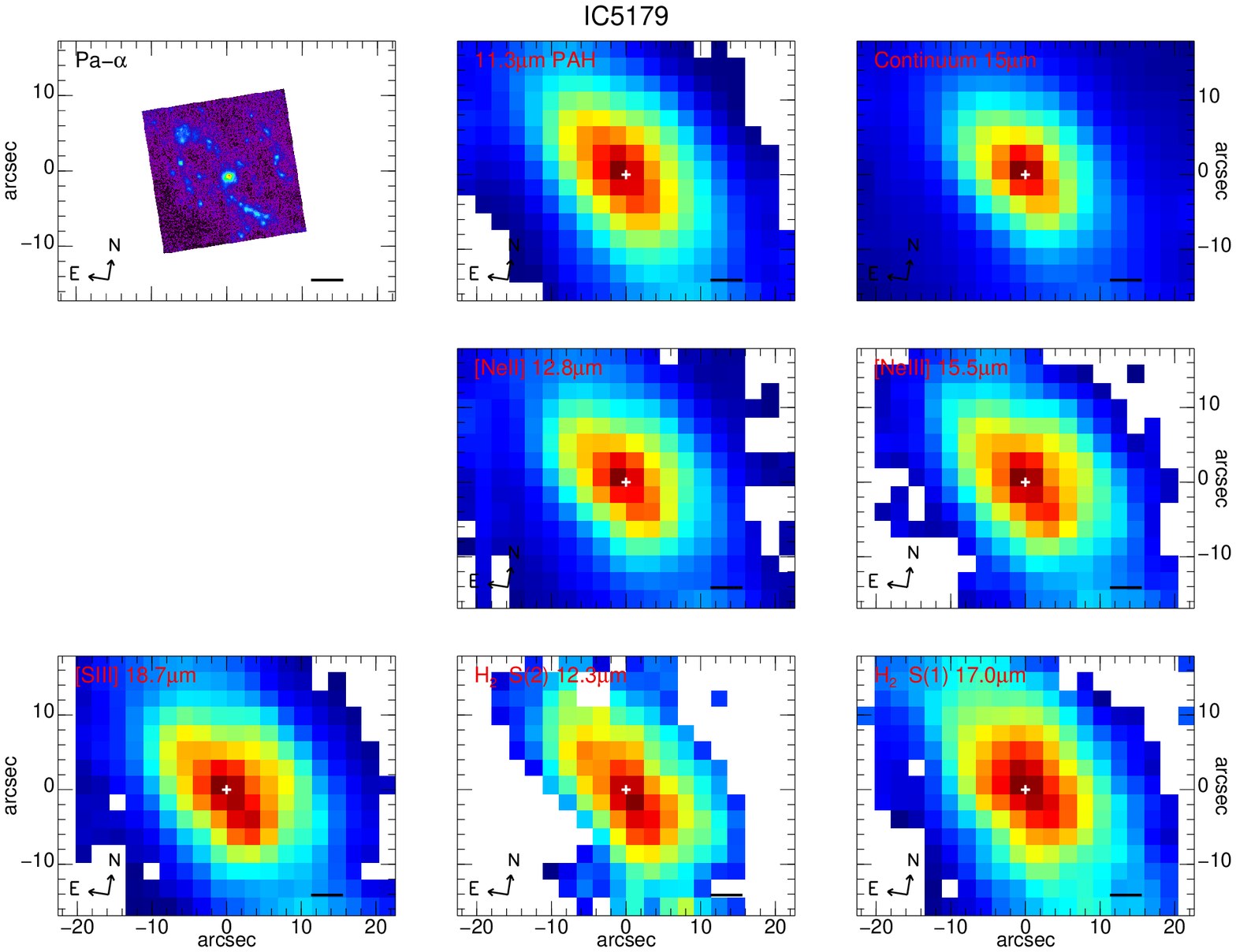}
\addtocounter{figure}{-1}
\caption{Continued.}
\end{figure*}

\begin{figure*}
\includegraphics[width=0.9\textwidth]{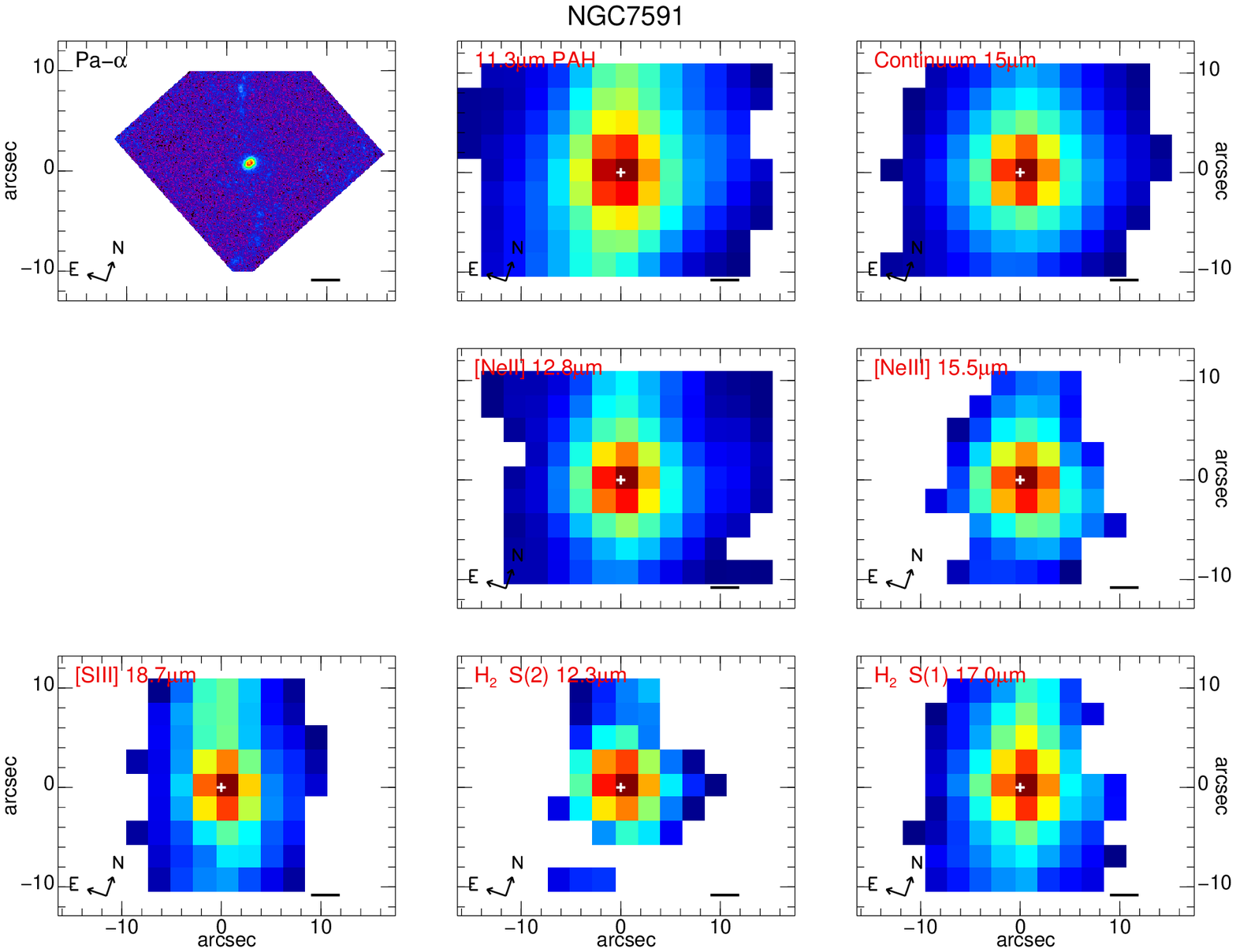}
\includegraphics[width=0.9\textwidth]{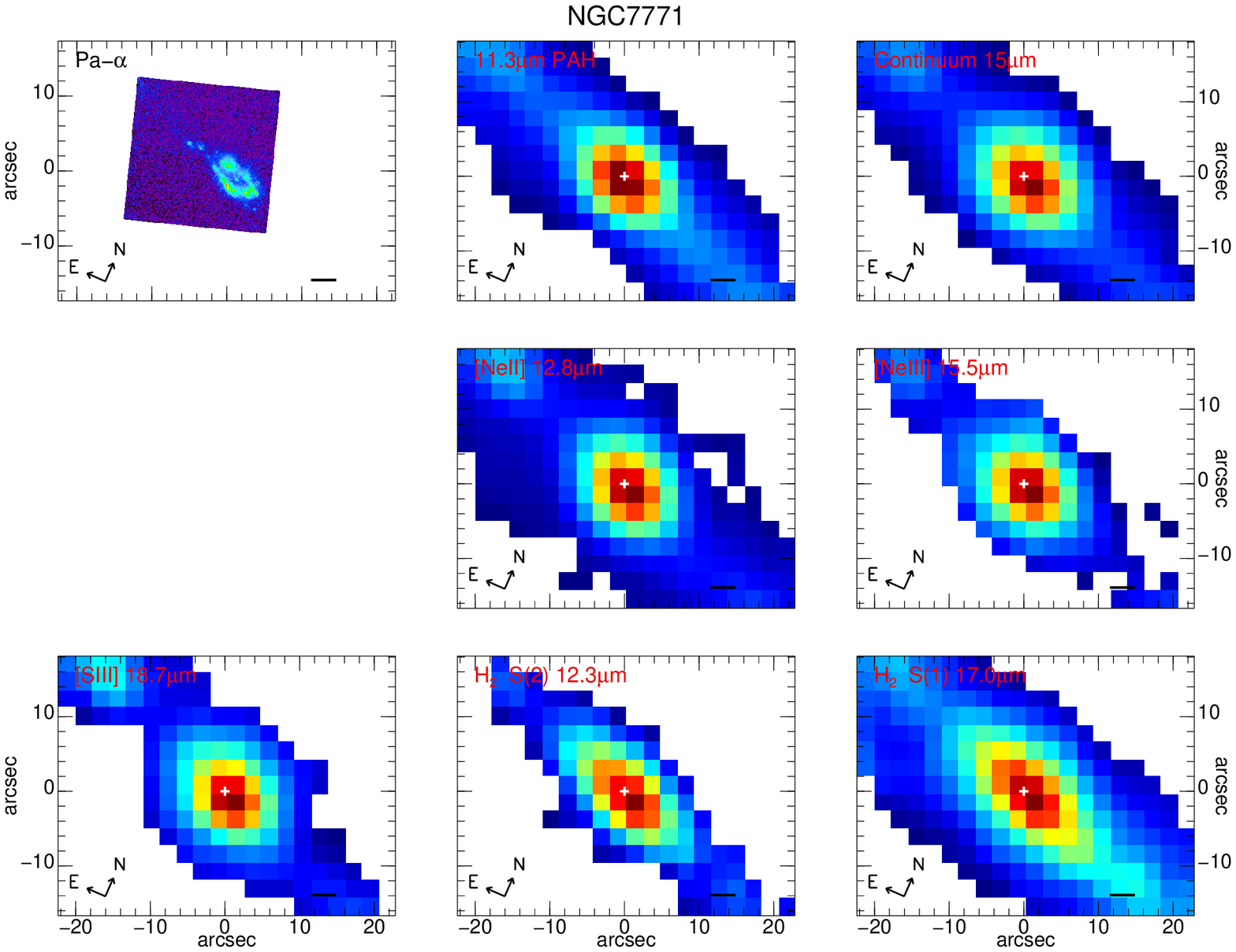}
\addtocounter{figure}{-1}
\caption{Continued.}
\end{figure*}

\begin{figure*}[!p]
\includegraphics[width=\textwidth]{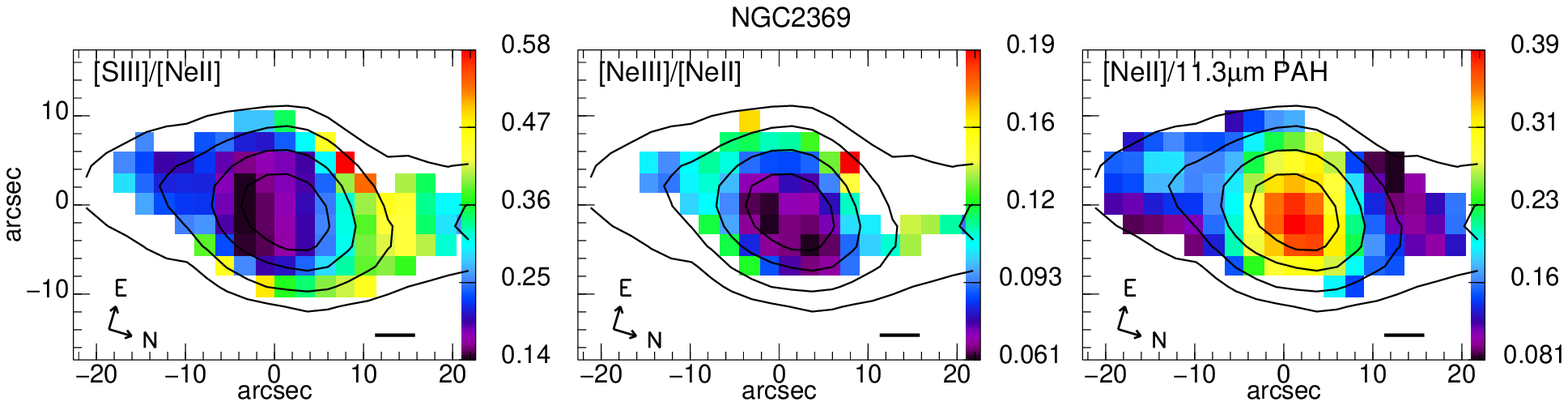}
\includegraphics[width=\textwidth]{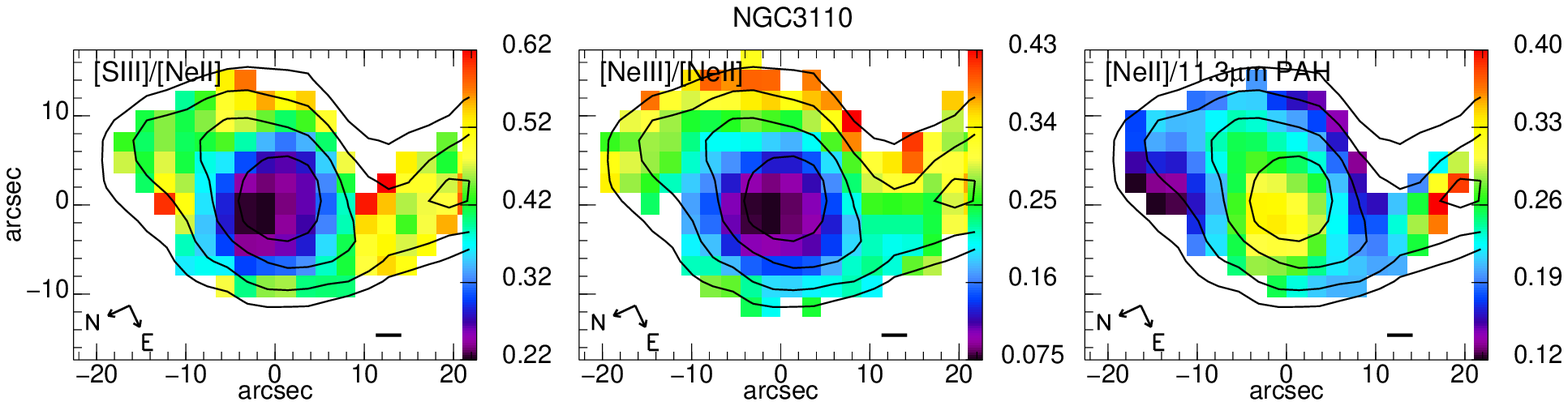}
\includegraphics[width=\textwidth]{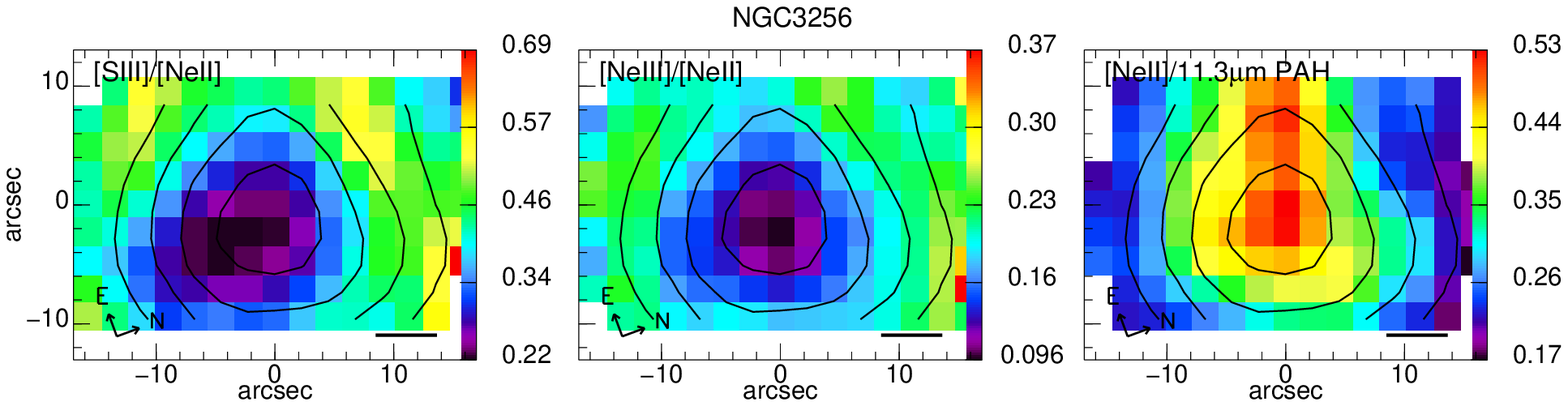}
\includegraphics[width=\textwidth]{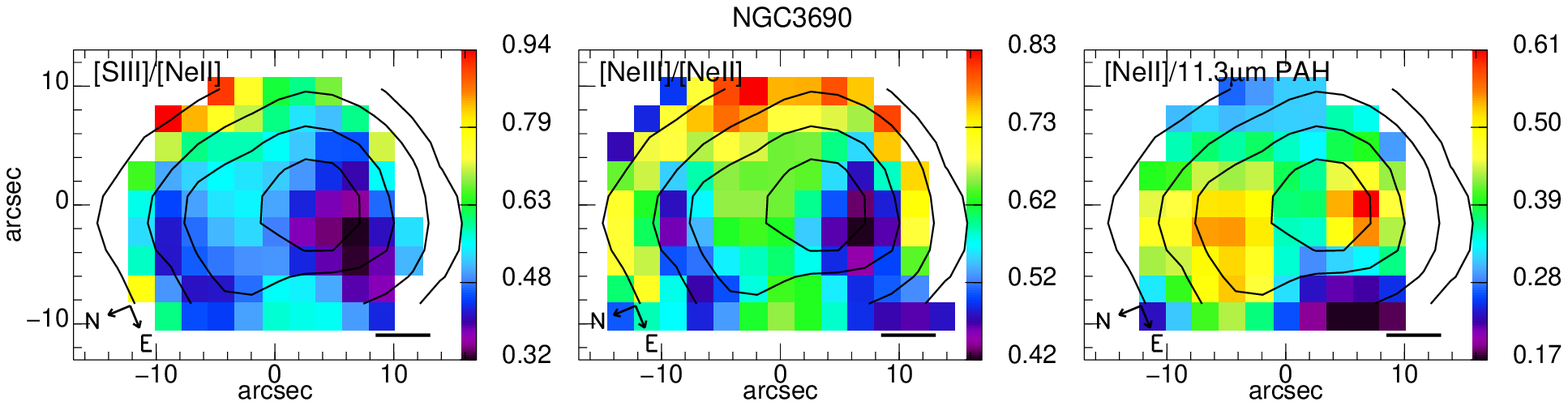}
\includegraphics[width=\textwidth]{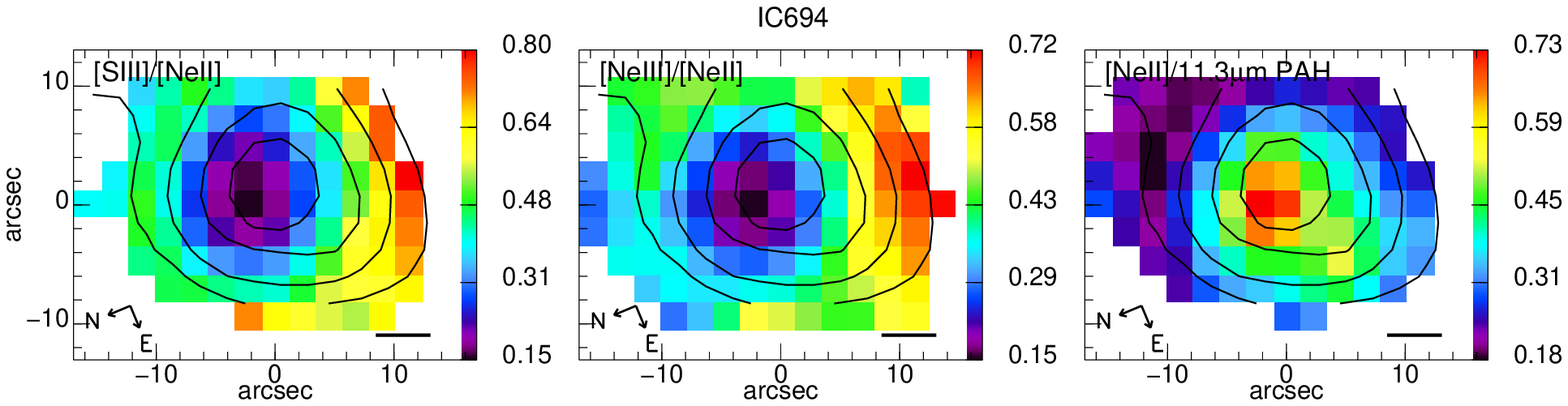}
\caption{Spitzer/IRS SH spectral maps of the \SIIIa\slash\Neii\ ratio, the \Neiii\slash\Neii\ ratio and the \Neii\slash\PAHonce . The 15.0\micron\ continuum contours are displayed to guide the eye. The image orientation is indicated on the maps for each galaxy. The scale represents 1 kpc. The ratio maps are shown in a linear scale.}
\label{fig_co_sh}
\end{figure*}

\begin{figure*}
\addtocounter{figure}{-1}
\includegraphics[width=\textwidth]{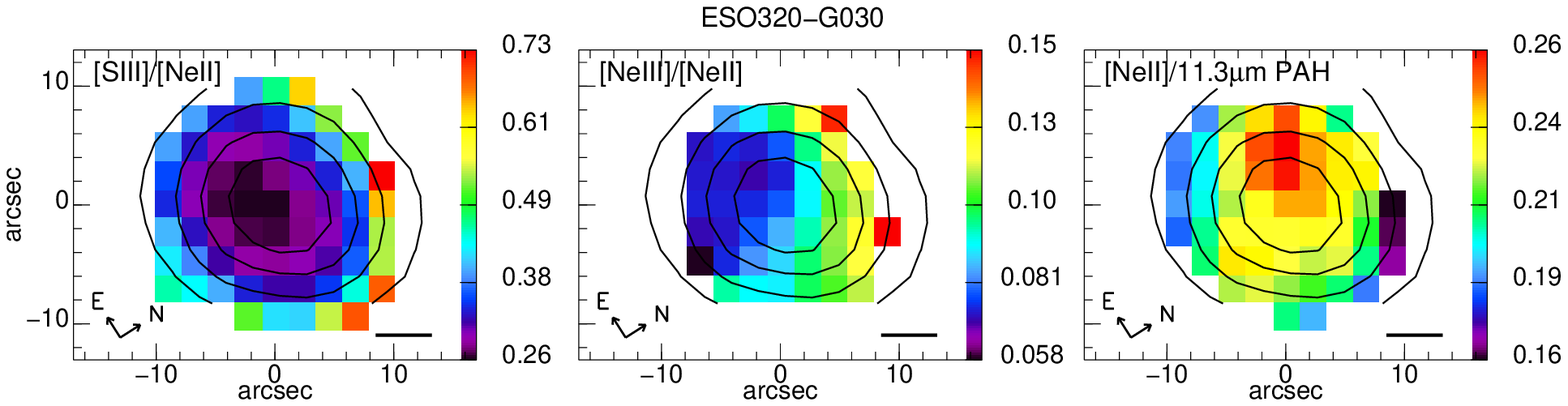}
\includegraphics[width=\textwidth]{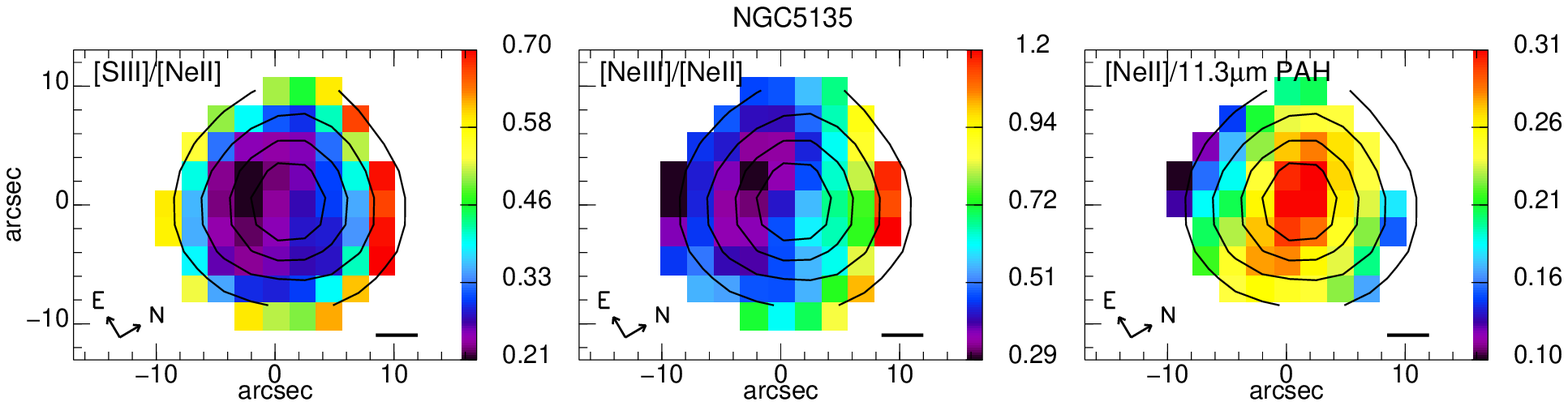}
\includegraphics[width=\textwidth]{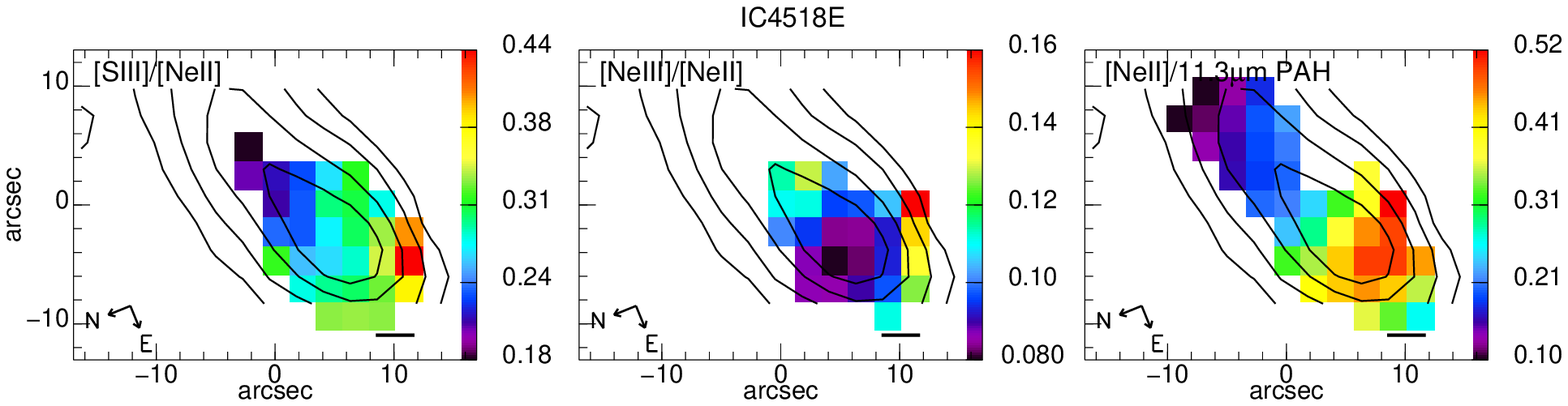}
\includegraphics[width=\textwidth]{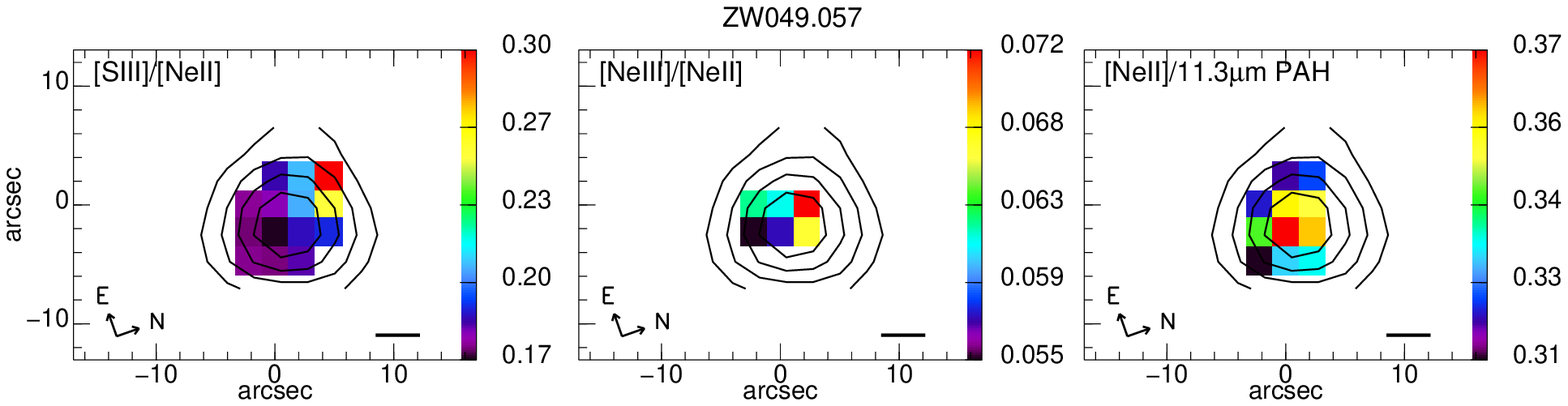}
\includegraphics[width=\textwidth]{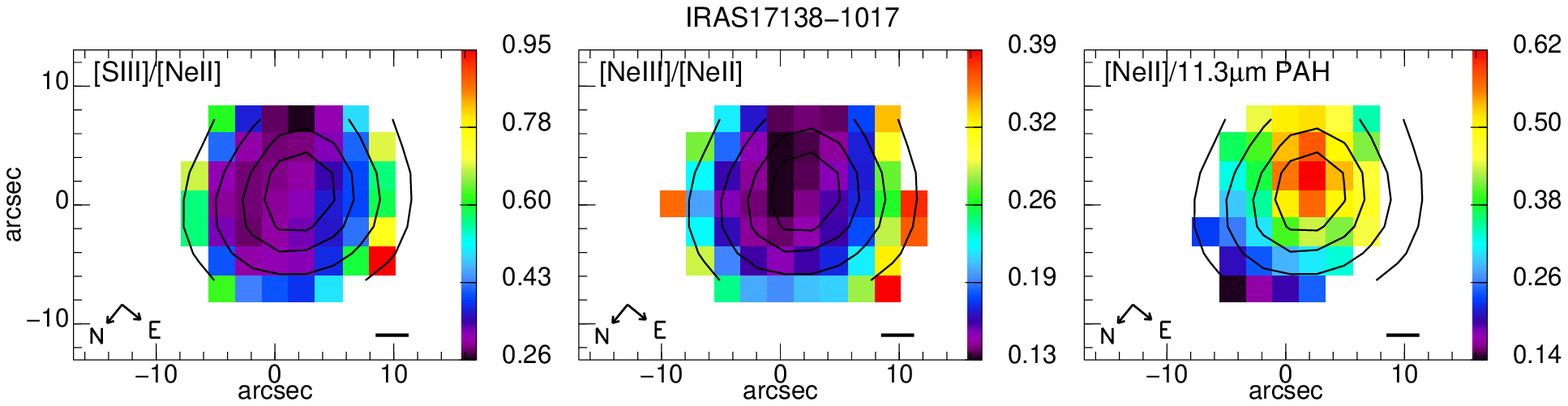}
\caption{Continued.}
\end{figure*}

\begin{figure*}
\addtocounter{figure}{-1}
\includegraphics[width=\textwidth]{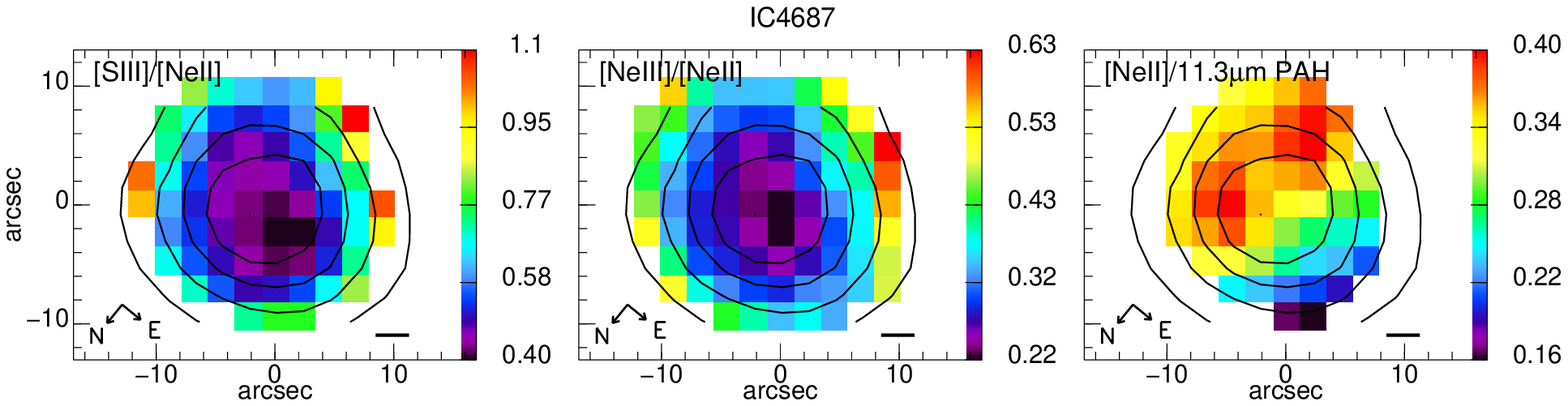}
\includegraphics[width=\textwidth]{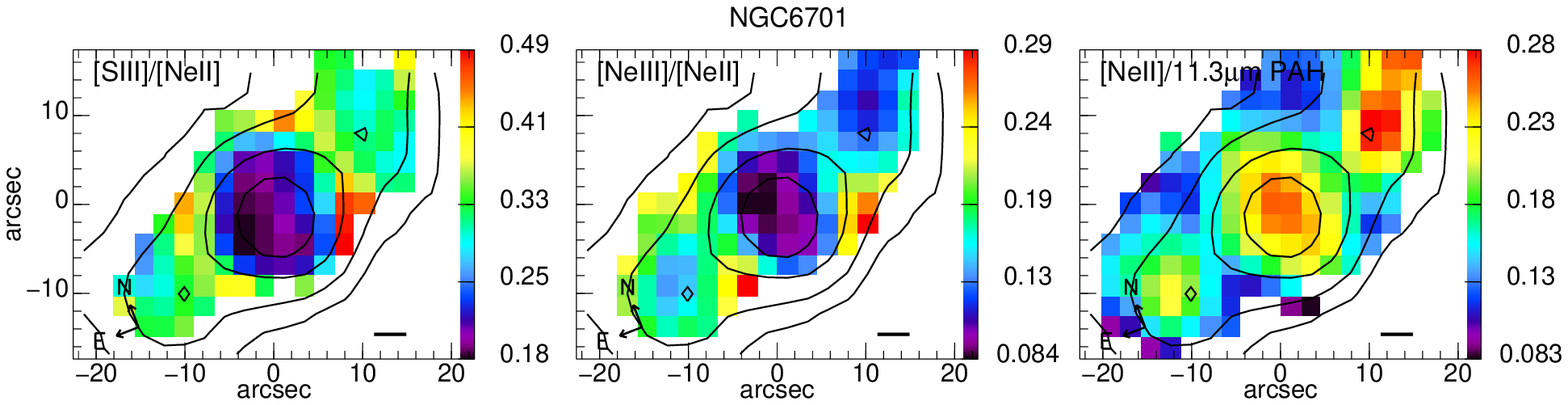}
\includegraphics[width=\textwidth]{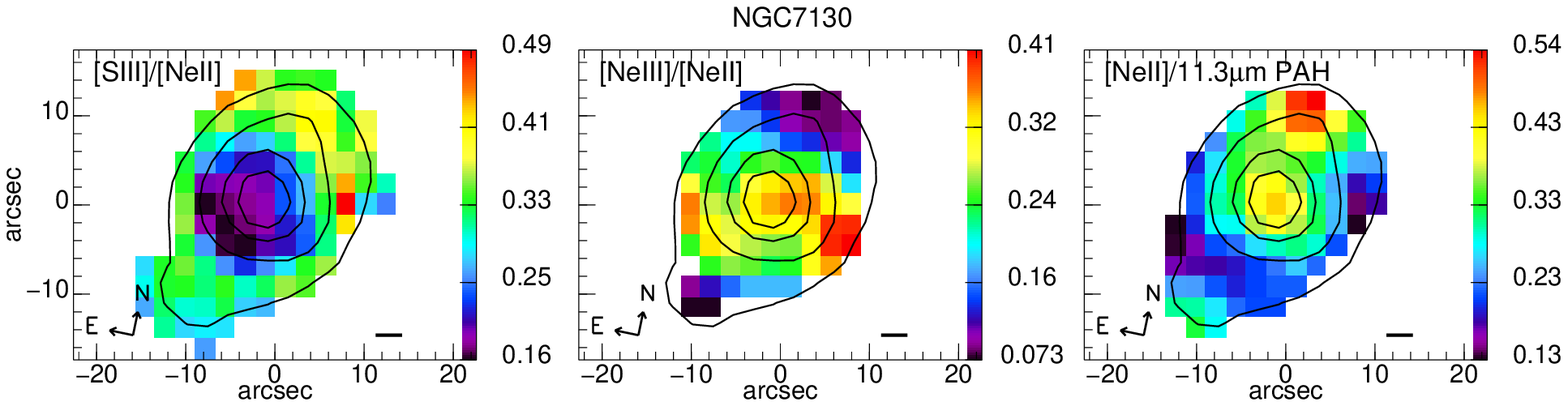}
\includegraphics[width=\textwidth]{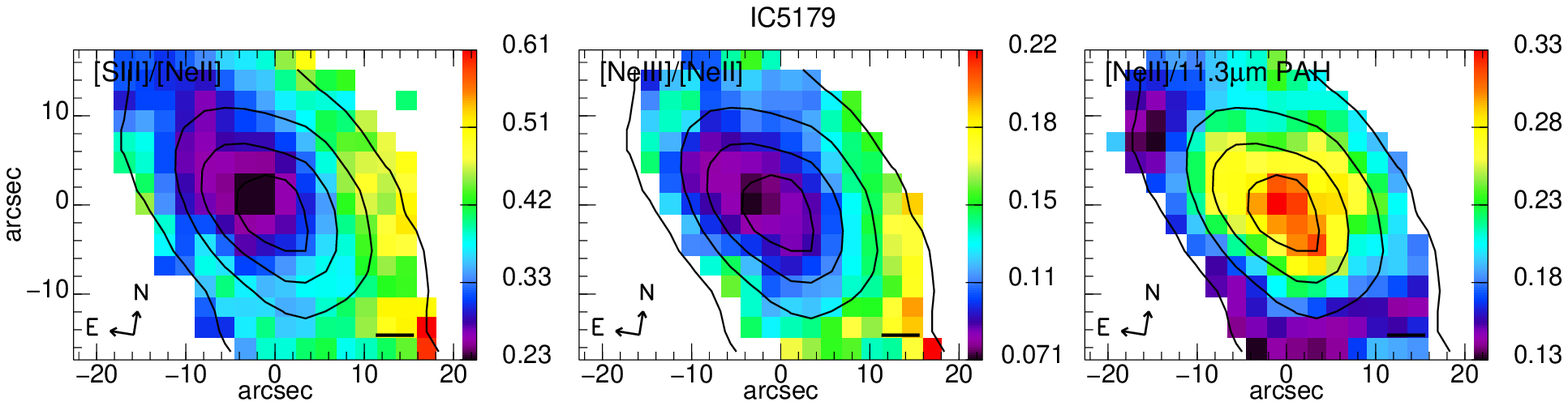}
\includegraphics[width=\textwidth]{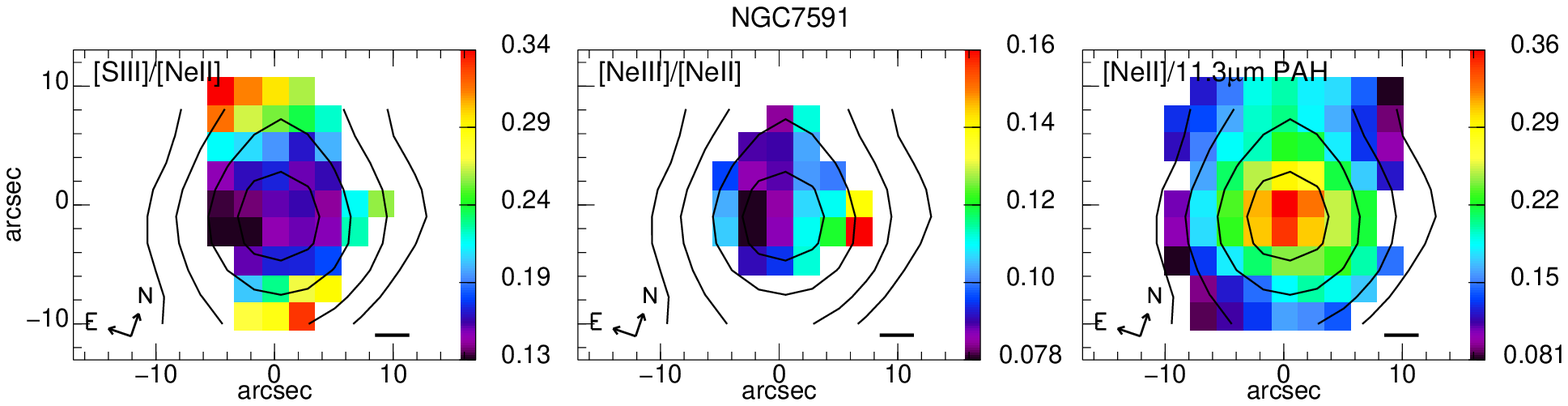}
\caption{Continued.}
\end{figure*}

\begin{figure*}
\addtocounter{figure}{-1}
\includegraphics[width=\textwidth]{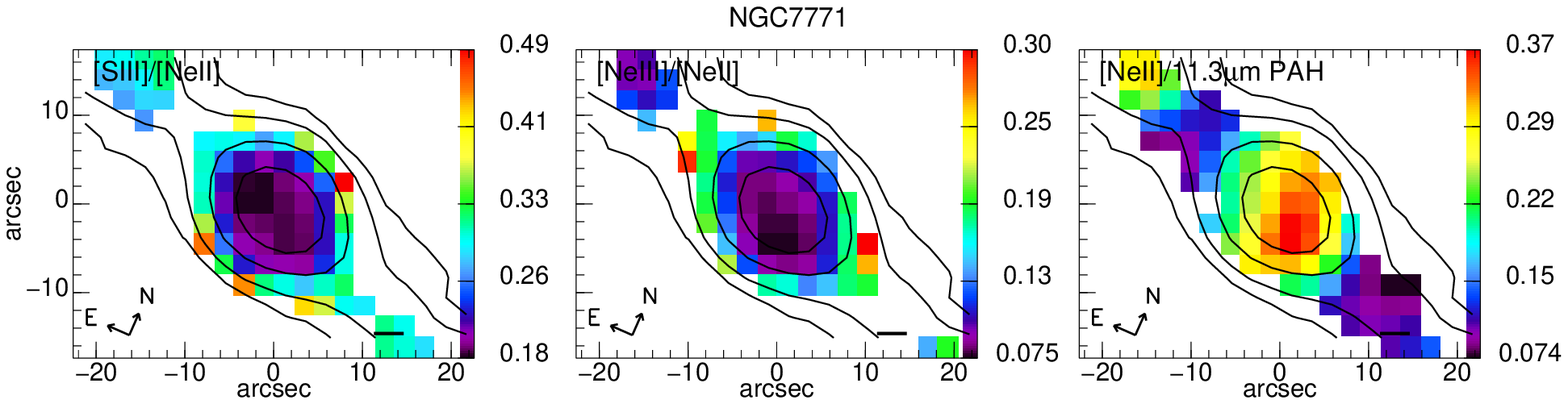}
\caption{Continued.}
\end{figure*}

\setcounter{figure}{7}
\begin{figure*}[!p]
\includegraphics[width=0.33\textwidth]{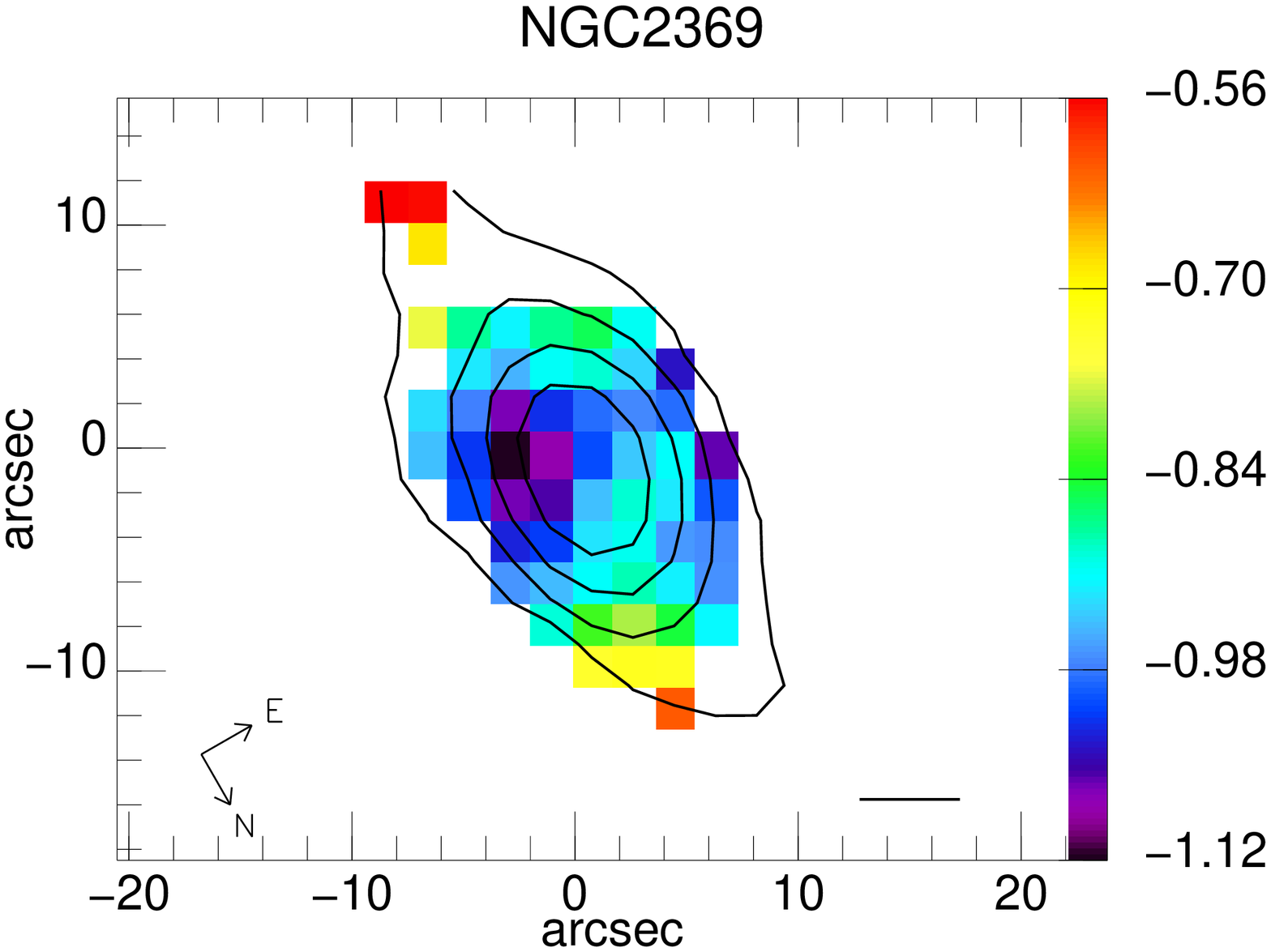}
\includegraphics[width=0.33\textwidth]{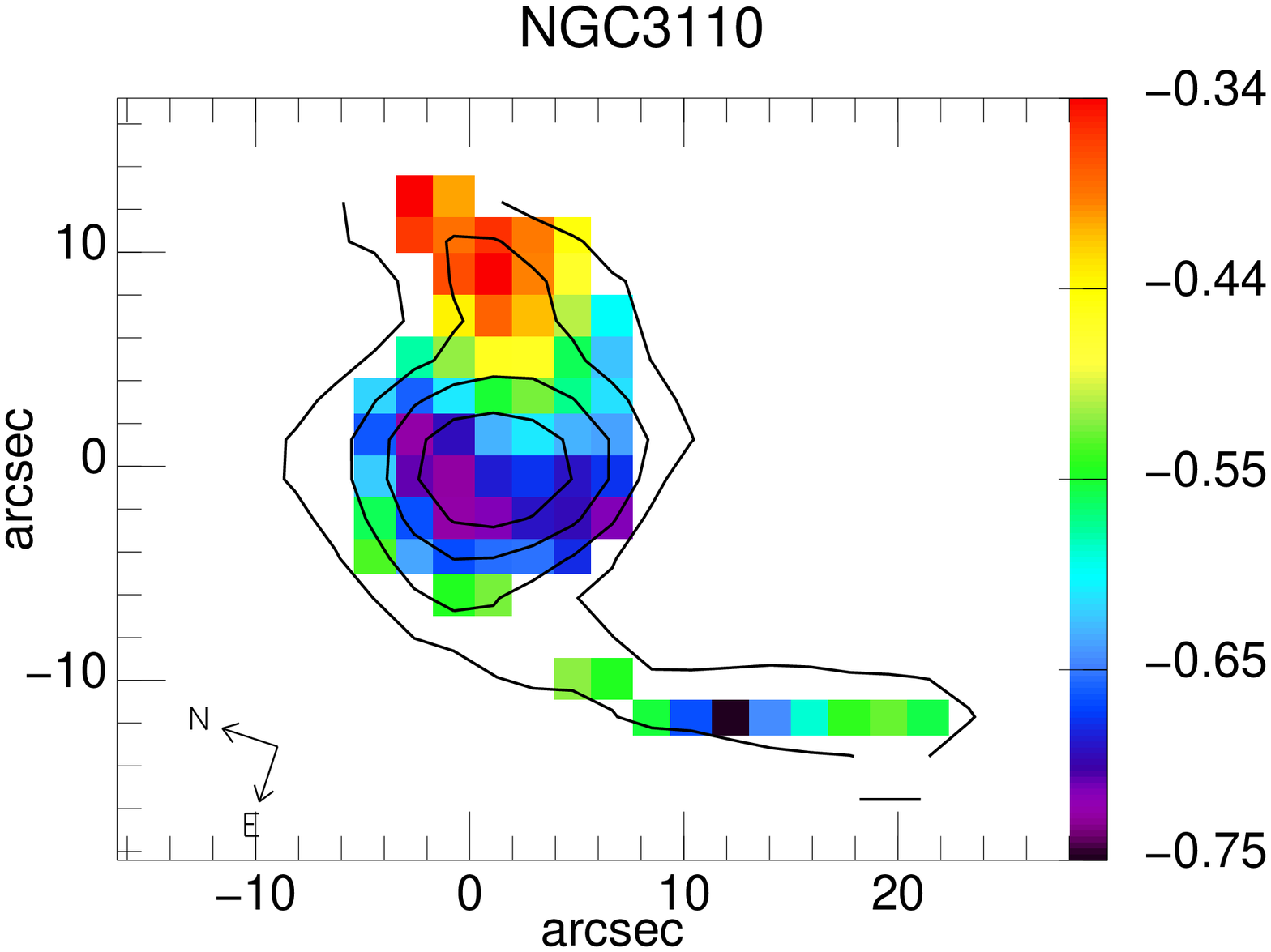}
\includegraphics[width=0.33\textwidth]{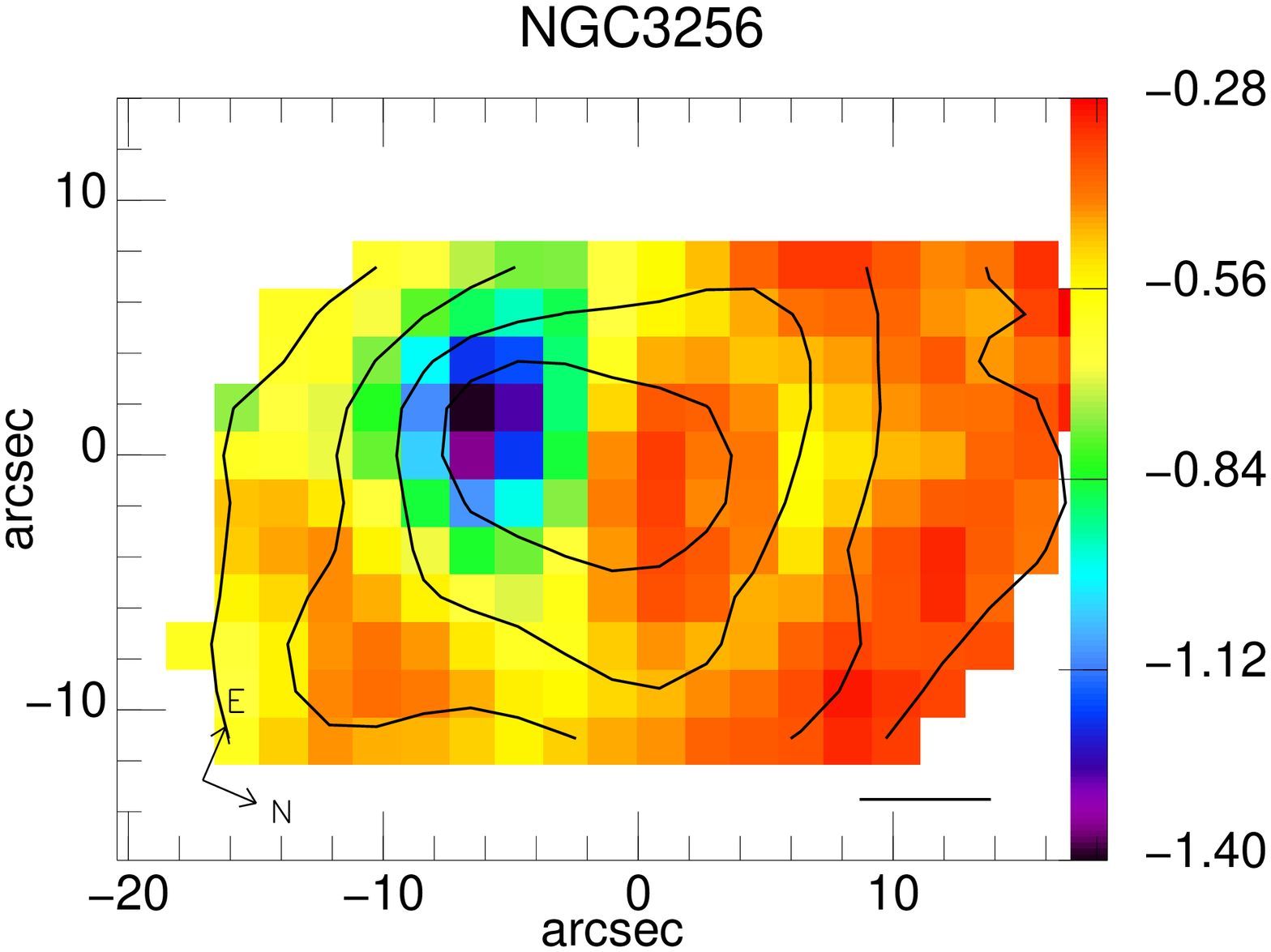}
\includegraphics[width=0.33\textwidth]{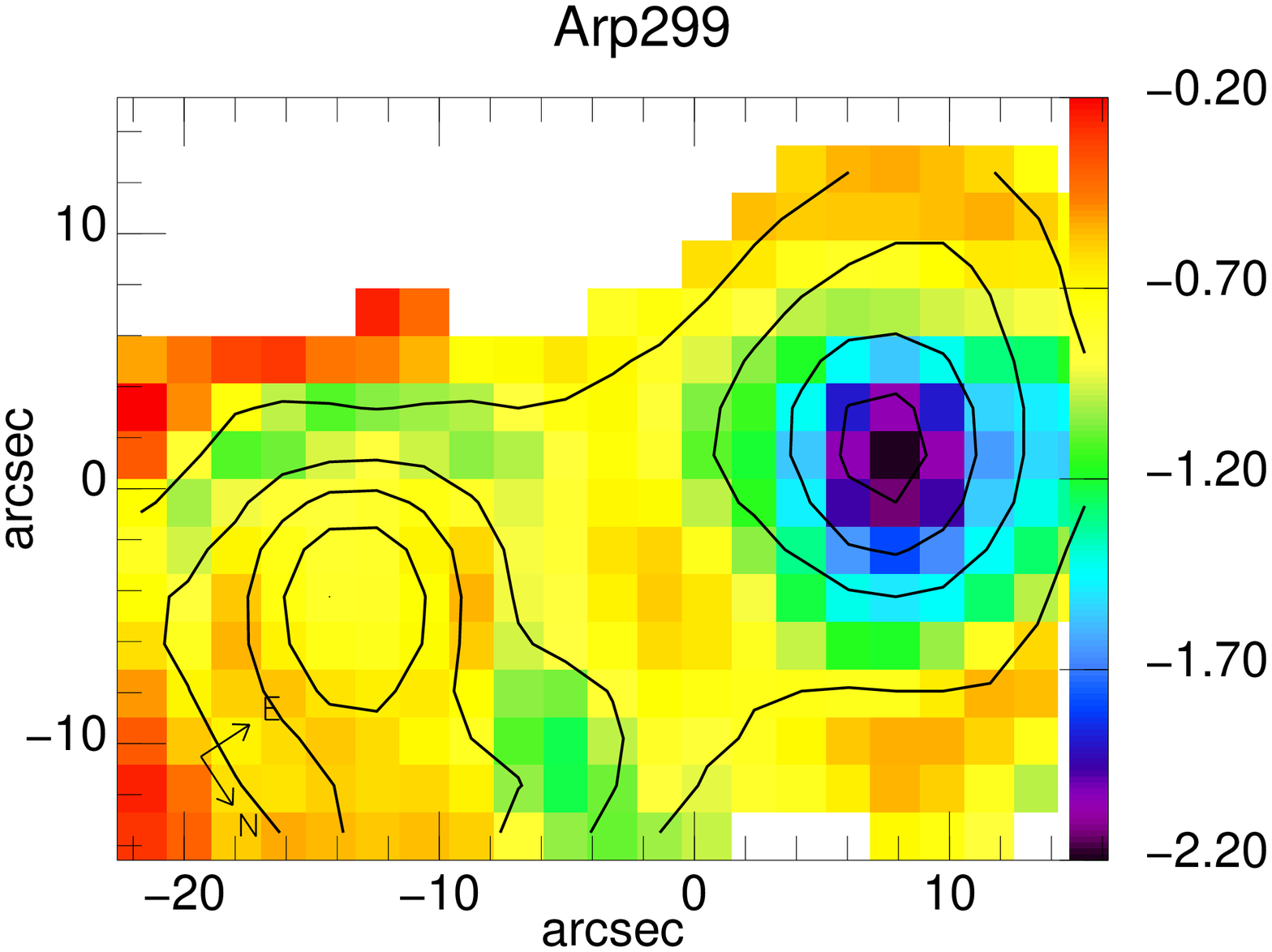}
\includegraphics[width=0.33\textwidth]{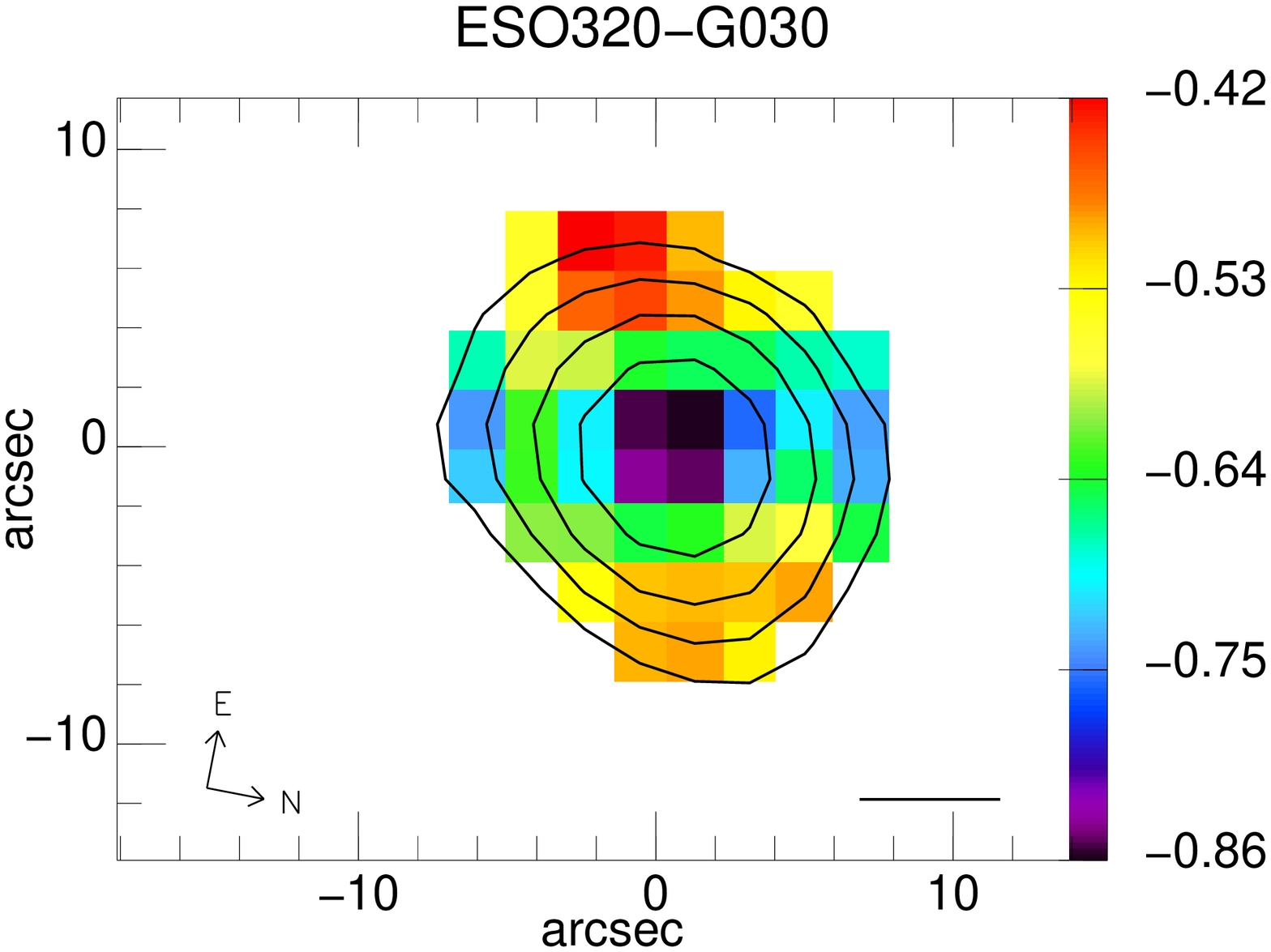}
\includegraphics[width=0.33\textwidth]{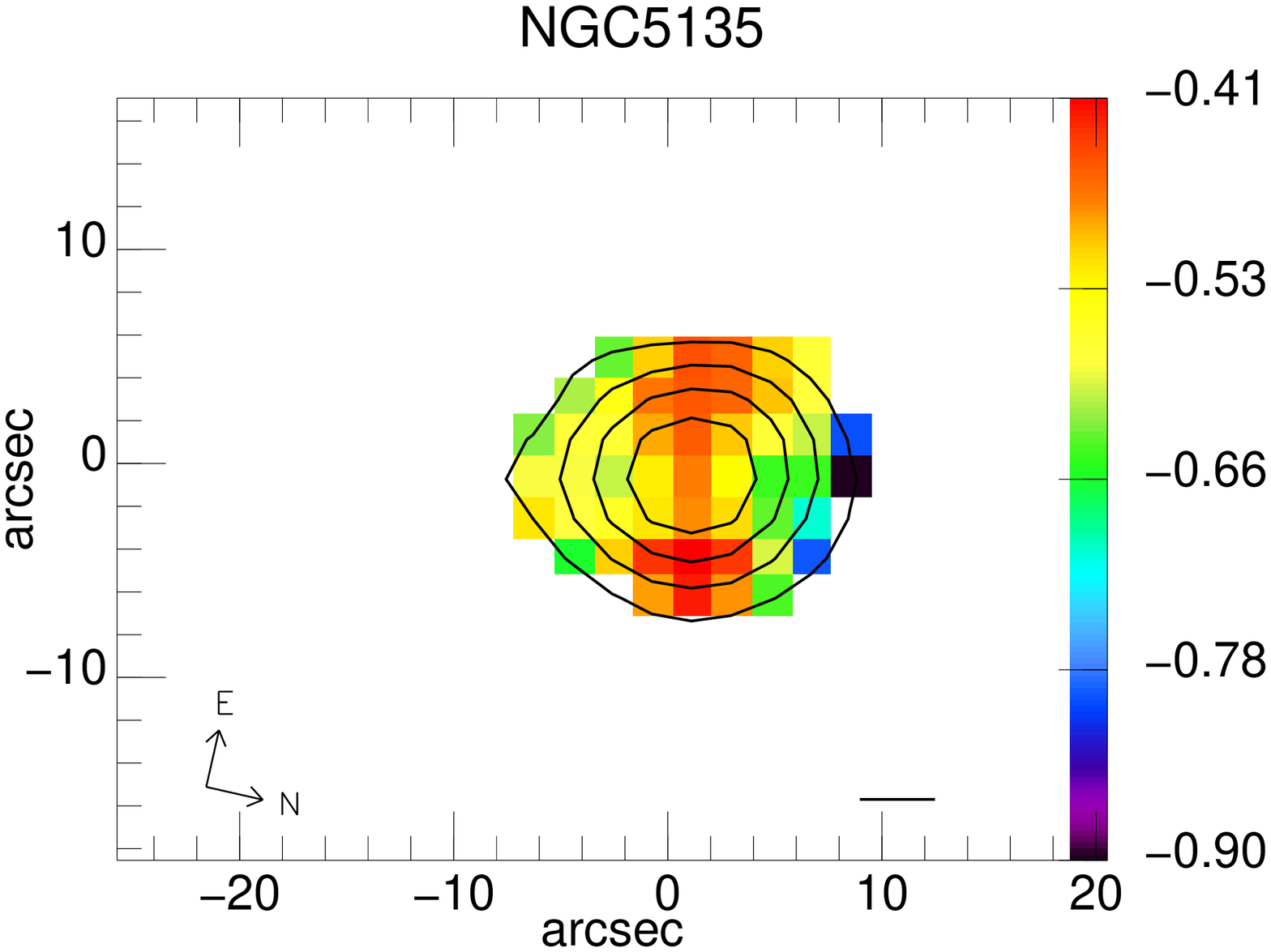}
\includegraphics[width=0.33\textwidth]{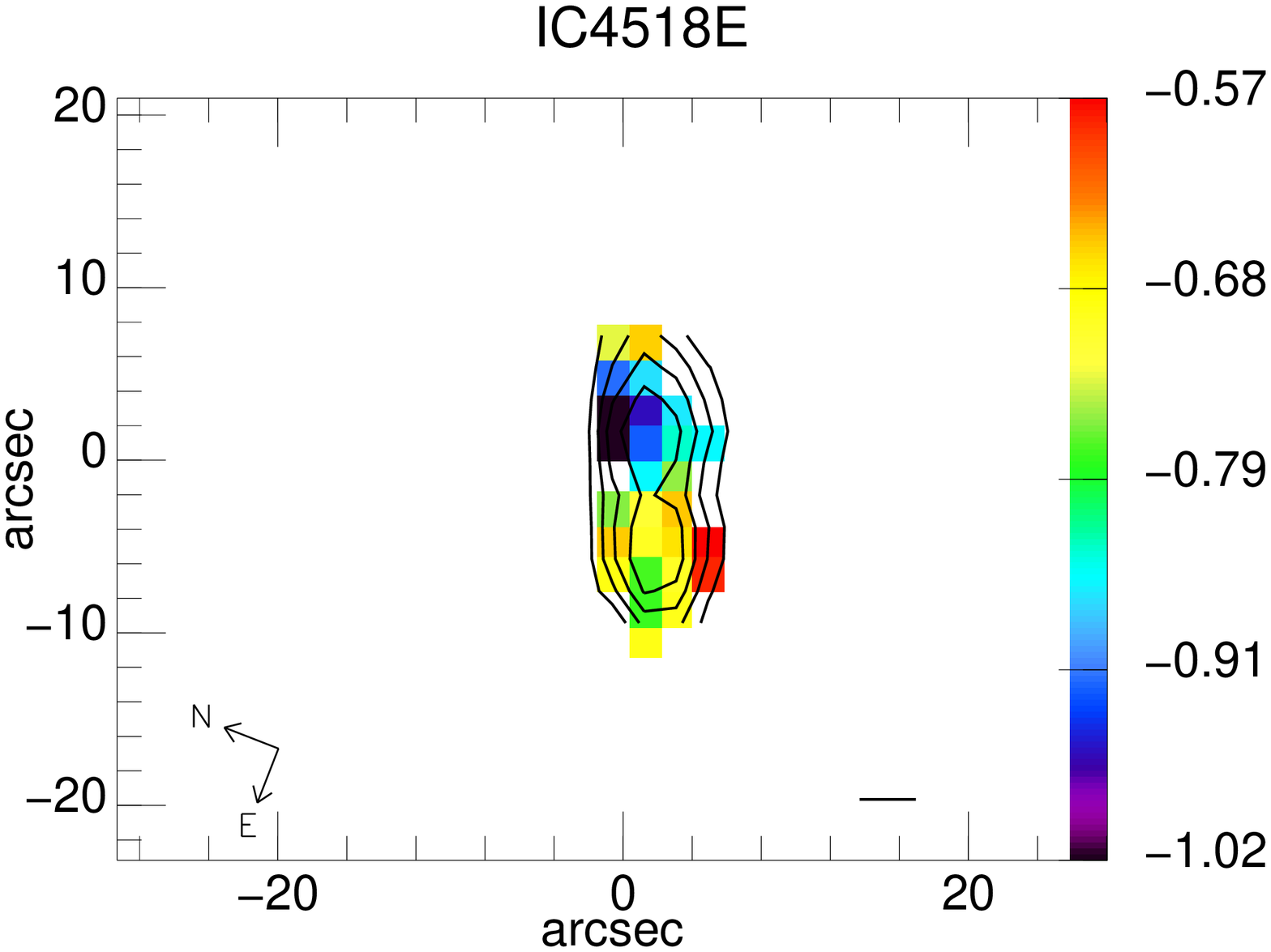}
\includegraphics[width=0.33\textwidth]{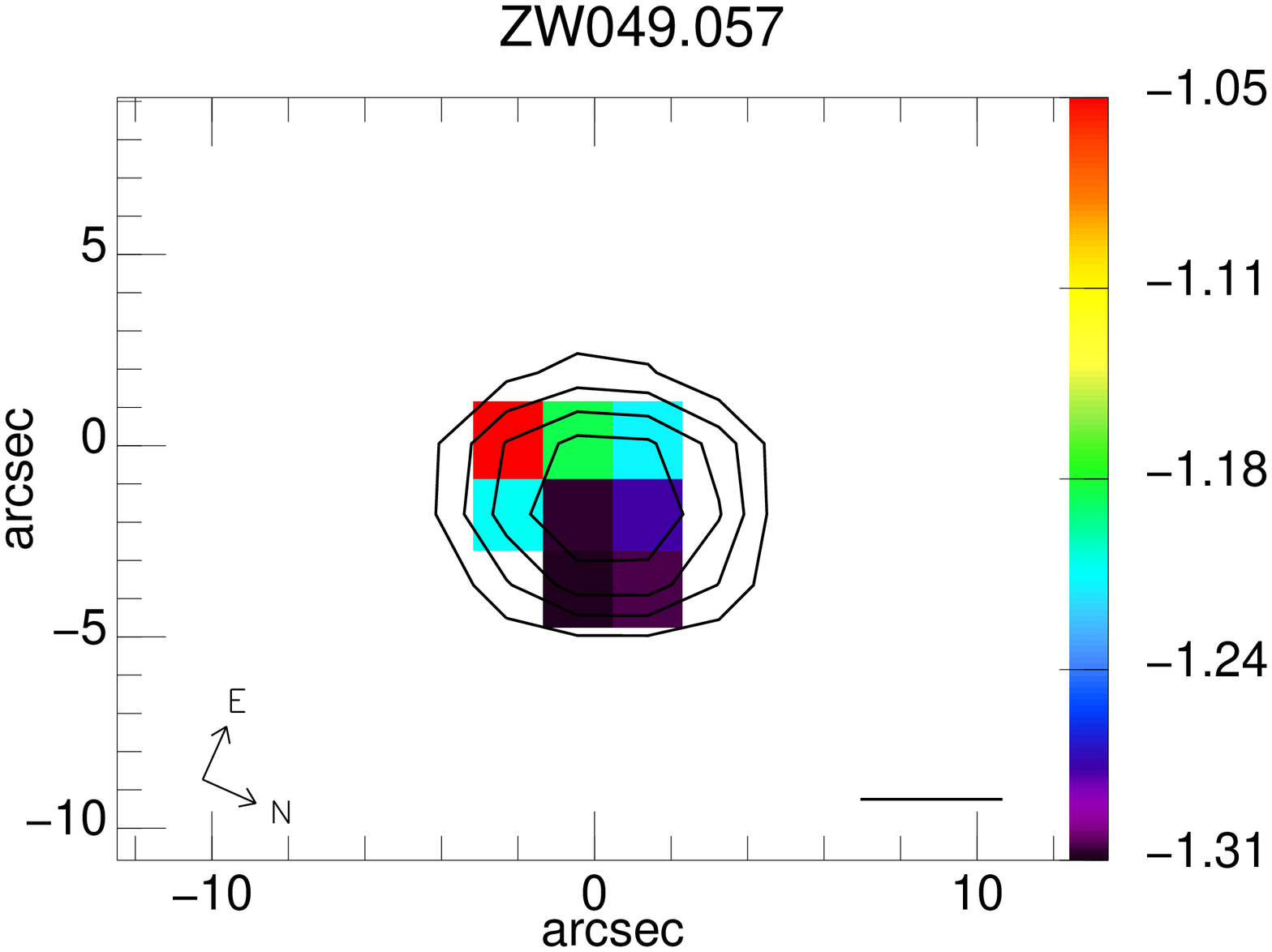}
\includegraphics[width=0.33\textwidth]{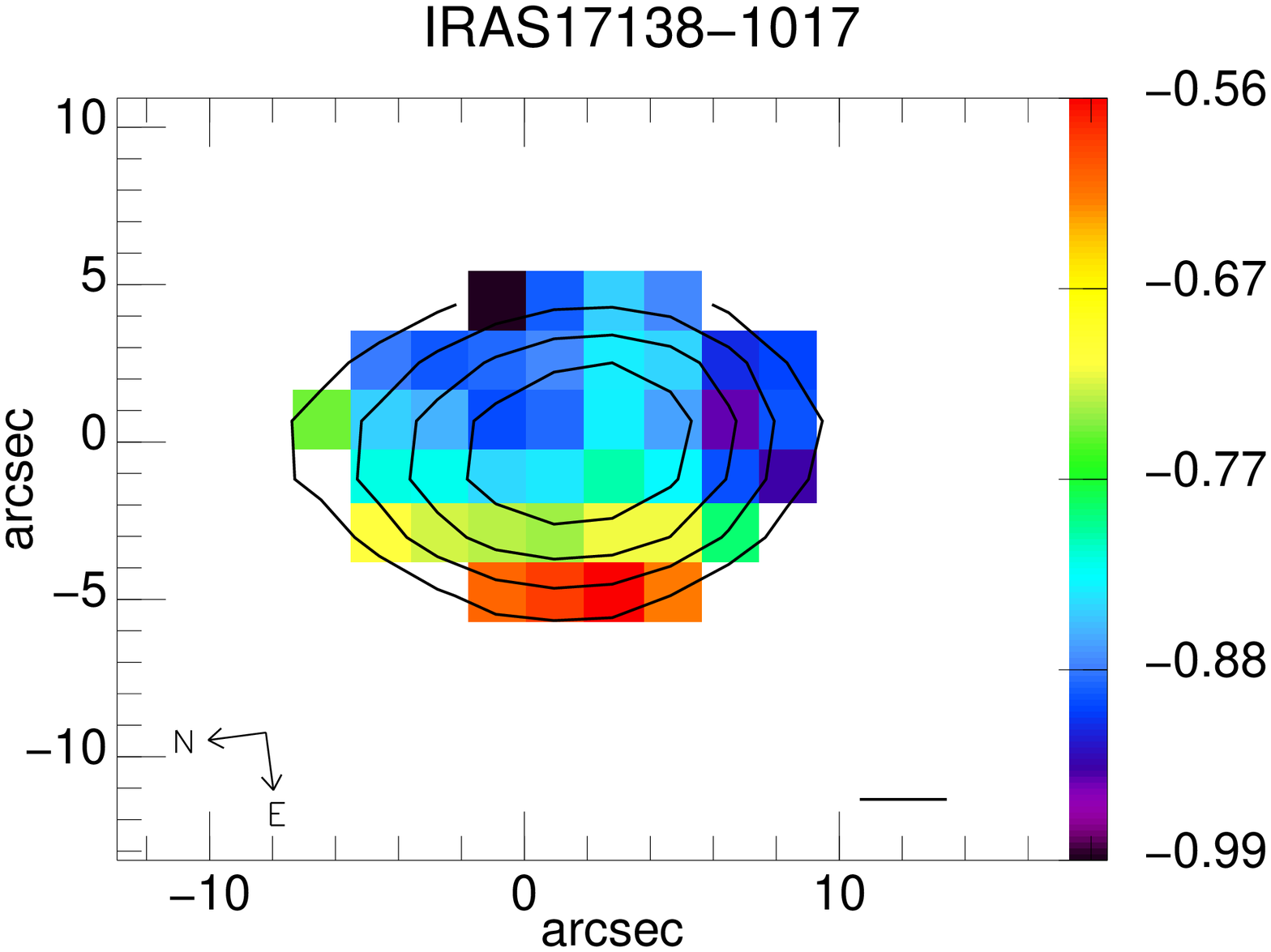}
\includegraphics[width=0.33\textwidth]{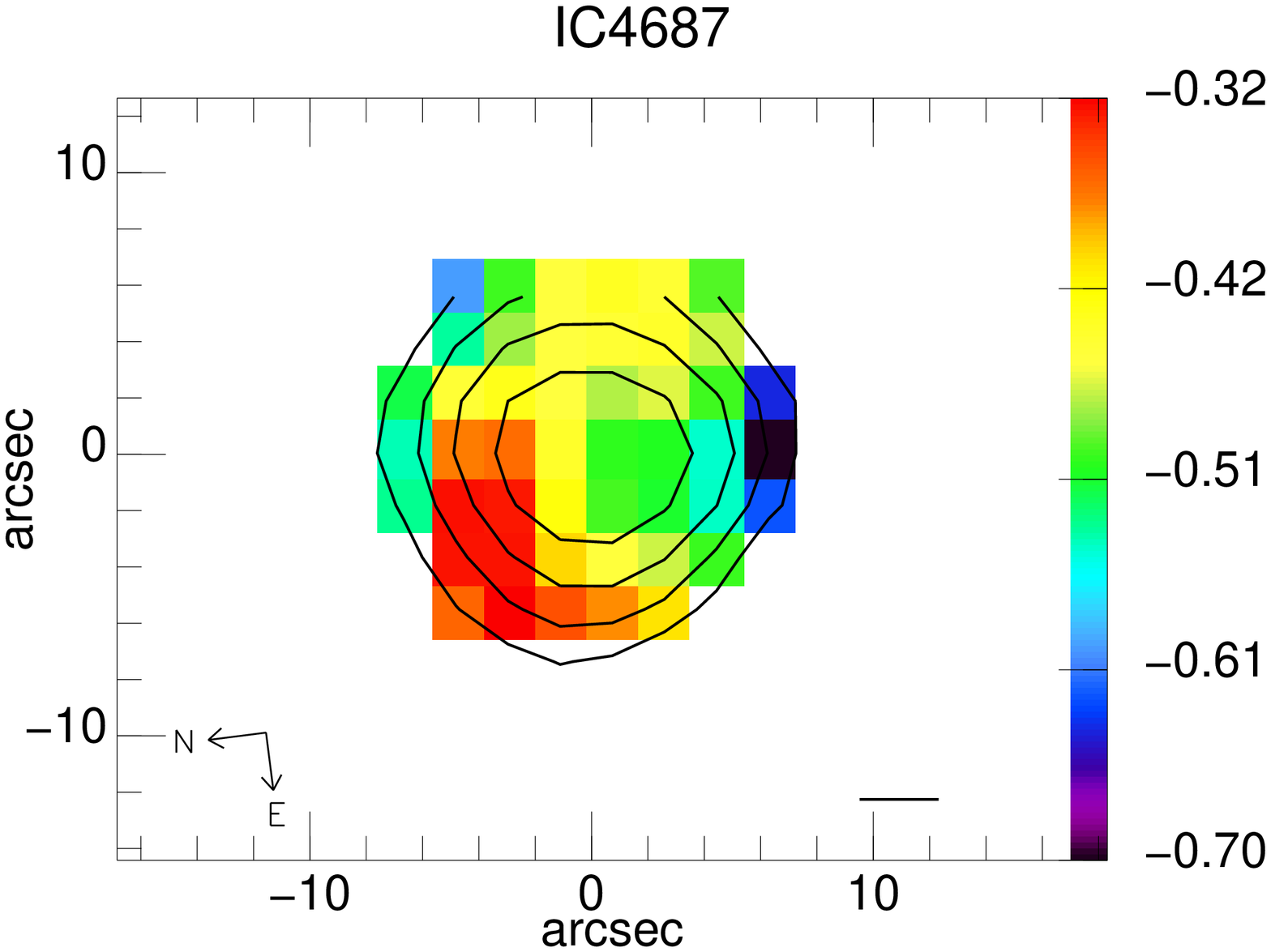}
\includegraphics[width=0.33\textwidth]{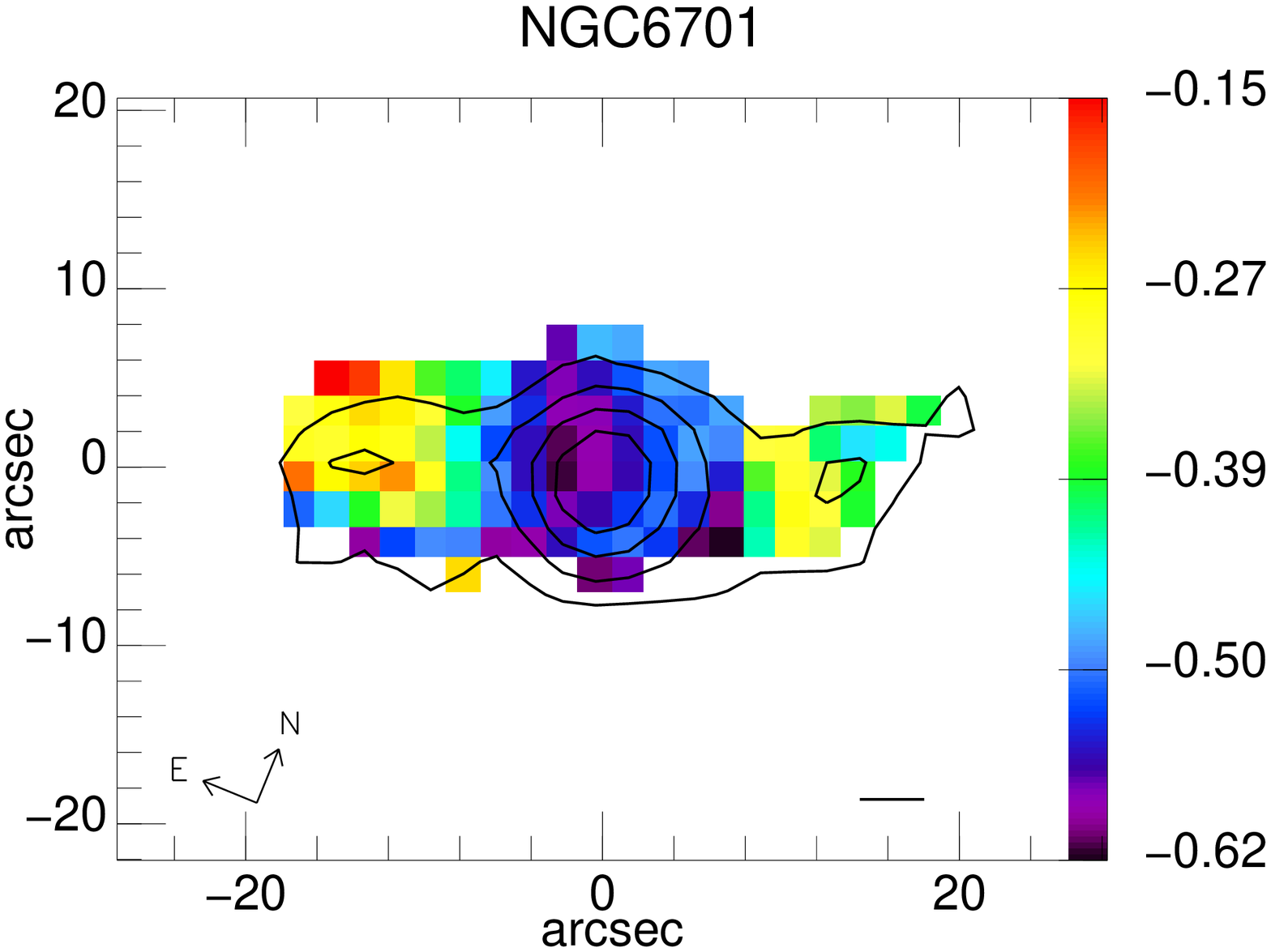}
\includegraphics[width=0.33\textwidth]{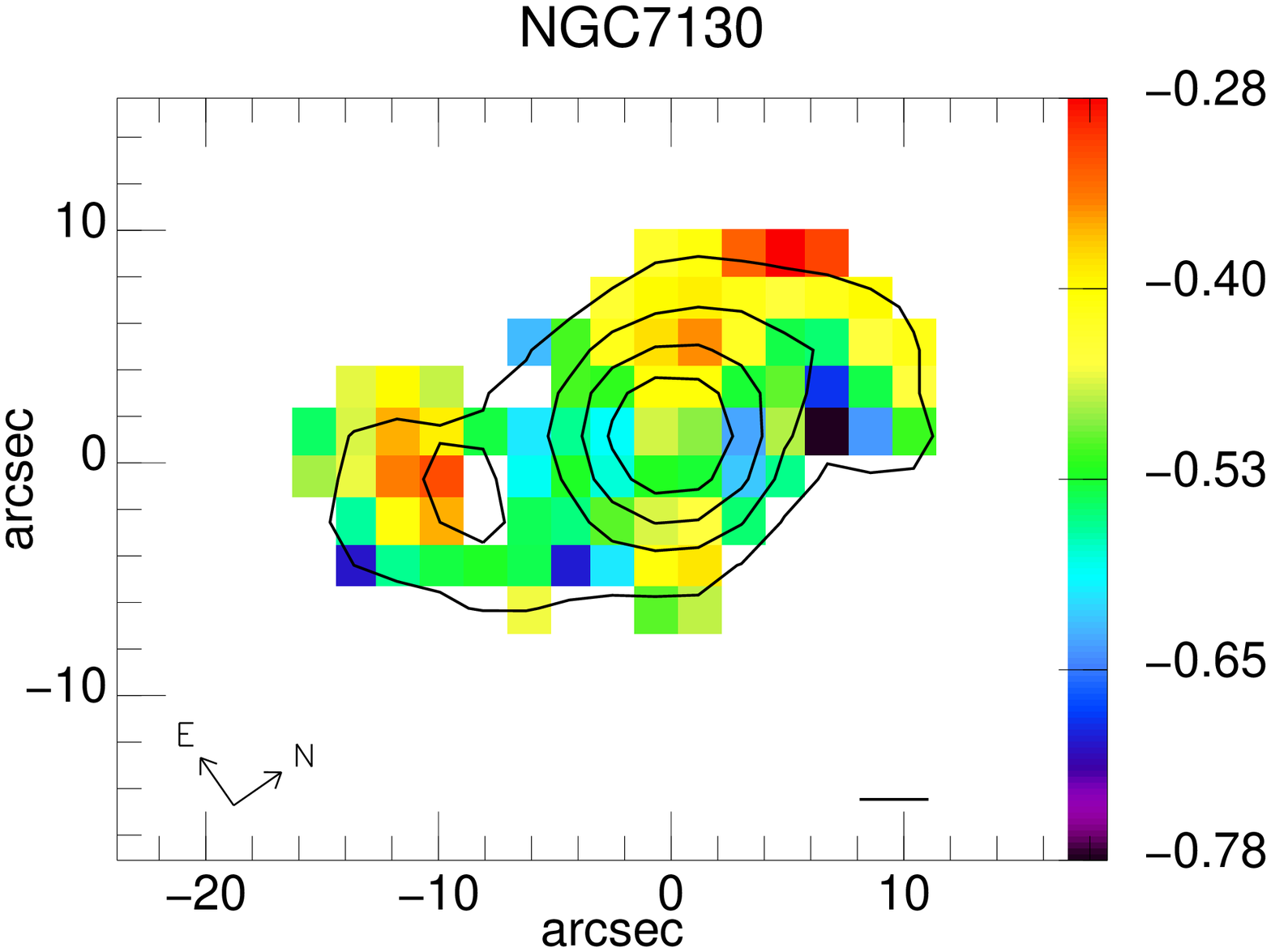}
\includegraphics[width=0.33\textwidth]{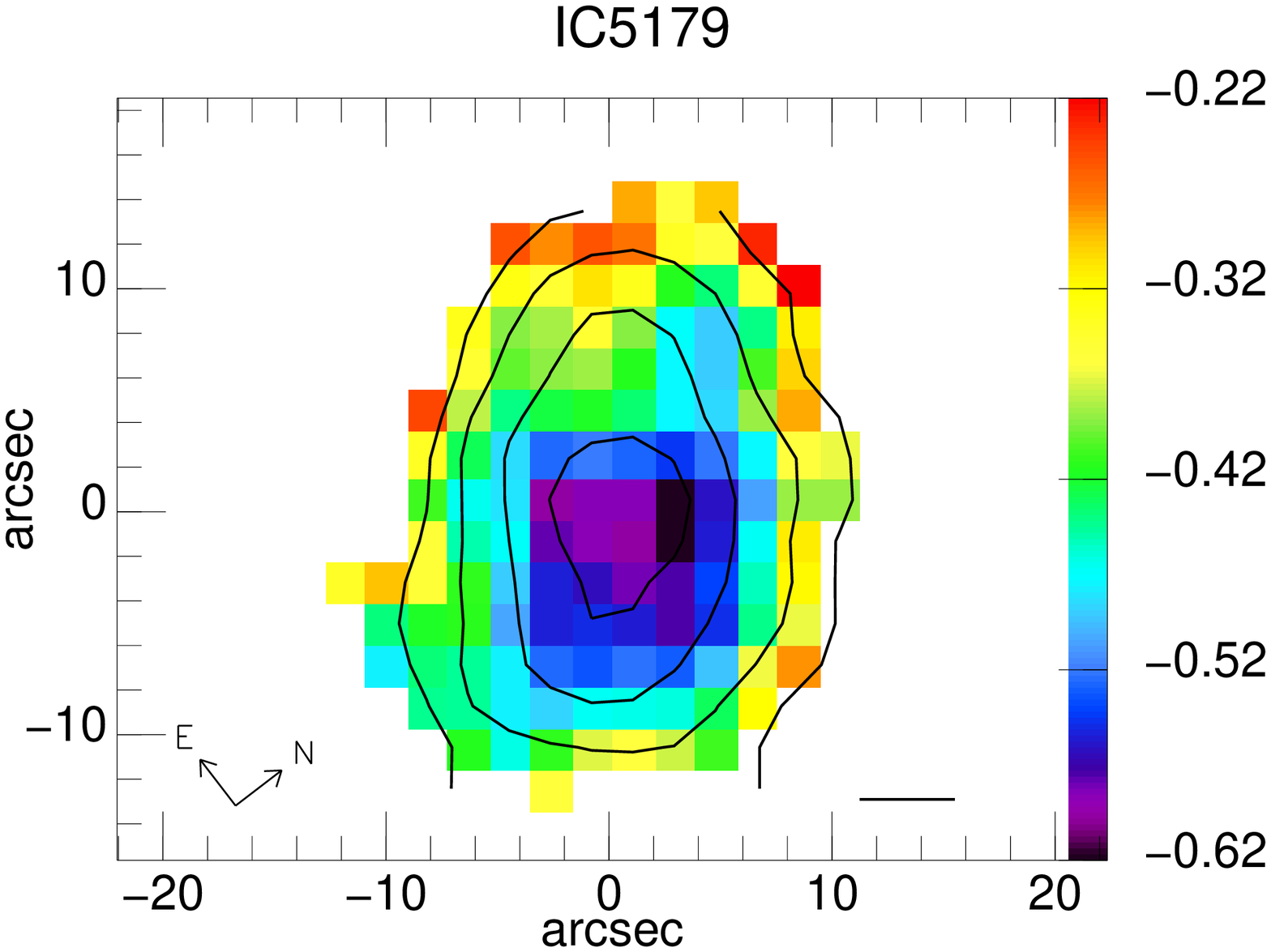}
\includegraphics[width=0.33\textwidth]{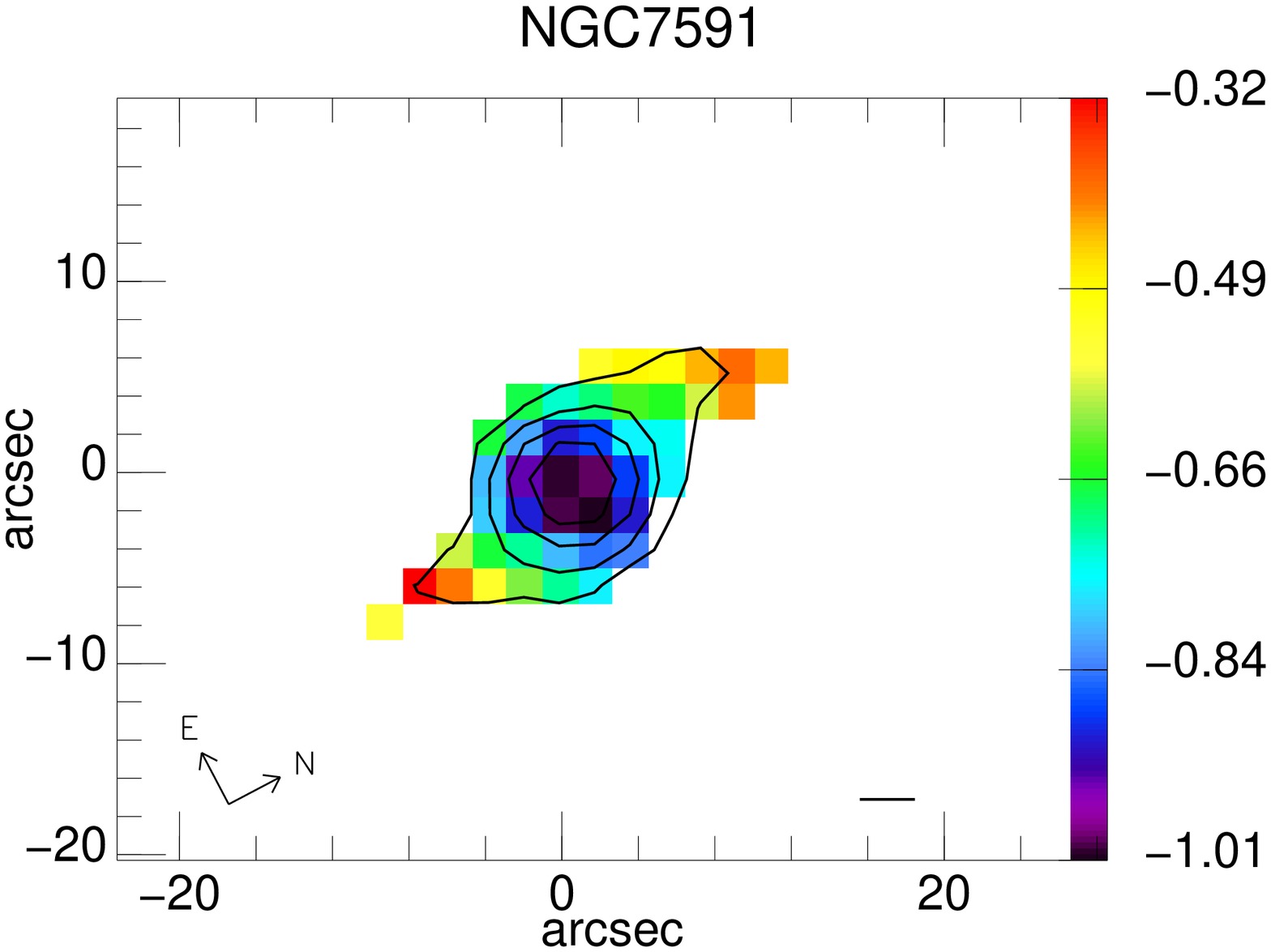}
\includegraphics[width=0.33\textwidth]{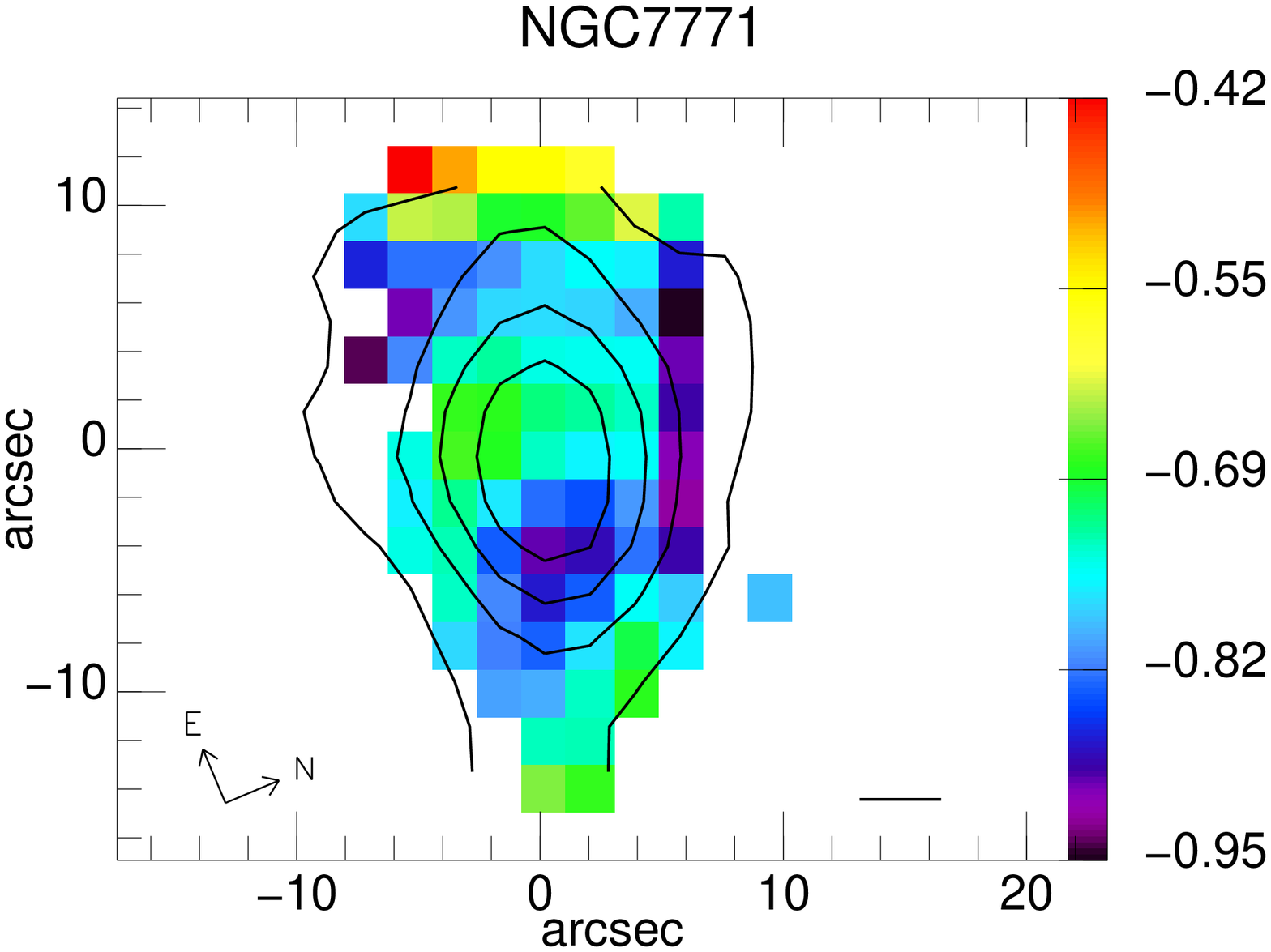}
\caption{Maps of the 9.7\micron\ silicate feature strength as defined in Section \ref{ss:silicate_feature_maps}. The 5.5\micron\ continuum contours are displayed to guide the eye. The image orientation is indicated on the maps for each galaxy. The scale represents 1 kpc. The maps are shown in a linear scale.}
\label{fig_map_sil}
\end{figure*}

\setcounter{figure}{9}
\begin{figure*}[!p]
\includegraphics[width=0.9\textwidth]{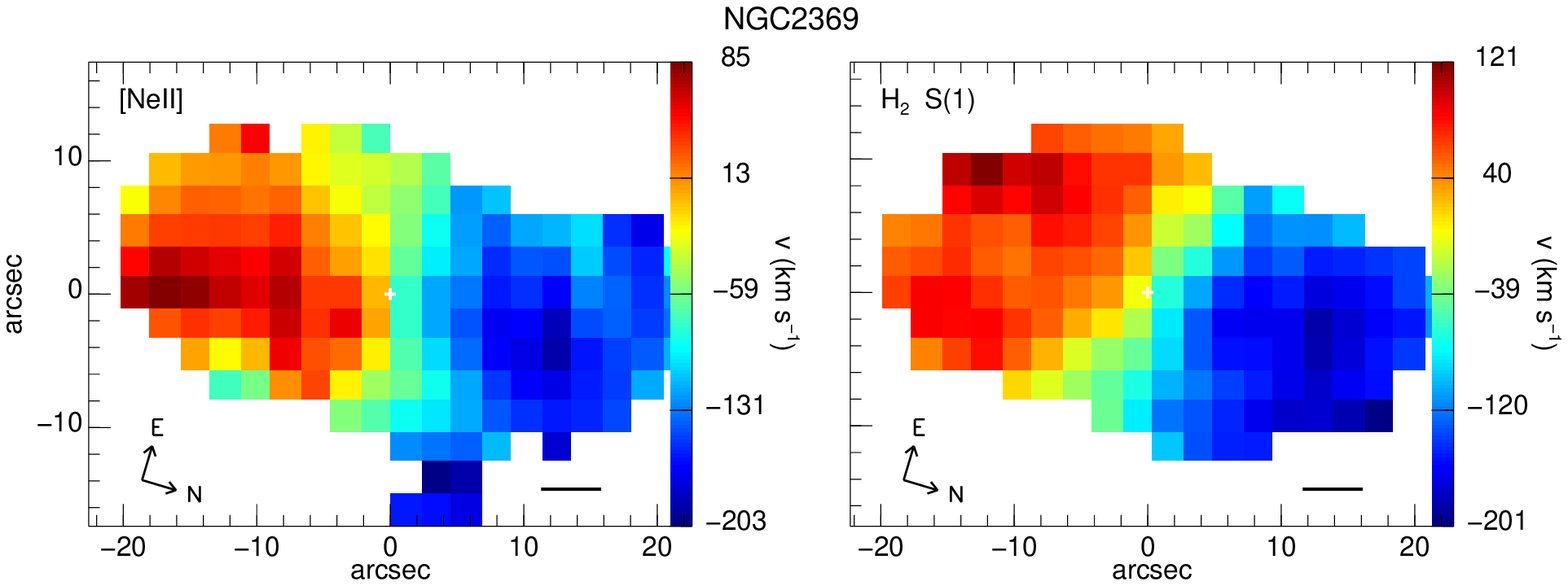}
\includegraphics[width=0.9\textwidth]{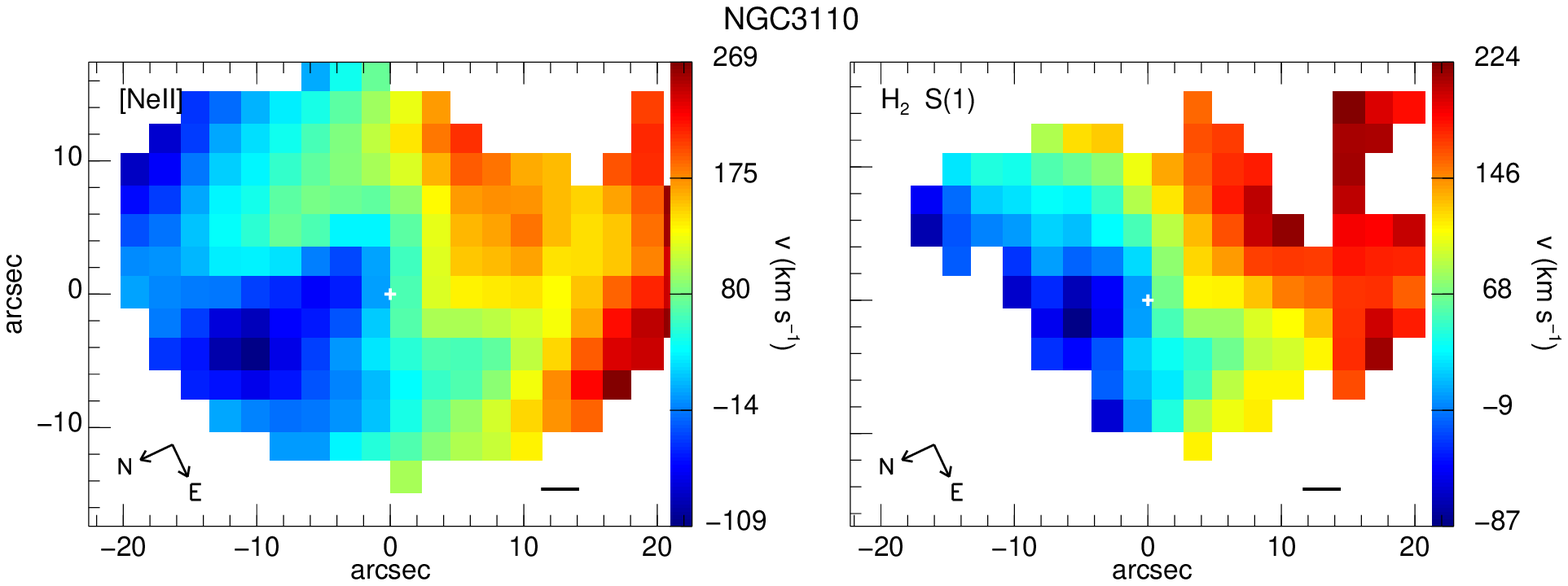}
\includegraphics[width=0.9\textwidth]{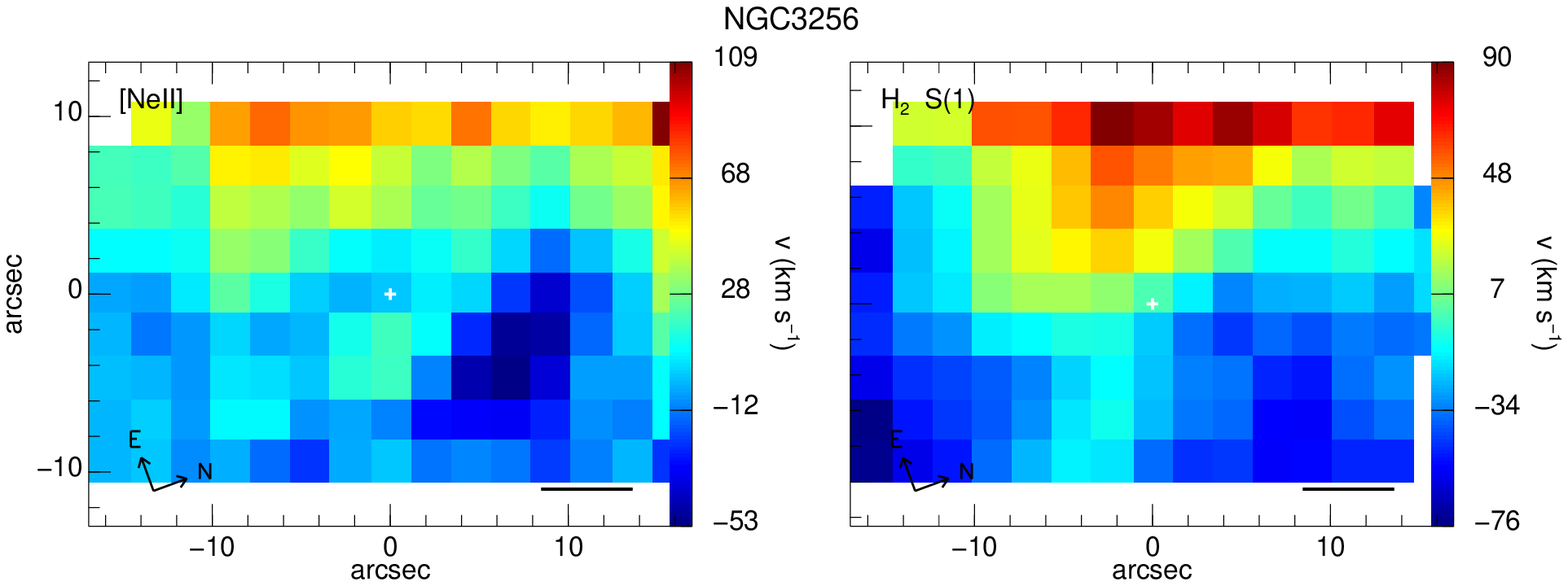}
\includegraphics[width=0.9\textwidth]{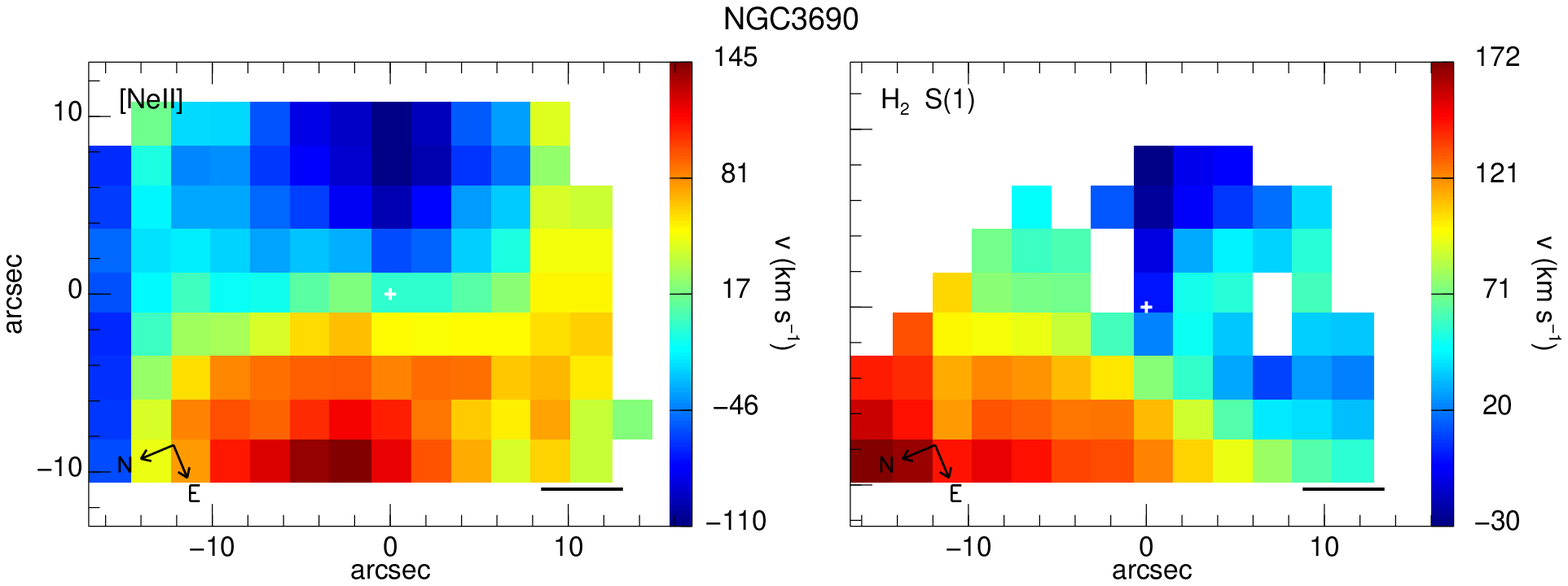}
\caption{Maps of the observed velocity field of \Neii and \Hm{1} at 17.0\micron.
The white cross marks the coordinates of the nucleus as listed in Table \ref{tbl_obs_map}. The image orientation is indicated on the maps for each galaxy. The scale represents 1 kpc.}
\label{fig_v}
\end{figure*}

\begin{figure*}
\addtocounter{figure}{-1}
\includegraphics[width=0.9\textwidth]{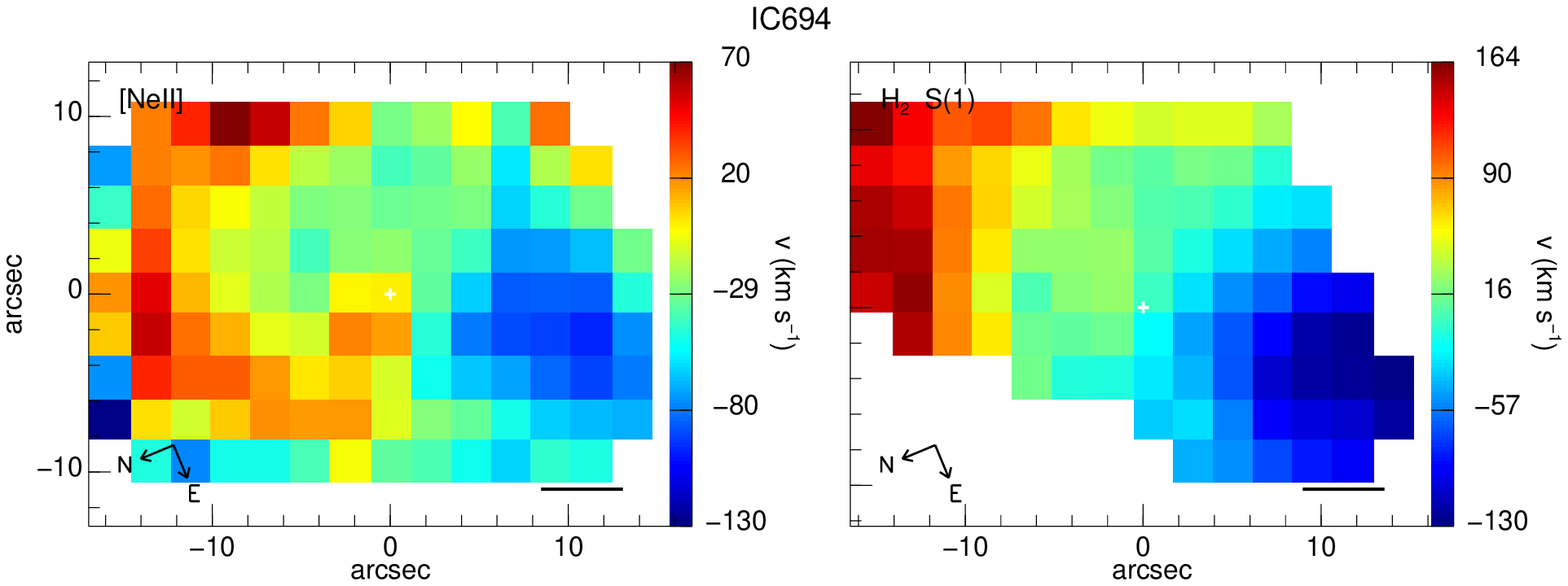}
\includegraphics[width=0.9\textwidth]{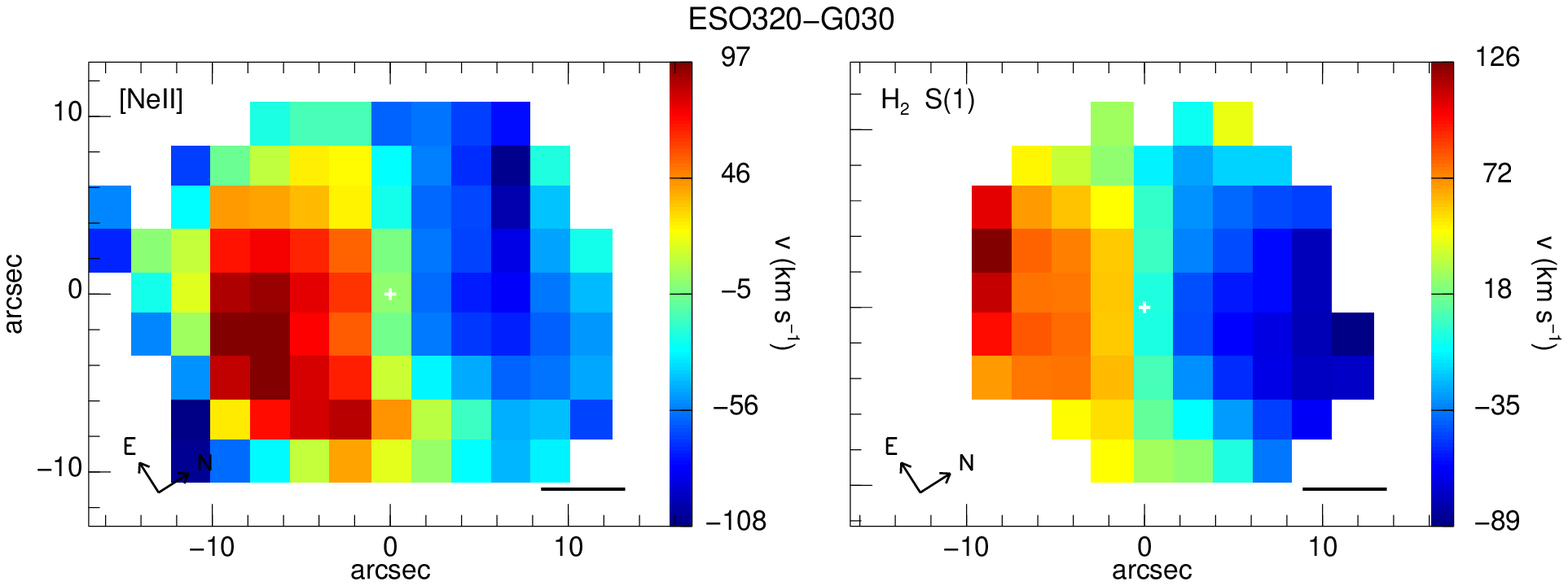}
\includegraphics[width=0.9\textwidth]{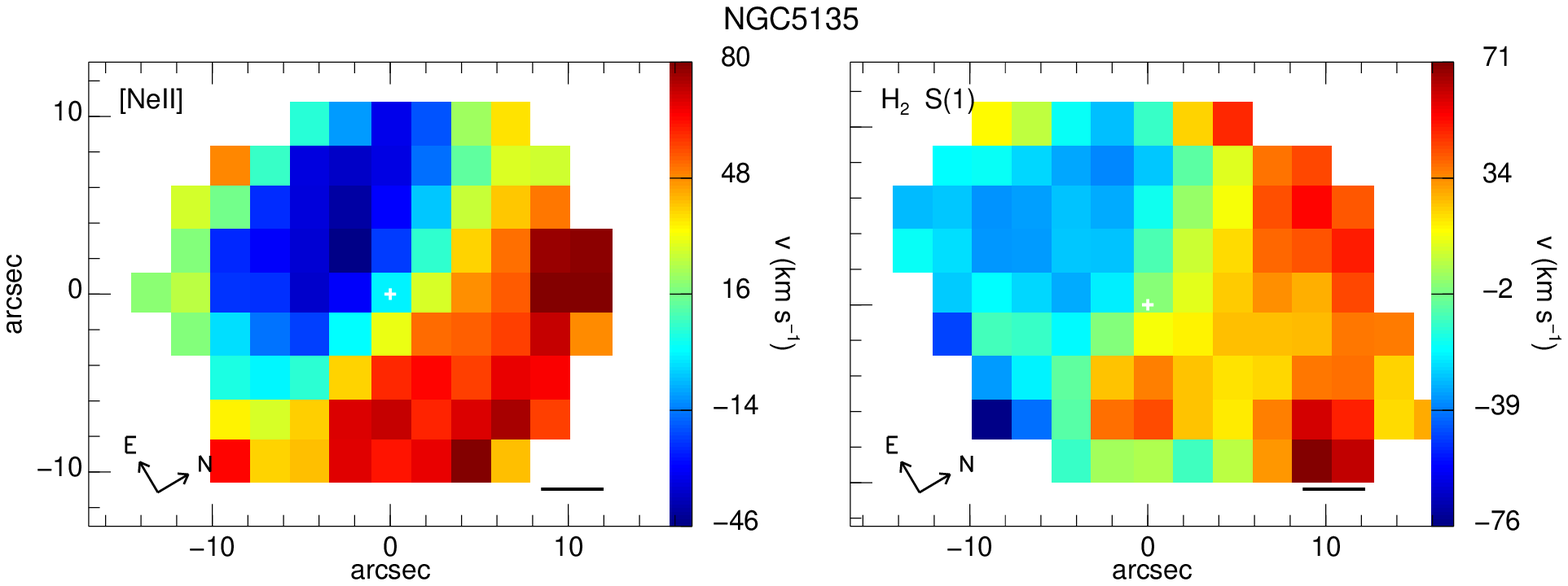}
\includegraphics[width=0.9\textwidth]{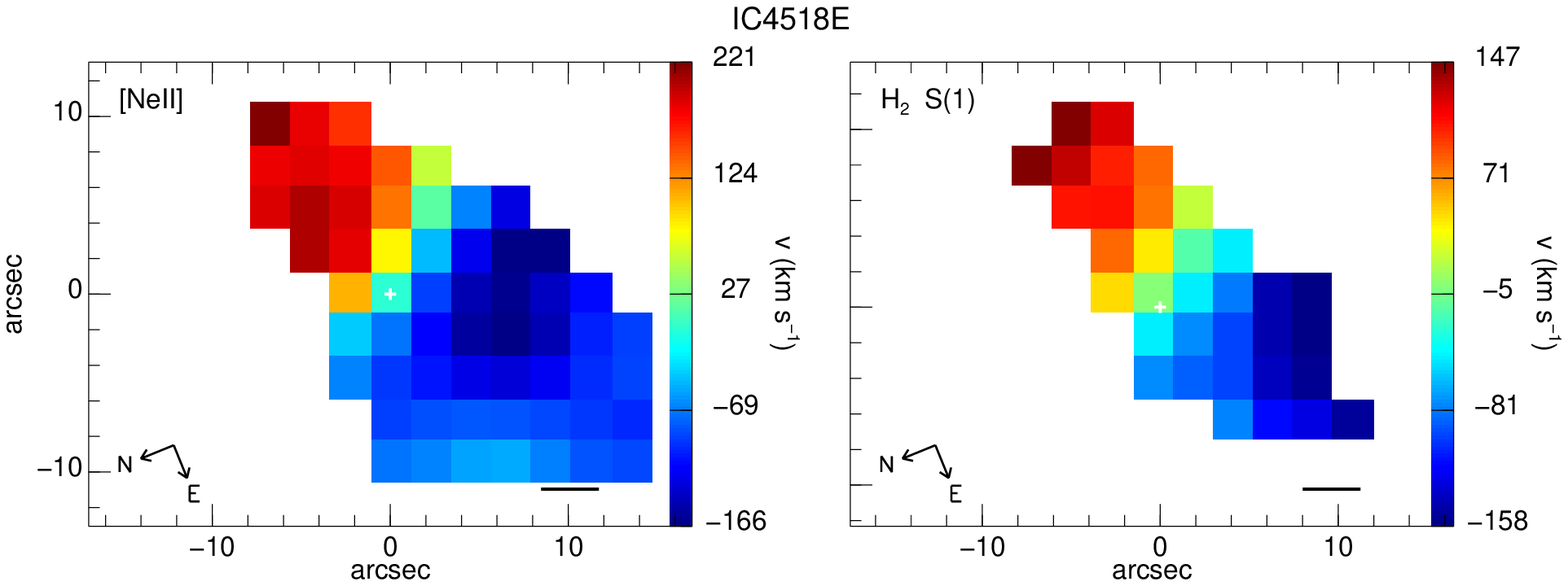}
\caption{Continued.}
\end{figure*}

\begin{figure*}
\addtocounter{figure}{-1}
\includegraphics[width=0.9\textwidth]{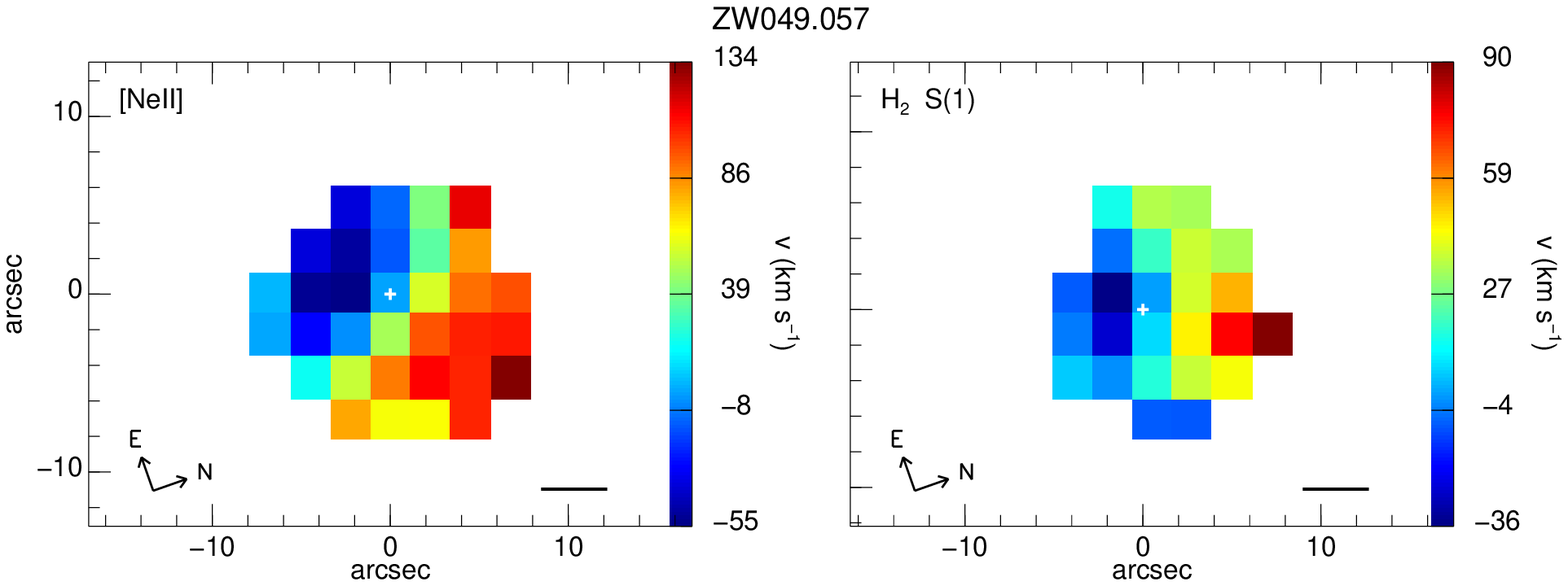}
\includegraphics[width=0.9\textwidth]{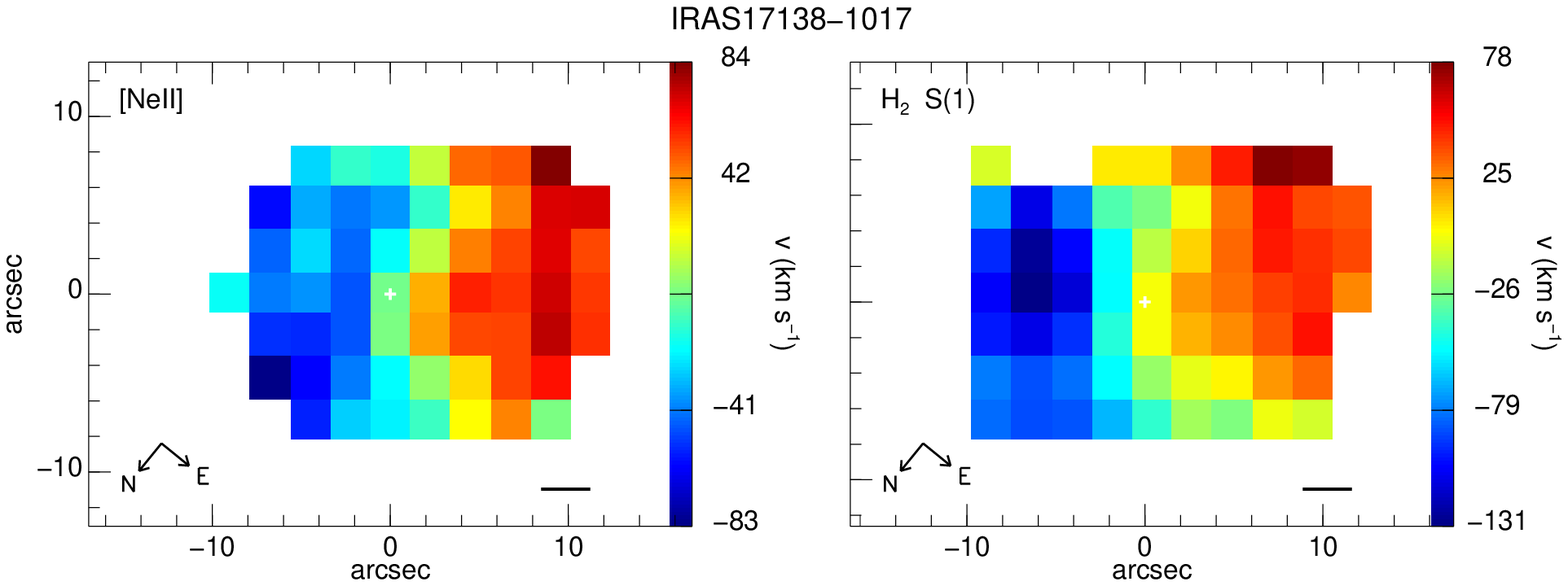}
\includegraphics[width=0.9\textwidth]{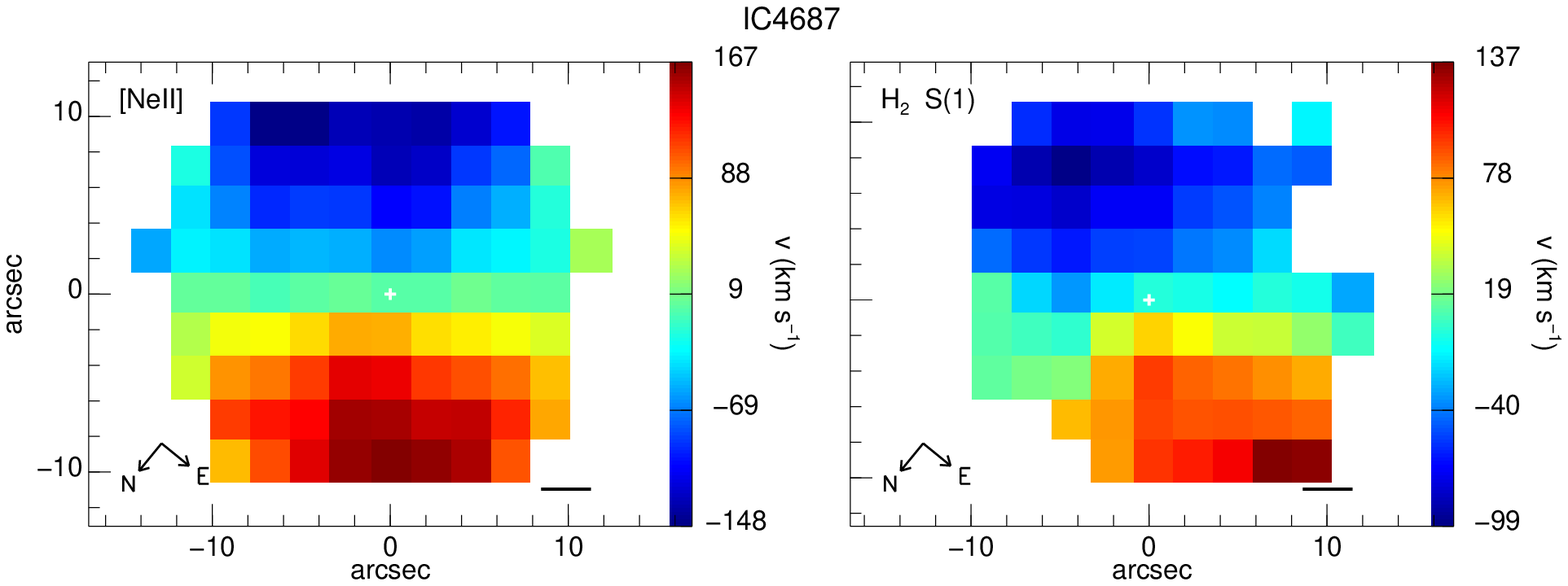}
\includegraphics[width=0.9\textwidth]{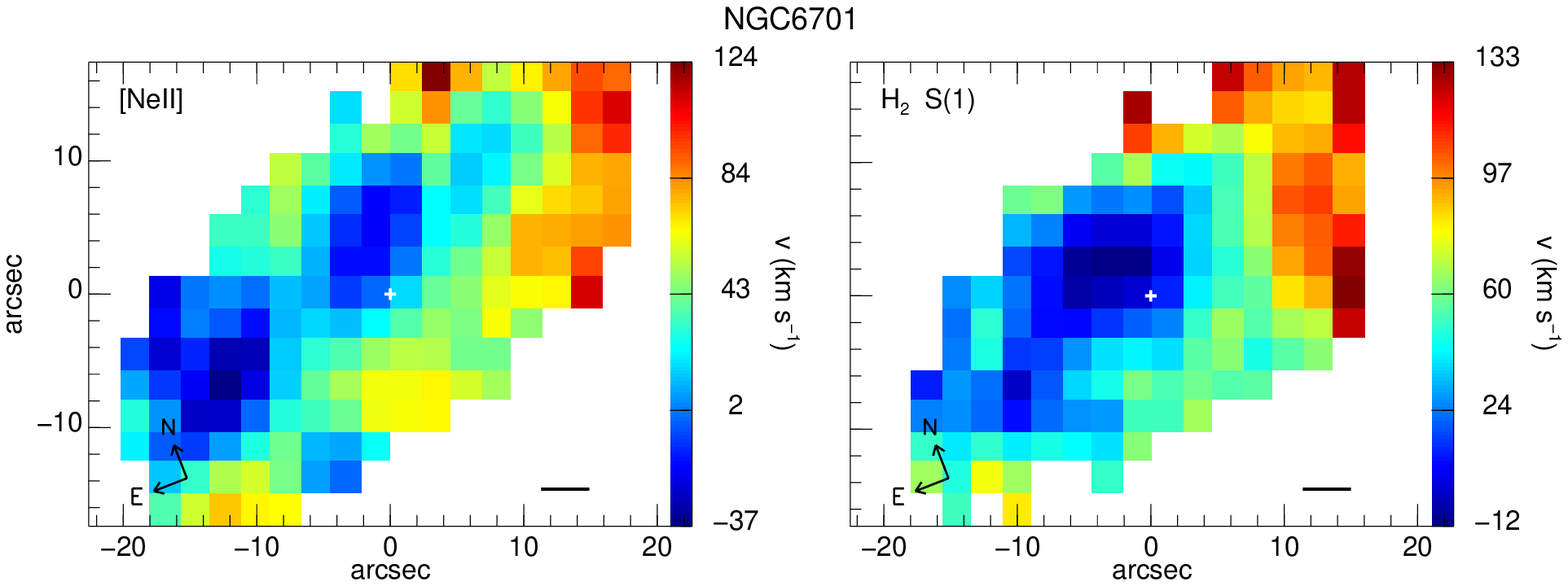}
\caption{Continued.}
\end{figure*}

\begin{figure*}
\includegraphics[width=0.9\textwidth]{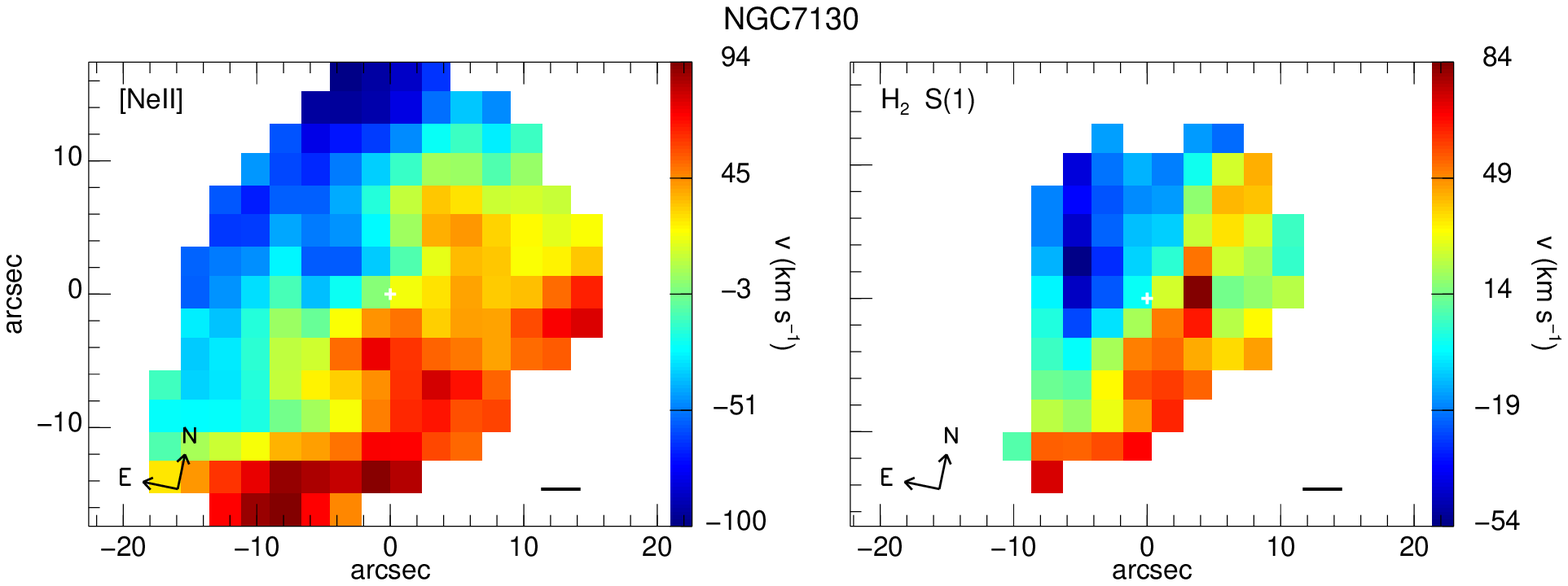}
\includegraphics[width=0.9\textwidth]{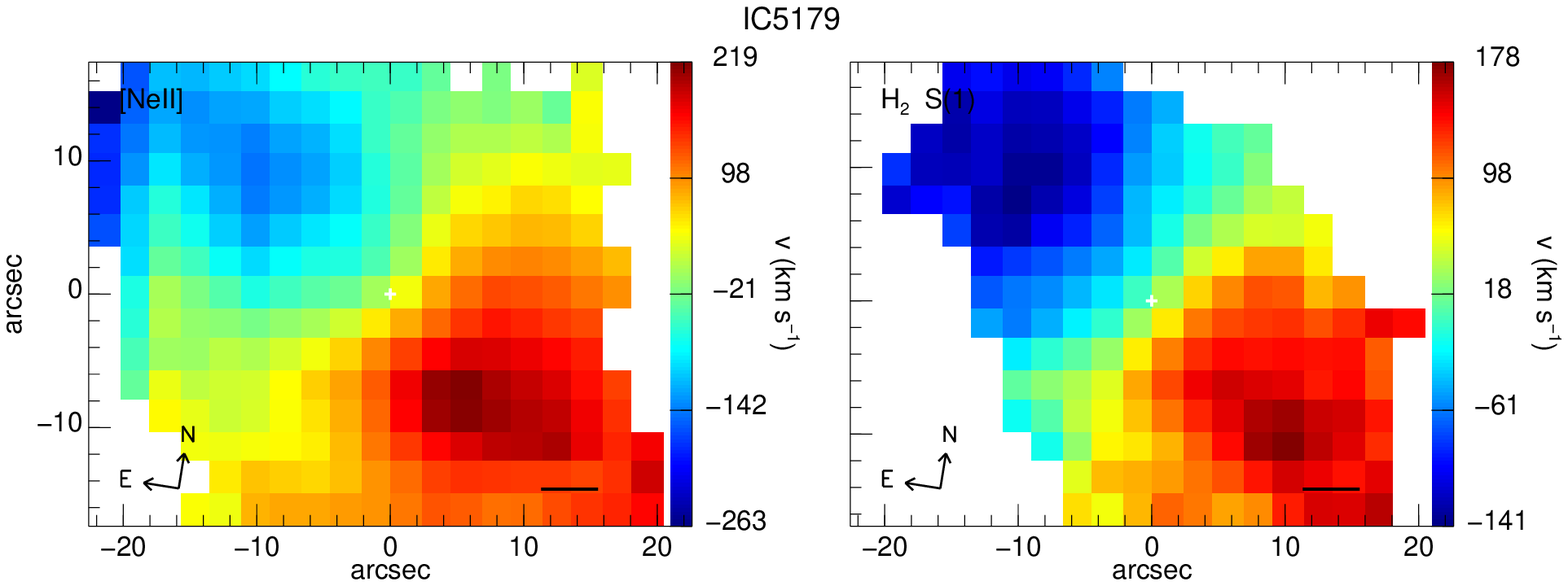}
\includegraphics[width=0.9\textwidth]{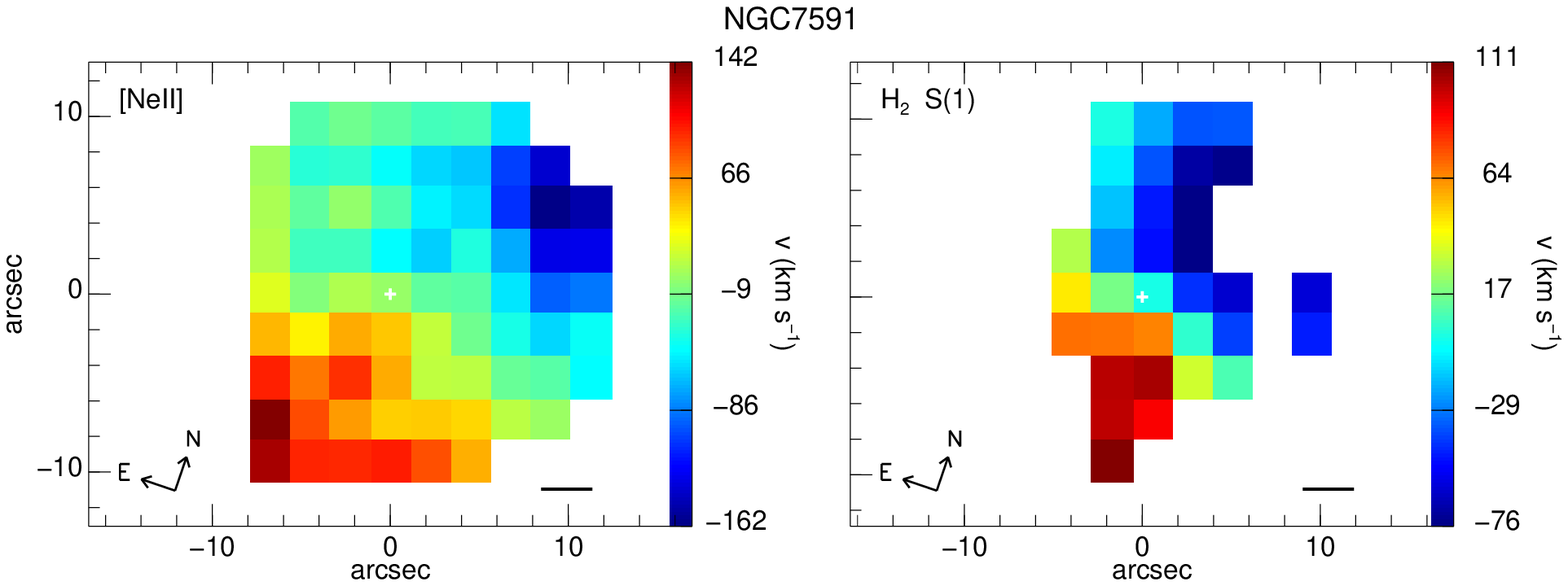}
\includegraphics[width=0.9\textwidth]{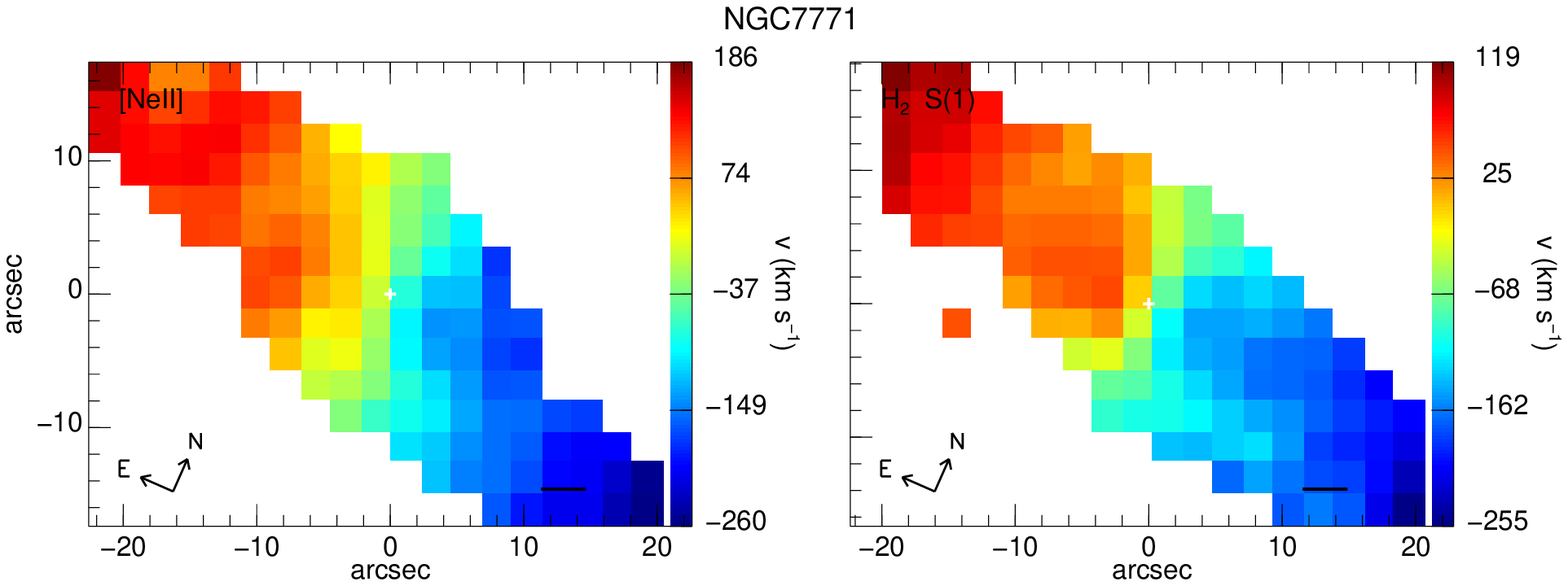}
\caption{Continued.}
\end{figure*}

\clearpage

\begin{appendix}
\section{The Point Spread Function of IRS Spectral Mapping Observations}\label{apxPSF}
When comparing maps of lines or features, at different wavelengths, the variation of the Point Spread Function(PSF) size can be significant.
The \Spitzer/IRS PSF is much less studied than that of the other \Spitzer\ instruments, although it is possible to generate synthetic IRS PSF using STinyTim program \citep{Krist2002}. This is useful for staring observations, but the spectral mapping observation mode introduces extra issues, such as the cube assembly from staring observations, wavelength calibration uncertainties, that cannot be easily taken into account using the modeled PSF. Because of that, a better way to characterize the IRS PSF of the spectral mapping observation mode is to use the mapping observations of calibration stars available in the \Spitzer\ archive.
Previously, \citet{Kospal08} studied the factor that should be applied to correct for flux losses when the slit is not centered properly on a point source. We extend this analysis to characterize the variations of the PSF.

To do this, we used the same data sets as \citet{Kospal08}, of spectral mapping observations of calibration stars. These observations were taken with the same instrument configuration used for the galaxies studied in this paper, but with much smaller steps between pointings, i.e. the calibration observations have a better spatial information of the PSF. Details of the observations are listed in Table \ref{tbl_log_cal}.

After obtaining BCD files processed by \Spitzer\ pipeline (version 15.3 for SH and SL modules, and version 17.2 for LH and LL modules), we built the spectral cubes with CUBISM \citep{SmithCUBISM}. The sky emission was subtracted using dedicated background observations. To take advantage of the better spatial information of the calibration observations, we set the pixel size in the final cube to be the size of the step in the perpendicular direction\footnote{In this Section we use the terms: ``perpendicular'' and ``parallel'' direction. They refer to the direction along the spectral dispersion axis and the spatial axis of the slit, respectively.} (see Table \ref{tbl_log_cal}). We did this for all IRS modules, however we only present the results for the SH, SL1 and SL2 modules, because the other modules (LH, LL1 and LL2), are heavily affected by fringing patterns and their PSF is not well reproduced by a Gaussian.

In order to study the PSF core variations we fitted each plane of the cube using a 2-D Gaussian. We allowed different values of the Full Width Half Maximum (FWHM) for each axis of the Gaussian, but the axis orientation is fixed and aligned with the parallel and perpendicular directions.
By doing this we obtained the FWHM and centroid of the PSF at every wavelength. These values are represented in Figure \ref{fig_PSF_modules}.

The undulating pattern of the FWHM and the centroid in the modules SL1 and SL2 is originated by several causes: (1) The spectral axis is not aligned with the detector, thus to extract the spectra is necessary to resample the original image. This resampling process produces this kind of periodic patterns. (2) The algorithm used to build the data cubes (see \citealt{SmithCUBISM} for details) does not try to correct for this effect. In any case, these variations are small: $<$10 \% for the FWHM and $<$0.2 arcsec for the centroid position.

The PSF core is slightly asymmetric for all modules. The FWHM in the perpendicular direction is larger than the FWHM in parallel direction, which has an almost constant value. A possible explanation to this is the way CUBISM builds the cubes. It assumes that there is no spatial information in the dispersion (or perpendicular) direction and the spatial information in that direction comes from different telescope pointings (for more details see \citealt{SmithCUBISM}).
The averaged values of the FWHM of each module are listed in Table \ref{tbl_fwhm_cal}.

The variation of the centroid position in SL modules is small. The 1$\sigma$ deviation around the average centroid position are 0.07 and 0.02 arcsec for the parallel and perpendicular direction respectively, that is a twentieth of the SL pixel size. These variations can be attributed to small telescope pointing errors and signal-to-noise limitation.
The position of the centroid in the SH module shows variations that cannot be explained just by instrumental pointing error or a limited signal-to-noise. The 1-$\sigma$ deviation in the perpendicular direction is 0.03 arcsec, but in the parallel direction the variation is one order of magnitude larger (0.2 arcsec) and shows a clear wavelength dependency. The most probable cause for this, is an error in the position and/or angle of the extraction aperture. However this effect (0.6 arcsec) is small when compared with the SH pixel size (2.26 arcsec).

We studied the variations of the diffraction rings in the SH module. We fitted the central row of the cube for each wavelength using a Gaussian for the PSF core and two Gaussians for the only  diffraction ring detected. These two Gaussians have the same FWHM and they are symmetrically placed around the PSF core position. Figure \ref{fig_PSF} shows a fit example.
The results of the fits are summarized in the Figure \ref{fig_PSF_ring}.
The FWHM of the diffraction ring, that does not depend on the wavelength, is 3.6 $\pm$ 0.3 arcsec. Instead the position of the ring depends linearly on wavelength.
The relative value of the diffraction ring peak respect to the PSF core peak, depends slightly on wavelength. At 10\micron\ it is 11\%\ of the core peak value and at 19 \micron\ it is 9\%\ .

When calculating feature ratios, the point spread function variations (FWHM and centroid), could originate artifacts. For the SL module these variations are less than one tenth of the pixel size and should not affect feature ratios. For the SH module the major source of uncertainty is the centroid position, but the difference is only one fourth of the pixel size. Since these variations are very small compared with the pixel size, they do not affect the calculated feature ratios.
Diffraction rings can produce artifacts in the feature ratios too. They cause inaccurate ratio values 3 or 4 pixels away from the center of the point source in the SH module. For extended sources the situation is more complex because it is not possible to know where the diffraction ring will appear. However, comparing the spatial profiles of the features and the underlying continuum, as we do in this paper, helps determine if the ratio values are real or artifacts.

\begin{deluxetable}{ccccc}
\tablewidth{0pt}
\tablecaption{Log of IRS mapping calibration observations\label{tbl_log_cal}}
\tablehead{
& & & \colhead{Step size perpendicular} & \colhead{Step size parallel} \\
\colhead{Module} & \colhead{AOR} & \colhead{Target} & \colhead{(arcsec)} & \colhead{(arcsec)}
}
\startdata
SL1, SL2 & 16295168 & HR 7341 & 0.6 & 15.5\\
LL1, LL2 & 16463104 & HR 6606 & 1.7 & 46.0\\
SH & 16294912 & HR 6688 & 0.8 & 0.8\\
SH & 16340224 & sky \\
LH & 16101888 & HR 2491 & 1.7 & 1.7 \\
LH & 116088320 & sky
\enddata
\end{deluxetable}

\begin{deluxetable}{ccccc}
\tablewidth{0pt}
\tablecaption{IRS PSF Full Width Half maximum\label{tbl_fwhm_cal}}
\tablehead{
 & \colhead{FWHM Perpendicular} & \colhead{FWHM Parallel} \\
\colhead{Module} & \colhead{(arcsec)} & \colhead{(arcsec)}
}
\startdata
SH & 5.16 $\pm$ 0.08 & 4.31 $\pm$ 0.37 \\
SL1 & 3.76 $\pm$ 0.07 & 3.27 $\pm$ 0.21 \\
SL2 & 3.88 $\pm$ 0.04 & 3.02 $\pm$ 0.17
\enddata
\end{deluxetable}
\setcounter{figure}{24}
\begin{figure}
\centering
\includegraphics[width=\textwidth]{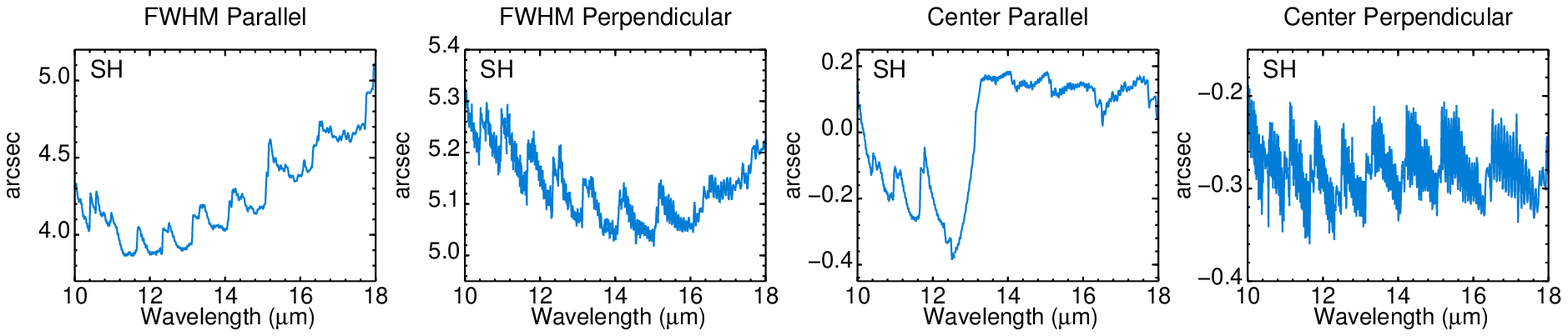} 
\includegraphics[width=\textwidth]{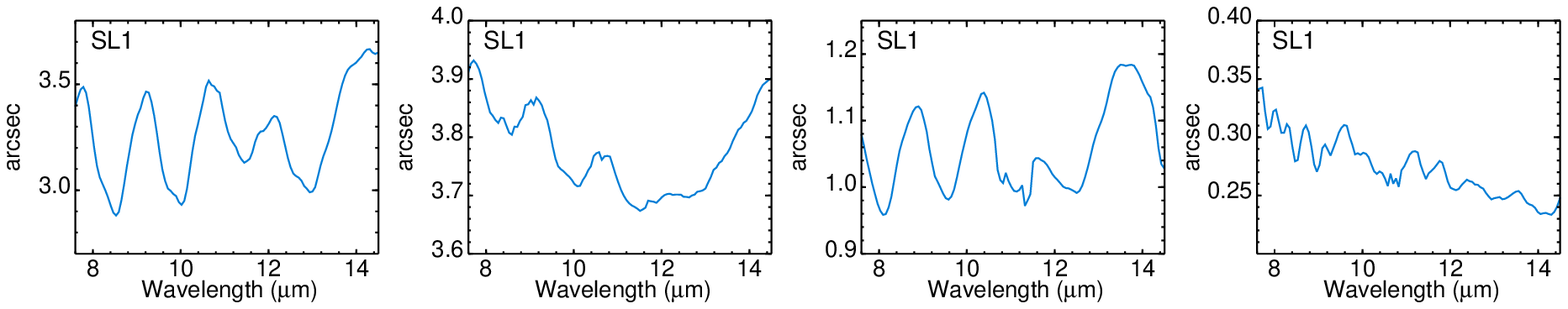} 
\includegraphics[width=\textwidth]{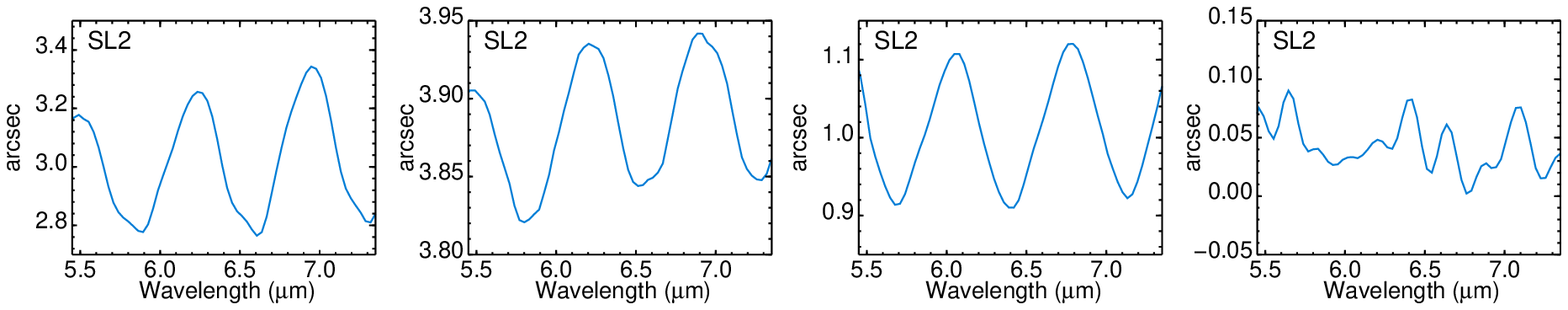} 
\caption{Full Width at Half Maximum (FWHM) and centroid position of the PSF of the modules SH, SL1 and SL2. Parallel and perpendicular directions are represented. We have calculated these values by fitting a 2-D Gaussian.}
\label{fig_PSF_modules}
\end{figure}

\begin{figure}
\centering
\includegraphics[width=0.45\textwidth]{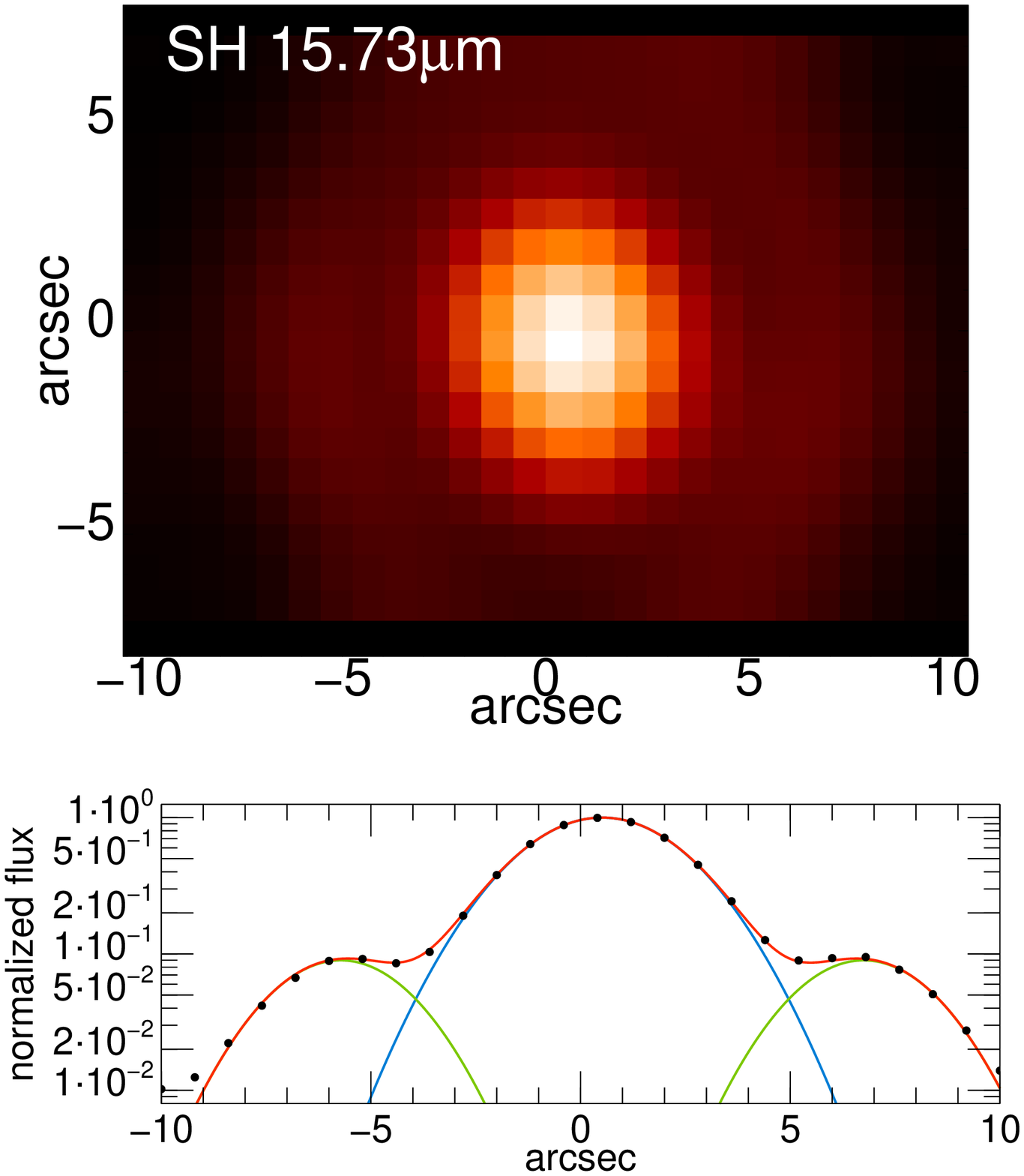}
\caption{Upper panel: PSF of the SH module at 15.73\micron\ . Lower Panel: Fit of the PSF profile at the central row. The blue line is the Gaussian used to fit the PSF core. The green line is the fit of the diffraction ring. The red line is the sum of both components.}
\label{fig_PSF}
\end{figure}

\begin{figure}
\centering
\includegraphics[width=0.45\textwidth]{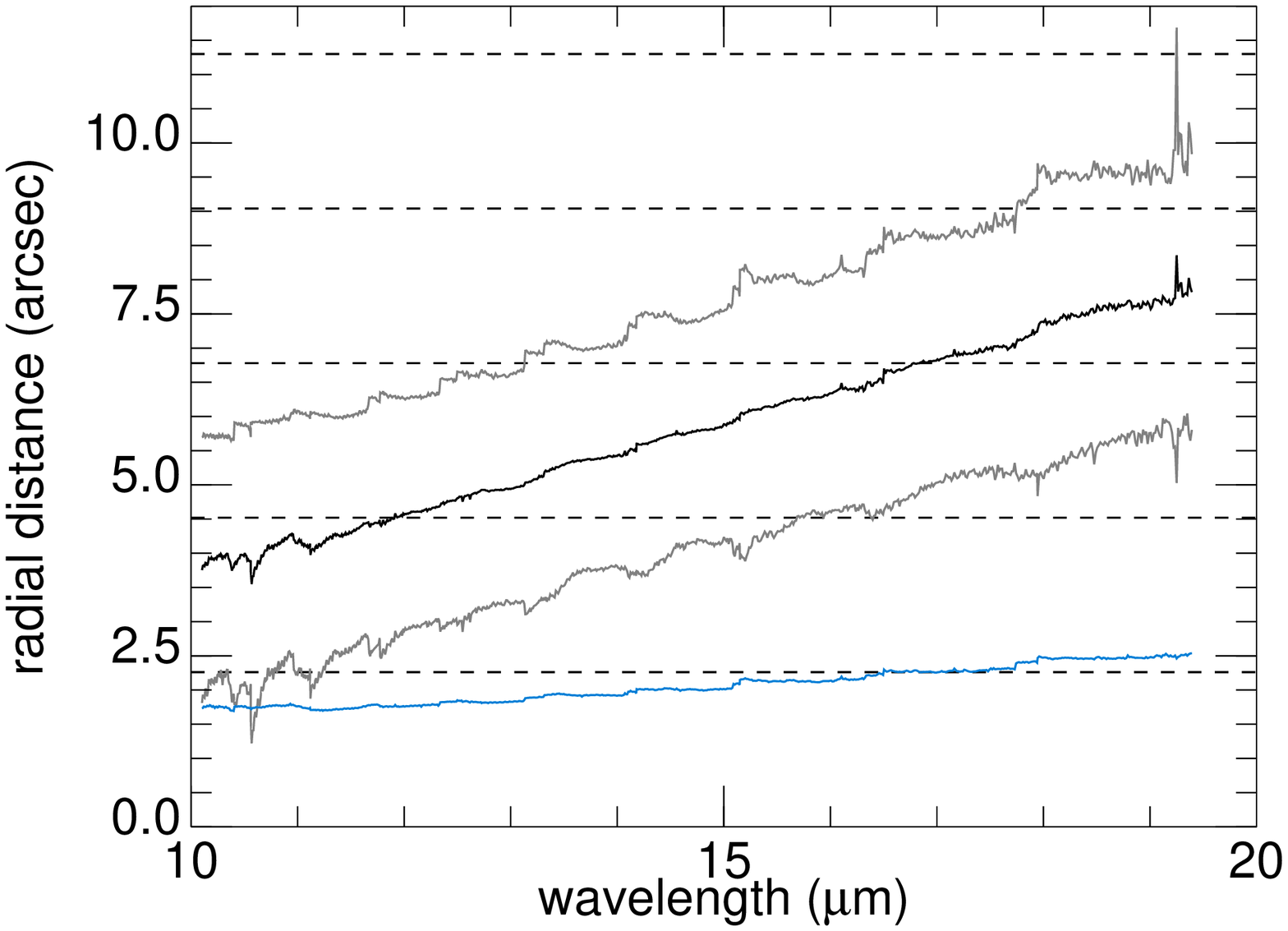}
\caption{Position of the diffraction ring in the SH module (black solid line). The position of the half-maximum of the diffraction ring is shown in gray. For comparison, the blue line is the half width at half maximum of the PSF core. The dashed black lines represent the pixel size of the SH module.}
\label{fig_PSF_ring}
\end{figure}

\clearpage
\section{Description of individual objects}\label{apxDescription}

\subsection{NGC~2369}
The MIR emission of NGC~2369 is dominated by the resolved nucleus. The molecular hydrogen (\Hm{1} and \Hm{2}) emission appears more extended along the galaxy disk than the \PAHonce\ emission. This causes a relatively large \Hm{1}\slash\PAHonce\ ratio in the regions around the nucleus.
The \Neii\slash\PAHonce\ ratio map reveals an \HII\ region $\sim$17 arcsec ($\sim$3.4 kpc projected) away from the nucleus to the south.
The nuclear silicate strength, $S_{\rm Si}\sim$0.9, is slightly larger than the average silicate strength of our LIRGs sample, probably because this galaxy is seen almost edge-on.

\subsection{NGC~3110}
It is a face-on galaxy with numerous \HII\ regions along the spiral arms\citep{Hattori04}. The \Neiii\ emission of this galaxy is especially bright and it resembles that of the H$\alpha$ emission. The \Neiii\slash\Neii\ ratio of the extranuclear \HII\ regions is $\sim$5 times larger than the nuclear ratio. The deepest silicate absorption is located at the nucleus.

\subsection{NGC~3256}
It is galaxy merger with two nuclei. The southern nucleus, located 5 arcsec away from the north nucleus, is highly obscured, although it is visible in the 5.5\micron\ continuum map. It has one of the deepest silicate absorption feature in our sample, $S_{\rm Si}\sim-1.4$. The silicate feature of the rest of the galaxy is similar to that found in the rest of the galaxies of our sample.
The PAH, continuum and atomic emission (except \SIV) peak at the north nucleus. The maximum of the \SIV\ emission is located $\sim$5 arcsec to the east of the north nucleus, and it is probably related to a star-forming region. This is the only extranuclear region (d$>\sim$1 kpc), apart from region C in Arp~299 \citep{AAH09Arp299}, where we detect the \SIV\ line. The \Hmol\ emission arises mainly from the southern nucleus and the surrounding regions.

\subsection{ESO~320-G030}
Due to the low spatial resolution of the IRS maps, we cannot distinguish any morphological detail in the emission maps this galaxy, although it is clearly spatially resolved. The maximum of the silicate feature absorption is located at the nucleus.

\subsection{NGC~5135}
This galaxy hosts a Sy2 nucleus surrounded by intense star formation (\citealt{Bedregal2009} and references therein), that we cannot separate due the limited spatial resolution of the IRS spectral maps. However the line ratios (e.g \Neiii\slash\Neii\ ratio) show some asymmetries that may be associated with the truncated ring-like structure of this galaxy. At our scale ($\sim$0.5 kpc), we do not see any noticeable variations in the PAH emission due to the AGN activity.
The silicate absorption feature is shallow (-0.47) and the maximum value is not located at the nucleus.

\subsection{IC~4518E}
This galaxy is part of an interacting pair (IC~4518E/IC~4518W). Most of the mid infrared emission (PAH, fine structure line and 15\micron\ continuum) emission is dominated by an \HII\ region located $\sim$7 arcsec to the south-east from the nucleus.
The nuclear bright is similar to that of the southwestern \HII\ region emission only for the molecular hydrogen (\Hm{1}) and the 5.5\micron\ continuum.
The deepest silicate absorption is found at the nucleus.

\subsection{Zw~049.057}
Due to the low signal-to-noise ratio of the observations of Zw~049.057, the maps of this galaxy contain very few pixels. The nuclear silicate absorption is, S$_{\rm Si}$= -1.2, which is large when compared to that of the other LIRGs.

\subsection{IRAS~17138$-$1017}
The limited spatial resolution of the emission maps does not allow us to identify any morphological details  in this galaxy. The maximum of the silicate absorption may not be located at the nucleus. Note however that the near-infrared \citep{AAH06s} and the mid-infrared \citep{Tanio2008} morphologies do not allow us to determine the position of the nucleus.

\subsection{IC~4687}
IC~4687 is part of an interacting system (IC~4686/IC~4687). Only IC~4687 was observed in the spectral mapping mode. In the emission line map it is not possible to isolate any region of interest, however the emission is spatially resolved. We detect the \SIV\ line, that seems to be associated to the nuclear \HII\ regions seen in the Pa$\alpha$ image. Two of these regions have slightly larger \Neii\slash\PAHonce\ ratio than the rest of the galaxy. The nuclear \Neiii\slash\Neii\ ratio of this galaxy is somewhat larger than those of the other \HII\ galaxies in our sample. The maximum of the silicate feature is not located at the nucleus.

\subsection{NGC~6701}
This LIRG is classified as composite(\HII\slash LINER) from optical spectroscopy \citep{AAH09PMAS}. We can identify two extranuclear \HII\ regions symmetrically located $\sim$10 arcsec away from the nucleus, one at each spiral arm. As is the case with the other extranuclear \HII\ regions, the molecular hydrogen of these two regions is relatively low when compared with the \Hmol\ nuclear emission. The silicate absorption is deeper in the nucleus than in the rest of the galaxy.

\subsection{NGC~7130}
This is a face-on galaxy whose nuclear activity is classified as Sy/LINER. We detect 2 extranuclear \HII\ regions located in the spiral arms. The brightest is located to the north, $\sim$10 arcsec from the nucleus. The other one is $\sim$14 arcsec to the south-east from the nucleus.
 The continuum and \Neiii\ emissions are concentrated at the nucleus. The PAH features and fine structure lines are bright in the spiral arms too. The minimum of the \Neiii\slash\Neii\ ratio is not located at the nucleus, as for the majority of the LIRGs in our sample. We detect nuclear \SIV\ emission. Also, this is one of the few galaxies in the sample whose \PAHonce\slash\PAHseis\ ratio minimum is not located at the nucleus. These facts may be related to the active nature of the nucleus of this galaxy.

\subsection{IC~5179}
This is a face-on spiral galaxy classified as \HII -like from optical spectroscopy, and it shows numerous \HII\ regions distributed in the spiral arms \citep{AAH06s}. These \HII\ regions are not resolved in the \Spitzer\ maps. They appear as a relatively uniform extended emission (PAH features and atomic lines) over $\sim$8 kpc. The continuum emission is slightly more concentrated towards the nucleus. The patterns in the ratio maps are possibly related to the intricate distribution of the \HII\ regions. The nucleus is the most obscured region of this galaxy.

\subsection{NGC~7591}
The nuclear activity of NGC~7591 is classified as composite \citep{AAH09PMAS}. Only in the low spectral resolution maps we can clearly distinguish the spiral arms of this face-on galaxy.
The continuum emission is concentrated at the nucleus, while we can see bright PAH and fine structure lines emission in the galaxy arms. The nucleus is the region with the deepest silicate absorption feature.

\subsection{NGC~7771}
The nucleus of this LIRG is classified as \HII -like from optical spectroscopy. It has a nuclear ring of star formation clearly seen in the Pa$\alpha$ images (Figure \ref{fig_map_sh}, see also \citealt{AAH06s} and \citealt{AAH09PMAS}), but it is not resolved by the IRS observations. This galaxy is seen almost edge-on and we can clearly observe the spiral arms in the line emission maps. We detect an \HII\ region in the eastern arm, $\sim$20 arcsec (6 kpc) away from the nucleus. The continuum is dominated by the nucleus. The \Hmol\ emission extends $\sim$5 kpc in the west galaxy arm with apparently no bright fine structure/PAH emission associated with it.
The silicate feature spatial distribution is somewhat irregular and we do not see any morphological detail. 

\end{appendix}


\clearpage

\begin{thebibliography}{74}
\expandafter\ifx\csname natexlab\endcsname\relax\def\natexlab#1{#1}\fi

\bibitem[{{Aalto} {et~al.}(1997){Aalto}, {Radford}, {Scoville}, \&
  {Sargent}}]{Aalto1997}
{Aalto}, S., {Radford}, S.~J.~E., {Scoville}, N.~Z., \& {Sargent}, A.~I. 1997,
  \apjl, 475, L107+

\bibitem[{{Alonso-Herrero} {et~al.}(2009{\natexlab{a}}){Alonso-Herrero},
  {Garc{\'{\i}}a-Mar{\'{\i}}n}, {Monreal-Ibero}, {Colina}, {Arribas},
  {Alfonso-Garz{\'o}n}, \& {Labiano}}]{AAH09PMAS}
{Alonso-Herrero}, A., {Garc{\'{\i}}a-Mar{\'{\i}}n}, M., {Monreal-Ibero}, A.,
  {Colina}, L., {Arribas}, S., {Alfonso-Garz{\'o}n}, J., \& {Labiano}, A.
  2009{\natexlab{a}}, \aap, 506, 1541

\bibitem[{{Alonso-Herrero} {et~al.}(2010){Alonso-Herrero}, {Pereira-Santaella},
  {Rieke}, {Colina}, {Engelbracht}, {P{\'e}rez-Gonz{\'a}lez},
  {D{\'{\i}}az-Santos}, \& {Smith}}]{AAH09ASR}
{Alonso-Herrero}, A., {Pereira-Santaella}, M., {Rieke}, G.~H., {Colina}, L.,
  {Engelbracht}, C.~W., {P{\'e}rez-Gonz{\'a}lez}, P.~G., {D{\'{\i}}az-Santos},
  T., \& {Smith}, J. 2010, Advances in Space Research, 45, 99

\bibitem[{{Alonso-Herrero} {et~al.}(2009{\natexlab{b}}){Alonso-Herrero},
  {Rieke}, {Colina}, {Pereira-Santaella}, {Garc{\'{\i}}a-Mar{\'{\i}}n},
  {Smith}, {Brandl}, {Charmandaris}, \& {Armus}}]{AAH09Arp299}
{Alonso-Herrero}, A., {Rieke}, G.~H., {Colina}, L., {Pereira-Santaella}, M.,
  {Garc{\'{\i}}a-Mar{\'{\i}}n}, M., {Smith}, J.-D.~T., {Brandl}, B.,
  {Charmandaris}, V., \& {Armus}, L. 2009{\natexlab{b}}, \apj, 697, 660

\bibitem[{{Alonso-Herrero} {et~al.}(2006){Alonso-Herrero}, {Rieke}, {Rieke},
  {Colina}, {P{\'e}rez-Gonz{\'a}lez}, \& {Ryder}}]{AAH06s}
{Alonso-Herrero}, A., {Rieke}, G.~H., {Rieke}, M.~J., {Colina}, L.,
  {P{\'e}rez-Gonz{\'a}lez}, P.~G., \& {Ryder}, S.~D. 2006, \apj, 650, 835

\bibitem[{{Alonso-Herrero} {et~al.}(2000){Alonso-Herrero}, {Rieke}, {Rieke}, \&
  {Scoville}}]{AAH2000}
{Alonso-Herrero}, A., {Rieke}, G.~H., {Rieke}, M.~J., \& {Scoville}, N.~Z.
  2000, \apj, 532, 845

\bibitem[{{Amram} {et~al.}(1994){Amram}, {Marcelin}, {Balkowski}, {Cayatte},
  {Sullivan}, \& {Le Coarer}}]{Amram1994}
{Amram}, P., {Marcelin}, M., {Balkowski}, C., {Cayatte}, V., {Sullivan}, III,
  W.~T., \& {Le Coarer}, E. 1994, \aaps, 103, 5

\bibitem[{{Armus} {et~al.}(2007){Armus}, {Charmandaris}, {Bernard-Salas},
  {Spoon}, {Marshall}, {Higdon}, {Desai}, {Teplitz}, {Hao}, {Devost}, {Brandl},
  {Wu}, {Sloan}, {Soifer}, {Houck}, \& {Herter}}]{Armus07}
{Armus}, L., {Charmandaris}, V., {Bernard-Salas}, J., {Spoon}, H.~W.~W.,
  {Marshall}, J.~A., {Higdon}, S.~J.~U., {Desai}, V., {Teplitz}, H.~I., {Hao},
  L., {Devost}, D., {Brandl}, B.~R., {Wu}, Y., {Sloan}, G.~C., {Soifer}, B.~T.,
  {Houck}, J.~R., \& {Herter}, T.~L. 2007, \apj, 656, 148

\bibitem[{{Armus} {et~al.}(2009){Armus}, {Mazzarella}, {Evans}, {Surace},
  {Sanders}, {Iwasawa}, {Frayer}, {Howell}, {Chan}, {Petric}, {Vavilkin},
  {Kim}, {Haan}, {Inami}, {Murphy}, {Appleton}, {Barnes}, {Bothun}, {Bridge},
  {Charmandaris}, {Jensen}, {Kewley}, {Lord}, {Madore}, {Marshall},
  {Melbourne}, {Rich}, {Satyapal}, {Schulz}, {Spoon}, {Sturm}, {U}, {Veilleux},
  \& {Xu}}]{Armus09}
{Armus}, L., {Mazzarella}, J.~M., {Evans}, A.~S., {Surace}, J.~A., {Sanders},
  D.~B., {Iwasawa}, K., {Frayer}, D.~T., {Howell}, J.~H., {Chan}, B., {Petric},
  A., {Vavilkin}, T., {Kim}, D.~C., {Haan}, S., {Inami}, H., {Murphy}, E.~J.,
  {Appleton}, P.~N., {Barnes}, J.~E., {Bothun}, G., {Bridge}, C.~R.,
  {Charmandaris}, V., {Jensen}, J.~B., {Kewley}, L.~J., {Lord}, S., {Madore},
  B.~F., {Marshall}, J.~A., {Melbourne}, J.~E., {Rich}, J., {Satyapal}, S.,
  {Schulz}, B., {Spoon}, H.~W.~W., {Sturm}, E., {U}, V., {Veilleux}, S., \&
  {Xu}, K. 2009, \pasp, 121, 559

\bibitem[{{Augarde} \& {Lequeux}(1985)}]{Augarde1985}
{Augarde}, R., \& {Lequeux}, J. 1985, \aap, 147, 273

\bibitem[{{Bedregal} {et~al.}(2009){Bedregal}, {Colina}, {Alonso-Herrero}, \&
  {Arribas}}]{Bedregal2009}
{Bedregal}, A.~G., {Colina}, L., {Alonso-Herrero}, A., \& {Arribas}, S. 2009,
  \apj, 698, 1852

\bibitem[{{Beir{\~a}o} {et~al.}(2008){Beir{\~a}o}, {Brandl}, {Appleton},
  {Groves}, {Armus}, {F{\"o}rster Schreiber}, {Smith}, {Charmandaris}, \&
  {Houck}}]{Beirao08}
{Beir{\~a}o}, P., {Brandl}, B.~R., {Appleton}, P.~N., {Groves}, B., {Armus},
  L., {F{\"o}rster Schreiber}, N.~M., {Smith}, J.~D., {Charmandaris}, V., \&
  {Houck}, J.~R. 2008, \apj, 676, 304

\bibitem[{{Brandl} {et~al.}(2006){Brandl}, {Bernard-Salas}, {Spoon}, {Devost},
  {Sloan}, {Guilles}, {Wu}, {Houck}, {Weedman}, {Armus}, {Appleton}, {Soifer},
  {Charmandaris}, {Hao}, {Higdon}, \& {Herter}}]{Brandl06}
{Brandl}, B.~R., {Bernard-Salas}, J., {Spoon}, H.~W.~W., {Devost}, D., {Sloan},
  G.~C., {Guilles}, S., {Wu}, Y., {Houck}, J.~R., {Weedman}, D.~W., {Armus},
  L., {Appleton}, P.~N., {Soifer}, B.~T., {Charmandaris}, V., {Hao}, L.,
  {Higdon}, J.~A.~M.~S.~J., \& {Herter}, T.~L. 2006, \apj, 653, 1129

\bibitem[{{Caputi} {et~al.}(2007){Caputi}, {Lagache}, {Yan}, {Dole},
  {Bavouzet}, {Le Floc'h}, {Choi}, {Helou}, \& {Reddy}}]{Caputi2007}
{Caputi}, K.~I., {Lagache}, G., {Yan}, L., {Dole}, H., {Bavouzet}, N., {Le
  Floc'h}, E., {Choi}, P.~I., {Helou}, G., \& {Reddy}, N. 2007, \apj, 660, 97

\bibitem[{{Casoli} {et~al.}(1999){Casoli}, {Willaime}, {Viallefond}, \&
  {Gerin}}]{Casoli1999}
{Casoli}, F., {Willaime}, M., {Viallefond}, F., \& {Gerin}, M. 1999, \aap, 346,
  663

\bibitem[{{Dale} {et~al.}(2006){Dale}, {Smith}, {Armus}, {Buckalew}, {Helou},
  {Kennicutt}, {Moustakas}, {Roussel}, {Sheth}, {Bendo}, {Calzetti}, {Draine},
  {Engelbracht}, {Gordon}, {Hollenbach}, {Jarrett}, {Kewley}, {Leitherer},
  {Li}, {Malhotra}, {Murphy}, \& {Walter}}]{Dale06}
{Dale}, D.~A., {Smith}, J.~D.~T., {Armus}, L., {Buckalew}, B.~A., {Helou}, G.,
  {Kennicutt}, Jr., R.~C., {Moustakas}, J., {Roussel}, H., {Sheth}, K.,
  {Bendo}, G.~J., {Calzetti}, D., {Draine}, B.~T., {Engelbracht}, C.~W.,
  {Gordon}, K.~D., {Hollenbach}, D.~J., {Jarrett}, T.~H., {Kewley}, L.~J.,
  {Leitherer}, C., {Li}, A., {Malhotra}, S., {Murphy}, E.~J., \& {Walter}, F.
  2006, \apj, 646, 161

\bibitem[{{Dale} {et~al.}(2009){Dale}, {Smith}, {Schlawin}, {Armus},
  {Buckalew}, {Cohen}, {Helou}, {Jarrett}, {Johnson}, {Moustakas}, {Murphy},
  {Roussel}, {Sheth}, {Staudaher}, {Bot}, {Calzetti}, {Engelbracht}, {Gordon},
  {Hollenbach}, {Kennicutt}, \& {Malhotra}}]{Dale2009}
{Dale}, D.~A., {Smith}, J.~D.~T., {Schlawin}, E.~A., {Armus}, L., {Buckalew},
  B.~A., {Cohen}, S.~A., {Helou}, G., {Jarrett}, T.~H., {Johnson}, L.~C.,
  {Moustakas}, J., {Murphy}, E.~J., {Roussel}, H., {Sheth}, K., {Staudaher},
  S., {Bot}, C., {Calzetti}, D., {Engelbracht}, C.~W., {Gordon}, K.~D.,
  {Hollenbach}, D.~J., {Kennicutt}, R.~C., \& {Malhotra}, S. 2009, \apj, 693,
  1821

\bibitem[{{Devost} {et~al.}(2004){Devost}, {Brandl}, {Armus}, {Barry}, {Sloan},
  {Charmandaris}, {Spoon}, {Bernard-Salas}, \& {Houck}}]{Devost04}
{Devost}, D., {Brandl}, B.~R., {Armus}, L., {Barry}, D.~J., {Sloan}, G.~C.,
  {Charmandaris}, V., {Spoon}, H., {Bernard-Salas}, J., \& {Houck}, J.~R. 2004,
  \apjs, 154, 242

\bibitem[{{D{\'{\i}}az-Santos} {et~al.}(2010){D{\'{\i}}az-Santos},
  {Alonso-Herrero}, {Colina}, {Packham}, {Levenson}, {Pereira-Santaella},
  {Roche}, \& {Telesco}}]{Tanio09}
{D{\'{\i}}az-Santos}, T., {Alonso-Herrero}, A., {Colina}, L., {Packham}, C.,
  {Levenson}, N.~A., {Pereira-Santaella}, M., {Roche}, P.~F., \& {Telesco},
  C.~M. 2010, \apj, 711, 328

\bibitem[{{D{\'{\i}}az-Santos} {et~al.}(2008){D{\'{\i}}az-Santos},
  {Alonso-Herrero}, {Colina}, {Packham}, {Radomski}, \& {Telesco}}]{Tanio2008}
{D{\'{\i}}az-Santos}, T., {Alonso-Herrero}, A., {Colina}, L., {Packham}, C.,
  {Radomski}, J.~T., \& {Telesco}, C.~M. 2008, \apj, 685, 211

\bibitem[{{Dopita} {et~al.}(2006){Dopita}, {Fischera}, {Sutherland}, {Kewley},
  {Leitherer}, {Tuffs}, {Popescu}, {van Breugel}, \& {Groves}}]{Dopita2006}
{Dopita}, M.~A., {Fischera}, J., {Sutherland}, R.~S., {Kewley}, L.~J.,
  {Leitherer}, C., {Tuffs}, R.~J., {Popescu}, C.~C., {van Breugel}, W., \&
  {Groves}, B.~A. 2006, \apjs, 167, 177

\bibitem[{{Draine} \& {Li}(2001)}]{Draine01}
{Draine}, B.~T., \& {Li}, A. 2001, \apj, 551, 807

\bibitem[{{Engelbracht} {et~al.}(2008){Engelbracht}, {Rieke}, {Gordon},
  {Smith}, {Werner}, {Moustakas}, {Willmer}, \& {Vanzi}}]{Engelbracht08}
{Engelbracht}, C.~W., {Rieke}, G.~H., {Gordon}, K.~D., {Smith}, J.-D.~T.,
  {Werner}, M.~W., {Moustakas}, J., {Willmer}, C.~N.~A., \& {Vanzi}, L. 2008,
  \apj, 678, 804

\bibitem[{{Farrah} {et~al.}(2007){Farrah}, {Bernard-Salas}, {Spoon}, {Soifer},
  {Armus}, {Brandl}, {Charmandaris}, {Desai}, {Higdon}, {Devost}, \&
  {Houck}}]{Farrah07}
{Farrah}, D., {Bernard-Salas}, J., {Spoon}, H.~W.~W., {Soifer}, B.~T., {Armus},
  L., {Brandl}, B., {Charmandaris}, V., {Desai}, V., {Higdon}, S., {Devost},
  D., \& {Houck}, J. 2007, \apj, 667, 149

\bibitem[{{Farrah} {et~al.}(2008){Farrah}, {Lonsdale}, {Weedman}, {Spoon},
  {Rowan-Robinson}, {Polletta}, {Oliver}, {Houck}, \& {Smith}}]{Farrah08}
{Farrah}, D., {Lonsdale}, C.~J., {Weedman}, D.~W., {Spoon}, H.~W.~W.,
  {Rowan-Robinson}, M., {Polletta}, M., {Oliver}, S., {Houck}, J.~R., \&
  {Smith}, H.~E. 2008, \apj, 677, 957

\bibitem[{{Galliano}(2006)}]{Galliano2006}
{Galliano}, F. 2006, ArXiv Astrophysics e-prints

\bibitem[{{Galliano} {et~al.}(2008){Galliano}, {Madden}, {Tielens}, {Peeters},
  \& {Jones}}]{Galliano2008}
{Galliano}, F., {Madden}, S.~C., {Tielens}, A.~G.~G.~M., {Peeters}, E., \&
  {Jones}, A.~P. 2008, \apj, 679, 310

\bibitem[{{Garc{\'{\i}}a-Mar{\'{\i}}n}
  {et~al.}(2006){Garc{\'{\i}}a-Mar{\'{\i}}n}, {Colina}, {Arribas},
  {Alonso-Herrero}, \& {Mediavilla}}]{GarciaMarin06}
{Garc{\'{\i}}a-Mar{\'{\i}}n}, M., {Colina}, L., {Arribas}, S.,
  {Alonso-Herrero}, A., \& {Mediavilla}, E. 2006, \apj, 650, 850

\bibitem[{{Genzel} {et~al.}(1998){Genzel}, {Lutz}, {Sturm}, {Egami}, {Kunze},
  {Moorwood}, {Rigopoulou}, {Spoon}, {Sternberg}, {Tacconi-Garman}, {Tacconi},
  \& {Thatte}}]{Genzel1998}
{Genzel}, R., {Lutz}, D., {Sturm}, E., {Egami}, E., {Kunze}, D., {Moorwood},
  A.~F.~M., {Rigopoulou}, D., {Spoon}, H.~W.~W., {Sternberg}, A.,
  {Tacconi-Garman}, L.~E., {Tacconi}, L., \& {Thatte}, N. 1998, \apj, 498, 579

\bibitem[{{Giveon} {et~al.}(2002){Giveon}, {Sternberg}, {Lutz}, {Feuchtgruber},
  \& {Pauldrach}}]{Giveon2002}
{Giveon}, U., {Sternberg}, A., {Lutz}, D., {Feuchtgruber}, H., \& {Pauldrach},
  A.~W.~A. 2002, \apj, 566, 880

\bibitem[{{Gordon} {et~al.}(2008){Gordon}, {Engelbracht}, {Rieke}, {Misselt},
  {Smith}, \& {Kennicutt}}]{Gordon08}
{Gordon}, K.~D., {Engelbracht}, C.~W., {Rieke}, G.~H., {Misselt}, K.~A.,
  {Smith}, J.-D.~T., \& {Kennicutt}, Jr., R.~C. 2008, \apj, 682, 336

\bibitem[{{Groves} {et~al.}(2004){Groves}, {Dopita}, \&
  {Sutherland}}]{Groves2004}
{Groves}, B.~A., {Dopita}, M.~A., \& {Sutherland}, R.~S. 2004, \apjs, 153, 75

\bibitem[{{Habart} {et~al.}(2003){Habart}, {Boulanger}, {Verstraete}, {Pineau
  des For{\^e}ts}, {Falgarone}, \& {Abergel}}]{Habart03}
{Habart}, E., {Boulanger}, F., {Verstraete}, L., {Pineau des For{\^e}ts}, G.,
  {Falgarone}, E., \& {Abergel}, A. 2003, \aap, 397, 623

\bibitem[{{Hattori} {et~al.}(2004){Hattori}, {Yoshida}, {Ohtani}, {Sugai},
  {Ishigaki}, {Sasaki}, {Hayashi}, {Ozaki}, {Ishii}, \& {Kawai}}]{Hattori04}
{Hattori}, T., {Yoshida}, M., {Ohtani}, H., {Sugai}, H., {Ishigaki}, T.,
  {Sasaki}, M., {Hayashi}, T., {Ozaki}, S., {Ishii}, M., \& {Kawai}, A. 2004,
  \aj, 127, 736

\bibitem[{{Higdon} {et~al.}(2006){Higdon}, {Armus}, {Higdon}, {Soifer}, \&
  {Spoon}}]{Higdon2006}
{Higdon}, S.~J.~U., {Armus}, L., {Higdon}, J.~L., {Soifer}, B.~T., \& {Spoon},
  H.~W.~W. 2006, \apj, 648, 323

\bibitem[{{Ho} \& {Keto}(2007)}]{Ho07}
{Ho}, L.~C., \& {Keto}, E. 2007, \apj, 658, 314

\bibitem[{{Hollenbach} \& {McKee}(1989)}]{Hollenbach1989}
{Hollenbach}, D., \& {McKee}, C.~F. 1989, \apj, 342, 306

\bibitem[{{Houck} {et~al.}(2004){Houck}, {Roellig}, {van Cleve}, {Forrest},
  {Herter}, {Lawrence}, {Matthews}, {Reitsema}, {Soifer}, {Watson}, {Weedman},
  {Huisjen}, {Troeltzsch}, {Barry}, {Bernard-Salas}, {Blacken}, {Brandl},
  {Charmandaris}, {Devost}, {Gull}, {Hall}, {Henderson}, {Higdon}, {Pirger},
  {Schoenwald}, {Sloan}, {Uchida}, {Appleton}, {Armus}, {Burgdorf},
  {Fajardo-Acosta}, {Grillmair}, {Ingalls}, {Morris}, \& {Teplitz}}]{HouckIRS}
{Houck}, J.~R., {Roellig}, T.~L., {van Cleve}, J., {Forrest}, W.~J., {Herter},
  T., {Lawrence}, C.~R., {Matthews}, K., {Reitsema}, H.~J., {Soifer}, B.~T.,
  {Watson}, D.~M., {Weedman}, D., {Huisjen}, M., {Troeltzsch}, J., {Barry},
  D.~J., {Bernard-Salas}, J., {Blacken}, C.~E., {Brandl}, B.~R.,
  {Charmandaris}, V., {Devost}, D., {Gull}, G.~E., {Hall}, P., {Henderson},
  C.~P., {Higdon}, S.~J.~U., {Pirger}, B.~E., {Schoenwald}, J., {Sloan}, G.~C.,
  {Uchida}, K.~I., {Appleton}, P.~N., {Armus}, L., {Burgdorf}, M.~J.,
  {Fajardo-Acosta}, S.~B., {Grillmair}, C.~J., {Ingalls}, J.~G., {Morris},
  P.~W., \& {Teplitz}, H.~I. 2004, \apjs, 154, 18

\bibitem[{{Imanishi} {et~al.}(2007){Imanishi}, {Dudley}, {Maiolino}, {Maloney},
  {Nakagawa}, \& {Risaliti}}]{Imanishi2007}
{Imanishi}, M., {Dudley}, C.~C., {Maiolino}, R., {Maloney}, P.~R., {Nakagawa},
  T., \& {Risaliti}, G. 2007, \apjs, 171, 72

\bibitem[{{K{\'o}sp{\'a}l} {et~al.}(2008){K{\'o}sp{\'a}l}, {{\'A}brah{\'a}m},
  {Apai}, {Ardila}, {Grady}, {Henning}, {Juh{\'a}sz}, {Miller}, \&
  {Mo{\'o}r}}]{Kospal08}
{K{\'o}sp{\'a}l}, {\'A}., {{\'A}brah{\'a}m}, P., {Apai}, D., {Ardila}, D.~R.,
  {Grady}, C.~A., {Henning}, T., {Juh{\'a}sz}, A., {Miller}, D.~W., \&
  {Mo{\'o}r}, A. 2008, \mnras, 383, 1015

\bibitem[{{Krist}(2002)}]{Krist2002}
{Krist}, J. 2002, {\it{Tiny Tim/SIRTF User's Guide}} (Pasadena, SSC)

\bibitem[{{Le Floc'h} {et~al.}(2005){Le Floc'h}, {Papovich}, {Dole}, {Bell},
  {Lagache}, {Rieke}, {Egami}, {P{\'e}rez-Gonz{\'a}lez}, {Alonso-Herrero},
  {Rieke}, {Blaylock}, {Engelbracht}, {Gordon}, {Hines}, {Misselt}, {Morrison},
  \& {Mould}}]{LeFloch2005}
{Le Floc'h}, E., {Papovich}, C., {Dole}, H., {Bell}, E.~F., {Lagache}, G.,
  {Rieke}, G.~H., {Egami}, E., {P{\'e}rez-Gonz{\'a}lez}, P.~G.,
  {Alonso-Herrero}, A., {Rieke}, M.~J., {Blaylock}, M., {Engelbracht}, C.~W.,
  {Gordon}, K.~D., {Hines}, D.~C., {Misselt}, K.~A., {Morrison}, J.~E., \&
  {Mould}, J. 2005, \apj, 632, 169

\bibitem[{{Lepp} \& {McCray}(1983)}]{Lepp1983}
{Lepp}, S., \& {McCray}, R. 1983, \apj, 269, 560

\bibitem[{{L{\'{\i}}pari} {et~al.}(2000){L{\'{\i}}pari}, {D{\'{\i}}az},
  {Taniguchi}, {Terlevich}, {Dottori}, \& {Carranza}}]{Lipari2000}
{L{\'{\i}}pari}, S., {D{\'{\i}}az}, R., {Taniguchi}, Y., {Terlevich}, R.,
  {Dottori}, H., \& {Carranza}, G. 2000, \aj, 120, 645

\bibitem[{{Markwardt}(2009)}]{MPFIT}
{Markwardt}, C.~B. 2009, ArXiv e-prints

\bibitem[{{Men{\'e}ndez-Delmestre} {et~al.}(2009){Men{\'e}ndez-Delmestre},
  {Blain}, {Smail}, {Alexander}, {Chapman}, {Armus}, {Frayer}, {Ivison}, \&
  {Teplitz}}]{MenendezDelmestre2009}
{Men{\'e}ndez-Delmestre}, K., {Blain}, A.~W., {Smail}, I., {Alexander}, D.~M.,
  {Chapman}, S.~C., {Armus}, L., {Frayer}, D., {Ivison}, R.~J., \& {Teplitz},
  H. 2009, ArXiv e-prints

\bibitem[{{Nardini} {et~al.}(2008){Nardini}, {Risaliti}, {Salvati}, {Sani},
  {Imanishi}, {Marconi}, \& {Maiolino}}]{Nardini2008}
{Nardini}, E., {Risaliti}, G., {Salvati}, M., {Sani}, E., {Imanishi}, M.,
  {Marconi}, A., \& {Maiolino}, R. 2008, \mnras, 385, L130

\bibitem[{{Peeters} {et~al.}(2004){Peeters}, {Spoon}, \&
  {Tielens}}]{Peeters2004}
{Peeters}, E., {Spoon}, H.~W.~W., \& {Tielens}, A.~G.~G.~M. 2004, \apj, 613,
  986

\bibitem[{{P{\'e}rez-Gonz{\'a}lez} {et~al.}(2005){P{\'e}rez-Gonz{\'a}lez},
  {Rieke}, {Egami}, {Alonso-Herrero}, {Dole}, {Papovich}, {Blaylock}, {Jones},
  {Rieke}, {Rigby}, {Barmby}, {Fazio}, {Huang}, \&
  {Martin}}]{PerezGonzalez2005}
{P{\'e}rez-Gonz{\'a}lez}, P.~G., {Rieke}, G.~H., {Egami}, E., {Alonso-Herrero},
  A., {Dole}, H., {Papovich}, C., {Blaylock}, M., {Jones}, J., {Rieke}, M.,
  {Rigby}, J., {Barmby}, P., {Fazio}, G.~G., {Huang}, J., \& {Martin}, C. 2005,
  \apj, 630, 82

\bibitem[{{Pope} {et~al.}(2007){Pope}, {Chary}, {Dickinson}, \&
  {Scott}}]{Pope2007}
{Pope}, A., {Chary}, R.-R., {Dickinson}, M., \& {Scott}, D. 2007, in
  Astronomical Society of the Pacific Conference Series, Vol. 380, Deepest
  Astronomical Surveys, ed. J.~{Afonso}, H.~C. {Ferguson}, B.~{Mobasher}, \&
  R.~{Norris}, 387--+

\bibitem[{{Rieke} \& {Lebofsky}(1985)}]{Rieke85}
{Rieke}, G.~H., \& {Lebofsky}, M.~J. 1985, \apj, 288, 618

\bibitem[{{Rieke} \& {Low}(1972)}]{Rieke72}
{Rieke}, G.~H., \& {Low}, F.~J. 1972, \apjl, 176, L95+

\bibitem[{{Rigby} {et~al.}(2008){Rigby}, {Marcillac}, {Egami}, {Rieke},
  {Richard}, {Kneib}, {Fadda}, {Willmer}, {Borys}, {van der Werf},
  {P{\'e}rez-Gonz{\'a}lez}, {Knudsen}, \& {Papovich}}]{Rigby2008}
{Rigby}, J.~R., {Marcillac}, D., {Egami}, E., {Rieke}, G.~H., {Richard}, J.,
  {Kneib}, J.-P., {Fadda}, D., {Willmer}, C.~N.~A., {Borys}, C., {van der
  Werf}, P.~P., {P{\'e}rez-Gonz{\'a}lez}, P.~G., {Knudsen}, K.~K., \&
  {Papovich}, C. 2008, \apj, 675, 262

\bibitem[{{Rigby} \& {Rieke}(2004)}]{Rigby2004}
{Rigby}, J.~R., \& {Rieke}, G.~H. 2004, \apj, 606, 237

\bibitem[{{Rigopoulou} {et~al.}(2002){Rigopoulou}, {Kunze}, {Lutz}, {Genzel},
  \& {Moorwood}}]{Rigopoulou02}
{Rigopoulou}, D., {Kunze}, D., {Lutz}, D., {Genzel}, R., \& {Moorwood},
  A.~F.~M. 2002, \aap, 389, 374

\bibitem[{{Roussel} {et~al.}(2007){Roussel}, {Helou}, {Hollenbach}, {Draine},
  {Smith}, {Armus}, {Schinnerer}, {Walter}, {Engelbracht}, {Thornley},
  {Kennicutt}, {Calzetti}, {Dale}, {Murphy}, \& {Bot}}]{Roussel07}
{Roussel}, H., {Helou}, G., {Hollenbach}, D.~J., {Draine}, B.~T., {Smith},
  J.~D., {Armus}, L., {Schinnerer}, E., {Walter}, F., {Engelbracht}, C.~W.,
  {Thornley}, M.~D., {Kennicutt}, R.~C., {Calzetti}, D., {Dale}, D.~A.,
  {Murphy}, E.~J., \& {Bot}, C. 2007, \apj, 669, 959

\bibitem[{{Rupke} {et~al.}(2008){Rupke}, {Veilleux}, \& {Baker}}]{Rupke2008}
{Rupke}, D.~S.~N., {Veilleux}, S., \& {Baker}, A.~J. 2008, \apj, 674, 172

\bibitem[{{Sanders} {et~al.}(2003){Sanders}, {Mazzarella}, {Kim}, {Surace}, \&
  {Soifer}}]{SandersRBGS}
{Sanders}, D.~B., {Mazzarella}, J.~M., {Kim}, D.-C., {Surace}, J.~A., \&
  {Soifer}, B.~T. 2003, \aj, 126, 1607

\bibitem[{{Sanders} \& {Mirabel}(1996)}]{Sanders96}
{Sanders}, D.~B., \& {Mirabel}, I.~F. 1996, \araa, 34, 749

\bibitem[{{Shaw} {et~al.}(2005){Shaw}, {Ferland}, {Abel}, {Stancil}, \& {van
  Hoof}}]{Shaw2005}
{Shaw}, G., {Ferland}, G.~J., {Abel}, N.~P., {Stancil}, P.~C., \& {van Hoof},
  P.~A.~M. 2005, \apj, 624, 794

\bibitem[{{Sirocky} {et~al.}(2008){Sirocky}, {Levenson}, {Elitzur}, {Spoon}, \&
  {Armus}}]{Sirocky08}
{Sirocky}, M.~M., {Levenson}, N.~A., {Elitzur}, M., {Spoon}, H.~W.~W., \&
  {Armus}, L. 2008, \apj, 678, 729

\bibitem[{{Smith} {et~al.}(2007{\natexlab{a}}){Smith}, {Armus}, {Dale},
  {Roussel}, {Sheth}, {Buckalew}, {Jarrett}, {Helou}, \&
  {Kennicutt}}]{SmithCUBISM}
{Smith}, J.~D.~T., {Armus}, L., {Dale}, D.~A., {Roussel}, H., {Sheth}, K.,
  {Buckalew}, B.~A., {Jarrett}, T.~H., {Helou}, G., \& {Kennicutt}, Jr., R.~C.
  2007{\natexlab{a}}, \pasp, 119, 1133

\bibitem[{{Smith} {et~al.}(2007{\natexlab{b}}){Smith}, {Draine}, {Dale},
  {Moustakas}, {Kennicutt}, {Helou}, {Armus}, {Roussel}, {Sheth}, {Bendo},
  {Buckalew}, {Calzetti}, {Engelbracht}, {Gordon}, {Hollenbach}, {Li},
  {Malhotra}, {Murphy}, \& {Walter}}]{Smith07}
{Smith}, J.~D.~T., {Draine}, B.~T., {Dale}, D.~A., {Moustakas}, J.,
  {Kennicutt}, Jr., R.~C., {Helou}, G., {Armus}, L., {Roussel}, H., {Sheth},
  K., {Bendo}, G.~J., {Buckalew}, B.~A., {Calzetti}, D., {Engelbracht}, C.~W.,
  {Gordon}, K.~D., {Hollenbach}, D.~J., {Li}, A., {Malhotra}, S., {Murphy},
  E.~J., \& {Walter}, F. 2007{\natexlab{b}}, \apj, 656, 770

\bibitem[{{Snijders} {et~al.}(2007){Snijders}, {Kewley}, \& {van der
  Werf}}]{Snijders07}
{Snijders}, L., {Kewley}, L.~J., \& {van der Werf}, P.~P. 2007, \apj, 669, 269

\bibitem[{{Soifer} {et~al.}(2000){Soifer}, {Neugebauer}, {Matthews}, {Egami},
  {Becklin}, {Weinberger}, {Ressler}, {Werner}, {Evans}, {Scoville}, {Surace},
  \& {Condon}}]{Soifer2000}
{Soifer}, B.~T., {Neugebauer}, G., {Matthews}, K., {Egami}, E., {Becklin},
  E.~E., {Weinberger}, A.~J., {Ressler}, M., {Werner}, M.~W., {Evans}, A.~S.,
  {Scoville}, N.~Z., {Surace}, J.~A., \& {Condon}, J.~J. 2000, \aj, 119, 509

\bibitem[{{Spoon} {et~al.}(2007){Spoon}, {Marshall}, {Houck}, {Elitzur}, {Hao},
  {Armus}, {Brandl}, \& {Charmandaris}}]{Spoon07}
{Spoon}, H.~W.~W., {Marshall}, J.~A., {Houck}, J.~R., {Elitzur}, M., {Hao}, L.,
  {Armus}, L., {Brandl}, B.~R., \& {Charmandaris}, V. 2007, \apjl, 654, L49

\bibitem[{{Surace} {et~al.}(2004){Surace}, {Sanders}, \&
  {Mazzarella}}]{Surace2004}
{Surace}, J.~A., {Sanders}, D.~B., \& {Mazzarella}, J.~M. 2004, \aj, 127, 3235

\bibitem[{{Thornley} {et~al.}(2000){Thornley}, {Schreiber}, {Lutz}, {Genzel},
  {Spoon}, {Kunze}, \& {Sternberg}}]{Thornley2000}
{Thornley}, M.~D., {Schreiber}, N.~M.~F., {Lutz}, D., {Genzel}, R., {Spoon},
  H.~W.~W., {Kunze}, D., \& {Sternberg}, A. 2000, \apj, 539, 641

\bibitem[{{Tommasin} {et~al.}(2008){Tommasin}, {Spinoglio}, {Malkan}, {Smith},
  {Gonz{\'a}lez-Alfonso}, \& {Charmandaris}}]{Tommasin08}
{Tommasin}, S., {Spinoglio}, L., {Malkan}, M.~A., {Smith}, H.,
  {Gonz{\'a}lez-Alfonso}, E., \& {Charmandaris}, V. 2008, \apj, 676, 836

\bibitem[{{Veilleux} {et~al.}(1995){Veilleux}, {Kim}, {Sanders}, {Mazzarella},
  \& {Soifer}}]{Veilleux1995}
{Veilleux}, S., {Kim}, D.-C., {Sanders}, D.~B., {Mazzarella}, J.~M., \&
  {Soifer}, B.~T. 1995, \apjs, 98, 171

\bibitem[{{Veilleux} {et~al.}(2009){Veilleux}, {Rupke}, {Kim}, {Genzel},
  {Sturm}, {Lutz}, {Contursi}, {Schweitzer}, {Tacconi}, {Netzer}, {Sternberg},
  {Mihos}, {Baker}, {Mazzarella}, {Lord}, {Sanders}, {Stockton}, {Joseph}, \&
  {Barnes}}]{Veilleux2009}
{Veilleux}, S., {Rupke}, D.~S.~N., {Kim}, D.-C., {Genzel}, R., {Sturm}, E.,
  {Lutz}, D., {Contursi}, A., {Schweitzer}, M., {Tacconi}, L.~J., {Netzer}, H.,
  {Sternberg}, A., {Mihos}, J.~C., {Baker}, A.~J., {Mazzarella}, J.~M., {Lord},
  S., {Sanders}, D.~B., {Stockton}, A., {Joseph}, R.~D., \& {Barnes}, J.~E.
  2009, \apjs, 182, 628

\bibitem[{{Velusamy} \& {Langer}(2008)}]{Velusamy08}
{Velusamy}, T., \& {Langer}, W.~D. 2008, \aj, 136, 602

\bibitem[{{Verma} {et~al.}(2003){Verma}, {Lutz}, {Sturm}, {Sternberg},
  {Genzel}, \& {Vacca}}]{Verma03}
{Verma}, A., {Lutz}, D., {Sturm}, E., {Sternberg}, A., {Genzel}, R., \&
  {Vacca}, W. 2003, \aap, 403, 829

\bibitem[{{Wu} {et~al.}(2006){Wu}, {Charmandaris}, {Hao}, {Brandl},
  {Bernard-Salas}, {Spoon}, \& {Houck}}]{Wu06}
{Wu}, Y., {Charmandaris}, V., {Hao}, L., {Brandl}, B.~R., {Bernard-Salas}, J.,
  {Spoon}, H.~W.~W., \& {Houck}, J.~R. 2006, \apj, 639, 157

\end{thebibliography}
\end{document}